\documentclass[12pt,a4paper]{book}
\usepackage{amsmath}
\usepackage{amssymb}
\usepackage{graphicx}
\usepackage{cite}
\usepackage{color}
\usepackage{url}
\usepackage{float}
\usepackage{latexsym}
\usepackage{bm}
\usepackage{multirow}
\usepackage{enumerate}
\usepackage{rotating}
\usepackage{changepage}

\newcommand{\ket} [1] {\vert #1 \rangle}
\newcommand{\bra} [1] {\langle #1 \vert}

\newcommand{\proj}[1]{\ket{#1}\bra{#1}}

\newcommand{\hr}[1]{\hat{\rho}_{#1}}

\newtheorem{theorem}{Theorem}

\newtheorem{definition}{Definition}

\definecolor{darkblue}{rgb}{0.2,0.2,0.6}
\usepackage[linktocpage,colorlinks=true,linkcolor=darkblue,citecolor=darkblue,urlcolor=darkblue]{hyperref}

\usepackage{setspace}
\usepackage[T1]{fontenc}
\usepackage{txfonts}
\usepackage[margin=3.0cm]{geometry}
\usepackage[english]{babel}
\usepackage{caption}
\begin{document}

\begin{center}
{\Large {\bf Some aspects of the interplay between bipartite correlations and quantum channels }}
\vskip 0.70cm
{\bf {\em By}} 
\vskip -0.2cm
{\bf {\large Krishna Kumar Sabapathy}}
\vskip 0.0cm
{\bf {\large PHYS10200605008}}
\vskip 0.5cm
{\bf {\large The Institute of Mathematical Sciences, Chennai}}
\vskip 2.6cm
{\bf {\em {\large A thesis submitted to the
\vskip 0.05cm
Board of Studies in Physical Sciences
\vskip 0.05cm
In partial fulfillment of requirements
\vskip 0.05cm
For the Degree of 
}}}
\vskip 0.05cm
{\bf {\large DOCTOR OF PHILOSOPHY}}
\vskip 0.1cm
{\bf {\em of}}
\vskip 0.1cm
{\bf {\large HOMI BHABHA NATIONAL INSTITUTE}}
\vfill
\includegraphics[height=3.5cm, width=3.5cm]{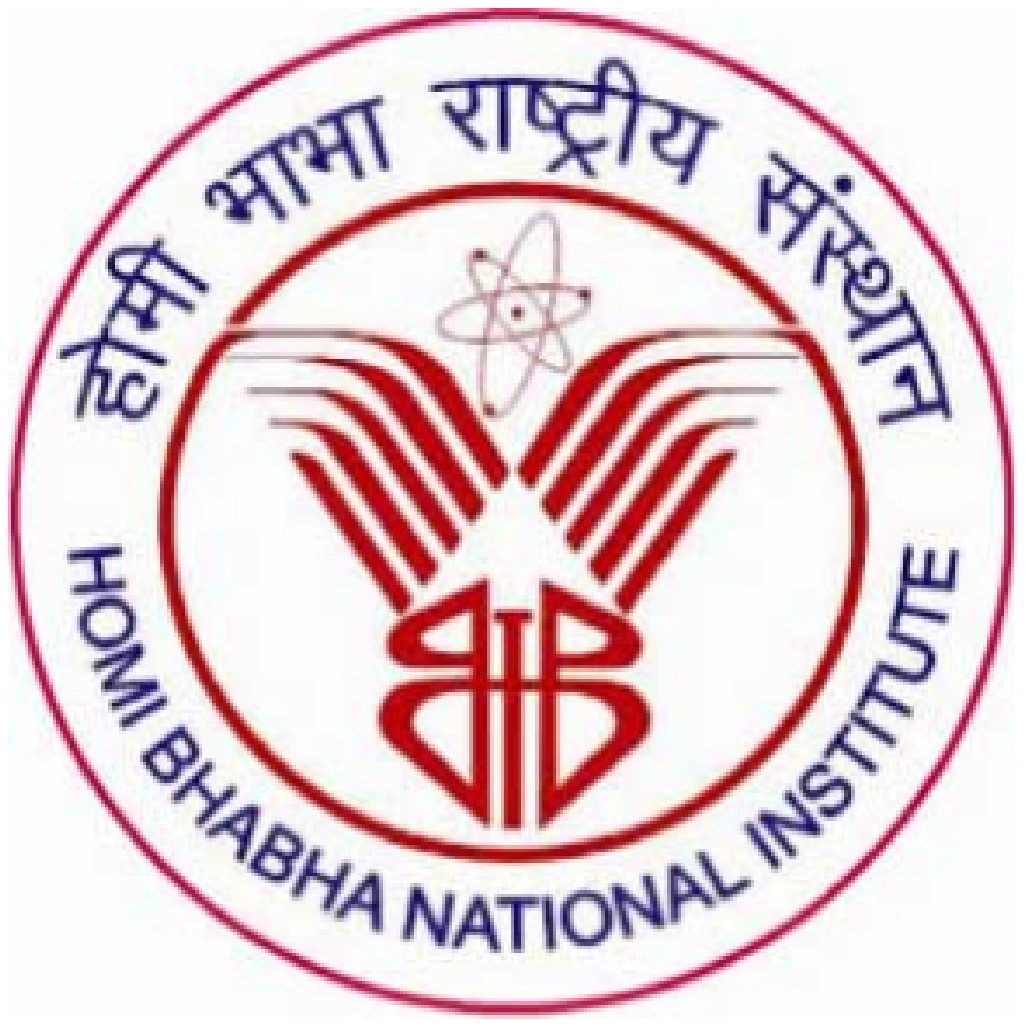}
\vfill
{\bf {\large July, 2013}}
\vfill
\end{center}

\newpage
\cleardoublepage
%
%
\centerline{{\bf{\LARGE Homi Bhabha National Institute}}}
\vskip 0.3cm
\centerline{{\bf {\large Recommendations of the Viva Voce Board}}}
\vskip 0.3cm
As members of the Viva Voce Board, we certify that we have read the
dissertation prepared by {\sf Krishna Kumar Sabapathy} entitled ``{\sf Some aspects of the interplay between bipartite correlations and quantum channels}'' and
recommend that it maybe accepted as fulfilling the dissertation
requirement for the Degree of Doctor of Philosophy.

\vskip 0.5cm
\underline{\hspace{12.0cm}} Date:
\vskip -0.1cm 
Chair/Guide\,: {\sf R. Simon} 
\vskip 0.7cm
\underline{\hspace{12.0cm}} Date:
\vskip -0.1cm 
Member 1\,: {\sf V. S. Sunder}
\vskip 0.6cm
\underline{\hspace{12.0cm}} Date:
\vskip -0.1cm 
Member 2\,: {\sf Sibasish Ghosh} 
\vskip 0.6cm
\underline{\hspace{12.0cm}} Date:
\vskip -0.1cm 
Member 3\,: {\sf S. R. Hassan}
\vskip 0.6cm
%
%
%
%
%
\underline{\hspace{12.0cm}} Date:
\vskip -0.1cm 
External Examiner\,: {\sf Andreas Winter}
%
\vfill
\hspace{0.7cm} Final approval and acceptance of this dissertation is
contingent upon the candidate's submission of the final copies of the
dissertation to HBNI.
\vskip -0.2cm
\hspace{0.7cm} I hereby certify that I have read this dissertation
prepared under my direction and recommend that it may be accepted as
fulfilling the dissertation requirement.
%
\vfill
{\bf Date:} 
\vskip 0.3cm 
{\bf Place:} \hfill Guide \hspace{1.0cm}
\newpage
\cleardoublepage
\centerline{{\bf {\large STATEMENT BY AUTHOR}}}
\vskip 1.00cm
This dissertation has been submitted in partial fulfillment of
requirements for an advanced degree at Homi Bhabha National Institute
(HBNI) and is deposited in the Library to be made available to borrowers
under rules of the HBNI.
\vskip 0.6cm
Brief quotations from this dissertation are allowable without special
permission, provided that accurate acknowledgement of the source is made.
Requests for permission for extended quotation from or reproduction of
this manuscript in whole or in part may be granted by the Competent
Authority of HBNI when in his or her judgement the proposed use of the
material is in the interests of scholarship. In all other instances,
however, permission must be obtained from the author.

\vskip 1.5cm


$~$\hspace{10.2cm} Krishna Kumar Sabapathy
\newpage
\cleardoublepage
~
\vskip 1.2cm
\centerline{{\bf{\large{DECLARATION}}}}
\vskip 1.2cm
I, hereby declare that the investigation presented in the thesis has been
carried out by me. The work is original and has not been submitted
earlier in whole or in part for a degree / diploma at this or any
other Institution / University.
\vskip 2.0cm
%
%
\rightline{Krishna Kumar Sabapathy \hspace{0.9cm}}
\newpage
\cleardoublepage
\centerline{{\bf {\large DEDICATIONS}}}
\vskip 0.5cm
%
\vspace{3cm}
\begin{center}
{\em To my parents and my well-wishers}
\end{center}
\newpage
\cleardoublepage
~
\vskip 1.0cm
\centerline{{\bf{\large ACKNOWLEDGEMENTS}}}
\vskip 0.5cm

This thesis has been written with an enormous amount of help from people around me. It is not without reason that Google scholar chose the tag line `Stand on the shoulders of giants'. This quote seems apt in the present context. 

It has been a long journey and I am grateful first and foremost to my adviser {\bf Prof. R. Simon}. Most of what I have learnt in my stay here is from various forms of interactions and discussions from him. I also wish to thank him for the interesting topics and subjects that I learnt and was exposed to in my study. It has been an honour to be able to get this opportunity and I am sure it has and continues to play an important role in my academic path. 

I would like to thank {\bf Prof. Mukunda} for the many inspiring lectures that he has delivered over the last few years that I had the privilege to attend and have greatly benefited from. I have learnt a lot from him right from my undergraduate days and I am greatly indebted to him for many things. There are also a lot of attributes of him that I aspire to be able to inculcate in my academic journey. 

An important role in the completion of this thesis was played by {\bf Dr. Solomon Ivan}. I have learnt a lot from him through many hours of lengthy discussions over the last few years. He has been very gracious in his time and helped me along through my stay here. I also wish to thank {\bf Dr. Sibasish Ghosh} who has introduced me to many interesting topics and has been a constant source of inspiration in pursuing my academic interests. He has also provided valuable inputs that have helped me in my work at various junctures. 

Finally, I would like to thank all my teachers at Matscience, my colleagues, and my batch mates who have helped me all along. I am also grateful to all the librarians who have been extremely helpful and patient with regards to many of my requests, and  without whom this thesis would not be completed.  I would also like to thank the administration and the persons incharge of the computing facilities, who have provided the support facilities and made my stay here as pleasant as possible.

\newpage
\cleardoublepage
~
\vskip 1.0cm
\centerline{{\bf{\large List of Publications arising from the thesis}}}
\vskip 0.5cm
{\bf Journal}
\begin{enumerate}[1.]
\item {\em Robustness of nonGaussian entanglement against noisy amplifier and attenuator environments}. {\bf K. K. Sabapathy}, J. S. Ivan, and R. Simon. \\ \href{http://prl.aps.org/abstract/PRL/v107/i13/e130501}{Physical Review Letters, {\bf 107}, 130501 (2011)}. 
\item {\em Nonclassicality breaking is the same as entanglement breaking for bosonic Gaussian channels}. J. S. Ivan, {\bf K. K. Sabapathy}, and R. Simon.  \\ \href{http://pra.aps.org/abstract/PRA/v88/i3/e032302}{Phys. Rev. A, {\bf 88}, 032302 (2013)}. 
\item  J. S. Ivan, {\bf K. K. Sabapathy}, and R. Simon. {\em  Operator-sum Representation for Bosonic Gaussian Channels}. 
\href{http://pra.aps.org/abstract/PRA/v84/i4/e042311}{Physical Review A, {\bf 84}, 042311 (2011)}.
\end{enumerate}

{\bf Others (communicated)}
\begin{enumerate}[1.]
\item  {\em Quantum discord plays no distinguished role in characterization of complete positivity: Robustness of the traditional scheme}. {\bf K. K. Sabapathy}, J. S. Ivan, S. Ghosh, and R. Simon. \href{http://arxiv.org/abs/1304.4857}{arXiv:1304.4857 [quant-ph]}.  
\item   {\em Quantum discord for two-qubit $X$-states: A comprehensive approach inspired by classical polarization optics }. {\bf K. K. Sabapathy} and R. Simon.\\ \href{http://arxiv.org/abs/1311.0210}{arXiv:1311.0210 [quant-ph]}.
\end{enumerate}

\newpage
\cleardoublepage
\vskip 3.0cm
\centerline{{\bf{\large List of corrections and changes as suggested by the Thesis and Viva Voce Examiners}}}
\vskip 0.5cm 
\vspace{2cm}
\noindent
\begin{enumerate}[1.]
\item Added Refs. 35, 37, 38, 40, 41, 44, and 51 
in the Introduction Chapter. 
\item  Added Refs. 147-150 on p.96 of Chapter 2.
\item The notions of critical noise and robustness of entanglement have been clarified in Definitions 1, 2, and 3 of Chapter 4.
\item Introduced Figs.\,\ref{fign0b}, \ref{fign0a}, and \,\ref{fign2} 
in Chapter 5 to bring out in a transparent manner the similarities and the differences between nonclassicality breaking and entanglement breaking channels. 
\item Corrected typos as suggested by the referees in all the Chapters.
\end{enumerate}

\vspace{5cm}

The corrections and changes suggested by the Thesis and Viva Voce Examiners  have been incorporated in the thesis. \\
\vspace{2cm}\\
~\hspace{5cm} Guide

\newpage
\cleardoublepage
\frontmatter
\pagestyle{plain}
\tableofcontents
\raggedbottom
\mainmatter
{
\addcontentsline{toc}{chapter}{Synopsis}
\doublespacing
\chapter*{Synopsis}

This thesis explores ways in which quantum channels and correlations (of both classical and quantum types) manifest themselves, and also studies the interplay between these two aspects in various physical settings. Quantum channels represent all possible evolutions of states, including measurements, allowed by quantum mechanics, while correlations are intrinsic (nonlocal) properties of  composite systems.

Given a quantum system with Hilbert space ${\cal H}_S$, states of this system are operators $\rho$ that satisfy $\rho = \rho^{\dagger}$, $\rho \geq 0$, and ${\rm Tr}\,\rho =1$. The set of all such operators (density matrices) constitute the (convex) state space  $\Lambda({\cal H}_S)$. Observables $\hat{O}$ are hermitian operators acting on ${\cal H}_S$.  Let the spectral resolution of $\hat{O}$ be $\hat{O}= \sum_j \lambda_j  P_j$. When $\hat{O}$ is measured, the $j^{\rm th}$ outcome corresponding to measurement operator $P_j$ (projection)  occurs with probability $p_j = {\rm Tr}(\rho \,P_j)$. One obtains a more general measurement scheme called POVM (positive operator valued measurement) when the projective measurement elements $P_j$ are replaced by positive operators $\Pi_j$ with $\sum_j \Pi_j=1\!\!1$, and the probabilities of outcomes are obtained in a similar manner\,: $p_j = {\rm Tr}(\rho \Pi_j)$.

If a system is isolated, then its dynamics is governed by the unitary Schr\"{o}dinger evolution. A unitary operator $U$ effects the following transformation $\rho \to \rho^{\,'} = U \,\rho\,U^{\dagger}$, $\rho$ and $\rho^{\,'} \in \Lambda({\cal H}_S)$. But if the system is in interaction with its environment, then the evolutions of the system of interest resulting from unitary evolutions of the composite are more general, but nevertheless described by linear maps acting on the state space, directly rather than through ${\cal H}_S$. 

Let $\Phi$ be a linear map that acts on states of the system. An obvious necessary requirement for $\Phi$ to be a valid evolution is that it takes states to states. We call a map that satisfies this condition as a {\em positive map}, i.e.,
\begin{align}
\Phi \text{ is positive} \Leftrightarrow \Phi(\rho_S)= \rho^{\,'}_S \in \Lambda({\cal H}_S),~
\text{i.e.,}~~ \Phi\left( \Lambda({\cal H}_S)\right) \subset  \Lambda({\cal H}_S).
\label{s1}
\end{align}
It turns out that not all positive maps are physical evolutions. For positive maps to be physical evolutions, there is a further requirement to be met. 

Let us consider a composite system in which the system is appended with an arbitrary ancilla or reservoir $R$. The Hilbert space of the composite system is ${\cal H}_S \otimes {\cal H}_R$, a tensor product of the individual subsystem Hilbert spaces. Let us denote the state space of this composite system by $\Lambda({\cal H}_S \otimes {\cal H}_R)$. 

It is both reasonable and necessary to require that local action of $\Phi$ takes states of the joint system also to states. In other words 
\begin{align}
&(\Phi \otimes 1\!\!1 )[\rho_{SR}]  = \rho_{SR}^{\,'} \in \Lambda({\cal H}_S \otimes {\cal H}_R), \nonumber\\
~ \text{i.e.,}~~ &[\Phi \otimes 1\!\!1]\left(\Lambda({\cal H}_S \otimes {\cal H}_R) \right) \subset \Lambda({\cal H}_S \otimes {\cal H}_R).
\label{s2}
\end{align} 
A positive map $\Phi$ that satisfies Eq.\,\eqref{s2}, is known as a {\em completely positive} (CP) trace-preserving (TP) map or a {\bf quantum channel}. 

It is known that every CP map can be realised in the following way. First, the systems states are elevated to product states on a larger Hilbert space (system + environment), with a fixed state of the environment\,: $\rho_S \to \rho_S \otimes \rho_R$, $\rho_R$ fixed. Then the product states are evolved by a joint unitary evolution, and finally the environment degrees of freedom are traced out to give the evolved system states. It turns out that this provides a suitable framework for the description of open quantum systems. 

An intrinsic property of composite systems that is of much importance is correlations between subsystems. One important aspect has been to segregate the classical and quantum contents of correlations. To this end, various measures and methods  have been proposed. Entanglement has been the most popular of these correlations owing to its inherent advantages in performing quantum computation and communication tasks\,\cite{syn-horo-rmp} and has been studied over the last few decades. 
But there are other correlations that are motivated from an information-theoretic or measurement perspective, which try to capture this classical-quantum boundary\,\cite{syn-modirmp}.  These include quantum discord, classical correlation, measurement induced disturbance, quantum deficit, and geometric variants of these measures. Of these, quantum discord and classical correlation have received enormous attention in recent years.  

Let us now consider a bipartite system with Hilbert space ${\cal H}_S \otimes {\cal H}_S$, where the two subsystem Hilbert spaces have been taken to be identical for simplicity.  A pure bipartite state is said to be separable if it can be written as a (tensor) product of states of the individual subsystems. Else, the pure state is said to be entangled.  While, a mixed state $\rho_{AB} \in \Lambda({\cal H}_S \otimes {\cal H}_S)$ is said to separable if it can be written as a convex combination of product states, i.e., 
\begin{align}
\rho_{AB} = \sum_j p_j\,\rho^A_j \otimes \rho^B_j.
\label{s3}
\end{align}
A state that cannot be written in this form is called an entangled state. The set  of separable states form a convex subset of the bipartite state space. The qualitative detection and quantitative estimation of entanglement have proved to be non-trivial. To this end, there have been many approaches that include Bell-type inequalities, entanglement witnesses, entropy based measures, distance (geometry) based measures, and criteria based on positive maps that are not completely positive.

{\bf Quantum discord} is a `beyond-entanglement' quantum correlation, since there exist separable states which return a non-zero value of quantum discord. A recent avenue has been to try and find advantages of these correlations, both in the theoretical and experimental domain, in respect of information precessing tasks. For example, some interesting applications of quantum discord in quantum computation, state merging, remote state preparation, and entanglement distillation have been reported. 

We may motivate the definition of quantum discord by first looking at the classical setting. Given a probability distribution $p(x,y)$ in two variables, the mutual information $I(x,y)$ is defined as   
\begin{align}
I(x,y) = H(x) - H(x|y), 
\label{s4}
\end{align}
where $H(\cdot)$ stands for the Shannon entropy  $H(x)= -\sum_x p(x)\, {\rm Log}[p(x)]$ and  $H(x|y)$ is the conditional entropy. Using Bayes rule we are lead to an equivalent expression for mutual information\,:
\begin{align}
I(x,y) = H(x) + H(y) - H(x,y). 
\label{s5}
\end{align}
The second expression \eqref{s5} for mutual information naturally generalises to the quantum setting when the bipartite probability distribution is replaced by a bipartite state $\rho_{AB}$ and the Shannon entropy $H(\cdot)$  by the von Neumann entropy $S(\cdot)$ of quantum states, and we have 
\begin{align}
I(\rho_{AB}) = S(\rho_A) + S(\rho_B) - S(\rho_{AB}).
\label{s6}
\end{align}
But the first expression \eqref{s4} for classical mutual information does {\em not} possess a straightforward generalization to the quantum case. In the quantum case, the conditional entropy is defined with respect to a measurement, where the measurement is performed on one of the subsystems, say subsystem B. Let us consider a POVM $\Pi^B = \{\Pi^B_j\}$ where $\Pi^B_j \geq 0$ and $\sum_j \Pi^B_j= 1\!\!1$. Then the conditional entropy post measurement is given by 
\begin{align}
S^A = \sum_j \, p_j \,S(\rho_j^A), 
\label{s7}
\end{align}
where the probabilities and  states post measurement are given by
\begin{align}
p_j &= {\rm Tr}\,(\Pi_j^B \, \rho_{AB}), \nonumber\\ 
\rho_j^A &= p_j^{-1}\, {\rm Tr}_B(\Pi_j^B \, \rho_{AB}).
\label{s8}
\end{align}
Let us denote by $S^A_{\rm min}$ the minimum of $S^A$ over all measurements or POVM's. 
The difference between these two classically equivalent expressions (optimized over all measurements) is called quantum discord ${\cal D}(\rho_{AB})$\,:
\begin{align}
{\cal D}(\rho_{AB}) &=    I(\rho_{AB}) - \left[ S(\rho_A) - S^A_{\rm min} \right],\nonumber\\
&=  S(\rho_B) - S(\rho_{AB}) + S^A_{\rm min}.
\label{s9}
\end{align} 
The quantity $C(\rho_{AB}) = S(\rho_A) - S^A_{\rm min}$ is defined as the {\bf classical correlation}. Thus, the mutual information which is supposed to capture the total correlation of a bipartite state is broken down into quantum discord, that captures the quantum correlations, and classical correlation $C(\rho_{AB})$. 

It is the interplay between correlations of bipartite states and their evolution through quantum channels that is the unifying theme of this thesis. We explore some aspects of this interplay in the different chapters. There are four broad topics that are covered in this thesis\,:
\begin{itemize}
\item Initial bipartite correlations and induced subsystem dynamics\,: Does initial correlation of the system-bath states provide a generalization of the folklore product realization of CP maps?
\item A geometric approach to computation of quantum discord for two-qubit $X$-states.
\item Robustness of nonGaussian vs Gaussian entanglement against noise\,: We demonstrate simple examples of nonGaussian states whose entanglement survives longer that  Gaussian entanglement under noisy channels
\item Is nonclassicality breaking the same thing as entanglement breaking? The answer is shown to be in the affirmative for bosonic Gaussian channels. 
\end{itemize}

In {\bf Chapter 1}, we provide a basic introduction to the concepts that are used in the thesis. 
In addition to setting up the notations, this Chapter helps to render the thesis reasonably self-contained. We describe the properties of bipartite correlations of interest to us, namely, classical correlation, quantum discord, and entanglement. 

We  briefly describe the notion of quantum channels. We discuss in  some detail the three well-established representations of CP maps\,\cite{syn-cp}. These are the operator-sum representation, the unitary representation, and the Choi-Jamiolkowski isomorphism between bipartite states and channels. We indicate how one can go from one representation to  another. 

We also indicate an operational way to check as to when a positive map be can called a CP map. We list some properties of channels and indicate when a channel is unital, dual, extremal, entanglement breaking, bistochastic, and so on. 

We then move on to a discussion of states and channels in the continuous variable setting, in particular the Gaussian case. Here, we begin by recapitulating the properties of Gaussian states. The phase space picture in terms of quasiprobability distributions is outlined and some basic aspects of the symplectic structure is recalled. We will be mainly concerned with the Wigner distribution and its associated characteristic function. Gaussian states are completely described in terms of the variance matrices and means. For these states, we describe the uncertainty principle, the canonical form of the variance matrix, Simon's criterion for detecting entanglement of two-mode Gaussian states, and a description of the more commonly used Gaussian states like the vacuum state, thermal state, squeezed state, and coherent state.

Then we proceed to a discussion of single-mode Bosonic Gaussian channels.  These  are trace-preserving completely positive  maps that take input Gaussian states to Gaussian states at the output. These channels play a fundamental role in continuous variable quantum information theory. We discuss their phase space description, the CP condition, and enumeration of their canonical forms. 

In the standard classification,  Bosonic Gaussian channels group themselves into five broad classes. Namely, the attenuator, amplifier, phase conjugation, singular channels and, finally, the classical noise channels. We then briefly describe the operator-sum representation\,\cite{syn-kraus10} for all single-mode Gaussian channels. 

Of particular importance to us is the analysis of quantum-limited channels. Quantum-limited channels are channels that saturate the CP condition
and hence do not contain extra additive classical noise over and above the minimum demanded by the uncertainty principle. We emphasise the fact that noisy channels can be factored as  product of a pair of noiseless or quantum-limited channels. The action in the Fock basis is brought out. 

The attenuator channel and the amplifier channel are of particular interest to us, and so we bring out some of its properties including the semigroup structure of the amplifier and the attenuator families of quantum-limited channels.  
The noisy versions of these channels can be easily obtained by composition of a pair of quantum-limited channels; 
this is explicitly shown for all channels and tabulated.  In particular, we 
 obtain a discrete operator-sum representation for the classical noise channel which may be contrasted with the familiar one in terms of a continuum of Weyl displacement operators. These representations lead to an interesting application which is pursued in Chapter 4. 
This Introductory chapter renders the passage to the main results of the thesis in subsequent chapters smooth. 

In {\bf Chapter 2} we consider the dynamics of a system that is in interaction with an environment, or in other words, the dynamics of an open quantum system. 

Dynamics of open quantum systems is fundamental to the study of any realistic or practical
application of quantum systems. Hence, there has been a rapidly growing interest in the understanding of 
various properties related to open quantum systems like its realization, control, and the role played by 
the noise in such dissipative systems, both in the theoretical and experimental domain. 
These studies have been motivated by applications to quantum computing, laser cooling,
quantum reservoir engineering, managing decoherence, and also to other fields like 
chemical reactions and energy transfer in molecules.

Here we study the induced dynamics of a system viewed as part of a larger composite system, when the system plus environment undergoes a unitary evolution. Specifically, we explore the effect of initial system-bath correlations on complete positivity of the reduced dynamics. 

{\bf Traditional (Folklore) Scheme}\,: 
In the folklore scheme, initial  system states $\rho_S$ 
 are elevated to  product states of the composite, for a {\em fixed} fiducial bath state 
$\rho_B^{\,\rm{fid}}$, through the assignment map    
$\rho_S \to \rho_S \otimes \rho_B^{\,\rm{fid}}$. These uncorrelated system-bath states are evolved under a 
joint unitary $U_{SB}(t)$ to  
$U _{SB}(t) \,\rho_S \otimes \rho_B^{\,\rm{fid}}\,U_{SB}(t)^{\dagger}$
and, finally, the bath degrees of freedom are 
traced out to obtain the time-evolved states of the system of interest\,: 
\begin{align}
\rho_S \to \rho_S(t) = \rm{Tr}_B\left[ U_{SB}(t)\, \rho_S 
 \otimes \rho_B^{\,\rm{fid}}\,U_{SB}(t)^{\dagger} \right].
\label{s10}
\end{align}
The resulting quantum dynamical process (QDP) ~$\rho_S \to 
\rho_S(t)$, parametrized by $\rho_B^{\,\rm{fid}}$ and $U_{SB}(t)$, is 
completely positive by construction.

Currently, however, the issue of system-bath initial correlations potentially affecting the 
reduced dynamics of the system has been attracting considerable interest. 
A specific, carefully detailed, and precise formulation of the issue 
of initial system-bath correlations possibly influencing the reduced dynamics was 
presented not long ago by  Shabani and Lidar (SL)\,\cite{syn-shabani09}.

{\bf Shabani-Lidar scheme}\,:
In sharp contrast to the folklore scheme, there is no assignment map in the SL scheme.  The distinguished bath state $\rho_B^{\,\rm{fid}}$ is replaced by 
a collection $\Omega^{SB}$ of (possibly correlated) system-bath 
 initial states $\rho_{SB}(0)$.
 The dynamics gets defined through
\begin{align} 
\rho_{SB}(0) \to \rho_{SB}(t) = 
U_{SB}(t)\,\rho_{SB}(0)\,U_{SB}(t)^{\dagger},
\label{s11}
\end{align}
 for all $\rho_{SB}(0) \in \Omega^{SB}$. 
 With reduced system states $\rho_S(0)$ and $\rho_S(t)$ defined through
the imaging or projection map   
$\rho_S(0) = \rm{Tr}_B\,\rho_{SB}(0)$ and $\rho_S(t) = \rm{Tr}_B 
\left[U_{SB}(t)\,\rho_{SB}(0)\,U_{SB}(t)^{\dagger} \right]$,   
this unitary dynamics of the composite induces on the system the QDP 
$\rho_S(0) \to \rho_S(t)$. 

Whether the SL QDP so described
is well-defined and completely positive is clearly an issue answered solely by 
the nature of the collection $\Omega^{SB}$. 
It is evident that the folklore scheme obtains  as a special case of the SL scheme.
This generalized formulation of QDP allows SL to transcribe the 
fundamental issue to this question: What are the necessary and sufficient 
conditions on the collection of initial states  so that the induced QDP  is  
 guaranteed to be CP {\em for all} joint unitaries? 
 
Motivated by the work of Rodriguez-Rosario et al.\,\cite{syn-rosario08}, SL advance the following resolution to this 
 issue: {\em The QDP is CP 
for all joint unitaries  if and only if the  quantum 
discord vanishes for all initial system-bath states $\in \Omega^{SB}$ , i.e., if and only if the 
initial system-bath correlations are purely classical}. 
The SL theorem has come to be counted among the more important recent results of quantum information theory, and it is 
paraded by many authors as one of the major achievements of quantum discord. Meanwhile, the very recent work of Brodutch et al.\,\cite{syn-brodutch13}, contests the claim of SL and asserts that vanishing quantum discord is 
sufficient but not necessary condition for complete positivity. 

Our entire analysis in Chapter 2 rests on two, almost obvious, necessary properties of the set of  initial system-bath states $\Omega^{SB}$ so that the resulting SL QDP would be well defined.  \\
\noindent
Property 1:  No state $\rho_S(0)$ can have two (or more) pre-images in $\Omega^{SB}$.\\
\noindent
Property 2: While every system state need not have 
a pre-image {\em actually enumerated} in $\Omega^{SB}$, 
the set of $\rho_S(0)$'s having pre-image in $\Omega^{SB}$ should be sufficiently large, such that the QDP can be extended by linearity to all states of the system, i.e., to the full state space of the system. 

Using these two requirements, we prove that {\em both the SL theorem and the assertion of Brodutch et al. are too strong to be tenable}. We labour to point out that rather than viewing this result as a negative verdict of the SL theorem, it is more constructive to view our result as demonstrating a kind of robustness of the traditional scheme. 

\noindent
In {\bf Chapter 3} we undertake a comprehensive analysis of the problem of computation of  correlations in two-qubit systems, especially the so-called $X$-states which have come to be accorded a distinguished status in this regard.  Our approach  exploits the very geometric nature of the problem, and clarifies some  issues regarding computation of correlations in $X$-states.  It may be emphasised that the geometric methods used here have been the basic tools of (classical) polarization optics for a very long time, and involve constructs like Stokes vectors, Poincar\'{e} sphere, and Mueller matrix\,\cite{syn-simon-mueller}. 

As noted earlier, the expressions for quantum discord and classical correlation are 
\begin{align}
{\cal D}(\rho_{AB}) &= S(\rho_B) - S(\rho_{AB}) + S^A_{\rm min},\nonumber\\
C(\rho_{AB}) &= S(\rho_A) - S^A_{\rm min}. 
\label{s12}
\end{align}
It is seen that the only term that requires an optimization is the conditional entropy post measurement, $S^A_{\rm min}$. Given a composite state $\rho_{AB}$, the other entropic quantities are immediately evaluated. Central to the simplicity and comprehensiveness of our analysis is the recognition that computation of $S^A_{\rm min}$ for two-qubit $X$-states is a {\em one-parameter optimization problem}, much against the impression given by a large section of the literature.

Our analysis begins by placing in context the use of the Mueller-Stokes formalism for estimating $S^A_{\rm min}$.  
Given a two-qubit  state $\rho_{AB}$, it can always be written as 
\begin{align}
\rho_{AB} = \frac{1}{4} \sum_{a,b=0}^3 M_{ab} \,\sigma_a \otimes \sigma_b^*,
\label{s13}
\end{align}
the associated $4 \times 4$ matrix $M$ being real; $\sigma_1$, $\sigma_2$, $\sigma_3$ are the Pauli matrices and $\sigma_0$ equals the unit matrix. Writing a POVM element on the B side as $K = \frac{1}{2} \sum_a S_a \sigma_a$, the output state of A  post measurement is obtained as the action of $M$ on the input Stokes vector $S^{\rm in}$ corresponding to the  POVM element. This may be compactly expressed as 
\begin{align}
S^{\rm out} = M S^{\rm in} 
\label{s14}
\end{align}
and has a form analogous to the input-output relation in polarization optics. In view of this analogy, we may call $M$ the Mueller matrix associated with $\hat{\rho}_{AB}$. 
In general $M$ need not correspond to a trace-preserving map, since the conditional output states need not be normalized. So they need to be normalised for calculating the conditional entropy. The manifold of these normalized conditional states is an ellipsoid, a convex subset of the Poincar\'{e} sphere, completely parametrised by the local unitarily invariant part of the $M$ matrix and, thereby, the local unitarily invariant part of the bipartite state $\rho_{AB}$. The boundary of this output ellipsoid corresponds to the images of all possible rank-one POVM's or light-like $S^{\rm in}$.

While this geometric picture a two-qubit state being fully captured by its Mueller matrix---or equivalently by this output ellipsoid---applies to every two-qubit state, $X$-states are distinguished by the fact that the centre $C$ of the output ellipsoid, the origin $O$ of the Poincar\'{e} sphere and $I$, the image of maximally mixed input $S^{\rm in}=(1,0,0,0)^T$ are all collinear and lie on one and the same principal axis of the ellipsoid. 

One realizes that the Mueller matrix of any $X$-state can, by local unitaries, be brought to a canonical form wherein the only nonvanishing off-diagonal elements are $m_{03}$ and $m_{30}$, and thus $X$-states form, in the canonical form, a five parameter family with $m_{11},\,m_{22},\,m_{33},\,m_{30}$, and $m_{03}$ as the canonical parameters ($m_{00}= {\rm Tr}\,\rho_{AB}=1$ identically). With this realization our entire analysis in Chapter 3 is geometric in flavour and content. The principle results of the Chapter may be summarized as follows\,:
\begin{itemize}
\item All $X$-states of vanishing discord are fully enumerated and contrasted with earlier results.
\item Computation of quantum discord of $X$-states is proved to be an optimization problem in one real variable.
\item It is shown that the optimal POVM never requires more that three elements.
\item In the manifold of $X$-states, the boundary between states requiring three elements for optimal POVM and those requiring just two is fully detailed. 
\end{itemize}
It may be stressed that our analysis in this Chapter is from first principles. It is comprehensive and geometric in nature. Our approach not only reproduces and unifies all known results in respect of $X$-states, but also brings out entirely new insights. 

In {\bf Chapter 4} we explore the connection between bipartite entanglement and local action of noisy channels in the context of continuous variable systems. Quantum entanglement in continuous variable systems has
proved to be a valuable resource in quantum information processes like
teleportation, cloning, dense
coding, quantum
cryptography, and quantum computation. 

These early developments in quantum information technology involving continuous 
variable (CV) systems largely concentrated on Gaussian states and 
Gaussian operations, mainly due to their
experimental viability within the current optical
technology. The symplectic group of linear 
canonical transformations is available as a handy and 
powerful tool in this Gaussian scenario, leading to an elegant
classification of permissible Gaussian processes or 
channels.

However,
the fact that states in the nonGaussian sector could offer advantage 
for several quantum information tasks has resulted more recently in 
considerable interest in nonGaussian 
states, both experimental and 
theoretical. 
 The use of nonGaussian resources for
teleportation, 
entanglement distillation, and its use in quantum
networks have been studied. So there 
has been interest to explore the essential differences between
Gaussian states and nonGaussian states as resources for performing
quantum information tasks.     

Allegra et al.\,\cite{syn-allegra} have studied the evolution of 
what they call {\em photon number entangled states} (PNES), 
\begin{align}
|\psi\rangle_{\rm PNES} = \sum_n\,c_n\,|n,\,n\rangle,
\label{s15}
\end{align}
in a  {\em noisy} attenuator environment. They conjectured based on  
numerical evidence that, for a given energy, Gaussian entanglement is 
more robust than  nonGaussian ones. Earlier Agarwal et 
al.\,\cite{syn-agarwal091} had shown that entanglement of the NOON state,
\begin{align}
|\psi\rangle_{\rm NOON} = \frac{1}{\sqrt{2}} (|n,0\rangle + |0,n\rangle),
\label{s16}
\end{align}
 is 
more robust than Gaussian entanglement in the {\em quantum limited} 
amplifier environment. Subsequently, Nha et al.\,\cite{syn-nha10}  
showed that nonclassical features, including entanglement, of several 
nonGaussian states survive a {\em quantum limited} amplifier 
environment much longer than Gaussian entanglement. 
Since the 
 conjecture of\,\cite{syn-allegra} refers to  noisy environment,   
 while the analysis of\,\cite{syn-agarwal091,syn-nha10} to the noiseless or 
quantum-limited case, the conclusions of the latter 
amount to neither confirmation nor 
refutation of the conjecture of\,\cite{syn-allegra}. In the meantime, Adesso 
argued\,\cite{syn-adesso} that the well known 
extremality\,\cite{syn-extremality} of Gaussian states implies 
`proof and rigorous validation' of the conjecture of\,\cite{syn-allegra}.

In the work described in Chapter 4 we employ the recently developed
Kraus representation of bosonic Gaussian channels\,\cite{syn-kraus10} to study 
analytically the behaviour of nonGaussian states in {\em noisy} 
attenuator or and amplifier environments. Both NOON states and a simple 
form of PNES are considered. Our results show conclusively that the 
conjecture of \cite{syn-allegra} is too strong to be maintainable, the `proof and rigorous validation' of \cite{syn-adesso} notwithstanding.   

An important point that emerges from this study is the fact that Gaussian entanglement resides entirely `in' the variance matrix or second moments, and hence 
disappears when environmental noise raises the variance matrix above the vacuum or 
quantum noise limit. That our chosen nonGaussian states survive these environments 
shows that their entanglement resides in the higher moments, in turn
demonstrating that their entanglement is genuine nonGaussian. Indeed,
the variance matrix of our PNES and NOON states for $N=5$ is six times `more noisy' than 
that of the vacuum state.

We study in {\bf Chapter 5} an interesting relationship between nonclassicality and entanglement in the context of bosonic Gaussian channels. We motivate and resolve the following issue\,: 
{\em which Gaussian channels have the property that their output 
is guaranteed to be classical independent of the input state}? 

We recall that the density operator $\hat{\rho}$ representing any state of 
radiation field is `{\em diagonal}' in the coherent state `{\em basis}'\,\cite{syn-ecg63},
and this happens because of the 
over-completeness property of the coherent state basis. 
An important notion that arises from the diagonal representation 
is the {\em classicality-nonclassicality divide}. 
Since coherent states are the most elementary 
of all quantum mechanical states exhibiting classical behaviour, any state
that can be written as a convex sum 
of these elementary classical states
is deemed classical. Any state which cannot be so written as a convex sum of
coherent states is deemed nonclassical. 

This classicality-nonclassicality 
divide leads to the following natural definition, inspired by the notion 
of entanglement breaking channels\,: we define a channel $\Gamma$ to be {\em nonclassicality breaking}
if and only if the output state $\hat{\rho}_{\rm out}= \Gamma (\hat{\rho}_{\rm in})$ is classical
{\em for every} input state $\hat{\rho}_{\rm in}$, i.e., if and only if the diagonal
`weight' function of every output state is a genuine probability distribution.

We first derive the {\em nonclassicality-based} canonical forms
for Gaussian channels\,\cite{syn-ncbc}. The available classification by Holevo and 
collaborators is {\em entanglement-based}, and so it is not suitable for our purpose, since
the notion of nonclassicality breaking has a more restricted invariance.
A nonclassicality breaking Gaussian channel $\Gamma$
preceded by any Gaussian unitary ${\cal U}({\cal S})$ is nonclassicality breaking
if and only if $\Gamma$ itself is nonclassicality breaking.
In contradistinction, the nonclassicality breaking aspect of $\Gamma$
and that of ${\cal U}({\cal S})\,\Gamma$ [$\Gamma$ followed the Gaussian unitary
${\cal U}({\cal S})$] are not equivalent in general. They are equivalent
if and only if ${\cal S}$ is in the 
intersection $Sp(2n,\,R) \cap SO(2n,\,R)$ of symplectic phase
space rotations, or passive elements in the quantum optical sense\,\cite{syn-simon94}. 
The canonical forms and 
the corresponding necessary and sufficient conditions for 
nonclassicality breaking, entanglement breaking and complete-positivity 
are listed in Table\,\ref{tables1}.

\begin{table}
\centering
\begin{tabular}{|c|c|c|c|c|}
\hline
&Canonical form & NB & EB & CP \\
\hline
I&$(\kappa\,{1\!\!1},\, {\rm diag}(a,b))$  &   $(a-1)(b-1) \geq \kappa^4$ 
& $ab\geq(1+\kappa^2)^2$  & $ab\geq (1-\kappa^2)^2$\\
\hline
II&$(\kappa\,\sigma_3,\,{\rm diag}(a,b))$ & $(a-1)(b-1) \geq \kappa^4$ 
& $ab\geq(1+\kappa^2)^2$  & $ab\geq (1+\kappa^2)^2$\\
\hline
III&$({\rm diag}( 1,0),\,Y)$, &$a,\,b \geq 1$, $a$, $b$ being & $ab\geq1$ & $ab \geq 1$ \\
&&eigenvalues of $Y$ & &  \\
&$({\rm diag}(0,0),\, {\rm diag}(a,b))$ &$a,\,b \geq 1$ & $ab\geq1$ & $ab \geq 1$ \\
\hline
\end{tabular}
\caption{Showing the nonclassicality breaking (NB), entanglement breaking (EB) and complete-positivity (CP) conditions for the three canonical forms.  \label{tables1}}
\end{table}

For all three canonical forms we show that a nonclassicality breaking
channel is necessarily entanglement breaking. There are channel 
parameter ranges wherein the channel is entanglement breaking
but not nonclassicality breaking, but the nonclassicality of the
output state is of a `weak' kind in the following sense\,:
For every entanglement breaking channel, there exists a particular
value of squeeze-parameter $r_{0}$, depending only on the channel
parameters and not on the input state, so that the entanglement breaking
channel followed by unitary squeezing of extent $r_0$ always results
in a nonclassicality breaking channel. It is in this precise sense 
that nonclassicality breaking channels and entanglement breaking channels
are essentially one and the same.

Squeezing is not the only form of nonclassicality. Our result
not only says that the output of an entanglement breaking 
channel could at the most have squeezing-type nonclassicality,
it further says that the nonclassicality of {\em all} output states can be
removed by a {\em fixed} unitary squeezing transformation.

Finally, in {\bf Chapter 6}, we briefly summarise the conclusions of each of the chapters, and explore possible  avenues and prospects for future directions of study.

\listoffigures
\addcontentsline{toc}{chapter}{List of Figures}

\listoftables
\addcontentsline{toc}{chapter}{List of Tables}

\part{Preliminaries}

\chapter{Introduction}
\section{States, observables, and measurements}
\footnote{This introduction Chapter has been written by generously borrowing from Prof. Simon's lectures delivered over the course of my doctoral work.}
Let us consider a quantum system $A$ with Hilbert space ${\cal H}_A$ of dimension $d_A$. 
{\em Pure states} $|\psi\rangle$ are equivalence classes of unit vectors (unit rays) in this Hilbert space\,:
\begin{align}
|\psi\rangle \in {\cal H}_A, ~~ \langle \psi| \psi \rangle =1, ~~ e^{i\alpha} |\psi\rangle \sim |\psi\rangle,~0 \leq \alpha < 2\pi.
\label{i1}
\end{align} 
A {\em mixed state} $\hr{A}$ is a statistical ensemble of pure states\,:
\begin{align}
\hr{A} = \sum_j p_j \,\proj{\psi_j},~~\sum_j p_j =1,~~p_j >0 ~\forall~ j. 
\label{i2}
\end{align}
Equivalently, one could view the state of a quantum system \eqref{i2} as a linear operator acting on ${\cal H}_A$ that satisfies the following defining properties\,:
\begin{itemize}
\item Hermiticity\,: $\hr{A} = \hr{A}^{\dagger}$.
\item Positivity\,: $\hr{A} \geq 0 $.
\item Unit trace condition\,:  ${\rm Tr}[\hr{A}] = 1$.
\end{itemize}
For any state $\hr{A}$, there exists a special decomposition called the {\em spectral decomposition}.  It states that any mixed state can be written as
\begin{align}
\hr{A} = \sum_{j=1}^{r} \lambda_j\, |\psi_j \rangle \langle \psi_j|,
\label{i3}
\end{align} 
where $\lambda_j$ are the eigenvalues (which are all positive) and $|\psi_j\rangle$ are the eigenvectors that satisfy $\langle \psi_j | \psi_k \rangle = \delta_{kj}$. Here, $r\leq d_A$ is the {\rm rank} of $\hr{A}$. The trace condition implies that $\sum_j \lambda_j =1$.

It is clear from the properties of a quantum state that the state space or collection of quantum states of a system form a convex set. 
We denote the state space of system $A$ by $\Lambda({\cal H}_A)$. The pure states are the extreme points of this convex set and mixed states are nonextremal.

A useful way to capture mixedness is through a quantity  known as {\em purity}. Purity is defined as ${\rm Tr}(\hr{A}^2)$. We see that for pure states
\begin{align}
{\rm Tr}\,\hr{A}^2 = {\rm Tr}\,\hr{A} = 1,
\label{i4}
\end{align}
while for mixed states  
\begin{align}
{\rm Tr}\,\hr{A}^2 < {\rm Tr}\,\hr{A} \text{ and so } {\rm Tr}\,\hr{A}^2 <1. 
\label{i5}
\end{align}
The least value of purity is assumed by the maximally mixed state, and evaluates to $1/d_A$.
Let us further denote by ${\cal B}({\cal H}_A)$ the complex linear space of  bounded linear operators on ${\cal H}_A$. The state space $\Lambda({\cal H}_A) \subset {\cal B}({\cal H}_A)$.

For illustration we consider the simplest quantum system, a qubit\,\cite{nielsenbook,vollbrecht00,peresbook}. For a qubit system $d_A=2$ and the states of the system can be represented as
\begin{align}
\hr{A} = \frac{1}{2} (1\!\!1 + \mathbf{a}.\boldsymbol{\sigma}),
\label{i6}
\end{align}
where $\mathbf{a} \in {\cal R}^3$, $|\mathbf{a}| \leq 1$ and $\boldsymbol{\sigma}=(\sigma_x,\,\sigma_y,\,\sigma_z)$ are the Pauli matrices. The state space here is a (solid) sphere of unit radius and is known as the Poincar\'{e} or Bloch sphere. In this case, and only in this case, are all boundary points extremals. One can write its spectral decomposition explicitly as
\begin{align}
\hr{A} = \frac{1+|\mathbf{a}|}{2}\left[ \frac{1}{2} (1\!\!1 + \hat{\mathbf{a}}.\boldsymbol{\sigma})\right] + \frac{1-|\mathbf{a}|}{2}\left[ \frac{1}{2} (1\!\!1 - \hat{\mathbf{a}}.\boldsymbol{\sigma})\right],
\label{i7}
\end{align} 
where the eigenvalues are 
\begin{align}
\frac{1}{2}(1\pm|\mathbf{a}|)
\label{i8}
\end{align} 
corresponding to eigenvectors
\begin{align}
\frac{1}{2} (1\!\!1 \pm \hat{\mathbf{a}}.\boldsymbol{\sigma}), ~~ \hat{\mathbf{a}} = \frac{\mathbf{a}}{|\mathbf{a}|}.
\label{i9}
\end{align}
It is easily seen that the eigenvectors \eqref{i9} are rank-one orthogonal projectors that lie on the boundary of the Bloch sphere and for a general mixed state $|\mathbf{a}|<1$. They are unique except for $\mathbf{a}=0$ in which case all states are eigenstates. 

Returning to the general case, a quantity of interest to us is the {\em von-Neumann} entropy of a state $\hr{A}$ and is given by
\begin{align}
S(\hr{A}) = -{\rm Tr}[ \hr{A}\, {\rm log}_2 (\hr{A})].
\label{i10}
\end{align}
From the spectral decomposition of $\hr{A}$ \eqref{i3}, it is easily seen that $S(\hr{A})$ is just the {\em Shannon entropy} of the probability distribution comprising the eigenvalues of $\hr{A}$, i.e., $S(\hr{A}) = -\sum_j \lambda_j \,{\rm log}_2 (\lambda_j) $. 




Having introduced the notion of states, we next consider the important concept of observables of a quantum system. \\

\noindent
{\bf Observables}\,:
Observables are physical variables of the system that are measurable. Observables $\hat{O}$ of a quantum system A are defined as   hermitian operators acting on ${\cal H}_A$.  Let the spectral resolution of a nondegenerate observable $\hat{O}$ be $\hat{O}= \sum_j \lambda_j  P_j$, where $\lambda_j$'s are the eigenvalues corresponding to the eigenvectors $P_j$. 
When $\hat{O}$ is measured in the state $\hr{A}$, the $j^{\rm th}$ outcome corresponding to measurement operator (one-dimensional projection) $P_j$ occurs with probability $p_j = {\rm Tr}(\hr{A} \,P_j)$. \\

\noindent
One obtains a more general measurement scheme called POVM (positive operator-valued measure) when the projective measurement elements $P_j$ are replaced by positive operators $\Pi_j$ with $\sum_j \Pi_j=1\!\!1$, and the probabilities of outcomes are obtained in a similar manner\,: $p_j = {\rm Tr}(\hr{A}\, \Pi_j)$. \\


\subsection{Composite systems} 
We next consider the case of composite systems. A composite system is one that has two (bipartite) or more  (multipartite) subsystems .  For our purpose, it suffices to concern ourselves with bipartite systems. 

The Hilbert space of a composite system is given by the tensor product of those of the individual subsystems. In other words ${\cal H}_{AB} = {\cal H}_A \otimes {\cal H}_B$. The state space of the composite system is denoted by $\Lambda({\cal H}_A \otimes {\cal H}_B)$. 

Let $\{|e_i\rangle \}_1^n$ and $\{|f_j\rangle \}_1^m$ be respective ONB in Hilbert spaces of subsystems $A$ and $B$. Then the collection of $mn$ vectors $\{ |e_i\rangle \otimes |f_j\rangle \}$ forms a basis in ${\cal H}_{AB}$. A product operator $A \otimes B$, with the matrix elements of $A$ given as
\begin{align}
A = \left[\begin{array}{cccc}
a_{11} & a_{12} & \cdots & a_{1n}\\
a_{21} & a_{22}&\cdots&a_{2n}\\
\vdots&&&\\
a_{n1} & a_{n2} & \cdots & a_{nn}
\end{array}
\right],
\label{i12a}
\end{align}
can be written as 
\begin{align}
A \otimes B = \left[\begin{array}{cccc}
a_{11} B & a_{12} B & \cdots & a_{1n} B \\
a_{21} B  & a_{22} B &\cdots&a_{2n} B \\
\vdots&&&\\
a_{n1} B & a_{n2} B & \cdots & a_{nn} B
\end{array}
\right].
\label{i12b}
\end{align}

A generic density operator of the composite system can be written as 
\begin{align}
\rho_{AB} = \left[\begin{array}{cccc}
A_{11} & A_{12} & \cdots & A_{1n}\\
A_{21} & A_{22}&\cdots&A_{2n}\\
\vdots&&&\\
A_{n1} & A_{n2} & \cdots & A_{nn}
\end{array}
\right].
\label{i13}
\end{align}
Here each matrix block $((A_{ij}))$ is a $m \times m$ matrix. 
The density matrix of subsystem $A$ is obtained through performing {\em partial trace} on subsystem $B$, i.e.,
\begin{align}
\hr{A} &= {\rm Tr}_B\,(\hr{AB})\nonumber\\
&= \sum_j \langle f^B_j |\hr{AB}| f^B_j \rangle,
\label{i14}
\end{align}
where $|f_j^B\rangle,\,j=1,2,\cdots,m$ is any ONB in ${\cal H}_B$. In terms of the matrix entries in Eq.\,\eqref{i13} we have
\begin{align}
\rho_A =  \left[\begin{array}{cccc}
{\rm Tr}\,A_{11} & {\rm Tr}\,A_{12} & \cdots & {\rm Tr}\,A_{1n}\\
{\rm Tr}\,A_{21} & {\rm Tr}\,A_{22}&\cdots&{\rm Tr}\,A_{2n}\\
\vdots&&&\\
{\rm Tr}\,A_{n1} & {\rm Tr}\,A_{n2} & \cdots & {\rm Tr}\,A_{nn}
\end{array}
\right].
\label{i15}
\end{align}
If instead the partial trace was performed over subsystem $A$ we have
\begin{align}
\rho_B = \sum_i A_{ii}.
\label{i16}
\end{align}

Another useful operation is the {\em partial transpose} operation. Performing the partial transpose on subsystem $B$ on matrix $\rho_{AB}$ in Eq.\,\eqref{i13}, we have
\begin{align}
\rho_{AB}^{T_B} =  \left[\begin{array}{cccc}
A_{11}^T & A^T_{12} & \cdots & A^T_{1n}\\
A^T_{21} &A^T_{22} &\cdots&A_{2n}^T\\
\vdots&&&\\
A^T_{n1} & A^T_{n2} & \cdots & A^T_{nn}
\end{array}
\right].
\label{i17}
\end{align}
We see that the transpose operation was performed on each of the sub-blocks of the composite state. If the transpose operation was performed on subsystem $A$ we have by Eq.\,\eqref{i13}
\begin{align}
\rho_{AB}^{T_A} =  \left[\begin{array}{cccc}
A_{11} & A_{21} & \cdots & A_{n1}\\
A_{12} & A_{22}&\cdots&A_{n2}\\
\vdots&&&\\
A_{1n} & A_{2n} & \cdots & A_{nn}
\end{array}
\right].
\label{i18}
\end{align}\\

Having introduced the notion of composite systems, we briefly discuss the connection between POVM's, projective measurements and composite systems. \\

\noindent
{\bf POVM}\,:
We have seen earlier that a POVM is a measurement scheme where the measurement elements $\Pi = \{\Pi_j \}$ are positive operators rather than projections. We now give a simple example in which a POVM results from a  projective measurement on a larger system. 

Consider a state $\hr{AB}$ of composite system of the form
\begin{align}
\hr{AB} = \hr{A} \otimes \proj{\psi_B}.
\label{i19}
\end{align}
Let $P = \{ P^{\,i}_{AB}\}$ be a collection of (one-dimensional) projection operators on the composite system $AB$ which is complete\,: $\sum_i P^i_{AB}=1\!\!1$. The probability of the result of the $i^{\rm th}$ measurement is given by
\begin{align}
p_i &= {\rm Tr} (\hr{AB}\,P^{\,i}_{AB}) \nonumber\\
&= {\rm Tr} (\hr{A} \otimes \proj{\psi_B}\,P^{\,i}_{AB})\nonumber\\
&= {\rm Tr}_A (\hr{A} \, \langle \psi_B|\,P^{\,i}_{AB}\,|\psi_B\rangle). 
\label{i20}
\end{align}
If we define $\Pi^A_i =  \langle \psi_B|\,P^{\,i}_{AB}\,|\psi_B\rangle$, we can write the above equation in a more suggestive form
\begin{align}
p_i = {\rm Tr}_A \, (\Pi^A_i \, \hr{A}). 
\label{i21}
\end{align}
The operators $\{\Pi^A_i \}$ are all positive and sum to identity on ${\cal H}_A$. Hence the set $\Pi = \{\Pi_i^A \}$ constitutes a POVM. We thus see  how a POVM results from the projective measurement on a larger system. 

The following theorem guarantees that there always exists a physical mechanism by which one can realise any given POVM\,\cite{peres90}. 
\begin{theorem}[Neumark]:
One can extend the Hilbert space ${\cal H}$ on which the POVM elements $\{\Pi_{j} \}$ act, in such a way that there exists in the extended space ${\cal K}$, a set of orthogonal projectors $\{P_j\}$ with $\sum_j P_j =1\!\!1_{\cal K}$, and such that $\Pi_j$ is the result of projecting $P_j$ from ${\cal K}$ to ${\cal H}$. 
\end{theorem}

Having collected some basic ideas relating to bipartite states, we next consider an important aspect of these bipartite states that we are interested in, which is correlation between the subsystems. 

\section{Correlations}
Correlations are intrinsic (nonlocal) properties of composite systems. Of the various quantifiers of correlations, we  first consider entanglement. A bipartite pure state $|\psi\rangle \in$ ${\cal H}_A \otimes {\cal H}_B$ is not entangled if and only if it is of the product form 
\begin{align}
|\psi\rangle = |u \rangle \otimes |v\rangle.
\label{i22}
\end{align}
A useful representation of pure bipartite states is the {\em Schmidt representation}\,\cite{nielsenbook}. The Schmidt form makes use of the  {\em singular value decomposition} theorem\,\cite{hornbook}\,: 
\begin{theorem}
An arbitrary complex $m \times n$ matrix $A$ of rank $k$ can be written in the form $A = V D W^{\dagger}$, where $V_{m \times m}$ and $W_{n \times n}$ are unitary, and $D$ diagonal (with $k$  entries which are positive and the rest zero). The non-zero diagonal entries of $D$ are the square roots of the eigenvalues of $AA^{\dagger} \sim A^{\dagger} A$. 
\end{theorem}

Let us write down a general pure state of system $AB$ as
\begin{align}
|\psi\rangle = \sum_{ij} c_{ij}\, |i\rangle \otimes |j\rangle,
\label{i23}
\end{align}
where $c_{ij}$ is a general complex (coefficient) matrix, where $\{|i\rangle \}$ and  $\{|j\rangle \}$ are the computational basis of systems $A$ and $B$ respectively. By applying suitable local unitaries, one can diagonalize any coefficient matrix $((c_{ij}))$ to bring it to a diagonal form as guaranteed by the singular value decomposition. We then have 
\begin{align}
|\psi\rangle = \sum^r_{j=1} \lambda_j\, |e_j^A\rangle \otimes |f_j^B\rangle,
\label{i24}
\end{align}
where $\{e_j \}$ and $\{f_j \}$ are orthonormal in ${\cal H}_A$, ${\cal H}_B$ respectively,  and the coefficients $\{\lambda_j \}$ are positive and $\sum_j \lambda_j^2=1$, as follows from the normalization condition. The number of terms in the above decomposition is called the {\em Schmidt rank} and can utmost take the value $r = {\rm min} (d_A,d_B)$. 
Further, if the Schmidt rank $r$ is one, then the pure state is a product state and therefore separable. If $r >1$, then the state $|\psi\rangle$ is entangled. 

A closely related and useful concept  is {\em purification}. Purification is an association of a generic mixed state of a system $A$ with a pure entangled state of a suitable composite system $AR$. To this end, 
let $\hr{A} = \sum_{j=1}^r \lambda_j\, |\psi_j\rangle \langle \psi_j|$ be the spectral decomposition of a mixed state $\hr{A}$. Let us append this system $A$ with a system $R$ with Hilbert space ${\cal H}_R$ of dimension equal to the  rank $r$ of $\hr{A}$. Let $\{|e\rangle^j_R\}_{1}^r$ be  an ONB for system $R$. Then  starting from a pure state written as
\begin{align}
|\psi\rangle_{AR} = \sum_j \sqrt{\lambda_j}\,|\psi_j \rangle_R \otimes |e^j\rangle_R,
\label{i25}
\end{align}
one obtains $\hr{A} = {\rm Tr}_R[\proj{\psi}_{AR}]$,  and  $|\psi\rangle_{AR}$ is called a purification of $\hr{A}$. In other words, it is seen that one recovers $\hr{A}$ by taking partial trace on subsystem $R$ of the pure state $|\psi\rangle_{AR}$. The purification $|\psi\rangle_{AR}$ has a local unitary freedom in system $R$ in the following sense. Any other choice of an ONB of system R also returns the same  state $\hr{A}$ under partial trace\,\cite{nielsenbook}. \\

Having considered separable and entangled pure states, a natural question would be to {\em quantify} the amount of entanglement in a pure state. 
A simple measure for the quantification of entanglement of bipartite pure states\,\cite{donald02} is given by the entropy of the reduced state, i.e.
\begin{align}
E(|\psi\rangle_{AB}) = S(\hr{A}), 
\label{i26}
\end{align}
where $\hr{A} = {\rm Tr}_B (|\psi\rangle_{AB} \langle \psi|)$ is obtained by partial trace over subsystem $B$. We see that for product pure states, the reduced state is a pure state and hence has zero entanglement. We note in passing that the entanglement measure in Eq.\,\eqref{i26} is symmetric with respect to the two subsystems in the following sense. One could have taken partial trace of system A instead of B in Eq.\,\eqref{i26}. Since $S(\hr{A}) = S(\hr{B})$, as can be easily seen from the Schmidt decomposition, the amount of entanglement is the same in either procedure.

We now wish to consider mixed states of the bipartite system $AB$.
\noindent
A {\em separable  mixed} state is one which can be written as\,\cite{werner89} 
\begin{align}
\hr{AB} = \sum_j p_j\, \hr{A}^{\,j} \otimes \hr{B}^{\,j},
\label{i27}
\end{align}
i.e., the bipartite density matrix can be written as a convex combination of product density matrices. If such a decomposition does not exist, then the state is said to be entangled.

\subsection{Entanglement detection}
The problem of studying entanglement for mixed states\,\cite{bennett96} turns out to be a non-trivial one. There have been many approaches to detect and quantify the entanglement\,\cite{terhal02,bruss02}. We now briefly review a few of the measures used often in the literature, and a few of these provide an operational method to detect  entanglement\,\cite{breuer06}. \\

\noindent
{\bf Partial transpose test}\,: 
Given a bipartite state $\hr{AB} $ whose matrix elements are written as
\begin{align}
\hr{AB} = \sum \rho_{jk;mn}\, |j\rangle \langle m| \otimes |k \rangle \langle n |,
\label{i28}
\end{align}
the partial transpose with respect to subsystem $B$ leads to
\begin{align}
\hr{AB}^{T_B} = \sum \rho_{jk;mn} \,|j\rangle \langle m| \otimes |n \rangle \langle k |,
\label{i29}
\end{align}
and that on subsystem $A$ gives
\begin{align}
\hr{AB}^{T_A} = \sum \rho_{jk;mn} \,|m\rangle \langle j| \otimes |k \rangle \langle n |.
\label{i30}
\end{align}
For $2\times 2$ and $2\times 3$ systems the lack of positivity of the state resulting from partial transpose provides a necessary and sufficient test to detect entanglement\,\cite{peres96,horodecki96}. If, say, $\hr{AB}^{T_B}$ is positive, then the state is separable; else it is entangled. The test however fails for higher dimensional systems as positivity under partial transpose (PPT) is not a sufficient condition for separability. In higher dimensions, if a state fails the partial transpose test, then it is entangled. But PPT  is not a sufficient condition for separability, and there exist entangled states which are PPT\,\cite{horodecki97,horodecki98}. \\

\noindent 
{\bf Positive maps}\,: A positive map $\Gamma$ is a linear map  on the space of bounded linear operators on a given Hilbert space which takes positive operators to positive operators. i.e.,
\begin{align}
\Gamma : {\cal B}({\cal H}_S) \to {\cal B}({\cal H}_S),~~  \Gamma (A) = A^{\,'} \geq 0 ~\forall~ A \geq 0.
\label{i31}
\end{align} 
It is immediately seen that the one-sided action of a positive map on a separable state takes it to a density operator. The necessary and sufficient condition for a state $\hr{AB}$ to be separable is that $[1\!\!1 \otimes \Gamma](\hr{AB}) \geq 0$ for all positive  maps $\Gamma$\,\cite{horodecki96}. It turns out that there is a very important subset of positive maps known as completely positive maps. The notion of completely positive maps will be discussed in a later Section. For detecting entanglement, it is positive maps that are not completely positive that are useful. The transpose map (used in the partial transpose test) considered above is an example of a positive map that is not completely positive.\\

\noindent
{\bf Entanglement witness}\,: A self-adjoint bipartite operator $W$ which has at least one negative eigenvalue and has nonnegative expectation on product states is called an entanglement witness\,\cite{horodecki96,terhal00b}. A state $\hr{AB}$ belongs to the set of separable states if it has a nonnegative mean value for all $W$, i.e.
\begin{align}
{\rm Tr}\,(W \,\hr{AB}) \geq 0 ~~\forall~~ W,
\label{i32}
\end{align}
where $W$ is an entanglement witness. For every entangled state $\hr{AB}$, there exist an entanglement witness $W$ such that ${\rm Tr}\,(W \,\hr{AB})) < 0$. We then say that the entanglement of $\hr{AB}$ is  witnessed by $W$.\\

\noindent
{\bf Reduction criterion}\,: Consider the following map that is known as the reduction map\,: 
\begin{align}
\Gamma(\hr{A}) = \frac{(1\!\!1 {\rm Tr}(\hr{A}) - \hr{A})}{d_A-1}.
\label{i32b}
\end{align}
The reduction separability criterion\,\cite{horodecki99,cerf99} states that a necessary condition for a state to be separable is that it satisfies 
\begin{align}
& ~~~~~[1\!\!1 \otimes \Gamma_B][\hr{AB}] \geq 0\nonumber\\
\Longrightarrow &~~~~~\hr{A} \otimes 1\!\!1 - \hr{AB} \geq 0.
\label{i33}
\end{align} 
The reduction criterion is weaker than the partial transpose test\,\cite{horodecki99}. \\

\noindent
{\bf Range criterion}\,: The range criterion\,\cite{horodecki97prl} states that if a state $\hr{AB}$ is separable then there exists a set of product vectors $\{|\psi_i^A \rangle \otimes |\phi_i^B\rangle\}$ such that it spans the range of $\hr{AB}$ and the set $\{|\psi_i^A \rangle \otimes |(\phi_i^B)^*\rangle\}$ spans that of $\hr{AB}^{T_B}$, where the complex conjugation is done in the same basis in which the partial transpose operation is performed. \\

\noindent
{\bf Unextendable product basis}\,: An unextendable product basis (UPB)\,\cite{upb} is a set $S_{u}$ of orthonormal product vectors such that there is no product vector that is orthogonal to all of them. Let us denote by $S_u^{\perp}$ the subspace that is orthogonal to the subspace spanned by vectors in $S_u$. Therefore, any vector in $S_u^{\perp}$ is entangled. By the range criterion, we have that any mixed state with support on this orthogonal space $S_u^{\perp}$ is entangled. Using this concept of UPB, one can construct entangled states that are PPT. \\

\noindent
{\bf 1-Entropic type}\,:
There is an entropic way to quantify the statement that `an entangled state gives more information about the total system than about the subsystems'. Indeed it was shown that the entropy of a subsystem can be greater that the entropy of the total system only when the state is entangled\,\cite{horodecki96b,horodecki96c}. In other words, for a separable state 
\begin{align}
S(\hr{A}) \leq S(\hr{AB}), \text{ and } S(\hr{B}) \leq S(\hr{AB}).
\label{i34}
\end{align}

\noindent
{\bf Majorization criterion}\,: 
A vector $\mathbf{x}$ is said to be majorized by $\mathbf{y}$\,\cite{majorbook}, denoted by $\mathbf{x} \prec \mathbf{y}$, both of dimension $d$, if 
\begin{align}
\sum_{j=1}^k x_j &\leq \sum_{j=1}^k y_j, ~~ \text{ for } ~~ k = 1, \cdots, d-1;\nonumber\\
\sum_{j=1}^d x_j &= \sum_{j=1}^d y_j,
\end{align}
it being assumed that the components are arranged in decreasing order. 
The majorization criterion states that if a  state $\hr{AB}$ is separable\,\cite{nielsen01}, then 
\begin{align}
\boldsymbol{\lambda}_{AB} \prec \boldsymbol{\lambda}_A \text{ and } \boldsymbol{\lambda}_{AB} \prec \boldsymbol{\lambda}_B,
\label{i35}
\end{align}
where $\boldsymbol{\lambda}_{(\cdot)}$ is the vector of eigenvalues of $\hr{(\cdot)}$ written in decreasing order. Therefore, we say that for a separable state the eigenvalues of the bipartite state is majorized by the ones of either reduced state. \\

\noindent
{\bf Realignment criterion}\,: The realignment map $R$ is defined as 
\begin{align}
[R(\rho_{AB})]_{ij;kl} = [\rho_{AB}]_{ik;jl}.
\label{i35b}
\end{align} 
The realignment criterion\,\cite{chen03} states that if a state $\hr{AB}$ is separable, then $||R(\hr{AB})||_1 \leq 1$. \\

\noindent
{\bf Bell-type}\,:
A Bell-type\,\cite{bell64} inequality is one which tries to capture entanglement through probabilities of outcomes of suitably chosen observables. An example of one such inequality in the two-qubit setting is the CHSH inequality\,\cite{chsh}\,:
\begin{align}
&|{\rm Tr}(\hat{O}_{CHSH} \, \hr{AB})| \leq 2,\nonumber\\
&\hat{O}_{CHSH} = A_1 \otimes (B_1 + B_2) + A_2 \otimes (B_1 - B_2).
\label{i36}
\end{align}
Here, $A_1$ and $A_2$ are respectively $\mathbf{a_1}.\boldsymbol{\sigma}$ and $\mathbf{a_2}.\boldsymbol{\sigma}$, $\mathbf{a}_1,\,\mathbf{a}_2 \in {\cal R}^3$. One similarly constructs operators $B_1,\,B_2$ with respect to two directions $\mathbf{b_1},\,\mathbf{b_2}$. Any state $\hr{AB}$ that violates this inequality is an entangled state. 

\subsection{Entanglement quantification}
In the previous Section we summarized a few ways to detect entanglement. We now briefly describe some measures of entanglement. We first begin by listing reasonable properties that any entanglement measure $E$ would be expected to satisfy\,\cite{vedral97,vedral98,horodecki00,vidal00}. 

\begin{itemize}
\item $E(\hr{AB})=0$ for $\hr{AB}$ separable.
\item $E$ is invariant under local unitary transformations: 
\begin{align}
E(U_{A}\otimes U_{B}\,\hr{AB}\,U_{A}^{\dagger}\otimes U_{B}^{\dagger})= E(\hr{AB}).
\label{i37}
\end{align}
\item $E$ is non-increasing under local operations and classical communications (LOCC): 
\begin{align}
E(\Gamma(\hr{AB})) \leq E(\hr{AB}), \text{ for } \Gamma \in \,{\rm LOCC}.  
\label{i38}
\end{align}
\item $E$ returns the value of von-Neumann entropy of the reduced states when evaluated on pure bipartite states, i.e., 
\begin{align}
E(|\psi_{AB}\rangle) = S({\rm Tr}_B(|\psi\rangle_{AB} \langle \psi|)).
\label{i39}
\end{align}
\item $E$ is subadditive over a general product of bipartite entangled  states, i.e., 
\begin{align}
E(\hr{AB} \otimes \hr{A^{\,'}B^{\,'}}) \leq  E(\hr{AB}) + E(\hr{A^{\,'}B^{\,'}}).
\label{i41}
\end{align} 
\item {\em Normalization}\,: $E(\hr{\rm max}) = {\rm Log}_2 d$ for a maximally entangled state $\hr{\rm max}$.
\end{itemize}
There are in addition some technical requirements that are also considered\,:
\begin{itemize}
\item $E$ is a convex function on the state space.
\item $E$ is continuous on the state space\,\cite{nielsen00}.
\end{itemize}
It is known that the measures to be considered below do not necessarily satisfy all the above mentioned properties, but are nevertheless used depending on the context. \\

\noindent
{\bf Entanglement of formation}\,: The expression for entanglement of formation is given by\,\cite{bennett96} 
\begin{align}
E_{F}(\hr{AB})  = \underset{\{p_j,|\psi^j\rangle\}}{{\rm min}} \sum_j \,p_j\, S({\rm Tr}_B(|\psi^j\rangle_{AB}\langle \psi^j|)),
\label{i42}
\end{align}
where $\hr{AB}=\sum_j p_j\,|\psi^j\rangle\langle \psi^j|$ is a pure state ensemble, and the optimization is over all possible convex pure state decompositions of the original bipartite mixed state $\hr{AB}$. 

This optimization has been solved analytically for very
few examples. These include two-qubit states\,\cite{wootters97,wootters98}, symmetric Gaussian states\,\cite{mm1}, general Gaussian states\,\cite{simoneof}, Gaussian entanglement of formation\,\cite{mm2}, werner states and O-O states\,\cite{voll01}, isotropic states\,\cite{terhal00}, some highly symmetric states\,\cite{cirac02}, flower states\,\cite{horo05}, examples in $16\times 16$ systems\,\cite{fei03}; special classes of states using the Koashi-Winter relation in two-qubit states\,\cite{saba13} and tripartite Gaussian states\,\cite{add1}; special examples using the Matsumoto-Shimono-Winter relation\,\cite{mswcmp}.   \\


\noindent
{\bf Entanglement cost}\,: The entanglement cost\,\cite{audenaert03,vidal02,brandao07} is defined as the asymptotic or regularized version of entanglement of formation\,\cite{hayden01}. In other words,
\begin{align}
E_C(\hr{AB}) = \underset{n \to \infty}{{\rm lim}} \frac{E_{F}(\hr{AB}^{\otimes n})}{n}.
\label{i43}
\end{align}
The entanglement cost has been evaluated for $3 \times 3$ anti-symmetric states\,\cite{yura03}, lower bounds for d-dimensional anti-symmetric states were obtained in\,\cite{christandl}, certain antisymmetric states with a non-identical bipartite separation\,\cite{matsumoto04}, examples of highly symmetric states\,\cite{cirac02}, and flower states\,\cite{horo05}.  \\


\noindent
{\bf Distillable entanglement}\,: The distillable entanglement\,\cite{rains99a,rains99b,lewenstein00,rains01} is a measure of how much entanglement can be extracted from an entangled state $\hr{AB}$\,\cite{bennett96a,bennett96b} in an asymptotic setting, i.e., 
\begin{align}
E_D(\hr{AB}) = {\rm sup} \left\{r: \underset{n \to \infty}{{\rm lim}} \left( \underset{\Gamma}{{\rm inf}} ||\Gamma(\hr{}^{\otimes n})- \Phi^+_{2^{ rn}} ||_1 \right)=0 \right\},
\label{i44}
\end{align}
where $\Gamma  $ is an LOCC operation, and $\Phi^+_{2^x}$ stands for $(\Phi^+)_2^{\otimes x}$, $\Phi^+_2$ being a Bell state. Here, $||A||_1$ stands for $\sum_{j} {\lambda}_j$, where $\lambda_j$'s are the singular values of $A$.   
Further, it is known that $E_D(\hr{AB}) \leq E_F(\hr{AB}) \leq E_C(\hr{AB})$, i.e., distillable entanglement is a lower bound for entanglement of formation\,\cite{bruss02}.\\

\noindent
{\bf Relative entropy of entanglement}\,:  The relative entropy\,\cite{audenaert01,vedral02,donald99} is just the `distance' of a given state $\hr{AB}$ to the closest separable state, i.e.
\begin{align}
E_{R} (\hr{AB}) = \underset{\sigma \in {\cal S}}{{\rm inf}}~~ S(\hr{AB}||\hat{\sigma}),
\label{i45}
\end{align}
where $\hat{\sigma}$ is an element of the set of separable states ${\cal S}$ and 
\begin{align}
S(\hat{a}||\hat{b}) = {\rm Tr}[\hat{a}\, ({\rm Log}\hat{a} - {\rm Log 
\hat{b}})],
\label{i46}
\end{align}
is known as the relative entropy between states $\hat{a},\,\hat{b}$. \\

\noindent
{\bf Squashed entanglement}\,: The expression for squashed entanglement of $\hr{AB}$ is given by\,\cite{christandl04} 
\begin{align}
E_{\rm sq}(\hr{AB})= \underset{\hr{ABE}}{{\rm inf}} \,\frac{1}{2} I(A:B|E), 
\label{i47}
\end{align}
where $I(A:B|E)= S_{AE} +S_{BE} - S_E - S_{ABE}$, $S_{X}$ denotes the  entropy of the state of system $X$ and the infimum is taken over all density matrices $\hr{ABE}$ such that $\hr{AB}= {\rm Tr}_E(\hr{ABE})$. $E_{\rm sq}$ enjoys many interesting properties like additivity over tensor products and superadditivity in general. $E_{\rm sq}$ is a lower bound of entanglement of formation and an upper bound on distillable entanglement\,\cite{christandl04}. \\

\noindent
{\bf Logarithmic negativity}\,: Logarithmic negativity is a straight-forward computable measure of entanglement\,\cite{werner02} often used in the literature. It is defined as the  logarithm of the sum of moduli of the eigenvalues of the partial transpose of a given bipartite state. The expression for logarithmic negativity is given by 
\begin{align}
E_N(\hr{AB}) = {\rm Log}_2 \left[ || \hr{AB}^{{\rm Tr}_B}||_1 \right]. 
\label{i48}
\end{align}

\subsection{Quantum discord, classical correlation and mutual information}
We now move on to a different set of correlations that are motivated more directly from a measurement perspective and capture a different sort of classical-quantum boundary as opposed to the separable-entangled boundary. Among these measurement-based correlations, the ones of primary interest to us are three closely related quantities namely classical correlation, quantum discord and mutual information. 

We may motivate the definition of quantum discord by first looking at the classical setting\,\cite{zurek01}. Given a probability distribution $p(x,y)$ in two variables, the mutual information $I(x,y)$ is defined as   
\begin{align}
I(x,y) = H(x) - H(x|y), 
\label{i49}
\end{align}
where $H(\cdot)$ stands for the Shannon entropy
\begin{align}
H(x)= -\sum_x p(x)\, {\rm Log}[p(x)]
\label{i50}
\end{align}
and $H(x|y)$ is the conditional entropy. Using Bayes rule we are lead to an equivalent expression for mutual information\,:
\begin{align}
I(x,y) = H(x) + H(y) - H(x,y). 
\label{i51}
\end{align}
This second expression in \eqref{i51} for mutual information naturally generalises to the quantum setting when the bipartite probability distribution is replaced by a bipartite state $\hat{\rho}_{AB}$ and the Shannon entropy $H(\cdot)$  by the von Neumann entropy $S(\cdot)$ of quantum states, and we have 
\begin{align}
I(\hat{\rho}_{AB}) = S(\hat{\rho}_A) + S(\hat{\rho}_B) - S(\hat{\rho}_{AB}).
\label{i52}
\end{align}
But the first expression \eqref{i49} for classical mutual information does {\em not} possess a straightforward generalization to the quantum case. In the quantum case, the conditional entropy is defined with respect to a measurement, where the measurement is performed on one of the subsystems, say subsystem B. Let us consider a POVM $\Pi^B = \{\Pi^B_j\}$ where $\Pi^B_j \geq 0$ and $\sum_j \Pi^B_j= 1\!\!1$. Then the (average) conditional entropy post measurement is given by 
\begin{align}
S^A = \sum_j \, p_j \,S(\hat{\rho}_j^A), 
\label{i53}
\end{align}
where the probabilities and  states post measurement are given by
\begin{align}
p_j &= {\rm Tr}_{AB}\,(\Pi_j^B \, \hat{\rho}_{AB}), \nonumber\\ 
\hat{\rho}_j^A &= p_j^{-1}\, {\rm Tr}_B(\Pi_j^B \, \hat{\rho}_{AB}).
\label{i54}
\end{align}
Let us denote by $S^A_{\rm min}$ the minimum of $S^A$ over all measurements or POVM's, i.e.,
\begin{align}
S^A_{\rm min} = \underset{\Pi}{{\rm min}} \sum_{j} p_j\, S(\hat{\rho}^A_j).
\label{samin}
\end{align} 
The difference between these two classically equivalent expressions (optimized over all measurements) is called quantum discord ${\cal D}(\hat{\rho}_{AB})$ and is given by the expression\,:
\begin{align}
{\cal D}(\hat{\rho}_{AB}) &=    I(\hat{\rho}_{AB}) - \left[ S(\hat{\rho}_A) - S^A_{\rm min} \right],\nonumber\\
&=  S(\hat{\rho}_B) - S(\hat{\rho}_{AB}) + S^A_{\rm min}.
\label{i55}
\end{align} 
The quantity in the square brackets above is defined as the classical correlation and is denoted by
\begin{align}
C(\hr{AB}) = S(\hr{A}) - S^A_{\rm min}.
\label{i56}
\end{align}
It is useful to keep in mind an alternate expression for mutual information which is given by
\begin{align}
I(\hr{AB}) = S(\hr{AB}||\hr{A} \otimes \hr{B}),
\label{i57}
\end{align}
where $S(\cdot||\cdot)$ is the relative entropy. We see that the mutual information is defined as the relative entropy between the given bipartite state and the (tensor) product  of its reductions. 
Thus, the mutual information which is supposed to capture the total correlation of a bipartite state is broken down into quantum discord ${\cal D}(\hat{\rho}_{AB})$, that captures the quantum correlations, and classical correlation $C(\hat{\rho}_{AB})$\,:
\begin{align}
I(\hr{AB}) = {\cal D}(\hr{AB}) +  C(\hr{AB})
\label{i57b}
\end{align}

There are many properties that are satisfied by quantum discord and classical correlations and we list some of them below\,\cite{modi-rmp}\,:

\begin{itemize}
\item Quantum discord and classical correlation are dependent on the subsystem on which the measurement is performed. Hence, they are not symmetric under exchange of the subsystems in general. 

\item Both classical correlation and quantum discord are non-negative quantities.

\item ${\cal D}(\hr{AB})$, $C(\hr{AB})$, and $I(\hr{AB})$ are invariant under local unitary transformations. This turns out to be useful for the computation of these quantities. 

\item $C(\hr{AB})=0$ only for a product state. 

\item $C(\hr{AB}) = E_F(\hr{AB})$ for any pure bipartite state $\hr{AB} = |\psi\rangle\langle \psi|$. 

\item $D(\hr{AB}) = E_F(\hr{AB})$ for any pure bipartite state $\hr{AB} = |\psi\rangle\langle \psi|$. In other words, both classical correlation and quantum discord reduce to the entanglement on pure bipartite states.

\item A state has vanishing quantum discord when 
\begin{align}
{\cal D}(\hr{AB}) =0 ~~\text{ or } ~~I(\hr{AB}) = C(\hr{AB}). 
\label{i58}
\end{align}
Further, any one-way zero discord state can be written as 
\begin{align}
\hr{AB} = \sum_i p_i\, \proj{i} \otimes \hr{b}^i,
\label{i59}
\end{align}
where $\{p_j\}$'s form a probability distribution and $\{|i\rangle \}$ is an orthonormal basis in the subsystem where the measurement was performed. In other words, a one-way zero discord state is invariant under some von-Neumann measurement on the subsystem. Such states are also known as classical-quantum states. 
\end{itemize}

As a simple example, we compute all the three correlations for a two-qubit Bell-state 
\begin{align}
|\psi^+\rangle = (|00\rangle + |11\rangle)/\sqrt{2}.
\label{i59b}
\end{align} 
For this pure state we have 
$I(|\psi\rangle) = 2{\rm Log}_2 2 =2$ bits. While $C(|\psi\rangle) = D(|\psi\rangle) = E(|\psi\rangle)= 1$ bit, illustrating $I(|\psi\rangle) = C(|\psi\rangle) + D(|\psi\rangle)$. 


\section{Positive maps and completely positive maps}
Having briefly considered correlations, we  now turn to another aspect of central importance to this thesis which is channels. 
We wish to know what are all the allowed physical evolutions of a given quantum system. If a system is isolated, then its dynamics is governed by the unitary Schr\"{o}dinger evolution. A unitary evolution $U$ effects the following transformation 
\begin{align}
\hr{A} \to \hr{A}^{\,'} = U \,\hr{A}\,U^{\dagger}, ~\hr{A} \text{ and } \hr{A}^{\,'} \in \Lambda({\cal H}_A). 
\label{i60}
\end{align}
But if the system is in interaction with its environment, then the evolutions of the system of interest resulting from unitary evolutions of the composite are more general, but nevertheless described by linear maps acting on the state space $\Lambda({\cal H}_A)$ directly rather than through its action on ${\cal H}_A$. 

Let $\Phi$ be a linear map that acts on states of the system, i.e., $\Phi\,: \Lambda({\cal H}_A) \to \Lambda({\cal H}_A) $. Writing this transformation in terms of the matrix elements, we have (dropping the system label $A$)\,:
\begin{align}
\rho \to \rho^{\,'} \,: ~~\Phi_{ij;k\ell} \, \rho_{k\ell} = \rho^{\,'}_{ij}.
\label{i61}
\end{align}
An obvious necessary requirement for $\Phi$ to be a valid evolution is that it takes states to states. So we require that the action of $\Phi$ preserve hermiticity, trace and positivity of the states. For the hermiticity of the output states we have that
\begin{align}
 \rho^{\,'}_{ij} = (\rho^{\,'}_{ji})^*,\nonumber\\
\text{i.e.,}~~ \Phi_{ij;\ell k}  = \Phi_{ji;k\ell}^*.
\label{i62}
\end{align}
The trace preserving condition manifests as 
\begin{align}
\sum_{i} \Phi_{ii;k\ell} = \delta_{k \ell}.
\label{i63}
\end{align}
Finally, the positivity condition can be written as 
\begin{align}
v_i^*\, \rho_{ij}^{\,'}\, v_j \geq 0 ~~ \forall ~ |v\rangle.
\end{align}
This property can be checked by assuming that $\rho = \proj{u}$, a pure state as input. We have 
\begin{align}
v_i^*\,\Phi_{ij;k\ell} \, \rho_{k\ell} \, v_j &\geq 0, \text{ for every } |u \rangle,\nonumber\\
\text{i.e., } ~   v_i^*\,\Phi_{ij;k\ell} \, u_k u^*_{\ell} \, v_j &\geq 0, \text{ for every } |u\rangle, |v\rangle.
\label{i64}
\end{align} 
It is instructive to write the matrix elements of $\Phi$ in terms of a new matrix we denote $\widetilde{\Phi}$,
\begin{align}
\widetilde{\Phi}_{ij ; k \ell} = {\Phi}_{ik;j \ell}. 
\label{i65}
\end{align}
The hermiticity preserving condition in Eq.\,\eqref{i62} now reads
\begin{align}
\widetilde{\Phi}_{i\ell;jk}  = \widetilde{\Phi}_{jk;i \ell}^*
\label{i66}
\end{align}
In other words, for the map $\Phi$ to be hermiticity preserving, we have that $\widetilde{\Phi}$ to be a hermitian matrix\,:
\begin{align}
\widetilde{\Phi} = \widetilde{\Phi}^{\dagger}.
\label{i67}
\end{align}
The trace condition in Eq.\,\eqref{i63} reads
\begin{align}
\sum_{i} \widetilde{\Phi}_{ik;i\ell} = \delta_{k \ell}.
\label{i68}
\end{align}
Finally, the positivity condition  of Eq.\,\eqref{i64} is then
\begin{align}
v_i^*\,\widetilde{\Phi}_{ik;j\ell} \, u_k u^*_{\ell} \, v_j &\geq 0,\nonumber\\
\Rightarrow \langle \psi | \widetilde{\Phi} | \psi \rangle &\geq 0, ~\forall~ |\psi\rangle = |u \rangle \otimes |v \rangle.
\label{i69}
\end{align}
We note that the positivity condition states that $\widetilde{\Phi}$ is positive over all {\em product} vectors. 

To summarize, we call a map that satisfies all the above conditions as a {\em positive map}, i.e.,
\begin{align}
\Phi \text{ is positive} ~\Leftrightarrow ~&\Phi(\rho_S)= \rho^{\,'}_S \in \Lambda({\cal H}_S),\nonumber\\
~ \text{i.e.,}~ &\Phi\left( \Lambda({\cal H}_S)\right) \subset  \Lambda({\cal H}_S).
\label{i70}
\end{align}
It turns out that not all positive maps are physical evolutions.  For positive maps to be physical evolutions, there is a further requirement to be met. 

Let us consider a composite system which consists of the original system  appended with an arbitrary ancilla or reservoir $R$. The Hilbert space of the composite system is ${\cal H}_S \otimes {\cal H}_R$, a tensor product of the individual subsystem Hilbert spaces. Let us denote the state space of this composite system by $\Lambda({\cal H}_S \otimes {\cal H}_R)$. 

To motivate the difference between positive and completely positive maps, we now give an example of a map that is positive but nevertheless nonpositive under local action, the transpose map.
The density operator for the Bell-state $|\psi^+\rangle$ in Eq.\,\eqref{i59b} is 
\begin{align}
\proj{\psi^+} = \frac{1}{2} \left( \begin{array}{cccc}
1 &0&0&1 \\
0&0&0&0 \\
0&0&0&0 \\
1&0&0&1 
\end{array} \right).
\label{i72}
\end{align} 
If we now apply the  transpose map locally on the B  system, by Eq.\,\eqref{i17}, we get 
\begin{align}
(\proj{\psi^+})^{\rm T_B} = \frac{1}{2} \left( \begin{array}{cccc}
1 &0&0&0 \\
0&0&1&0 \\
0&1&0&0 \\
0&0&0&1 
\end{array} \right).
\label{i73}
\end{align} 
We see that the above operator has negative eigenvalues. Therefore, the local action of the transpose map does not give a positive operator of the larger system, even though transpose map by itself is a positive map.

It is both reasonable and necessary to require that local action of $\Phi$ takes states of the joint system also to states for $\Phi$ to be a physical evolution.  In other words 
\begin{align}
&(\Phi \otimes 1\!\!1 )[\rho_{SR}]  = \rho_{SR}^{\,'} \in \Lambda({\cal H}_S \otimes {\cal H}_R),\nonumber\\
~ \text{i.e.,} ~~&[\Phi \otimes 1\!\!1]\left(\Lambda({\cal H}_S \otimes {\cal H}_R) \right) \subset \Lambda({\cal H}_S \otimes {\cal H}_R).
\label{i74}
\end{align} 
A positive map $\Phi$ that satisfies Eq.\,\eqref{i74}, is known as a {\em completely positive} (CP) trace-preserving (TP) map or a {\bf quantum channel}. 

An important class of positive maps from entanglement perspective is the so called {\em decomposable} maps\,\cite{horodecki96}. A positive map $\Phi$ is said to be decomposable if 
\begin{align}
\Phi = \Phi_1 + \Phi_2 \circ T,
\label{i75}
\end{align}
where $\Phi_1$ and $\Phi_2$ are both completely positive maps and $T$ is the transpose map. One application of the decomposability notion is that positive maps that are decomposable are `weaker' than the transpose map in the detection of entanglement. 
In other words, one would like to look for maps that are not decomposable to detect entangled states which are PPT\,\cite{simon06}. 

We have seen above that a completely positive map is positive under local action on the system, appended with a system $R$ of any dimension. It turns out that there is an operational criterion that captures this aspect. 

Consider a composite system whose Hilbert space is given by ${\cal H}_S \otimes {\cal H}_S$, where ${\cal H}_S$ is the system Hilbert space. We denote the maximally entangled state of the composite system by\,:
\begin{align}
|\psi\rangle_{\rm max} = \frac{1}{\sqrt{d}} \sum_{i=1}^{d}|ii\rangle.
\label{i76}
\end{align}
Let us denote $d\, |\psi\rangle \langle \psi|_{\rm max}$ by $\hat{\sigma}$, which is explicitly written down as 
\begin{align}
\hat{\sigma} = \sum_{i,j=1}^{d} |i \rangle \langle j| \otimes |i \rangle \langle j|.
\label{i77}
\end{align}
We then have the following theorem regarding completely positive maps\,\cite{choi75}\,:
\begin{theorem}[Choi]
A positive map $\Phi$ is completely positive if and only if $[\Phi \otimes 1\!\!1] (\hat{\sigma}) \geq 0$. 
\end{theorem}

The matrix $D_{\Phi} = [\Phi \otimes 1\!\!1] (\hat{\sigma})$ is known as the dynamical matrix\,\cite{ecg61}. We now discuss the properties and representation of CP maps in some detail in the following Section.

\section{Representations of CP maps}
There are three well-known representations of completely positive maps. These are the unitary (Stinespring) dilation\,\cite{stinespring}, the operator sum representation (OSR)\,\cite{ecg61,kraus71}, and the Choi-Jamiolkowski isomohpishm (CJI)\,\cite{choi75,jamiolkowski72}. 

\subsection{Unitary representation}
It is known that every CP map can be realised in the following way. Let us consider a composite system constructed from the system plus an environment $R$ with Hilbert space ${\cal H}_R$. First, the systems states are elevated to product states on the larger Hilbert space (system + environment), with a fixed state of the environment\,: 
\begin{align}
\hr{S} \to \hr{S} \otimes |\psi\rangle_R\langle\psi|, ~~|\psi_R\rangle \text{ fixed}. 
\label{i78}
\end{align} 
Then the product states are evolved by a joint unitary evolution $U_{SR}$\,:
\begin{align}
\hr{S} \otimes |\psi\rangle_R\langle\psi| \to U_{SR} \,( \hr{S} \otimes |\psi\rangle_R\langle\psi|)\, U_{SR}^{\dagger}.
\label{i79}
\end{align}
Finally, the environment degrees of freedom are traced out to give the evolved system states\,:
\begin{align}
\Phi(\hr{S}) = \hr{S}^{\,'} = {\rm Tr}_R \, \left[ U_{SR} \,( \hr{S} \otimes |\psi\rangle_R\langle\psi|)\, U_{SR}^{\dagger}\right]. 
\label{i80}
\end{align}
We see that the map $\Phi\,: \hr{S} \to \hr{S}^{\,'}$ is completely specified by the triplet $({\cal H}_R, \, U_{SR},\,|\psi\rangle_R)$. A schematic diagram for the unitary representation is shown in Fig.\,\ref{figi1}.
\begin{figure}
\begin{center}
\scalebox{0.9}{\includegraphics{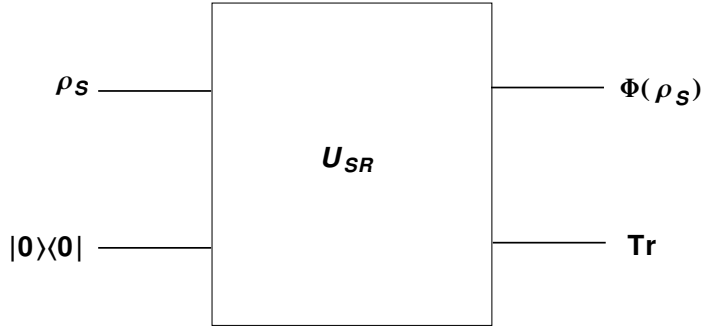}}
\end{center}
\caption{
Showing the unitary realization of any quantum channel. The initial state is appended with a fixed environment state denoted $|0\rangle \langle 0|$ and the composite is evolved through a joint unitary $U_{SR}$. Then the environment degrees are ignored to obtain the evolved system state. 
\label{figi1}}
\end{figure}
It is immediately clear that this representation is not unique as can be seen from the following example.  Performing a unitary transformation on system $R$ and appropriately changing the fixed pure state of the environment, we obtain the same CP map $\Phi$. In other words
\begin{align}
U_{SR} \to U_{SR} (1\!\!1_S \otimes U_R), ~~ |\psi_R\rangle \to U_R^{\dagger}|\psi_R\rangle,
\label{i81}
\end{align}
will result in the same map. The map $\Phi$ is trace-preserving by construction. The unitary representation can also be written in the following form\,:
\begin{align}
\Phi(\hr{S}) = {\rm Tr}_R\,( V_{SR}\, \hr{S}\, V_{SR}^{\dagger}),
\label{i82}
\end{align}
where $V_{SR} = U_{SR} |\psi\rangle_R$ is an isometry from ${\cal H}_S \to {\cal H}_S \otimes {\cal H}_{R}$. We recall that an isometry $V\,: {\cal H}_A \to  {\cal H}_A \otimes {\cal H}_B$ is a linear operator such that $V^{\dagger} V = 1\!\!1_A$, which is satisfied by $V_{SR}$ defined in Eq.\,\eqref{i82}.   

\subsection{Operator sum representation}
An alternative and equivalent representation of a quantum channel is known as the operator sum representation (OSR). Every channel $\Phi$ can be expressed as 
\begin{align}
\Phi({\cal H}_S) = \sum_k A_k\, \hr{S} \,A_k^{\dagger},
\label{i83}
\end{align} 
where the operators $A_k$ are called Kraus operators. The trace-preserving condition reads 
\begin{align}
\sum_{k} A_k^{\dagger} A_K = 1\!\!1.
\label{i84}
\end{align}
Let us now consider a new set of Kraus operators given by $\tilde{A}_k = V_{kj} A_{j}$. Let us impose the trace-preserving condition
\begin{align}
&\sum_k \tilde{A}_k^{\dagger} \tilde{A}_K = 1\!\!1,\nonumber\\
\text{i.e., }~ &\sum_k \sum_{ij}  A_{i}^{\dagger} \,V_{ki}^* V_{kj}\, A_{j} = 1\!\!1.
\label{i85}
\end{align}
In other words, we require 
\begin{align}
\sum_k &V_{ki}^* V_{kj} = \delta_{ij},\nonumber\\
 ~\Rightarrow~ &V^{\dagger} V = 1\!\!1,
\label{i86}
\end{align}
i.e., $V$ is required to be an isometry by the trace preserving condition in Eq.\,\eqref{i84}. 

Let us denote by $\tilde{\Phi}$ the channel corresponding to the new set of Kraus operators $\{\tilde{A}_k\}$. The operator sum representation is then given by 
\begin{align}
\tilde{\Phi}(\hr{S}) &= \sum_k \tilde{A}_k \, \hr{S} \, A_k^{\dagger}\nonumber\\
&=\sum_k \sum_{ij} V_{kj} A_{j}\,\hr{S} \, A_{i}^{\dagger} V_{ki}^* \nonumber\\
&= \sum_{ij} \sum_{k} V_{kj} V_{ki}^*\,A_{j}\,\hr{S} \, A_{i}^{\dagger}\nonumber\\
&= \sum_{ij} \delta_{ij} A_{j}\,\hr{S} \, A_{i}^{\dagger}\nonumber\\
&= \sum_j A_j\, \hr{S}\,A_j^{\dagger}.
\label{i87}
\end{align}
We see that $\tilde{\Phi} = \Phi$, i.e., there is a isometry freedom in the definition of the operator sum representation. In other words, if two sets of Kraus operators are related by an isometry, then the corresponding channels $\Phi, \, \widetilde{\Phi}$ defined through OSR, will represent one and the same map.

\subsection{Choi-Jamiolkowski representation}
The third representation is the Choi-Jamiolkowski state corresponding to a given CP map $\Phi$. Consider the  composite system $AB$ whose Hilbert space is given by ${\cal H}_S \otimes {\cal H}_S$. We will make use of the maximally entangled pure state given in Eq.\,\eqref{i76}. The Choi-Jamiolkowski state is obtained from the one-sided action of the CP  map $\Phi$. We have 
\begin{align}
\Gamma_{\Phi} = (\Phi \otimes 1\!\!1)\frac{\hat{\sigma}}{d}.  
\label{i88}
\end{align}
The state $\Gamma_{\Phi}$ associated with $\Phi$ gives a complete description of the CP map.  

$\Phi$ is trace-preserving only if ${\rm Tr}_A(\Gamma_{\Phi}) = 1\!\!1/d$. We note that the dynamical matrix $D_{\Phi}$ is related to the Choi-Jamiolkowski state by\,: $D_{\Phi} = d\, \Gamma_{\Phi}$. The CJ-representation turns out to be useful in obtaining the operator sum representation of $\Phi$,  as will be detailed in the next Section.

\subsection{Connecting the three representations}
We will now briefly describe how the three representations are interconnected, and how one can go from one representation to the another. \\

\noindent
{\bf Unitary $\to$ OSR}\,:\\
\noindent
Let ${\cal E} =\{ |e^j_R\rangle \}$ be an orthonormal basis for system $R$. We first begin with the unitary representation and perform the trace in basis ${\cal E}$. We have 
\begin{align}
\Phi(\hr{S}) &=  {\rm Tr}_R \, \left[ U_{SR} \,\left( \hr{S} \otimes |\psi\rangle_R\langle\psi|\right)\, U_{SR}^{\dagger}\right],\nonumber\\
&= \sum_j \langle e^j_R| \, \left[ U_{SR} \,\left( \hr{S} \otimes |\psi\rangle_R\langle\psi|\right)\, U_{SR}^{\dagger}\right] \, |e^j_R\rangle\nonumber\\
&=\sum_j ( \langle e^j_R|  U_{SR} |\psi\rangle_R) \, \hr{S}\, ({}_R \langle\psi| U_{SR}^{\dagger}  |e^j_R\rangle)
\label{i89}
\end{align}
Let us now define operators $A_k\,: {\cal H}_S \to {\cal H}_S$ where
\begin{align}
A_k &= \langle e^k_R| U_{SR} | \psi_R\rangle. 
\label{i90}
\end{align}
Then the expression for $\Phi(\hr{S})$ in Eq.\,\eqref{i89} reduces to 
\begin{align}
\Phi(\hr{S}) = \sum_j A_k\,\hr{S}\,A_k^{\dagger},
\label{i91}
\end{align}
which is the operator sum representation. To check the trace condition we evaluate 
\begin{align}
\sum_j A_j^{\dagger} A_j &= \sum_j  \langle \psi_R| U^{\dagger}_{SR} | e^j_R\rangle \langle e^j_R| U_{SR} | \psi_R\rangle\nonumber\\
&= \langle \psi_R| U^{\dagger}_{SR}\, \left(\sum_j | e^j_R\rangle \langle e^j_R|\right)\, U_{SR}\, | \psi_R\rangle\nonumber\\
&= \langle \psi_R| U^{\dagger}_{SR} U_{SR} | \psi_R\rangle \nonumber\\
&=  \langle \psi_R| 1\!\!1_S \otimes 1\!\!1_R  | \psi_R\rangle \nonumber\\
&= 1\!\!1_S,
\label{i92}
\end{align}
as expected. Had we instead chosen some other complete basis to evaluate the partial trace in the unitary representation, we would have obtained another operator sum representation for the same map $\Phi$, connected to the original one by an isometry as seen earlier in Eq.\,\eqref{i87}.\\

\noindent
{\bf OSR $\to$ Unitary}\,:\\
\noindent
We will describe how to obtain the unitary representation starting from the operator sum representation. Let us begin with
\begin{align}
\Phi(\hr{S}) = \sum_{k=1}^{r} A_k \, \hr{S}\,A_k^{\dagger},
\label{i93}
\end{align}
where $r$ denotes the number of Kraus operators in the operators sum representation. Let us consider a composite system with Hilbert space ${\cal H}_R \otimes {\cal H}_S$, with system $R$ of dimension $r$. 
Let us arrange the operators $A_k$ in a suggestive form to obtain a bipartite operator $V\,: {\cal H}_S \to {\cal H}_R \otimes {\cal H}_S$ defined as 
\begin{align} 
V &= \sum_k |k\rangle \otimes A_k \nonumber\\
&= \begin{pmatrix}
A_1 \\
A_2 \\
\vdots\\
A_r
\end{pmatrix},
\label{i94}
\end{align}
and $\{ |k\rangle \}_{1}^{r}$ is an ONB for system $R$. 
It is immediate that $V$ is an isometry as can be seen from the fact that 
\begin{align}
V^{\dagger} V &= \sum_{j,k}^{r}  \langle j | k \rangle ~ A_j^{\dagger} A_k \nonumber\\
&= \sum_j A_j^{\dagger} A_k \delta_{jk}\nonumber\\
&= 1\!\!1_S.
\label{i95}
\end{align}
We note that $V$ is a $dr \times d$ matrix. We already see that the operator sum representation can be written as 
\begin{align}
\Phi(\hr{S}) &= {\rm Tr}_R V\,\hr{S}\,V^{\dagger} \nonumber\\
&= {\rm Tr}_R \left( \sum_k  |k\rangle \otimes A_k\right)\,\hr{S}\,\left( \sum_j \langle j|  \otimes A_j^{\dagger}\right)  \nonumber\\
&= \sum_{jk} {\rm Tr}_R (|k\rangle \langle j|) \otimes A_k\,\hr{S}\,A_j^{\dagger}  \nonumber\\
&= \sum_{jk} A_k\,\hr{S}\,A_j^{\dagger}\, \delta_{jk}\nonumber\\
&= \sum_k^{r} A_k \, \hr{S}\,A_k^{\dagger}.
\label{i96}
\end{align}
The isometry $V$ which is $dr \times d$ matrix can be appropriately completed to a unitary $dr \times dr$ matrix. For convenience, we can make the choice $U_{SR}|1\rangle_R$ = $V$, where $|1\rangle_R$ is the first vector of the computational basis in system $R$. Then one recovers operator $V$ as given in Eq.\,\eqref{i94}. In this way, we obtain the unitary representation of the CP map. \\

\noindent
{\bf CJ $\to ~ \Phi$}\,:\\
\noindent
The action of $\Phi$ on a state $\hr{S}$ can be written down from the dynamical matrix $D_{\Phi}$ associated with $\Phi$. We have
\begin{align}
\Phi(\hr{}) = {\rm Tr}_R\,(D_{\Phi}\,\hr{}^T). 
\label{i97}
\end{align}
This expression can be verified in a straight-forward manner\,:
\begin{align}
{\rm Tr}_R\,(D_{\Phi}\,[1\!\!1 \otimes \hr{}^T]) &= \sum_{ij}\,[\Phi \otimes 1\!\!1_R]  {\rm Tr}_R\, (|i \rangle \langle j| \otimes |i \rangle \langle j|\,\hr{}^T)\nonumber\\
&= \sum_{ij}\,  \left( \Phi[|i \rangle \langle j|]\, \langle j |\hr{}^T|i \rangle \right),\nonumber\\
&=  \sum_{ij}\,  \left( \Phi[|i \rangle \langle j|]\, \rho_{ij} \right)\nonumber\\
&= \Phi(\hr{}).
\label{i98}
\end{align}

\noindent
{\bf CJ $\to$ OSR}\,:\\
\noindent
Here we outline a simple procedure to obtain the operator sum representation from the CJ state or dynamical matrix. 
Let us first write down a decomposition of the dynamical matrix $D_{\Phi}$ into pure states, the spectral resolution being a special choice with orthonormal vectors\,:
\begin{align}
D_{\Phi} = \sum_j \proj{\psi_j}.
\label{i99}
\end{align}
Note that the vectors $\{|\psi_j\rangle \}$ are not normalized. Let us write down the vectors $\{|\psi_j \rangle \}$ as 
\begin{align}
|\psi_j\rangle = \sum_{mn} c^j_{mn}\, |m \rangle \otimes |n\rangle,
\label{i100}
\end{align}
where $c^{j}_{mn}$ is the coefficient matrix for every vector $|\psi_j\rangle$. To each of these vectors we associate an operator $\widetilde{K}^j$ using the Jamiolkowski isomorphism\,\cite{jamiolkowski72}  which is defined as 
\begin{align}
\widetilde{K}^j = \sum_{mn} c^j_{mn}\, |m \rangle \langle n|.
\label{i101}
\end{align}
In other words, we flip the second ket of the vector to a bra to obtain the associated operator. We see that the isomorphism associates a vector $|v\rangle \in {\cal H}_A \otimes {\cal H}_A$ to a linear operator $V\,: {\cal B}({\cal H}_A) \to {\cal B}({\cal H}_A)$. 

Let $\widetilde{\Phi}$ be the map whose Kraus operators are the $\widetilde{K}^j$'s. 
Consider the one-sided action of the CP map $\widetilde{\Phi}$\,:
\begin{align}
D_{\widetilde{\Phi}} = [\widetilde{\Phi} \otimes 1\!\!1](\hat{\sigma}) &= \sum_{ij}\, [\widetilde{\Phi} \otimes 1\!\!1] [ |i \rangle \langle j| \otimes |i \rangle \langle j|]\nonumber\\
&= \sum_{ij}\,\sum_{k} \widetilde{K}^k\,    |i \rangle \langle j| \,(\widetilde{K}^k)^{\dagger} \otimes |i \rangle \langle j|\nonumber\\
&= \sum_{ij}\,\sum_k \sum_m c^k_{mi}\, |m\rangle \langle n|\, (c^k_{nj})^* \otimes |i\rangle \langle j|\nonumber\\
&= \sum_k \left(\sum_{mi} c^k_{mi} |m i\rangle \right)\, \left(\sum_{nj} (c^k_{nj})^* \langle nj|\right)\nonumber\\
&= \sum_k \proj{\psi_k} = D_{\phi},
\label{i102}
\end{align} 
proving the assertion. By Eq.\,\eqref{i102}, the association from the vector to the operator is made transparent by the following identity\,:
\begin{align}
\sum_{ij} \left[\widetilde{K}^k \otimes 1\!\!1 \right]\, \left(|i \rangle \langle j| \otimes |i \rangle \langle j|\right)\, \left[(\widetilde{K}^k)^{\dagger} \otimes 1\!\!1 \right]= \proj{\psi_k},
\label{i103}
\end{align}

\begin{figure}
\begin{center}
\scalebox{1}{\includegraphics{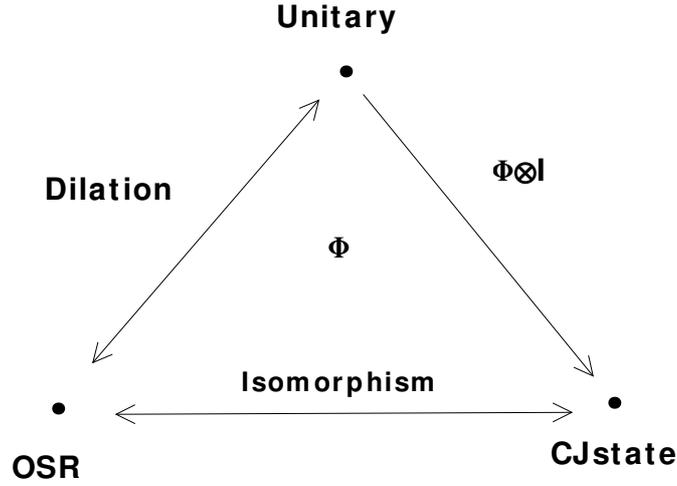}}
\end{center}
\caption{
Showing a schematic diagram for the various ways in which the three representations of CPTP maps, namely, the unitary representation, the operator sum representation and the Choi-Jamiolkowski representation are related. 
\label{figi2}}
\end{figure}

A schematic diagram with the connections between the various representations is shown in Fig.\,\ref{figi2}. A few remarks are in order with respect to obtaining the operator sum representation from the CJ state. Note that we first began with the dynamical matrix for which the trace is not unity, and in fact ${\rm Tr}(\Gamma_{\Phi}) = d_A$. This facilitated the obtaining of the Kraus operators directly. That the resulting CP map is trace-preserving is a consequence of the fact that ${\rm Tr}_A \Gamma_{\Phi}= 1\!\!1/d$, i.e., ${\rm Tr}_A D_{\phi}=1\!\!1$.

We see that the rank of the CJ state  $\Gamma_{\Phi}$  or dynamical matrix $D_{\Phi}$ corresponding to a channel $\Phi$ gives the minimum number of Kraus operators in the operator sum representation. We will call the operator sum representation of a channel {\em minimal} when the number of Kraus operators is the minimum number possible.  Let us denote the rank of $\Gamma_{\Phi}$ as $r$. So we have that 
\begin{align}
r \leq d^2.
\label{i104}
\end{align}
In other words, the maximum number of Kraus operators in the minimal representation is $d^2$. One way to obtain this minimal representation is to consider the spectral decomposition of the $\Gamma_{\Phi}$. We have seen earlier that any isometry on the set of operators also gives rise to equivalent operator sum representations. These are precisely the various rank-one decompositions of $\Gamma_{\Phi}$. 
Finally, we see that the unitary representation requires an ancilla system $R$ of dimension $r \leq d^2$ to realize any channel acting on a system with Hilbert space dimension $d$.

\subsection{Properties of CP maps}
We  now provide a useful guide to the various properties of CPTP maps in a suitable representation. \\

\noindent
{\bf Dual}\,: Given a channel $\Phi$ with Kraus operators $\{A_k\}$, the {\em dual} $\Phi^{\,'}$ is defined as the CP map that has an operator sum representation with Kraus operators  $\{\tilde{A}_k = {A}_k^{\dagger}\}$, i.e., 
\begin{align}
\Phi^{\,'} (\hr{S}) = \sum_k \tilde{A}_k\,\hr{S}\,\tilde{A}_k^{\dagger}.
\label{i105}
\end{align}
Since $\Phi$ is trace-preserving, it implies that
\begin{align}
\sum_k A_k^{\dagger} A_k = \sum_k \tilde{A}_k \tilde{A}_k^{\dagger} = 1\!\!1.
\label{i106}
\end{align}

\noindent
{\bf Unital}\,: A unital CP map is one that takes the identity operator to itself, i.e.,
\begin{align}
\Phi(1\!\!1) = 1\!\!1.
\label{i107}
\end{align}
From the operator sum representation of $\Phi$, we have the following condition on the Kraus operators\,: 
\begin{align}
\sum_k A_k\, A_k^{\dagger} = 1\!\!1.
\label{i108}
\end{align}
Therefore, the dual of a quantum channel is a unital CP map as can be seen by Eq.\,\eqref{i105}. An alternative way to see this fact is by considering the Choi-Jamiolkoski state $\Gamma_{\Phi}$. We have that a CP map $\Phi$ is unital if and only if 
\begin{align}
{\rm Tr}_B\, (\Gamma_{\Phi})= \frac{1\!\!1}{d_A}, ~~ {\rm Tr}_B\, (D_{\Phi})= 1\!\!1.
\label{i109}
\end{align}
\\

\noindent
{\bf Bistochastic}\,: A channel $\Phi$ that is also unital is called a bistochastic map. So the conditions in terms of the Kraus operators are given by\,:
\begin{align}
&\sum_k A_k^{\dagger} A_k = 1\!\!1, \nonumber\\
\text{ and } ~ & \sum_k A_k A_k^{\dagger} = 1\!\!1.
\label{i110}
\end{align}
Alternately, a channel $\Phi$ is bistochastic if and only if
\begin{align}
{\rm Tr}_A \,(\Gamma_{\Phi}) = \frac{1\!\!1}{d}, \text{ and } ~~{\rm Tr}_B \,(\Gamma_{\Phi}) = \frac{1\!\!1}{d}.
\label{i111}
\end{align}

\noindent
{\bf Random unitary}\,: A random unitary channel\,\cite{audenaert08} is a channel which is a convex combination of unitary channels. In other words, the operator sum representation of a random unitary channel can be given in the form\,: 
\begin{align}
\Phi(\hr{S}) = \sum_k p_k\, U_k\,\hr{S}\,U_k^{\dagger}.
\label{i112}
\end{align}
By definition, a random unitary channel is bistochastic. But not every bistochastic channel is random unitary. Examples of channels which are bistochastic but not random unitary were provided in the finite-dimensional setting in\,\cite{kummerer87,landau93,wolf09} and for continuous variable systems in\,\cite{kraus10}.\\

\noindent
{\bf Extremal}\,: The set of quantum channels acting on a given Hilbert space forms a convex set, i.e.,
\begin{align}
\Phi = p_1 \Phi_1 + (1-p_1) \Phi_2,
\label{i113}
\end{align}
is also a CPTP map when $\Phi_1,\,\Phi_2$ are channels. An extremal channel is one that cannot be written as a convex combination of other quantum channels. A simple example of an extremal map is a unitary channel, i.e.,
\begin{align}
\Phi(\hr{S}) = U \, \hr{S}\,U^{\dagger}. 
\label{i114}
\end{align}
By definition, a random unitary channel is not extremal. A theorem by Choi\,\cite{choi75} gives a way to check if a channel is extremal or not.  

\begin{theorem}[Choi]
A CPTP map $\Phi$ with minimal operator sum representation $\Phi(\hr{S}) = \sum_k A_k\,\hr{S}\,A_K^{\dagger}$ is extremal if and only if the  operators $\{ {\cal A}_{kj} = A_k^{\dagger} A_j\}$ are linearly independent. 
\end{theorem}
If the number of Kraus operators is $r$, then the number of operators $\{{\cal A}_{kj}\}$ is $r^2$. Since we require linear independence of $\{ {\cal A}_{kj} \}$ for an extremal channel, we have $r^2 \leq d^2$. In other words, the operator sum representation of an extremal channel can have utmost $d$ Kraus operators in the minimal representation. \\


\noindent
{\bf Entanglement-breaking}\,: A  channel $\Phi\,: \Lambda({\cal H}_S) \to \Lambda({\cal H}_S)$ is said to be entanglement-breaking\,\cite{horodecki03} if its one-sided action takes every bipartite state $\hr{SR} \in \Lambda({\cal H}_S \otimes {\cal H}_R)$ to a separable state for an arbitrary system $R$. Much like the CP condition, there is an operational way to check whether a channel $\Phi$ is entanglement-breaking or not. We have the following theorem\,\cite{horodecki03}\,:
\begin{theorem}[Horodecki-Shor-Ruskai]
A channel is entanglement-breaking iff 
\begin{align}
\Gamma_{\Phi} = \frac{1}{d} [\Phi \otimes 1\!\!1]\,(\hat{\sigma}).
\label{i115}
\end{align}
 is separable. Further, every entanglement-breaking channel has an operator sum representation in which every Kraus operator is  rank-one.
\end{theorem} 
That there exists an operator sum representation having rank-one elements for every entanglement-breaking channel is a consequence of the fact that every separable state has a decomposition in terms of products of projectors. 

Having assembled the basic notions of correlations and channels of interest to us, we next consider some preliminaries regarding continuous variable systems.

\section{Single mode of radiation}
Let us consider as our quantum system a  single-mode of a radiation field (a harmonic oscillator)\,\cite{simon94}. The Hilbert space is the space of all (complex) square integrable functions $\psi$ over one real variable, the configuration space, and is denoted by ${\cal L}^2({\cal R})$\,:
\begin{align}
\psi \in {\cal L}^2({\cal R})  \longleftrightarrow \int dx \, |\psi(x)|^2 < \infty. 
\label{j1} \end{align}
The creation and annihilation operators, denoted by $\hat{a},\,\hat{a}^{\dagger}$ of the quantum system satisfy the standard bosonic commutation relation\,:
\begin{align}
[\hat{a},\,\hat{a}^{\dagger}] = 1.
\label{j2} \end{align}
In terms of the hermitian position and momentum variables, these ladder operators have the expression 
\begin{align}
\hat{a} = \frac{\hat{q}+i\hat{p}}{\sqrt{2}},~~ \hat{a}^{\dagger} = \frac{\hat{q}-i\hat{p}}{\sqrt{2}},
\label{j3} \end{align}
and the equivalent commutation relation reads
\begin{align}
[\hat{q},\hat{p}]=i,
\label{j4} \end{align}
where we have set $\hbar =1$.
Let us arrange the operators $\hat{q},\,\hat{p}$ as a column vector\,:
\begin{align}
\hat{\xi} = \begin{pmatrix}
~\hat{q}~\\
~\hat{p}~
\end{pmatrix}.
\label{j5} \end{align}
Then the commutation relations, using Eq.\,\eqref{j4}, can be compactly written as 
\begin{align}
[\hat{\xi}_i,\,\hat{\xi}_j] = i\, \beta_{ij},
\label{j6} \end{align}
where 
\begin{align}
\beta = \begin{pmatrix}
0 & 1 \\
-1 & 0
\end{pmatrix}.
\label{j7} \end{align}
Consider a linear transformation on $\hat{q}$ and $\hat{p}$ specified by a $2 \times 2$ real matrix $S$\,:
\begin{align}
\hat{\xi} \to \hat{\xi}^{\,'} = S \hat{\xi}.
\label{j8} \end{align}
Since the new variables also need to satisfy the canonical commutation relations of\,\eqref{j6}, we have that 
\begin{align}
S \,\beta\,S^T = \beta.
\label{j9} 
\end{align}
In other words, $S$ is an element of the symplectic group $Sp(2,{\cal R})$. 

An important aspect to note is that these linear transformations are induced by unitary evolutions generated by Hamiltonians that are quadratic in the mode operators\,\cite{simon94}. In other words any $\hat{H} = \sum_{ij}\, h_{ij} \,\hat{\xi}_i\,\hat{\xi}_j$, $((h_{ij}))$ being real symmetric,  the corresponding unitary transformation $\hat{U} = e^{-i  \hat{H}}$ induces 
\begin{align}
\hat{U}^{\dagger} \, \hat{\xi}\,\hat{U} = S(h)\,\hat{\xi}, ~~ S(h) \in Sp(2,{\cal R}).
\label{j10} 
\end{align}

{\bf Passive transformations}\,:\\
A subset of transformations of particular interest to us are the what are known as passive transformations\,\cite{simon94}.  Passive transformations are those symplectic transformations that are phase-space rotations as well. We only consider the single-mode case for simplicity. We denote the collection of passive transformations on single-mode systems by $K(1)$. We have\,
\begin{align}
K(1) = \left\{ S|\, S \in Sp(2,{\cal R}) \cap SO(2,{\cal R})\right\}.
\label{j41} \end{align}
It turns out that $K(1)=SO(2,{\cal R})$ is isomorphic to $U(1)$. All the above properties suitably generalise to the multi-mode case. Passive transformations conserve photon number and play an important role in the definition of squeezing\,\cite{simon94}.

\section{Phase space distributions}
The study of phase space distributions can be motivated from  the possibility of using these functions as a `weight' functions in an integral representation of a given operator\,\cite{cahill69a,cahill69b}. Before we describe the notion of phase space distributions, we now briefly discuss an important class of operators known as the Weyl displacement operators. 

For each complex number $\alpha \in {\cal C}$, there is an associated operator ${\cal D}(\alpha)$, which is defined as  
\begin{align}
{\cal D}(\alpha) = \exp[\alpha \hat{a}^{\dagger}- \alpha^* \hat{a}].
\label{j11} \end{align}
The operators $\{ {\cal D}(\alpha) \}$ are known as the displacement operators. 
From the definition we see that ${\cal D}(\alpha)$ is unitary and ${\cal D}^{\dagger}(\alpha) = {\cal D}(-\alpha) = {\cal D}(\alpha)^{-1}$. The operators ${\cal D}(\alpha)$ are called displacement operators for the following reason\,:
\begin{align}
&{\cal D}(\alpha)^{\dagger}\,\hat{a} \, {\cal D}(\alpha) = \hat{a} + \alpha\nonumber\\
&{\cal D}(\alpha)^{\dagger}\, \hat{a}^{\dagger} \, {\cal D}(\alpha) = \hat{a}^{\dagger} + \alpha^*
\label{j12} \end{align}
The composition of two displacement operators with independent arguments gives\,:
\begin{align}
{\cal D}(\alpha) {\cal D}(\beta) = \exp\left[\frac{1}{2}(\alpha \beta^* - \alpha^* \beta)\right]\,{\cal D}(\alpha + \beta).
\label{j13} \end{align}
We finally mention the orthogonality property\,:
\begin{align}
{\rm Tr}\,[{\cal D}(\alpha)\,{\cal D}^{-1}(\beta)] = \pi \delta^{(2)}(\alpha - \beta).
\label{j14} \end{align}
It may be `visually' seen from the definition \eqref{j11} that ${\cal D}(\alpha)$ is simply the `quantized' version of the plane wave $\exp[\alpha z^* - \alpha^* z]$ over the classical $q-p$ phase-space, with $z=(q+ip)/\sqrt{2}$. It should thus come as no surprise that the collection $\{{\cal D}(\alpha), \alpha \in {\cal C} \}$ satisfy a completeness relation corresponding to the completeness of the plane waves (Fourier integral theorem). Consequently, the displacement operators ${\cal D}(\alpha)$ form a basis for expansion of generic operators acting on ${\cal H}={\cal L}^2({\cal R})$\,\cite{cahill69a}. 


The displacement operators can be expressed in various ways to correspond to various ordering schemes.  Ordering refers to the order in which the ladder operators are written in the polynomial expansion of the displacement operators. Two particular choices of ordering are the normal ordering and the anti-normal ordering. The normal ordering of the displacement operator ${\cal D}(\alpha)$ is given by the expression 
\begin{align}{\cal D}(\alpha) = \exp[{-|\alpha|^2/2}]\, \exp[{\alpha \hat{a}^{\dagger}}] \, \exp[{-\alpha^* \hat{a}}],
\label{j15} \end{align}
and the anti-normal ordering  by 
\begin{align}
{\cal D}(\alpha) = \exp[|\alpha|^2/2]\,  \exp[{-\alpha^* \hat{a}}]\,\exp[{\alpha \hat{a}^{\dagger}}]. 
\label{j16} \end{align} 
The expression in \eqref{j11} corresponds to Weyl or symmetric ordering. The s-ordered displacement operator, $s \in [-1,1]$, denoted by ${\cal D}(\alpha;s)$ is defined as
\begin{align}
{\cal D}(\alpha;s) &= \exp[s|\alpha|^2/2]\,{\cal D}(\alpha).
\label{j17} \end{align}
So normal ordering corresponds to $s=1$, anti-normal ordering to $s=-1$, and the symmetric or Weyl ordering corresponds to the case $s=0$. 
The $s$-ordered monomial $\{ (\hat{a}^{\dagger})^n \hat{a}^m\}_s$  is defined as 
\begin{align}
{\cal D}(\alpha;s)= \sum_{n,m=0}^{\infty} \,\{ (\hat{a}^{\dagger})^n \hat{a}^m\}_s \, \frac{\alpha^n(-\alpha^*)^m}{n!m!},
\label{j18} 
\end{align}
or equivalently we have 
\begin{align}
\{ (\hat{a}^{\dagger})^n \hat{a}^m\}_s = \left.\frac{\partial^{n+m}{\cal D}(\alpha;s)}{\partial \alpha^n \partial (-\alpha^*)^m}\right|_{\alpha =0}.
\label{j19} \end{align}

The $s$-ordered displacement operators facilitate the definition of the $s$-ordered characteristic function. The $s$-ordered characteristic function associated with a given density operator $\hr{}$ is defined as 
\begin{align}
\chi_s(\hr{},\alpha) &= {\rm Tr}[{\cal D}(\alpha;s) \, \hr{}].
\label{j20} \end{align}
From the completeness of the displacement operators, we have the following representation or inverse relation for any operator $\hr{}$\,\cite{cahill69a}:
\begin{align}
\hr{} = \int \frac{d^2 \alpha}{\pi}\, \chi_s(\hr{},\alpha)\,{\cal D}^{-1}(\alpha;s).
\label{j21} \end{align}
The $s$-ordered quasiprobability associated with a state $\hr{}$ is defined as the two-dimensional Fourier transform of the corresponding $s$-ordered characteristic function. We have
\begin{align}
W_s(\hr{},\xi) = \int \frac{d^2\alpha}{\pi}\, e^{\,\xi \alpha^*  - \xi^* \alpha}\,\chi_s(\hr{},\alpha).
\label{j22} \end{align}
We now briefly detail three frequently used quasiprobabilities from among the one-parameter family of quasiprobabilities, namely, the Wigner, the $Q$ and the $\phi$ distributions\,\cite{cahill69b}. \\

\noindent
{\bf Wigner function}\,: \\
The Wigner function  $W_0(\hr,\alpha)$ associated with a given state $\hr{}$ results as the symmetric-ordered ($s=0$) quasiprobability\,:
\begin{align}
W(\hr{},\alpha) \equiv W_0(\hr,\alpha) =  \int \frac{d^2\xi}{\pi}\, e^{\alpha \xi^* - \xi \alpha^*}\,\chi_0(\hr{},\xi).
\label{j23} \end{align}
The Wigner function (and indeed every s-ordered quasiprobability) is real and normalized in accordance with the hermiticity and trace condition of a density operator $\hr{}$ \,:
\begin{align}
W(\hr{},\alpha) &= W(\hr{},\alpha)^*,\nonumber\\
 \int \frac{d^2 \alpha}{\pi} \,W(\hr{},\alpha) &=1.
\label{j24} \end{align}
The Wigner representation is particularly useful for evaluating expectation values of operators written in the symmetric ordered form\,:
\begin{align}
{\rm Tr}\,[\hr{}\,\{(\hat{a}^{\dagger})^n \hat{a}^m \}_0] = \int \frac{d^2 \alpha}{\pi} \,W(\hr{},\alpha) \,( \alpha^*)^n \alpha^m. 
\label{j25} \end{align}
We see that the symmetric-ordered operators are just replaced by the c-number equivalents and the density operator is replaced by the associated Wigner function.

A useful property of the Wigner function is the ease with which symplectic transformations reflect in the Wigner description. We had seen earlier that unitary transformations generated by Hamiltonians quadratic in mode operators, lead to a symplectic transformation of the mode operators. We have in the Wigner picture\,:
\begin{align}
\hr{} &\to \hr{}^{\,'} = U(S) \, \hr{} \, U(S)^{\dagger}\nonumber\\
&\Longleftrightarrow  \hat{\xi} \to \hat{\xi}^{\,'} = S \hat{\xi} \nonumber\\
&\Longleftrightarrow W(\hr{},\xi) \to W(\hr{}^{\,'},\xi) =  W(\hr{},S^{-1}\xi)\nonumber\\
&\Longleftrightarrow \chi(\hr{},\xi) \to \chi(\hr{}^{\,'},\xi)= \chi(\hr{},S^{-1}\xi)\nonumber\\
&\Longleftrightarrow V \to V^{\,'} = S\, V \,S^T,
\label{j40} \end{align}
where in the last line of Eq.\,\eqref{j40}, $V$ stands for the variance matrix associated with $\hr{}$. We will consider the notion of variance matrix in the next Section. 
We will repeatedly appeal to the above transformations in phase space as well as the corresponding transformations at the level of the variance matrix in the following Sections. \\

\noindent
{\bf Husimi function}\,: \\
\noindent
The Husimi $Q$-function is the anti-normal ordered quasiprobability and is defined as 
\begin{align}
Q(\hr{},\alpha) \equiv W_{-1}(\hr{},\alpha) = \int \frac{d^2\xi}{\pi}\, e^{\alpha \xi^* - \xi \alpha^*}\,\chi_{-1}(\hr{},\xi).
\label{j26} \end{align}
It can be shown that the $Q$-function can alternately be written as\,\cite{cahill69b}\,: 
\begin{align}
Q(\hr{},\alpha) = \langle \alpha | \hr{} | \alpha \rangle.
\label{j27} \end{align}
We see that the $Q$-function is always pointwise positive irrespective of the state $\hr{}$ and its numerical value is bounded from above by 1, i.e., $Q(\hr{},\alpha) \leq 1$. Further, $Q$ is normalised\,:
\begin{align}
\int \frac{d^2 \alpha}{\pi}\,Q(\hr{},\alpha) =1.
\label{j28} \end{align} 
We see that the $Q$-function is a probability distribution over the complex plane. We mention in passing that though every $Q$-function is a probability distribution, the converse however is not true.  The $Q$-function facilitates the computation of the ensemble averages of anti-normally ordered operators analogous to how the Wigner function was useful for computing ensemble averages of symmetric-ordered operators. \\

\noindent
{\bf Diagonal `weight' function}\,:\\
\noindent
The third important quasiprobability we shall be interested in is the normal-ordered distribution corresponding to $s=1$. The quasiprobability corresponding to $s=1$ is called the Sudarshan-Glauber diagonal weight function denoted by $\phi$\,\cite{ecg63,glauber63}.
The diagonal weight $\phi(\hr{},\alpha)$ associated with a density matrix $\hr{}$ is defined as 
\begin{align}
\phi(\hr{},\alpha) \equiv W_{1}(\hr{},\alpha) = \int \frac{d^2\xi}{\pi}\, e^{\alpha \xi^* - \xi \alpha^*}\,\chi_{1}(\hr{},\xi).
\label{j29} \end{align}
Every density matrix $\hr{}$ can be expressed in the `diagonal' form in the (over-complete) coherent state basis as 
\begin{align}
\hr{} = \int \frac{d^2 \alpha}{\pi} \, \phi(\alpha) \proj{\alpha}.
\label{j30} \end{align}
We note that coherent states form a complete non-orthogonal set.
From the trace condition of $\hr{}$, we have that
\begin{align}
\int \frac{d^2 \alpha}{\pi} \, \phi(\alpha) =1.
\label{j31} \end{align}
Unlike the $Q$-function and the Wigner function which are well-behaved on the complex plane, the diagonal weight function can be highly singular. Finally, we note that the diagonal function $\phi$ helps to easily evaluate the ensemble averages of normally-ordered operators which is of much interest from an experimental perspective\,\cite{mandelwolfbook}.

\section{Gaussian states}
We now begin with a brief discussion on the notion of a variance matrix associated with a state $\hr{}$\,\cite{simon87,simon88,simon94}. Consider the $2\times 2$ matrix of operators $\hat{\xi}\,\hat{\xi}^T$. The following identity associated with this operator matrix is obtained by using the commutation and anti-commutation relations of the mode operators\,:
\begin{align}
2\, (\hat{\xi}\hat{\xi}^T)_{ij} &= 2\, \hat{\xi}_i\,\hat{\xi}_j\nonumber\\
&=\{\hat{\xi}_i,\hat{\xi}_j \} + [\hat{\xi}_i,\hat{\xi}_j]\nonumber\\
&= \{\hat{\xi}_i,\hat{\xi}_j \} + i\, \beta_{ij}.
\label{j32} \end{align}
Taking the expectation value in the state $\hr{}$, we have
\begin{align}
2 \langle \hat{\xi}\hat{\xi}^T \rangle_{ij} =  {\rm Tr}\,(\{\hat{\xi}_i,\hat{\xi}_j \}\,\hr{}) + i\, \beta_{ij}.
\label{j33} \end{align}
Let us assume without loss of generality that state is one for which the means are zero. We now define the variance matrix $V$ of a given state $\hr{}$ as 
\begin{align}
V_{ij} = {\rm Tr}\,(\{\hat{\xi}_i,\hat{\xi}_j \}\,\hr{}).
\label{j34} \end{align}
The matrix $V$ is real, symmetric and positive definite. These properties of the variance matrix would also hold for a classical probability distribution. However, for the quantum case there is an additional condition that $V$ has to satisfy for it to be a valid variance matrix. This additional constraint is the uncertainty principle\,\cite{simon94}\,:
\begin{align}
V + i\, \beta \geq 0.
\label{j35} \end{align}

It is known that every variance matrix can be diagonalised by a symplectic transformation\,\cite{simon94,simon98}. For the single mode case, by choosing a suitable symplectic transformation $S$, the variance matrix can be diagonalised, i.e.,
\begin{align}
V \to V_{\rm can} &= S\, V\, S^T \nonumber\\
&= \begin{pmatrix}
\kappa & 0\\
0&\kappa
\end{pmatrix},
\label{j36} \end{align}
where $\kappa$ is called the symplectic eigenvalue of $V$. In this canonical form, the uncertainty principle of Eq.\,\eqref{j35}  reads
\begin{align}
\kappa \geq 1.
\label{j37} \end{align}

We now describe a particularly important class of states known as Gaussian states\,\cite{simon88,simon94,krp1,krp2}. We assume that the state has zero first moments [this can be achieved by a rigid  phase space translation that is effected by the action of a suitable (unitary) displacement operator]. 
A Gaussian state is one whose Wigner function is a Gaussian function\,:
\begin{align}
W(\hr{},\alpha) = \frac{1}{2\sqrt{{\rm Det} V}} \,\exp[{-\frac{1}{2} \alpha^T\, V^{-1}\,\alpha}],
\label{j38} \end{align}
where the complex number $\alpha = x+iy$ can also be viewed as the vector $(x,y)^T$.
Equivalently, Gaussian states are states whose corresponding (symmetric-ordered) characteristic function is a Gaussian function. 
In other words we have
\begin{align}
\chi(\hr{},\xi) = \exp[{-\frac{1}{2} \xi^T\beta V\beta^T \xi}], 
\label{j39} \end{align}
where $V$ is the variance matrix associated with the state $\hr{}$. We wish to emphasis that a Gaussian state is completely specified by its first (means) and second moments (variances).

Simple examples of pure Gaussian states include the vacuum state or the ground state of the harmonic oscillator $|0 \rangle$,  coherent states $|\alpha \rangle =D(\alpha)|0\rangle$, squeezed state $S(\eta)|0 \rangle$, which is obtained by the action of the squeeze transformation $S(\eta)$ on the vacuum state. The thermal state is an example of a mixed Gaussian state. 
 
From the canonical form of the variance matrix in Eq.\,\eqref{j36}, we infer that by applying a suitable symplectic transformation, the variance matrix of any pure Gaussian state can be brought to the identity matrix, while any mixed Gaussian state can be brought to  the form $\kappa 1\!\!1_{2 \times 2}$, where $\kappa > 1$ is the symplectic eigenvalue.


\subsection{Two-mode systems}
The Hilbert space of the two-mode system is ${\cal L}^2({\cal R}) \otimes {\cal L}^2({\cal R}) = {\cal L}^2({\cal R}^2)$ and consists of vectors that are square integrable over a two-plane. As in the single mode case, we arrange the quadrature operators $\hat{q}_1,\,\hat{p}_1$, and $\hat{q}_2,\,\hat{p}_2$ associated with the modes as a column vector
\begin{align}
\hat{\xi} = (\hat{q}_1,\,\hat{p}_1,\,\hat{q}_2,\,\hat{p}_2)^T.
\label{j42} \end{align}
Then the canonical commutation relations read
\begin{align}
[\hat{\xi}_i,\,\hat{\xi}_j] &= i\, \Omega_{ij},\nonumber\\
\text{ where, }~~~ \Omega &= \beta \oplus \,\beta.
\label{j43} \end{align}
The mode operators $\hat{a}_1,\,\hat{a}_2$ are defined for each mode in the standard way.

A  Gaussian state of a two-mode system (with zero mean) is completely described by a $4 \times 4$ variance matrix which satisfies the uncertainty relation\,:
\begin{align}
V + i\, \Omega \geq 0.
\label{j44} \end{align}

Of importance to us is the detection of entanglement of two-mode Gaussian states. It turns out that there is necessary and sufficient criteria for detecting entanglement of two-mode Gaussian states\,\cite{simon00}. For this we require the use of the  transpose map $\Lambda$. The transpose map transcribes on the Wigner function faithfully into a mirror reflection of the underlying phase space. In other words, we have\,:
\begin{align}
\hat{\xi} \to \hat{\xi}^{\,'} = \Lambda \hat{\xi} = (\hat{q},\,-\hat{p}).
\label{j45} \end{align}
We now state the following necessary and sufficient condition for detecting entanglement of two-mode Gaussian states\,\cite{simon00}. 

\begin{theorem}[Simon]
A two-mode Gaussian state with variance matrix $V$ is separable if and only if the local application of the transpose map by one and only one of the parties leads to a valid variance matrix. The state is entangled otherwise.
\end{theorem}
The mirror reflection corresponding to partial transpose $\widetilde{\Lambda}$, with the transpose performed on the second mode, can be written as $\widetilde{\Lambda} = {\rm diag}(1,1,1,-1)$. The separability criterion can then be written as the additional requirement
\begin{align}
V + i \,\widetilde{\Omega} \geq 0, \text{ where }~ \widetilde{\Omega} = \widetilde{\Lambda}\,\Omega\,\widetilde{\Lambda},
\label{j46} \end{align}
over and above the uncertainty principle \,\eqref{j44}.

\section{Gaussian channels}
Having outlined the basic and fundamental properties of Gaussian states, we now consider the notion of Gaussian channels. Before we begin with the description of Gaussian channels, we wish to motivate the notion of Gaussian channels from the analogous classical setting. 

It is well known in classical probability theory that 
a Gaussian probability distribution denoted by
\begin{align}
{\cal P}_{\rm G}(\boldsymbol{\xi}) = \frac{1}{\sqrt{ (2\pi)^n {\rm Det V}}} ~ e^{-\frac{1}{2} \boldsymbol{\xi}^{\,\rm T} V^{-1} \boldsymbol{\xi}}
\label{j47} \end{align} 
remains Gaussian under all affine 
transformations of the form $\boldsymbol{\xi} \rightarrow A\boldsymbol{\xi} + \boldsymbol{b}$ and
convolutions with Gaussian distributions. The affine transformation 
$\boldsymbol{\xi} \rightarrow A\boldsymbol{\xi}$, induces the following  
transformation on the characteristic function, i.e.,
\begin{align}
\chi_{\rm G}(\boldsymbol{x}) &\rightarrow \chi_G(B \boldsymbol{x}), ~~B = (A^{-1})^T.
\label{j48} \end{align}
So we see that the translation by the vector $\boldsymbol{b}$ reflects as a linear phase factor in the characteristic 
function, and the homogeneous transformation $A$ reflects as
a corresponding homogeneous transformation $B=(A^{-1})^T$ on $\chi_{\rm G}(\boldsymbol{x})$.  
There are no restrictions on $A$ and $\boldsymbol{b}$ for a
Gaussian probability to be taken to a Gaussian probability under such a
transformation. 

The analogue of Gaussian probability distributions in quantum mechanics
are Gaussian Wigner distributions. 
It is true that
a Gaussian Wigner function is taken to a Gaussian probability under
all affine transformations. But to remain a valid Wigner distribution, additional
constraints have to be satisfied in the form of the
uncertainty principle in Eq.\,\eqref{j35} which we detail below.

The action of any  Gaussian channel on system A 
may be realized through the action of a  Gaussianity preserving unitary on a suitably
enlarged  system\,: 
\begin{align}
\rho_A \rightarrow \rho_A^{\,'} = \text{Tr}_B\left[ U_{AB} \,(\rho_A
  \otimes \rho_B) \, U_{AB}^{\dagger} \right]. 
\label{j49} \end{align}  
Here $\rho_{B}$ is a Gaussian state of the ancilla B, and $U_{AB}$ is a
linear canonical transformation on the enlarged composite system
consisting of the system of interest A and
the ancilla B. That all Gaussian channels can indeed be realized in
this manner has been shown by the work of Holevo and coauthors
\cite{holevo01,caruso06,holevo07,caruso08,caruso11}. 

For arbitrary input state with symmetric-ordered characteristic function 
$\chi_{W}(\xi;\rho)$, we have resulting from \eqref{j49}
\begin{align}
\chi^{\rm in}(\xi;\rho) \rightarrow \chi^{\rm out}(\xi;\rho) = 
\chi(X \xi;\rho) \exp\left[-\frac{\xi^T
    Y \xi}{2} \right],
\label{j51} 
\end{align}
where $X$ and $Y$ are real matrices, and $Y$ being positive definite. The pair $(X,Y)$ are completely specified by 
the unitary representation\,\eqref{j49}. 

So the action of a Gaussian channel thus 
manifests simply as a linear transformation on the variance matrix $V$. Under
the action of a Gaussian channel described by $(X, Y)$\,\cite{holevo01}\,:
\begin{align}
V \rightarrow V' = X^T V X +Y.
\label{j50} 
\end{align} 

Suppose we are instead given a general $(X,Y)$ which effect the transformation in\,\eqref{j50}.
For $V^{\,'}$ to be a valid variance matrix for arbitrary input, $(X,Y)$ have to satisfy a constraint 
that is a consequence of the uncertainty principle, which we detail below.

Let us consider the one-sided action of a Gaussian map described by $(X,Y)$  on a two-mode squeezed vacuum state with squeeze parameter $r$. The two-mode squeezed vacuum state is represented in the Fock basis as  
\begin{align}
|\psi_{r}\rangle ={\rm sech}\, r \sum_{k=0}^{\infty}({\rm tanh}\, r)^{k}|k,\,k\rangle,
\label{j51b}
\end{align} 
and its variance matrix is given by
\begin{align}
V_{\rm out}(r)=\left(
\begin{matrix}
c_{2r}  & 0 & s_{2r}&0 \\
0& c_{2r} & 0 & - s_{2r}\\ 
s_{2r} & 0 & c_{2r} & 0 \\
0& -s_{2r} & 0 & c_{2r}
\end{matrix} 
\right),
\label{51c}
\end{align}
where $c_{2r}=\cosh\, 2r$, $s_{2r}= \sinh\, 2r$.

The result of this one-sided action by the map $(X,Y)$ is a two-mode mixed Gaussian state specified by variance matrix
\begin{align}
V_{\rm out}(r)=\left(
\begin{matrix}
c_{2r} (X^TX) +Y && s_{2r}(X^T \sigma_3) \\
s_{2r} (\sigma_3 X) && c_{2r} ({1\!\!1}_2)
\end{matrix} 
\right),
\label{j52} \end{align}
$\sigma_3$ being the standard Pauli matrix.
It is clear that $V_{\rm out}(r)$ should obey the mandatory uncertainty principle
\begin{align}
V_{\rm out}(r) + i\, \Omega \geq 0, 
\label{j53} 
\end{align}
for all values of squeezing. In fact, this requirement in terms of the uncertainty principle 
is both a necessary and sufficient condition on $(X, Y)$ to correspond to a Gaussian
channel, and it may be restated in the form\,\cite{caruso06,wolf07}
\begin{align}
Y + i\,\Omega \geq i\, X^T\,\Omega\,X.
\label{j54} \end{align} 

\subsection{Canonical forms for quantum-limited and noisy channels} 
Given a Gaussian channel $\Gamma$ we can construct, `quite
cheaply', an entire family of Gaussian channels by simply preceding
and following $\Gamma$ with unitary (symplectic) Gaussian channels
$U(S_1), U(S_2)$ corresponding respectively to symplectic matrices
$S_1,\,S_2$. Therefore in classifying Gaussian 
channels it is sufficient to classify these orbits or double cosets
and, further, we may identify each orbit with the `simplest' looking
representative element of that orbit (the canonical form). Since 
\begin{align}
U(S_1) \,\Gamma \, U(S_2) \, :\, \chi(\xi) \rightarrow \chi(S_2\,X\,S_1\,
\xi) \exp[-\frac{1}{2} \xi^T\, S_1^T\, Y\, S_1\, \xi],
\label{j55} \end{align}
the task actually reduces to enumeration of the orbits of $(X,Y)$ under the
transformation $(X,Y) \rightarrow (X^{\,'}, Y^{\,'}) = (S_2\,X\,S_1,\,
S_1^T\,Y\,S_1)$.

We wish to make one important remark regarding Gaussian channels.
The injection of an arbitrary amount of classical
(Gaussian) noise into the state is obviously a Gaussian channel\,:
$\chi(\xi) \rightarrow \chi(\xi) \, \exp[-a\, |\xi|^2/2 ], \,
a>0$. It is called the classical noise channel. Now, given a Gaussian
channel we may follow it up with a classical noise channel to obtain
another Gaussian channel. A Gaussian channel will be said to be {\em
  quantum-limited} if it cannot be realized as another Gaussian
channel followed by a classical noise channel. Conversely, the most general
Gaussian channel is a quantum-limited Gaussian channel followed by a classical
noise channel, and it follows that quantum-limited channels are the
primary objects which need to be classified into orbits.

In other words, for a given $X$, the minimal $Y$, say $Y_0$, that saturates the inequality in\,\eqref{j54}
represents the threshold Gaussian noise that needs to be added to $\chi(X\xi)$
to make atonement for the failure of $X$ to be a symplectic matrix,
and thus rendering the map completely positive; if $X$ happens to be
a symplectic  matrix, then the corresponding minimal $Y_0=0$. And $Y \neq 0$ whenever
$X$ is not a symplectic matrix.

In the single-mode case where $(X,Y)$ are $2 \times 2$ matrices, $S_1,
S_2 \in Sp(2,R)$ can be so chosen that $X^{\,'}$ equals a multiple of
identity, a multiple of $\sigma_3$, or $(1\!\!1 + \sigma_3)/2$ while $Y^{\,'}$
equals a multiple of identity or $(1\!\!1 + \sigma_3)/2$. Thus the canonical
form of a Gaussian channel $X,Y$ is fully determined by the rank and
determinant of $(X,Y)$, and classification of
{\em quantum-limited bosonic  Gaussian channels} \cite{caruso06,
  holevo07} is shown in Table\,\ref{tablej1}

\begin{table}
\centering
\begin{tabular}{|c|c|c|c|}
\hline
Quantum-limited& $X$ & $Y_0$ & Noisy Channel\\
Channel $(X,Y_0)$ &&&~~~$Y= Y_0 + a 1\!\!1$~~~\\
\hline
~~${\cal D}(\kappa;0)$ ~~~~&~~~~$ -\kappa \sigma_3$ ~~~~&~~~~$
(1+\kappa^2) 1\!\!1$ \,~ $\kappa > 0$& ~${\cal D}(\kappa;a)$~\\
~~${\cal C}_1(\kappa;0)$ ~~~~&~~~~$ \kappa 1\!\!1$ ~~~~&~~~~$ (1-\kappa^2)1\!\!1$
\,~ $0 \leq \kappa < 1$&  ~${\cal C}_1(\kappa;a)$~\\
~~${\cal C}_2(\kappa;0)$ ~~~~&~~~~$ \kappa 1\!\!1$ ~~~~&~~~~$ (\kappa^2-1) 1\!\!1$
\,~ $\kappa > 1$ &  ~${\cal C}_2(\kappa;a)$~\\
~~${\cal A}_1(0)$ ~~~~&~~~~$0$ ~~~~&~~~~$1\!\!1$&~${\cal A}_1(a)$~\\
~~${\cal A}_2(0)$ ~~~~&~~~~$(1\!\!1 + \sigma_3)/2$ ~~~~&~~~~$1\!\!1$&~${\cal A}_2(a)$~\\
~~${\cal B}_2(0)$ ~~~~&~~~~$ 1\!\!1$ ~~~~&~~~~$0$& ~${\cal B}_2(a)$~\\
~~${\cal B}_1(0)$ ~~~~&~~~~$1\!\!1$ ~~~~&~~~~$ 0$& ~${\cal B}_1(a)$~\\
\hline
\end{tabular}
\caption{Showing the quantum-limited bosonic Gaussian channels. The noisy versions of these channels are obtained by replacing $Y_0$ by $Y= Y_0 + a 1\!\!1$ and so $Y >Y_0$.\label{tablej1}}
\end{table}

By following the above listed quantum-limited channels by injection of
classical noise of magnitude $a$ we get respectively ${\cal
  D}(\kappa;a)$, ${\cal C}_1(\kappa;a)$, ${\cal C}_2(\kappa;a)$,
${\cal A}_1(a)$, ${\cal A}_2(a)$, and  ${\cal B}_2(a)$; the last case
${\cal B}_1(a)$ is special in that it is obtained from ${\cal B}_1(0)$
by injection of noise into {\em just one quadrature}\,: $\chi(\xi)
\rightarrow \chi(\xi) \, \exp[-a\, \xi^T (1\!\!1 + \sigma_3) \xi/4 ]$. 

It is clear in the case of ${\cal D}(\kappa;0)$ that $X= -\kappa
\sigma_3$ corresponds to (scaled) phase conjugation or matrix
transposition of the density operator. And the phase conjugation is the most famous
among positive maps which are not CP \cite{peres96, horodecki96, simon00}; it is the injection of additional
classical noise of magnitude (not less than) $1+\kappa^2$, represented
by $Y_0$, that mends it into a CP map. It may be noted that the quantum-limited end of both the ${\cal B}_1$
and ${\cal B}_2$ families is the trivial identity channel.

The reason for the special emphasis on quantum-limited channels in our
enumeration of the Holevo classification is this\,: every noisy
Gaussian channel [except ${\cal B}_1(a)$] can be realized, as we shall
see later, as the composite of a pair of quantum-limited channels.  This fact will be exploited to study an application in Chapter 4. 

\subsection{Operator sum representation}
We now briefly touch upon the operator sum representation of single-mode bosonic Gaussian channels\,\cite{kraus10}. The operator sum representation was obtained by considering the unitary representation of Gaussian channels. The system is first appended with a fixed Gaussian environment state (vacuum state for example), then the joint system is evolved through a two-mode Gaussian unitary transformation, and finally the environment mode is traced out in a suitable basis (Fock states for example) to obtain the resulting Kraus operators. 

The Kraus operators thus constructed in \cite{kraus10} for all the quantum-limited channels and the classical noise channels is presented in Table\,\ref{tablej2}. We see that in this representation, the beamsplitter, amplifier and phase conjugation channels have a discrete index Kraus representation whereas the classical noise channels and the singular channels have a continuous index Kraus representation. 
 
\begin{table}
\centering
\begin{tabular}{|c|c|c|}
\hline
Channel & Kraus operators & OSR \\
\hline
~~${\cal D}(\kappa;0)$ &$T_{\ell}(\kappa) = \sum_{n=0}^{\ell} (\sqrt{1+\kappa^2})^{-(n+1)}
 (\sqrt{1+\kappa^{-2}})^{-(\ell-n)} \sqrt{{}^{\ell}C_n}   |\ell-n
 \rangle \langle n |$&$\sum_{\ell=0}^{\infty} T_{\ell}(\kappa)\, (\cdot)\,T_{\ell}(\kappa)^{\dagger}$\\
~~${\cal C}_1(\kappa;0)$ &$
B_{\ell}(\kappa)= \sum_{m=0}^{\infty} \sqrt{{}^{m+\ell} C_{\ell}}\,
(\sqrt{1-\kappa^2})^{\ell}\, {\kappa}^{m}  
| m \rangle \langle m+\ell|$& $\sum_{\ell=0}^{\infty} B_{\ell}(\kappa)\, (\cdot)\,B_{\ell}(\kappa)^{\dagger} $\\
~~${\cal C}_2(\kappa;0)$ &$A_{\ell}(\kappa)= \kappa^{-1} \sum_{m=0}^{\infty} \sqrt{{}^{m+\ell}
  C_{\ell}} \left(\sqrt{1-\kappa^{-2}}\right)^{\ell} (\kappa^{-1})^{m}  
| m +\ell \rangle \langle m |$&$ \sum_{\ell=0}^{\infty} A_{\ell}(\kappa)\,(\cdot)\, A_{\ell}(\kappa)^{\dagger}$\\
~~${\cal A}_1(0)$ &$B_k = |0\rangle \langle k |$& $\sum_k B_k \,(\cdot)\,B_k^{\dagger}$\\
\hline
~~${\cal A}_2(0)$ &$V_q =  | q/\sqrt{2} ) \, \langle q |$&$ \int dq\, V_q \,( \cdot)\,V_q^{\dagger}$\\
~~${\cal B}_2(a)$ &$D_{\alpha} = (\pi a)^{-1/2}\,\exp[-|\alpha|^2/2a]\,{\cal D}(\alpha) $&$\int d^2 \alpha\,D_{\alpha}\,(\cdot)\,D_{\alpha}^{\dagger}$\\
~~${\cal B}_1(a)$ &$Z_q \equiv (\pi
a)^{-1/4} \exp[-q^2/2a]\, {\cal D}(q/\sqrt{2})$&$\int dq \,Z_q \,(\cdot)\, Z_q^{\dagger}$\\
\hline
\end{tabular}
\caption{Showing the OSR of the quantum-limited bosonic Gaussian channels and the classical noise channels.\label{tablej2}}
\end{table}

Of particular interest to us is the action of the beamsplitter and amplifier channels on the Fock basis. We first consider the quantum-limited beamsplitter channel. We wish to consider the action of the channel on the operator basis consisting of the Fock operators $\{|m\rangle \langle n| \}$. From Table\,\ref{tablej1}, we see that the action of the quantum-limited beamsplitter channel on a general operator $|m\rangle \langle n|$ is given by\,: 
\begin{align}
|m \rangle \langle n | & \rightarrow \sum_{\ell=0}^{\infty} B_{\ell}(\kappa)
|m\rangle \langle n| B_{\ell}^{\dagger}(\kappa) \nonumber \\ 
& = \sum_{{\ell}=0}^{\text{min}\{m,n \}} \sqrt{{}^m C_{\ell}\, {}^n C_{\ell}}\,
{({1-\kappa^2})}^{\ell} {\kappa}^{m+n-2{\ell}} 
|m-{\ell} \rangle \langle n-{\ell}|. 
\label{j56} \end{align}
We see that the resulting operator is of finite rank. Further, for an input Fock state $|n\rangle \langle n|$, the output consists of all Fock state projectors up to the value $n$.

A similar analysis of the action of the quantum-limited amplifier channel on the operator $|m\rangle \langle n|$ leads to\,:
\begin{align}
{|m \rangle \langle n|} &\rightarrow \sum_{\ell=0}^{\infty} {A}_{\ell}(\kappa) |m \rangle \langle n|
{A}_{\ell}^{\dagger}(\kappa) \nonumber \\ 
&={\kappa}^{-2} {\kappa}^{-(n+m)} \sum_{\ell=0}^{\infty} 
\sqrt{{}^{n+\ell}C_{\ell} \, {}^{m+\ell}C_{\ell}} \,(1-\kappa^{-2})^{\ell}\,|m+{\ell}\rangle \langle n+{\ell} |.
\label{j57} 
\end{align} 
In contrast to the quantum-limited beamsplitter case, we see that the output operator is of infinite rank. 



\subsection{Semigroup property}
It is clear from Table\,\ref{tablej1} (action in phase space) that successive actions of
two quantum-limited beamsplitter channels with parameter values ${\kappa}_1, {\kappa}_2$ is a quantum-limited beamsplitter 
channel whose parameter ${\kappa}$ equals the product ${\kappa}_1{\kappa}_2$ of the 
individual channel parameters \,:
\begin{align}
{\cal C}_1(\kappa_1)\,:  ~~\chi_W(\xi) \rightarrow \chi{\,'}_W(\xi) &= \chi_W({\kappa}_1  \,\xi) ~
\exp{[-(1-{\kappa}_1^{2}) |\xi|^2/2]}, \nonumber \\
{\cal C}_1(\kappa_2)\,:  ~~ \chi{\, '}_W(\xi) \rightarrow \chi{\,''}_W(\xi) &= \chi{\, '}_W({\kappa}_2  \,\xi) ~
\exp{[-(1-{\kappa}_{2}^{2}) |\xi|^2/2]} \nonumber \\
&= \chi_W({\kappa}_1{\kappa}_2  \,\xi) ~\exp{[-(1-{{\kappa}}_{1}^{2} {{\kappa}}_{2}^{2}
  )|\xi|^2/2]}. 
\label{j58} \end{align} 
It is instructive to see how this semigroup property emerges in the 
Kraus representation. Let $\{B_{\ell_1}(\kappa_1)\}$ and
$\{B_{\ell_2}(\kappa_2) \}$ be the Kraus operators of the two
channels. The product of two Kraus operators
$B_{{\ell}_1}{({\kappa_1})}$, $B_{{\ell}_2}{({\kappa_2})}$, one
from each set, is 
\begin{align}
B_{{\ell}_1}({\kappa}_1) B_{{\ell}_2}({\kappa}_2) = &\sum_{m =0}^{\infty}\,
\sqrt{{}^{{\ell}_1+{\ell}_2} C_{\ell_1}}
\left(\sqrt{1-\kappa_1^2}\right)^{\ell_1}
\left(\sqrt{1-\kappa_2^2}\right)^{\ell_2} \nonumber \\
&\times  \sqrt{{}^{m + {\ell}_1+{\ell}_2} C_{{\ell}_1+{\ell}_2}}
\, ({\kappa}_1{\kappa}_2)^{m} \,{\kappa}_2^{{\ell}_1}  |m 
  \rangle\langle m + {\ell}_1 + {\ell}_2|. 
\label{j59} \end{align}
The action of the product channel on the input operator 
$|r \rangle \langle r+\delta|$ is 
\begin{align}
&\sum_{{\ell}_1,{\ell}_2} B_{{\ell}_1}({\kappa}_1) B_{{\ell}_2}({\kappa}_2) |r \rangle \langle r+\delta|
B_{{\ell}_2}({\kappa}_2)^{\dagger} \, B_{{\ell}_1}({\kappa}_1)^{\dagger} \nonumber \\
&= \sum_{{\ell}_1,{\ell}_2,m,n} \sqrt{{}^{{\ell}_1+{\ell}_2} C_{\ell_1}}  \left(\sqrt{1-\kappa_1^2}\right)^{\ell_1} \left(\sqrt{1-\kappa_2^2}\right)^{\ell_2}
\sqrt{{}^{m + {\ell}_1+{\ell}_2} C_{{\ell}_1+{\ell}_2}}
\,({\kappa}_1{\kappa}_2)^{m} {\kappa}_2^{{\ell}_1}  
  \nonumber \\
&\,\,~~~\,\,\,\,\times  \sqrt{{}^{{\ell}_1+{\ell}_2} C_{\ell_1}} \, \left(\sqrt{1-\kappa_1^2}\right)^{\ell_1} \left(\sqrt{1-\kappa_2^2}\right)^{\ell_2}
\sqrt{{}^{n + {\ell}_1+{\ell}_2} C_{{\ell}_1+{\ell}_2}}
\,({\kappa}_1{\kappa}_2)^{n} {\kappa}_2^{{\ell}_1} 
\nonumber\\ 
& ~~~~~~\times |m
  \rangle\langle m + {\ell}_1 + {\ell}_2 |r \rangle \langle
  r+\delta| n+ {\ell}_1 + {\ell}_2
  \rangle\langle n |.
\label{j60} \end{align}
Denoting ${\ell}_1+{\ell}_2 = {\ell}$, the expression on the RHS of Eq.\,\eqref{j60} becomes 
\begin{align}
\text{RHS  }= \sum_{{\ell}=0}^{r} \sum_{{\ell}_1=0}^{{\ell}} \sum_{m,n=0} ^{\infty}
&{}^{{\ell}}C_{{\ell}_1} {\kappa}_2^{2{\ell}_1}
{({1-\kappa_1^2})}^{{\ell}_1} 
{({1-\kappa_2^2})}^{({\ell} -{\ell}_1)}
({\kappa}_1{\kappa}_2)^{m+n} \nonumber\\
&\times \sqrt{{}^{{\ell}+m} C_{\ell} {}^{{\ell}+n} C_{\ell}} 
\, \delta_{r,m+\ell} \, \delta_{r+\delta, n+\ell} \, 
|m \rangle \langle n |.
\label{j61} \end{align}
The sum over ${\ell}_1$ is the binomial expansion of
$[({{1-\kappa_1^2})}{\kappa}_2^2+{({1-\kappa_2^2})}]^{\ell} =
(1-\kappa_1^2\kappa_2^2)^{\ell}$ and, in addition, we have the
constraints 
$m + {\ell} =r$ and $n + {\ell} = r+ \delta$. With this the
expression (\ref{j61}) reduces to 
\begin{align}
\text{RHS }=\sum_{{\ell}=0}^{r} \, (1-\kappa_1^2\kappa_2^2)^{\ell} \, \sqrt{ {}^{r} C_{\ell}\,
  {}^{r+\delta} C_{\ell}} \, ({\kappa}_1{\kappa}_2)^{2r-2{\ell} + \delta} |r-{\ell} \rangle
\langle r-{\ell} + \delta|. 
\label{j62} \end{align} 
Comparing Eqs.\,\eqref{j62} and (\ref{j56}) we find that the expression in
(\ref{j62}) is precisely the action of a quantum-limited
attenuator channel with parameter ${\kappa}_1{\kappa}_2$. In other words, we have that 
\begin{align}
\sum_{{\ell}_1,{\ell}_2 =0}^{\infty} B_{{\ell}_1}({\kappa}_1) B_{{\ell}_2}({\kappa}_2) |r \rangle \langle r+\delta|
B^{\dagger}_{{\ell}_2}({\kappa}_2) B^{\dagger}_{{\ell}_1}({\kappa}_1)
= \sum_{\ell=0}^{\infty} \,B_{\ell}({\kappa}_1{\kappa}_2) |r \rangle
\langle r+\delta|  B_{\ell}^{\dagger}({\kappa}_1{\kappa}_2).
\label{j63} \end{align}  
\noindent
An identical result can be similarly obtained for the behaviour of $|r
+ \delta \rangle \langle r|$, and thus we have proved the semigroup
property 
\begin{align}
{\cal C}_1(\kappa_1)\circ {\cal C}_1(\kappa_2) = {\cal C}_1(\kappa_1\kappa_2).
\label{j64} \end{align}

We now analyze the composition of two quantum-limited amplifier channels, as in the beamsplitter channel case.
It follows from the very definition of the amplifier channel that the
composition of two quantum-limited amplifier channels
 with parameters $\kappa_1$ and $\kappa_2$ is also a quantum-limited amplifier channel with
 parameter $\kappa = \kappa_1\kappa_2 >1$\,:
\begin{align}
{\cal C}_2(\kappa_2) \circ {\cal C}_2(\kappa_1): \, \chi_W(\xi) \rightarrow \chi^{\,'}_W(\xi) =\chi_W(\kappa_1 \kappa_2 \,\xi) ~
\exp{[-(\kappa_{1}^{2} \kappa_{2}^{2}  -1 ) |\xi|^2/2]}.
\label{j65a} \end{align}
That is, 
\begin{align}
{\cal C}_2(\kappa_2) \circ {\cal C}_2(\kappa_1) = {\cal C}_2(\kappa_1
\kappa_2) = {\cal C}_2(\kappa_1) \circ {\cal C}_2(\kappa_2). 
\label{j65} \end{align}
It will be instructive to examine how this fact emerges from the structure of
the Kraus operators. Let the set 
$\{ A_{\ell_1}(\kappa_1)\}$ be the Kraus operators of the first amplifier and
let $\{ A_{\ell_2}(\kappa_2)\}$ be that of the second. Then the product of
a pair of Kraus operators, one from each set, is  
\begin{align}
A_{\ell_1}(\kappa_1) A_{\ell_2}(\kappa_2) &= (\kappa_1 \kappa_2)^{-1} \sqrt{{}^{\ell_1+\ell_2}C_{\ell_1}}
\sum_{n=0}^{\infty} \sqrt{{}^{n+\ell_1+\ell_2}C_{\ell_1+\ell_2}}
\left(\sqrt{1-\kappa_1^{-2}}\right)^{\ell_1}
 \nonumber \\ 
~~&~~~~~~~ \times\left(\sqrt{1-\kappa_2^{-2}} \right)^{\ell_2} (\kappa_1\kappa_2)^{-n} \kappa_1^{-\ell_2} |n+\ell_1+\ell_2 \rangle \langle n|. 
\label{j66} \end{align}  
Thus, under the successive action of these two amplifier channels the operator
$|j \rangle \langle j+\delta|$ goes to
\begin{align}
&\sum_{{\ell}_1,{\ell}_2} A_{{\ell}_1}(\kappa_1) A_{{\ell}_2}(\kappa_2) |j \rangle \langle j+\delta|
A_{{\ell}_2}(\kappa_2)^{\dagger} A_{{\ell}_1}(\kappa_1)^{\dagger} \nonumber \\
&=(\kappa_1 \kappa_2)^{-2} \sum_{{\ell}_1,{\ell}_2}\sum_{n=0}^{\infty}\sum_{m=0}^{\infty} 
{}^{{\ell}_1+{\ell}_2}C_{{\ell}_1}\,  \left({1-\kappa_1^{-2}}\right)^{\ell_1} \left({ 1-\kappa_2^{-2}}\right)^{\ell_2} (\kappa_1\kappa_2)^{-(n+m)}
\kappa_1^{-2{\ell}_2} \nonumber \\ 
& ~~~~~~~~\times
\,\sqrt{{}^{n+{\ell}_1+{\ell}_2}C_{{\ell}_1+\ell_2}\,{}^{m+{\ell}_1+{\ell}_2}C_{{\ell}_1+\ell_2}} 
|n+{\ell}_1+{\ell}_2 \rangle \langle n| j \rangle \langle j+\delta| m \rangle
\langle m +{\ell}_1 +{\ell}_2|.  
\label{j67} \end{align}
Denoting ${\ell}_1+{\ell}_2 = {\ell}$, the RHS of the expression in Eq.\,\eqref{j67}
reduces to  
\begin{align}
&(\kappa_1 \kappa_2)^{-2} \,  \sum_{{\ell}=0}^{\infty} \sum_{{\ell}_1=0}^{{\ell}}
\sum_{n=0}^{\infty}\sum_{m=0}^{\infty} \,
{}^{{\ell}}C_{{\ell}_1}  \, (1-\kappa^{-2}_1)^{{\ell}_1} \, 
(\kappa_1^{-2} (1-\kappa_2^{-2}))^{({\ell}- {\ell}_1)}
 \, (\kappa_1\kappa_2)^{-(n+m)} \nonumber \\ 
 &~~~~\times \sqrt{{}^{n+{\ell}}C_{{\ell}}\, {}^{m+{\ell}}C_{{\ell}}} \,
\, \delta_{m,j+\delta} \, \delta_{n,j}\,
|n+{\ell}\rangle \langle n+\delta +{\ell}|. 
\label{j68} \end{align}
As in the beamsplitter case, the summation over the index $\ell_1$ is
a binomial expansion, and 
the expression in Eq.\,\eqref{j68} reduces to
\begin{align}
 (\kappa_1 \kappa_2)^{-2}  \sum_{{\ell}=0}^{\infty} \,
(1-\kappa_1^{-2}\kappa_2^{-2} )^{\ell} \,
(\kappa_1\kappa_2)^{-(j+j+\delta)} \,
\sqrt{{}^{j+\ell}C_{{\ell}}\,{}^{j+\ell+\delta}C_{{\ell}}} \, |j+{\ell}
\rangle \langle j+{\ell}+\delta|.
\label{j69} \end{align}
Comparing Eqs.\,\eqref{j69} and \eqref{j57}, we see that the latter
is the Kraus representation for a single quantum-limited amplifier
channel. That is, 
\begin{align}
\sum_{{\ell}_1,{\ell}_2} A_{{\ell}_1}(\kappa_1) A_{{\ell}_2}(\kappa_2) |j \rangle \langle j+\delta|
A_{{\ell}_2}(\kappa_2)^{\dagger} A_{{\ell}_1}(\kappa_1)^{\dagger} 
=  \sum_{\ell}  A_{\ell}(\kappa_1\kappa_2)|j \rangle \langle j+\delta|
A_{{\ell}}(\kappa_1\kappa_2)^{\dagger}.
\label{j70} \end{align}
A similar behaviour holds for $|j+\delta \rangle \langle j|$ as well. 
And this is what we set out to demonstrate. \\


\subsection{Noisy channels from quantum-limited ones}
Our considerations so far have been in respect of quantum-limited
channels. We turn our attention now to the case of noisy channels. It
turns out that every noisy channel, except ${\cal B}_1(a)$ which
corresponds to injection of classical noise in just one quadrature,
can be realised (in a non-unique way) as composition of two
quantum-limited channels, so that the Kraus operators are products of
those of the constituent quantum-limited channels.

We have noted already in the previous subsection that the composition of two quantum-limited 
attenuator (or amplifier) channels is again a quantum-limited
attenuator (or amplifier) channel. This special semigroup property
however does not obtain under composition for other quantum-limited
channels. In general, composition of two quantum-limited channels 
results in a channel with additional classical noise. We will now consider pairs of quantum-limited channels 
from Table\,\ref{tablej1} and construct the Kraus operators of the resulting noisy channel.

\subsubsection{The composite ${\cal C}_2(\kappa_2;0) \circ {\cal C}_1(\kappa_1;0), ~
  \kappa_2 > 1, ~ \kappa_1 < 1$ \label{comp1}}
It is clear from the very definition of these channels through their action
on the characteristic function that the composite ${\cal
  C}_2(\kappa_2;0) \circ {\cal C}_1(\kappa_1;0) $ is a noisy
amplifier, a classical noise channel, or 
a noisy attenuator depending on the numerical value of $\kappa_2\kappa_1$\,:
it equals ${\cal C}_1(\kappa_2\kappa_1; 2(\kappa_2^2-1))$ for
$\kappa_2\kappa_1 < 1$, ${\cal B}_2(2(\kappa_2^2-1))$ for
$\kappa_1\kappa_2=1$, and ${\cal C}_2(\kappa_2\kappa_1;
2\kappa_2^2(1-\kappa_1^2))$ for 
$\kappa_2\kappa_1 > 1 $, as may be readily read off from Table \ref{tablej3}. 
 
{\scriptsize
\begin{table}
\begin{center}
\begin{tabular}{|c|c|c|c|c|}
\hline
\hline
$X ,  Y \rightarrow$  & ${\cal D}(\kappa_1;0)$  & ${\cal C}_1(\kappa_1;0)$ &
${\cal C}_2 (\kappa_1;0)$ & ${\cal A}_2(0)$ \\
$\downarrow$~~~~~~~~ &&&&\\
\hline 
  & ${\cal C}_1(\kappa_2\kappa_1;2\kappa_2^2(1+\kappa_1^2))$,
& &  & \\
${\cal D}(\kappa_2;0)$ &for $\kappa_2\kappa_1 < 1$.&${\cal D}(\kappa_2\kappa_1; 2\kappa_2^2(1- \kappa_1^2)$&${\cal
  D}(\kappa_2\kappa_1;0 )$&${\cal A}_2(2\kappa_2^2) $\\
&${\cal C}_2(\kappa_2 \kappa_1;2(1+\kappa_2^2))$, &&&\\
&for $\kappa_2\kappa_1 >1 $.&&&\\
&${\cal B}_2(2(1+\kappa_2^2))$,&&&\\
&for $\kappa_2\kappa_1 =1 $.&&&\\
\hline
  &  &  & ${\cal
   C}_1(\kappa_2\kappa_1;2\kappa_2^2(\kappa_1^2-1))$,&\\
 ${\cal C}_1(\kappa_2;0)$ & ${\cal D}(\kappa_2\kappa_1;0)$&${\cal
   C}_1(\kappa_2\kappa_1;0)$&for $\kappa_2\kappa_1 < 1$.& ${\cal A}_2(0)$\\ 
&&&${\cal C}_2(\kappa_2\kappa_1;2(1-\kappa_2^2))$,&\\ 
&&&for $\kappa_2 \kappa_1 >1 $.&\\
&&&${\cal B}_2(2(1-\kappa_2^2)) $&\\
&&&for $\kappa_2 \kappa_1 = 1 $&\\
\hline
&
& ${\cal C}_1(\kappa_2\kappa_1;2(\kappa_2^2-1))$,  &  &\\
${\cal C}_2(\kappa_2;0)$ & ${\cal
  D}(\kappa_2\kappa_1;2(\kappa_2^2-1))$&for $\kappa_2\kappa_1 <
1$.&${\cal C}_2(\kappa_2 
\kappa_1;0)$&${\cal A}_2(2(\kappa_2^2-1)) $\\ 
&&${\cal C}_2(\kappa_2\kappa_1; 2\kappa_2^2(1-\kappa_1^2))$, &&\\
&&for $\kappa_2 \kappa_1 > 1$.&&\\
&&${\cal B}_2(2(\kappa_2^2-1))$.&&\\
&&for $\kappa_2 \kappa_1 =1$.&&\\
\hline
${\cal A}_2(0)$ &${\cal A}_2(\sqrt{\kappa_1^2+2} -1)$&${\cal
  A}_2(\sqrt{2-\kappa_1^2}-1)$&${\cal A}_2(\kappa_1-1)$&${\cal
  A}_2(\sqrt{2}-1)$\\ 
\hline 
\hline
\end{tabular}
\end{center}
\caption{Showing the composition $X \circ Y$ of quantum-limited channels
  $X,Y$ {\em assumed to be in their respective canonical forms
    simultaneously}. The composition results, in several cases, in
  noisy channels thereby enabling description of noisy Gaussian
  channels, including the classical noise channel ${\cal B}_2(a)$, in terms of {\em
    discrete sets} of linearly independent Kraus operators.  \label{tablej3}}
\end{table}
}

The Kraus operators for the composite is given by the set
$\{A_m(\kappa_2) B_{n}(\kappa_1)\} $ with $(m,n)$ running
independently over the range $0 \leq m,n < \infty$. 
By computing the products $A_m(\kappa_2) B_n(\kappa_1)$, we have, 
\begin{align}
A_{\ell + \delta}(\kappa_2) B_{\ell}(\kappa_1) &= \sum_{j=0}^{\infty}
g_1(\delta)_{\ell j} | j + \delta \rangle \langle j|, \nonumber \\
A_{\ell}(\kappa_2) B_{\ell+\delta}(\kappa_1) &= \sum_{j=0}^{\infty}
\tilde{g}_1(\delta)_{\ell j} | j \rangle \langle j+\delta|,\nonumber \\
g_1(\delta)_{\ell j} &= \kappa_2^{-1} \,(\kappa_2^{-1}\kappa_1)^{j-\ell}\, \sqrt{{}^{j+\delta}C_{\ell +
    \delta} \, {}^{j}C_{\ell}}\nonumber\\
&~~~~~~~~\times \left(\sqrt{1-\kappa_1^2}\right)^{\ell} \,
\left(\sqrt{1-\kappa_2^{-2}}\right)^{\ell+ \delta}, \text{  for } j\geq \ell, \nonumber\\
&= 0, \text{  for } j < \ell;\nonumber \\ 
\tilde{g}_1(\delta)_{\ell j} &= \kappa_2^{-1} \,(\kappa_2^{-1}\kappa_1)^{j-\ell}\, \sqrt{{}^{j+\delta}C_{\ell +
    \delta} \, {}^{j}C_{\ell}}  \nonumber\\
&~~~~~~~~~~\times \left(\sqrt{1-\kappa_2^{-2}}\right)^{\ell} \,
\left(\sqrt{1-\kappa_1^{2}}\right)^{\ell + \delta}, \text{  for } j\geq \ell,
\nonumber\\ 
&= 0, \text{  for } j < \ell.
\label{j71} 
\end{align}

\subsubsection{ The composite ${\cal C}_1(\kappa_2;0) \circ {\cal C}_2(\kappa_1;0), ~
  \kappa_2 < 1, ~ \kappa_1 > 1$ \label{comp2}}
Again the composite ${\cal C}_1(\kappa_2;0) \circ {\cal C}_2(\kappa_1;0)$ is 
a noisy amplifier, a classical noise channel, or a noisy attenuator
depending on the numerical value of
$\kappa_2\kappa_1$ and the details may be read off from Table \ref{tablej3}. The
Kraus operators for the composite ${\cal C}_1(\kappa_2;0) \circ {\cal
C}_2(\kappa_1;0)$ are given by $\{B_m(\kappa_2) A_n(\kappa_1) \}$, $0
\leq m,n < \infty$. We have 
\begin{align}
&B_{\ell + \delta}(\kappa_2) A_{\ell}(\kappa_1) = \sum_{j=0}^{\infty}
g_2(\delta)_{\ell j} |j \rangle \langle j+ \delta|, \nonumber \\
&B_{\ell}(\kappa_2) A_{\ell+\delta}(\kappa_1) = \sum_{j=0}^{\infty}
\tilde{g}_2(\delta)_{\ell j} |j +\delta \rangle \langle j|, \nonumber \\ 
&g_2(\delta)_{\ell j}= \kappa_1^{-1}\,\sqrt{{}^{j+\ell+\delta}C_{\ell + \delta} \,
  {}^{j+\delta+\ell}C_{\ell}} \, \kappa_1^{-(j+\delta)}  \left(\sqrt{1-\kappa^{-2}_1}\right)^{\ell}\,
\kappa_2^{j} \left(\sqrt{1-\kappa^2_2}\right)^{\ell+ \delta},\nonumber \\
&\tilde{g}_2(\delta)_{\ell j}= \kappa_1^{-1}\,\sqrt{{}^{j+\ell+\delta}C_{\ell + \delta} \,
  {}^{j+\delta+\ell}C_{\ell}} \, \kappa_1^{-j}  \left(\sqrt{1-\kappa^{-2}_1}\right)^{\ell+\delta}\,
\kappa_2^{j+\delta} \left(\sqrt{1-\kappa^2_2}\right)^{\ell}.
\label{j72} \end{align}

\subsubsection{The composite $ {\cal D}(\kappa_2) \circ {\cal
    D}(\kappa_1), ~ \kappa_2 > 0,\,   \kappa_1 > 0$ \label{comp3}}
Similar to the earlier two cases, the composite ${\cal D}(\kappa_2;0)
\circ {\cal D}(\kappa_1;0)$ is a noisy amplifier, a classical noise
channel, or a noisy attenuator depending
on the numerical value of $\kappa_2\kappa_1$, as in the earlier two cases, and the details can be read off
from Table \ref{tablej3}. It may be noted, again from Table \ref{tablej3},
that this case tends to be more noisy 
than the earlier two cases.

The Kraus operators for this composite are given by
$\{T_m(\kappa_2)T_n(\kappa_1)\}, ~ 0 \leq\, m,n < \infty $. 
The products $T_m(\kappa_2)T_n(\kappa_1)$ have the form 
\begin{align}
&T_{\ell + \delta}(\kappa_2)T_{\ell}(\kappa_1) = \sum_{j=0}^{\infty}
g_3(\delta)_{\ell j} |j+\delta \rangle \langle j|, \nonumber \\ 
&T_{\ell}(\kappa_2)T_{\ell + \delta}(\kappa_1) = \sum_{j=0}^{\infty}
\tilde{g}_3(\delta)_{\ell j} |j+\delta \rangle \langle j|, \nonumber \\  
&g_3(\delta)_{\ell j} = \left(\sqrt{1+\kappa_1^2}\right)^{-1} 
\left(\sqrt{1+\kappa_2^2}\right)^{-1}\,
\sqrt{{}^{\ell}C_{j}\, {}^{\ell+\delta}C_{j}} \,
\left[\sqrt{(1+\kappa_2^2)(1+\kappa_1^{-2})}  \right]^{-(\ell -j)}
\,\nonumber \\
&~~~~~~~~~~~~~~\times \left[\sqrt{(1+\kappa_1^2)(1+\kappa_2^{-2})}\right]^{-j} \,
\left(\sqrt{1+\kappa_2^{-2}}\right)^{-\delta}, ~~ \text{for } j\leq \ell, \nonumber \\ 
&~~~~~~~~~~=0, ~~\text{for } j>\ell, \nonumber \\ 
&\tilde{g}_3(\delta)_{\ell j} = \left(\sqrt{1+\kappa_1^2}\right)^{-1}
\left(\sqrt{1+\kappa_2^2}\right)^{-1}\, 
\sqrt{{}^{\ell}C_{j}\, {}^{\ell+\delta}C_{j}} \,
\left[\sqrt{(1+\kappa_2^2)(1+\kappa_1^{-2})}\right]^{-(\ell -j)}
\,\nonumber \\ 
&~~~~~~~~~~~~~~~~~\times \left[\sqrt{(1+\kappa_1^2)(1+\kappa_2^{-2})}\right]^{-j} \,
\left(\sqrt{1+\kappa_1^2}\right)^{-\delta}, ~~ \text{for } j\leq \ell, \nonumber \\ 
&~~~~~~~~~~=0, ~~\text{for } j>\ell. 
\label{j73} \end{align} 

\subsubsection{The composite  ${\cal D}(\kappa_2;0) \circ {\cal C}_1(\kappa_1;0), ~ 
  \kappa_2 > 0, \,  \kappa_1  < 1$ \label{comp4}}
Kraus operators of this composite, which always corresponds to a noisy
transpose channel (see Table \ref{tablej3}), are $\{T_m(\kappa_2) B_n(\kappa_1) \}, ~ 0 \leq m,n
< \infty$. We have 
\begin{align}
T_m(\kappa_2)B_n(\kappa_1) &= \sum_{j=0}^{\infty} \xi^{j}_{mn} |m-j \rangle
\langle n+j|\nonumber\\
\xi^j_{mn} &= \left(\sqrt{1+\kappa_2^2}\right)^{-1} \sqrt{{}^mC_j\,
  {}^{n+j}C_j} \, \left(\sqrt{1+\kappa_2^2}\right)^{-j}\, 
\left(\sqrt{1+\kappa_2^{-2}}\right)^{-(m-j)}\nonumber\\
&~~~~~~\times \kappa_1^j \left(\sqrt{1-\kappa_1^2}\right)^{n}, \text{ for }
j \leq m; \nonumber \\
&=0, \, \text{ for } j>m.
\label{j74} \end{align}  


\subsubsection{The composite ${\cal C}_1(\kappa_2;0) \circ {\cal D}(\kappa_1;0), ~ 
  \kappa_1 > 0,\,  \kappa_2  < 1$ \label{comp5}}
This composite channel corresponds to a {\em quantum-limited} transpose
channel (see Table \ref{tablej3}).
The Kraus operators $\{B_m(\kappa_2) T_n(\kappa_1) \}, ~ 0 \leq m,n
< \infty $ (which as a set should be equivalent to
$\{T_{\ell}(\kappa_2\kappa_1) \}, \, 0 \leq \ell < \infty$), are 
\begin{align}
B_m(\kappa_2) T_n(\kappa_1) &= \sum_{j=m}^{n}\, \xi^j_{mn} |j-m
\rangle \langle n-j|, \nonumber \\
\xi^j_{mn} &= \sqrt{{}^{j}C_{m}\,
  {}^{n}C_{j}}\, \left(\sqrt{1-\kappa_2^2}\right)^{m}\, \kappa_2^{j-m}\,
\left(\sqrt{1+\kappa_1^2}\right)^{-(n-j+1)}\nonumber\\
&~~~~~ \times \,\left(\sqrt{1+\kappa_1^{-2}}\right)^{-j}, \text{ for } n\geq m ;
\nonumber \\
&= 0, \text{ for } n<m.  
\label{j75} \end{align}

\subsubsection{The composite ${\cal C}_2(\kappa_2;0) \circ {\cal D}(\kappa_1;0), ~ 
  \ \kappa_2  > 1 ,\, \kappa_1 > 0$ \label{comp6}} 
This composite channel corresponds, for all $\kappa_1, \kappa_2$,  to
a noisy transpose channel, 
similar to the case of ${\cal D}(\kappa_2;0)\circ {\cal C}_1(\kappa_1;0)$
considered earlier. The Kraus operators $\{A_m(\kappa_2) T_n(\kappa_1)
\}, ~ 0 \leq m,n < \infty$ have the form 
\begin{align}
A_m(\kappa_2) T_n(\kappa_1) &= \sum_{j=0}^n \, \xi^j_{mn} |j+m \rangle
\langle n-j|, \nonumber \\
\xi^j_{mn} &= \kappa_2^{-1}\left(\sqrt{1+\kappa_1^2}\right)^{-1} 
\, \sqrt{{}^{m+j}C_{j}
  \, {}^{n}C_{j}} \, \left(\sqrt{1-\kappa_2^{-2}}\right)^{m} \, \kappa_2^{-j}\,\nonumber\\
&~~~~~~~ \times \left(\sqrt{1+\kappa_1^2}\right)^{-(n-j)}  
\left( \sqrt{1+\kappa_1^{-2}}\right)^{-j}, \text{ for } j\leq n;
\nonumber \\
&=0, \text{ for } j>n. 
\label{j76} \end{align}

\subsubsection{The composite ${\cal D}(\kappa_2;0)\circ {\cal C}_2(\kappa_1;0), ~ \kappa_2
   > 0,\, \kappa_1 > 1$ \label{comp7}}
This composite is a {\em quantum-limited} transpose channel (see Table \ref{tablej3}),
with Kraus operators $\{T_m(\kappa_2) A_n(\kappa_1) \}, ~ 0 \leq m,n <
\infty$. The product Kraus operators are computed as 
\begin{align}
T_m(\kappa_2) A_n(\kappa_1) &= \sum_{k=0}^m \xi^k_{mn} |m-k \rangle \langle k-n|,\nonumber\\
\xi^k_{mn} & = (\sqrt{1+\kappa_2^2})^{-(k+1)} (\sqrt{1+\kappa_2^{-2}})^{-(\ell-k)} \sqrt{{}^{\ell}C_k}  \nonumber\\
&~~~\times  \kappa_1^{-1}  \sqrt{{}^{k}C_{n}} \left(\sqrt{1-\kappa_1^{-2}}\right)^{n} (\kappa_1^{-1})^{k-n}, \text{ for } k>n,\nonumber\\
&=0, \text{ for } k<n.
\label{j77}
\end{align}


\noindent
{\bf Remark}\,: We wish to make a final remark regarding the Kraus operators for the  composite channels obtained as the product of the Kraus operators of the quantum-limited channels as detailed above. The Kraus operators for the composites in Eqs.\,\eqref{j71}, \eqref{j72}, \eqref{j73}, \eqref{j74}, \eqref{j76} were shown to be linearly independent in Ref.\,\cite{kraus10}. However, the Kraus operators in \eqref{j75} and \eqref{j77}, are linearly dependent. Nevertheless, they still give rise to a valid operator sum representation for the corresponding composite channels.

\section{Entanglement-breaking bosonic Gaussian channels}
We now consider the important notion of entanglement-breaking bosonic Gaussian channels. 
We recall\,\eqref{i115} that a channel
$\Gamma$ {\em acting on system $S$} is entanglement-breaking if the bipartite 
output state $(\Gamma \otimes {1\!\!1}_E)\,(\hat{\rho}_{SE})$ is separable
for every input state $\hat{\rho}_{SE}$, 
the ancilla system $E$ being arbitrary\,\cite{horodecki03}. 

A bosonic Gaussian channel is said to be entanglement-breaking if its one-sided action on a two-mode state is separable for all input bipartite states. It turns out that for single-mode bosonic Gaussian channels, the entanglement-breaking condition can be written down compactly by resorting to Simon's criterion\,\cite{simon00}.

A single-mode bosonic Gaussian channel $\Phi$ is said to EB if and only if $T\circ \Phi$ is also a channel, where $T$ stands for the transpose operation. By Eq.\,\eqref{j54} we have that
\begin{align}
&\Lambda\,Y\,\Lambda + i\,\beta \geq i\, \Lambda\,X^T\,\beta\, X\,\Lambda\nonumber\\
\Longrightarrow &~~~ Y - i\, \beta \geq   i \,X^T\,\beta\, X,
\label{j78}
\end{align}
where, as noted in Eq.\,\eqref{j45}, $\Lambda$ transcribes for the transpose map. We add that this requirement is in addition to the constraint satisfied by $(X,Y)$ for $\Phi$ to be a channel in Eq.\,\eqref{j54}. 
Further, we note that if a given channel $(X,Y)$ is EB, then adding additional classical noise will also result in an EB channel.

\begin{table}
\centering
\begin{tabular}{|c|c|c|}
\hline
Channel &  EB region & Y  \\
\hline
${\cal C}_{1}(\kappa;\alpha)$ & $\alpha \geq 2\kappa^2$ & $Y \geq  (\kappa^2+1)\, 1\!\!1$\\
${\cal C}_{2}(\kappa;\alpha)$ & $\alpha \geq 2$ & $Y \geq  (\kappa^2+1)\, 1\!\!1$ \\
${\cal D}(\kappa;\alpha)$ & $\alpha \geq 0$ & $Y \geq  (\kappa^2+1)\, 1\!\!1$ \\
${\cal A}_{2}((1+\sigma_2)/2;\alpha)$ & $\alpha \geq 0$ & $Y \geq 1\!\!1$ \\
\hline
\end{tabular}
\caption{Showing the EB bosonic Gaussian channels.\label{tablej4}}
\end{table}

Using the criterion provided in Eq.\eqref{j78}, we classify the EB Gaussian channels\,\cite{holevo08} for each of the canonical forms and tabulate then in Table\,\ref{tablej4}. We see that the quantum-limited phase conjugation channels ${\cal D}(\kappa;0)$  and singular channels ${\cal A}_2(0)$ are already entanglement breaking. Hence, these classes of channels are always entanglement breaking irrespective of the noise. One other quantum-limited channel that is EB is the $\kappa=0$ end of the attenuator channel ${\cal C}_1(\kappa;0)$, i.e. ${\cal C}_1(0,0)$. The noisy channels ${\cal C}_1(\kappa;\alpha)$ and ${\cal C}_2(\kappa;\alpha)$ are EB for $\alpha \geq 2\kappa^2$ and $2$ respectively. We will explore more properties of EB Gaussian channels in Chapter 5.

\part{Main Results}

\chapter{CP maps and initial correlations}

\section{Introduction}
Open quantum systems are systems that are in interaction with its environment. Therefore, 
open quantum systems play a very fundamental role in the study of every realistic or practical
application of quantum systems. There has been a rapid growth in the understanding of 
various properties related to open quantum systems like its realization, control, and the role played by 
the noise in such dissipative systems, both in the theoretical and experimental domain. 

Recent studies on various aspects of control of open quantum systems has appeared\,\cite{lloyd00,lloyd01,altafini03,altafini04,potz05,potz06,potz06b,manko03,clark05,hu06,romano06,rabitz10,rabitz07,rabitz07b}. 
These studies have been motivated by applications to quantum computing\,\cite{tarasov02,lidar05,knight00}, laser cooling\,\cite{itano79,navin04},
quantum reservoir engineering\,\cite{zoller96,morigi10}, managing decoherence\,\cite{gringo05,gringo05b,lidar05b,viola05,morton05}, and also to other fields like 
chemical reactions and energy transfer in molecules\,\cite{tim04,herek02,levis01,assion98}. There has also been 
a study of experimental aspects of environment induced decoherence in various physical scenarios including 
atomic systems\,\cite{pfau94,myatt00,kimble05,diehl10,diehl11}, spin networks\,\cite{igor06}, and molecular physics\,\cite{arndt03,arndt04}. 

A related recent avenue has been to exploit the dissipation into the environment. Here, theoretical studies 
of basic tasks in quantum information theory like state preparation\,\cite{zoller08,diehl08,cirac09,cirac11,sorenson11}, distillation\,\cite{cirac11b}, storage\,\cite{cirac11c},
cooling\,\cite{eisert12}, and including their experimental aspects\,\cite{blatt11,polzik12}, are performed by 
engineering the system-environment coupling. Further, the issue of timing in such dissipative quantum information processing
was addressed in\,\cite{eisert13}.  

In this chapter we study the induced dynamics of the system resulting from the dynamics of an open quantum system. In particular, we explore the role of the initial system-bath states, especially in respect of a possible connection to quantum discord, as brought out in recent literature. From the various manifestations of open quantum systems listed above, it is pertinent 
to understand this aspect of realization of an open quantum system.


Every physical system is in interaction with its environment, {\em the bath}, to a smaller or 
larger degree of strength. Therefore, the joint unitary dynamics or unitary Schr\"{o}dinger evolutions 
of the system and bath induces a dissipative non-unitary dynamics for the system\,\cite{breuerbook}.
We now briefly recapitulate the folklore scheme or Stinepring dilation\,\cite{stinespring,gks,lindblad76,krpd1,krpd2}.
The Hilbert spaces ${\cal H}_S$ and ${\cal H}_B$ of 
the system and the bath are of dimensions $d_S$, $d_B$ respectively. The ($d^2_S-1$)-dimensional (convex) 
state space $\Lambda_S$ is a subset of ${\cal B}({\cal H}_S)$. We also denote the collection of initial-system bath states by $\Omega^{SB} \subset {\cal B}({\cal H}_S \otimes {\cal H}_B)$,
the convex hull of $\Omega^{SB}$ being denoted $\overline{\Omega^{SB}}$. The definition of $\Omega^{SB}$ will become clear in a subsequent Section.  

\subsection{Folklore scheme}
The folklore scheme (see Fig.\,\ref{figl1}) 
 for realizing  open system 
dynamics is to first elevate the  system states $\rho_S$ to the (tensor)  
products $\rho_S \otimes \rho_B^{\,\rm{fid}}$, for a {\em fixed} fiducial bath state 
$\rho_B^{\,\rm{fid}}$. Then these composite uncorrelated system-bath states are evolved under a 
joint unitary $U_{SB}(t)$, and finally the bath degrees of 
freedom are traced out to obtain the evolved states $\rho_S(t)$ of the system\,:
\begin{align}
\rho_S \rightarrow \rho_S \otimes \rho_B^{\,\rm{fid}} 
&\to U_{SB}(t) \,(\rho_S \otimes \rho_B^{\,\rm{fid}})\,U_{SB}(t)^{\dagger}\nonumber \\ 
\to \rho_S(t) &= \rm{Tr}_B\left[ U_{SB}(t) \,(\rho_S 
 \otimes \rho_B^{\,\rm{fid}})\,U_{SB}(t)^{\dagger} \right].  
\label{l1}
\end{align} 
The resulting quantum dynamical process (QDP) $\rho_S \to 
\rho_S(t)$, parametrized by $\rho_B^{\,\rm{fid}}$ and $U_{SB}(t)$, is provably 
completely positive (CP)\,\cite{ecg61,kraus71,choi75,lindblad76,gks}. 

\begin{figure}
\begin{center}
\scalebox{1.5}{
\includegraphics{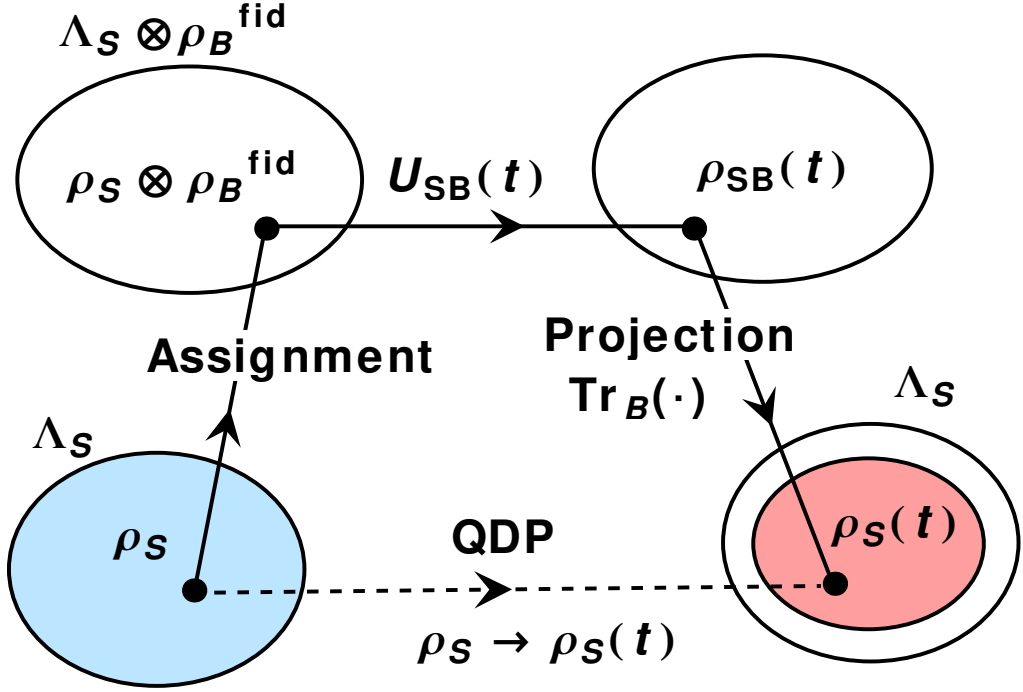}}
\end{center}
\caption{
Showing the folklore scheme. 
In the folklore scheme, initial  system states $\rho_S$ 
 are elevated to  product states of the composite, for a {\em fixed} fiducial bath state 
$\rho_B^{\,\rm{fid}}$, through the assignment map    
$\rho_S \to \rho_S \otimes \rho_B^{\,\rm{fid}}$. These uncorrelated system-bath states are evolved under a 
joint unitary $U_{SB}(t)$ to  
$U _{SB}(t) \,\rho_S \otimes \rho_B^{\,\rm{fid}}\,U_{SB}(t)^{\dagger}$
and, finally, the bath degrees of freedom are 
traced out to obtain the time-evolved states 
$\rho_S(t) = \rm{Tr}_B\left[ U_{SB}(t) \rho_S 
 \otimes \rho_B^{\,\rm{fid}}\,U_{SB}(t)^{\dagger} \right]$
 of the system. 
The resulting quantum dynamical process (QDP) $\rho_S \to 
\rho_S(t)$, parametrized by $\rho_B^{\,\rm{fid}}$ and $U_{SB}(t)$, is 
completely positive by construction. Initial system states are identified by the blue region and the final states by the red.
 \label{figl1}} 
\end{figure}

Indeed if $\rho_B^{\,\rm{fid}} = |\psi_B \rangle \langle \psi_B|$, 
and $\{|v_B\rangle \}$ is a complete basis for system $B$, then the operator-sum 
representation for the QDP can be written as 
\begin{align}
\rho_S(t) = \sum_k A_k(t)\, \rho_S \, A_k^{\dagger}(t),
\label{l2}
\end{align} 
where $A_k(t)$ are the sum-operators which are given by 
\begin{align}
A_k(t) = \langle v^k_B | U_{SB}(t)|\psi_B\rangle.
\label{l3}
\end{align}
If instead we have a mixed state $\rho_B^{\,\rm{fid}}$, then the operator-sum
representation will just be a convex combination of maps resulting from 
each pure state in, say, the spectral resolution of $\rho_B^{\,\rm{fid}}$.

While every CP map can be thus realized with 
uncorrelated initial states of the composite, there has been 
various studies in literature that explore more general 
realizations of CP maps\,\cite{park,raggio,parkreply,pechukas,alickireply,pechukasreply,buzek01,hayashi03,jordan04,shaji05}.  Possible effects of system-bath initial correlations on the  
reduced dynamics for the system has been the subject of several recent 
studies\,\cite{rosario08,karol08,shabani09,shabani09b,guzik10,usha11,cracken13,brodutch13,buscemi13}. Some of these works look at the connection between the concept of quantum discord and the complete positivity of the reduced dynamics\,\cite{rosario08,shabani09,brodutch13}; these are of much interest to us.

\subsection{SL scheme}
A specific, carefully detailed, and precise formulation of the issue 
  of initial system-bath correlations possibly influencing the reduced dynamics was 
presented not long ago by Shabani and Lidar\,\cite{shabani09}. In this 
formulation (see Fig.\,\ref{figl2}), the distinguished bath state $\rho_B^{\,\rm{fid}}$ is replaced by 
a collection of (possibly correlated) system-bath 
 initial states $\Omega^{SB} \in {\cal B}({\cal H}_S 
\otimes {\cal H}_B)$.
 The dynamics gets defined through a joint unitary $U_{SB}(t)$\,:
\begin{align} 
\rho_{SB}(0) \to \rho_{SB}(t) &= 
U_{SB}(t)\,\rho_{SB}(0)\,U_{SB}(t)^{\dagger}, 
 ~~~~\forall~ 
\rho_{SB}(0) \in \Omega^{SB}. 
\label{l4}
\end{align} 
This composite dynamics induces on the system the QDP 
\begin{align} 
\rho_S(0) \to \rho_S(t), 
\label{l5}
\end{align} 
with $\rho_S(0)$ and $\rho_S(t)$ defined through this natural imaging from $\Omega^{SB}$ to the system state space $\Lambda_S$\,:  
\begin{align} 
\rho_S(0) = \rm{Tr}_B\,\rho_{SB}(0),~\rho_S(t) = \rm{Tr}_B \rho_{SB}(t). 
\label{l6}
\end{align} 

\begin{figure}
\begin{center}
\scalebox{1.5}{\includegraphics{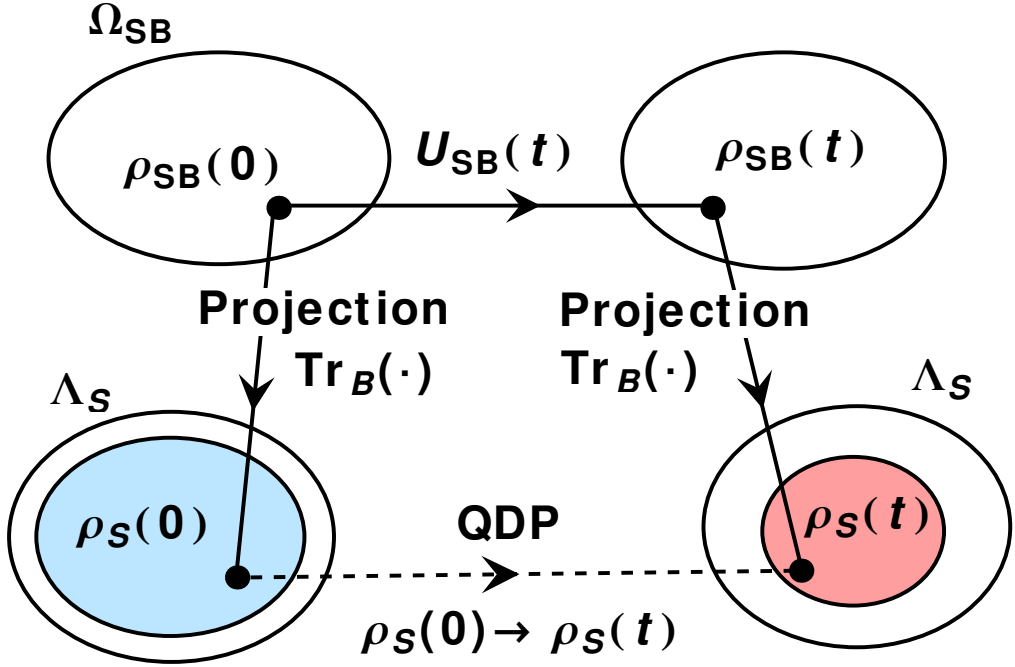}}
\end{center}
\caption{
Showing the SL scheme. In sharp contrast to the folklore scheme, there is no assignment map in the SL scheme.  The distinguished bath state $\rho_B^{\,\rm{fid}}$ is replaced by 
a collection $\Omega^{SB}$ of (possibly correlated) system-bath 
 initial states $\rho_{SB}(0)$.
 The dynamics gets defined through 
$\rho_{SB}(0) \to \rho_{SB}(t) = 
U_{SB}(t)\,\rho_{SB}(0)\,U_{SB}(t)^{\dagger}$
 for all $\rho_{SB}(0) \in \Omega^{SB}$. 
 With reduced system states $\rho_S(0)$ and $\rho_S(t)$ defined through
the imaging or projection map   
$\rho_S(0) = \rm{Tr}_B\,\rho_{SB}(0)$ and $\rho_S(t) = \rm{Tr}_B 
\left[U_{SB}(t)\,\rho_{SB}(0)\,U_{SB}(t)^{\dagger} \right]$,   
this unitary dynamics of the composite induces on the system the QDP 
$\rho_S(0) \to \rho_S(t)$. 
As before, initial system states are identified by the blue region and the final states by the red.
\label{figl2}}
\end{figure}
It is clear the folklore scheme is a particular case of the SL scheme corresponding to  
$\Omega^{{SB}} = \{\,\rho_S \otimes \rho_B^{\,\rm{fid}}\,|\, 
\rho_B^{\,\rm{fid}}={\rm fixed}\,\}$.

This generalized formulation of QDP allows SL to transcribe the 
fundamental issue to this question: What are the necessary and sufficient 
conditions on the collection $\Omega^{SB}$ so that the induced QDP $\rho_S(0) 
\to \rho_S(t)$ in Eq.\,\eqref{l5} is  
 guaranteed to be CP {\em for all} joint unitaries 
$U_{SB}(t)$? Motivated by the work of Rodriguez-Rosario et al.\,\cite{rosario08}, 
and indeed highlighting it 
 as {\em `a recent breakthrough'}, SL advance the following resolution to this 
 issue\,: 
\begin{theorem}[Shabani-Lidar]: The QDP in Eq.\,\eqref{l5} is CP 
for all joint unitaries $U_{SB}(t)$ if and only if the  quantum 
discord vanishes for all $\rho_{SB} \in \Omega_{SB}$, i.e., if and only if the 
initial system-bath correlations are purely classical. 
\end{theorem}
Whether the Shabani-Lidar QDP so described
is well-defined and completely positive is clearly an issue answered solely by 
the nature of the collection $\Omega^{SB}$. 

\section{Properties of SL $\Omega^{SB}$}
 In order that the QDP in Eq.\,\eqref{l5} be {\em well defined} in the first place, the set $\Omega^{SB}$ 
 should necessarily satisfy the following two properties; since our entire analysis rests 
critically on these properties, we begin by motivating them.

\subsection{Property 1}
 No state $\rho_S(0)$ can have two (or more) pre-images in $\Omega^{SB}$.
To see this fact unfold
assume, to the contrary, that 
\begin{align}
\rm{Tr}_B\, \rho_{SB}(0) &=\rm{Tr}_B\, \rho^{\,'}_{SB}(0),  
~~ \rho_{SB}(0) \neq \rho^{\,'}_{SB}(0),\nonumber\\
 & {\rm for~ two~ states~} \rho_{SB}(0),~~ \rho^{\,'}_{SB}(0) \in \Omega^{SB}.
\label{l7}
\end{align}
Clearly, the difference $\triangle \rho_{SB}(0) = \rho_{SB}(0) - \rho^{\,'}_{SB}(0) \neq 0$
should necessarily meet the property $\rm{Tr}_B \triangle \rho_{SB}(0)=0$. Let
$\{\lambda_u \}_{u=1}^{d_S^{\,2}-1}$ be a set of orthonormal hermitian
traceless $d_S \times d_S$ matrices so that together with the unit matrix
$\lambda_0 = 1\!\!1_{d_S \times d_S }$ these matrices form a hermitian basis for ${\cal
B}({\cal H}_S)$, the set of all $d_S \times d_S$  (complex) matrices.
Let $\{\gamma_v
\}_{v=1}^{d_B^{\,2}-1}$, $\gamma_0= 1\!\!1_{d_B \times d_B}$ be a similar
basis for ${\cal B}(\cal H_B)$. The $(d_Sd_B)^2$ tensor products
$\{\lambda_u \otimes \gamma_v \}$ form a basis for ${\cal B}({\cal H}_S
\otimes {\cal H}_B )$, and  $\triangle \rho_{SB}(0)$ can be written in the form
\begin{align}
\triangle \rho_{SB}(0) = \sum_{u=0}^{d_S^{\,2}-1}\;
\sum_{v=0}^{d_B^{\,2}-1} C_{uv}\, \lambda_u \otimes \gamma_v, ~ C_{uv} \text{ real}.
\label{l8}
\end{align}
Now,
the property $\rm{Tr}_B\, \triangle \rho_{SB}(0)=0$ is strictly equivalent to
the requirement that the expansion coefficient $C_{u0} = 0$ for all $u= 0,\,1,\,\cdots\, d_S^{\,2}-1$.
 Since the $[(d_S
d_B)^2-1]$-parameter unitary group $SU(d_Sd_B)$ acts {\em irreducibly} on the $[(d_S
d_B)^2-1]$-dimensional subspace of ${\cal B}({\cal H}_S \otimes {\cal H}_B)$
consisting of all traceless $d_Sd_B$-dimensional matrices [\,this is the adjoint
representation of $SU(d_Sd_B)$\,], there exists an $U_{SB}(t) \in SU(d_Sd_B)$
which takes $\triangle \rho_{SB}(0) \ne 0$ into a matrix whose
 expansion coefficient $C_{u0} \neq 0$ for
some $u$. That is, if the initial $\triangle \rho_{SB}(0)\ne 0$ then
one and the same system state $\rho_S(0)$ will evolve into two distinct
\begin{align}
\rho_S(t) =& \rm{Tr}_B\left[ U_{SB}(t) \rho_{SB}(0) U_{SB}(t)^{\dagger}\right], 
     \nonumber\\
 \rho^{\,\prime} _S(t) =& 
\rm{Tr}_B\left[U_{SB}(t)\rho^{\,\prime} _{SB}(0)U_{SB}(t)^{\dagger}\right],
\label{l9}
\end{align} 
for some $U_{SB}(t)$, rendering the QDP
in equation\,\eqref{l5} one-to-many, and hence ill-defined.

\subsection{Property 2}
  While every system state $\rho_S(0)$ need not have 
a pre-image {\em actually enumerated} in $\Omega^{SB}$, 
 the set of $\rho_S(0)$'s having pre-image should be sufficiently large.  
Indeed, Rodriguez-Rosario et al.\,\cite{rosario08} have
 rightly emphasised that it should be {\em `a large enough set of
states such that the QDP in Eq.\,\eqref{l5} can be extended by linearity to all
states of the system'}. It is easy to see that 
if $\Omega^{SB}$ fails this property, 
 then the very issue of CP would make no sense. For, in carrying out verification
 of CP property, the QDP  would be required to act, 
as is well known\,\cite{choi75}, on 
 $\{|j\rangle \langle k|\}$ for $j,\,k = 1,2,\,\cdots\,d_S$; i.e., 
on generic complex $d_S$-dimensional
square matrices, and not just on positive or hermitian matrices alone.
Since the basic issue on hand is to check if the QDP as a map on
${\cal B}({\cal H}_S)$ is CP or not, it is essential that it be well
 defined (at least by linear extension)
on the entire {\em complex} linear space ${\cal B}({\cal H}_S)$.

\section{Main Result}
With the two properties of $\Omega^{SB}$ thus motivated, we proceed to prove our main result.  We 
 `assume', {\em for the time being}, that every pure state $| \psi \rangle$ of 
the system has a pre-image in $\Omega^{SB}$. This assumption may appear, at first sight, 
to be a drastic one. But we show later that it entails indeed {\em no loss of generality}.

 It is evident that, for every pure state $|\psi\rangle$, the pre-image 
 in $\Omega^{SB}$ has to necessarily 
 assume the (uncorrelated) product form $|\psi \rangle \langle \psi| \otimes 
\rho_B$ , $\rho_B$ being a state of the 
bath which could possibly 
depend on the system state $|\psi \rangle$.
 

Now, let $\{|\psi_k \rangle\}_{k=1}^{d_S}$ be an orthonormal basis in ${\cal H}_S$ 
and let $\{|\phi_{\alpha}\rangle\}_{\alpha=1}^{d_S}$ be another orthonormal 
basis related to the former through a complex Hadamard unitary matrix $U$. 
 Recall that a unitary U is Hadamard if $|U_{k\alpha}|= 1/\sqrt{d_S}$, independent 
of $k,\alpha$. 
For 
instance, the characters of the cyclic group of order $d_S$ 
 written out as a $d_S \times d_S$ matrix is Hadamard. 
The fact that the $\{|\psi_k\rangle \}$ basis and 
the $\{|\phi_{\alpha}\rangle \}$ basis are related by a Hadamard means that 
 $|\langle \psi_k |\phi_{\alpha} \rangle|$ 
is independent of both $k$ and $\alpha$, and hence equals $1/\sqrt{d_S}$ uniformly.
 We may refer to such a pair as {\em relatively unbiased bases}. 
 
 
Let $|\psi_k \rangle \langle \psi_k| \otimes O_k$ be the pre-image of $|\psi_k 
\rangle \langle \psi_k|$ in $\Omega^{SB}$ and $|\phi_{\alpha} \rangle \langle 
\phi_{\alpha}|\otimes \widetilde{O}_{\alpha}$ that of $|\phi_{\alpha} \rangle 
\langle \phi_{\alpha}|$, $k,\alpha = 1,2,\cdots,d_S$. Possible dependence of 
the bath states $O_k$ on $|\psi_k \rangle$ and $\widetilde{O}_{\alpha}$ on 
$|\phi_{\alpha}\rangle$ has not been ruled out as yet.  
Since the maximally mixed state of the system
 can be expressed in 
two equivalent ways as ${d_S^{-1}} \sum_k |\psi_k \rangle \langle \psi_k| = 
{d_S^{-1}} \sum_{\alpha} |\phi_{\alpha} \rangle \langle \phi_{\alpha}|$, 
 {\em uniqueness} of its pre-image in $\Omega^{SB}$ (Property 1) demands
\begin{align}
\sum_{k=1}^{d_S} |\psi_k \rangle \langle \psi_k|\otimes O_k 
= \sum_{\alpha=1}^{d_S} |\phi_{\alpha} \rangle \langle \phi_{\alpha}| 
  \otimes \widetilde{O}_{\alpha}. 
\label{l10}
\end{align} 
Taking projection of both sides on 
$|\psi_j \rangle \langle \psi_j|$, and using the Hadamard property $|\langle \psi_j|\phi_{\alpha} 
\rangle|^2 = {d_S^{-1}}$, we have 
\begin{align} 
O_j = \frac{1}{d_S} 
\sum_{\alpha=1}^{d_S} \widetilde{O}_{\alpha}, ~~ j=1,2,\cdots,d_S, 
\label{l11}
\end{align} 
while projection on $|\phi_{\beta} \rangle \langle \phi_{\beta}|$ leads to 
\begin{align} 
\widetilde{O}_{\beta} = \frac{1}{d_S} \sum_{k=1}^{d_S} O_k, ~~ 
\beta=1,2,\cdots,d_S. 
\label{l12}
\end{align} 
The $2 d_S$ constraints of Eqs.\,\eqref{l11},\,\eqref{l12} together imply that $O_j= \widetilde{O}_{\beta}$ 
uniformly for all $j,\,\beta$.  
Thus the pre-image of $ |\psi_k \rangle 
\langle \psi_k|$ is $ |\psi_k \rangle \langle \psi_k| \otimes 
\rho_B^{\,\rm{fid}}$ and that of $|\phi_{\alpha} \rangle \langle 
\phi_{\alpha}|$ is $ |\phi_{\alpha} \rangle \langle \phi_{\alpha}| \otimes 
\rho_B^{\,\rm{fid}}$, for all $k,\alpha$, for some {\em fixed fiducial 
 bath state} $\rho_B^{\,\rm{fid}}$.  
And, perhaps more importantly, the 
pre-image of the maximally mixed state $d_S^{-1} 1\!\!1$ 
 necessarily equals $d_S^{-1} 1\!\!1 \otimes \rho_B^{\,\rm{fid}}$ as well.

Taking another pair of relatively unbiased bases $\{|\psi^{\,'}_k \rangle\}$, 
$\{ |\phi^{\,'}_{\alpha}\rangle\}$ one similarly concludes that the pure 
states $ |\psi^{\,'}_k \rangle \langle \psi^{\,'}_k|$, $ |\phi^{\,'}_{\alpha} 
\rangle \langle \phi^{\,'}_{\alpha}|$ too have pre-images $|\psi^{\,'}_k 
\rangle \langle \psi^{\,'}_k| \otimes \rho_B^{\,\rm{fid}}$, 
$|\phi^{\,'}_{\alpha} \rangle \langle \phi^{\,'}_{\alpha}| \otimes 
\rho_B^{\,\rm{fid}}$ respectively, with the same 
fixed fiducial bath state $\rho_B^{\,\rm{fid}}$ as before. This is so, since the 
maximally mixed state is {\em common} to both sets.   

Considering in this manner enough  
 number of pure states or projections $|\psi \rangle \langle \psi|$ sufficient to span---by linearity---the entire system 
state space $\Lambda_S$, and hence ${\cal B}({\cal H}_S)$,
and using the fact that convex sums goes to corresponding convex sums under
pre-imaging,
 one readily concludes that {\em every element} $\rho_{SB}(0)$ of $\Omega^{SB}$ 
(irrespective of whether ${\rm Tr}_B\,\rho_{SB}(0)$ is pure or mixed) 
 {\em necessarily} needs to be 
of the product form $\rho_S(0) \otimes \rho_B^{\,\rm{fid}}$, 
for some {\em fixed}  
bath state $\rho_B^{\,\rm{fid}}$. But this is exactly the folklore 
realization of non-unitary dissipative dynamics, to surpass 
which was the primary goal of the SL scheme.  We have thus proved our principal result: 
\begin{itemize} 
\item[] No 
initial correlations---{\em even classical ones}---are 
permissible in the SL scheme. That is, quantum discord is 
no less destructive as far as CP property of QDP 
is concerned. 
\end{itemize}
 It is true that we have proved this result 
under an assumption but, as we show below, 
this assumption entails no loss of generality at all. 


As we have noted, if at all a pure state $\rho_S(0)= |\psi \rangle \langle \psi|$ has a pre-image in 
$\Omega^{SB}$ it would necessarily be of the product form $ |\psi \rangle 
\langle \psi| \otimes \rho_B$, for some (possibly $|\psi\rangle$-dependent) 
bath state $\rho_B$. While this is self-evident and is independent of SL, 
 it is instructive to view it  as a consequence of 
the {\em necessary condition part} of SL theorem. Then our 
principal conclusion above can be rephrased to say that validity of SL theorem for 
pure states of the system readily leads to the folklore product-scheme as the {\em only solution} within the SL framework.   
This interesting aspect comes through  in an even  
more striking manner in our proof below   
  that our earlier `assumption' is one without loss of generality. 

\subsection{Assumption entails no loss of generality}

Let us focus, to begin with, on the convex hull  $\overline{\Omega^{SB}}$ of
$\Omega^{SB}$ rather than the full (complex) linear span of $\Omega^{SB}$   
to which we are entitled. Let us further allow for the possibility that the image 
of $\overline{\Omega^{SB}}$ under the convexity-preserving 
linear map $\rho_{SB}(0) \to 
{\rm Tr}_B \rho_{SB}(0)$ fills not the entire (convex) state 
space $\Lambda_S$---the $(d^{\,2}_S-1)$-dimensional 
generalized Bloch sphere---of the system, but 
only a portion thereof, possibly a very small part. Even so, in order that 
our QDP in equation\,\eqref{l5} be well-defined, this portion would 
{\em occupy a non-zero volume} of the $(d^{\,2}_S-1)$-dimensional 
state space of the system (Property 2).

Let us consider one set of all mutually commuting elements of the 
system state space $\Lambda_S$. If 
the full state space were available under the imaging 
$\rho_{SB}(0) \to 
{\rm Tr}_B \rho_{SB}(0)$ of $\overline{\Omega^{SB}}$, 
then the resulting mutually commuting images would have  
filled the  entire $(d_S-1)$-simplex, 
the classical state space of a $d_S$-level system, 
this being respectively the triangle and the tetrahedron 
when $d_S=3,4$\,\cite{simonpramana,goyal}. 
Since the full state space is not assumed to be available, these commuting 
elements possibly fill only a, perhaps very small but nevertheless of nontrivial measure, 
proper convex subset of the $(d_S-1)$-simplex, 
 depicted in Fig.\,\ref{figl3} as region $R$ for the case $d_S=3$, (qutrit).
\begin{figure}
\begin{center}
\scalebox{1.4}{\includegraphics{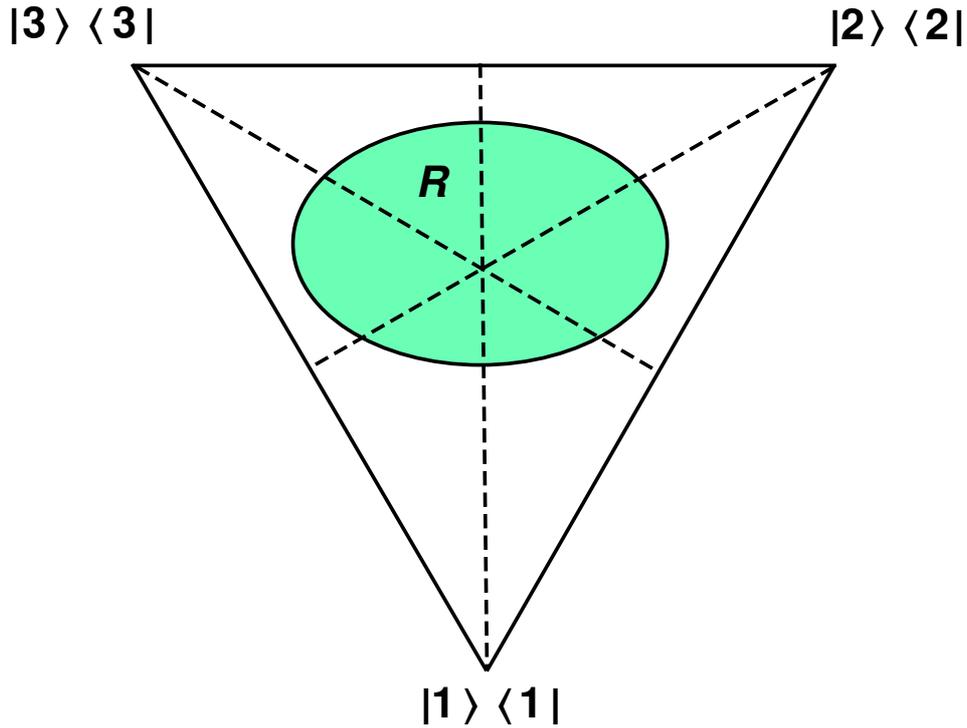}}
\end{center}
\caption{Depicting, for the case $d_S=3$ (qutrit), the image of $\overline{\Omega^{SB}}$ under 
     ${\rm Tr}_B (\cdot)$ in the plane spanned by the commuting (diagonal)  
      $\lambda$-matrices $(\lambda_3,\lambda_8)$.
 \label{figl3}} 
\end{figure}

Elements of these simultaneously diagonal density matrices 
of the system can be expressed as convex 
sums of orthogonal pure states or one-dimensional 
projections. 
For a generic element in this region, the spectrum is 
non-degenerate, and hence the projections are unique and commuting, being the 
eigenstates of $\rho_S(0)$, and correspond to the $d_S$ vertices of the ($d_S-1$)-simplex. In the case of qutrit, it is pictorially seen in Fig.\,\ref{figl3} that only the points  
 on the three dotted lines correspond to doubly degenerate density 
matrices and the centre alone is 
triply degenerate, rendering transparent the fact that 
being nondegenerate is a generic attribute of  region $R$.  

Now  consider the pre-image $\rho_{SB}(0)$ in $\overline{\Omega^{SB}}$ of 
such a generic 
non-degenerate $\rho_S(0) \in R$. Application of the SL requirement 
of vanishing discord (again, only the necessity part of 
the SL theorem) to this $\rho_{SB}(0)$ implies, by definition\,\cite{zurek01,modirmp}, that
  the pre-image  has  the form
\begin{align}
\rho_{SB}(0) = \sum_{j=1}^{d_S} p_j |j \rangle \langle j| \otimes \rho_{Bj}(0), 
\label{l13}
\end{align}
 where the probabilities $p_j$ and the pure states $|j \rangle \langle j|$
  are uniquely determined (in view of nondegeneracy) by the spectral resolution  
\begin{align} 
\rho_S(0) = {\rm 
Tr}_B\,\rho_{SB}(0) = \sum_{j=1}^{d_S} p_j |j \rangle \langle j| . 
\label{l14}
\end{align} 
And $\rho_{Bj}(0)$'s are bath states, possibly dependent on $|j \rangle 
\langle j|$  as indicated by the label $j$ in $\rho_{Bj}(0)$.  
  These considerations  hold for every  nondegenerate element of  
  region $R$ of probabilities $\{\,p_j\,\}$. 
   In view of generic nondegeneracy,  
 the requirement\,\eqref{l13}  
 implies that each of the $d_S$ pure states $|j \rangle \langle j|$ has 
 pre-image of the form $|j \rangle \langle j|\otimes \rho_{Bj}(0)$ in the {\em linear
span} of the pre-image of R---at least   
 as seen by the QDP\,\eqref{l5}. That is, $\rho_{Bj}(0)$'s can have no 
dependence on the probabilities $\{\,p_j\,\}$.

Since every pure state of the system 
constitutes one of the vertices of some  
$(d_S-1)$-simplex in $\Lambda_S$ comprising one set of all mutually commuting density 
operators $\rho_S(0)$, the conclusion that a pure 
state effectively has in the linear span of $\Omega^{SB}$ a pre-image, and one necessarily of the product form, 
{\em applies to every pure state}, showing that the `assumption' 
in our earlier analysis indeed  {\em entails no loss of generality}. 

\section{Conclusion}
To summarize, it is clear that the dynamics described 
 by \\
$\rho_{SB}(0) \to \rho_{SB}(t) = U_{SB}(t)\,\rho_{SB}(0)\,U_{SB}(t)^{\dagger}$, $\rho_{SB}(0) \in \Omega^{SB}$  would {\em `see'} only the full (complex) linear span of  
 $\Omega^{SB}$, and {\em not so much the actual enumeration}
 of $\Omega^{SB}$  as such. 
 But as indicated by the imaging (projection) map $\rho_{SB}(0) \to \rho_S(0) = \rm{Tr}_B\,\rho_{SB}(0)$, the 
only elements of this linear span 
which are immediately relevant for the QDP are those which are 
hermitian, positive semidefinite, and have unit trace; these are precisely the 
elements of  $\overline{\Omega^{SB}}$, the convex hull of $\Omega^{SB}$.
 Since no system state can have two or more pre-images (see Property 1), 
 in order to render the QDP in\,\eqref{l5} 
 well defined these relevant elements are forced to 
constitute a {\em faithful linear embedding}, in  ${\cal B}({\cal H}_S \otimes {\cal H}_B)$, 
of (a nontrivial convex subset of) the system's state space. 
 In the SL scheme of things, this leaves us with just the folklore embedding
 $\rho_S(0)\to \rho_{SB}(0)=\rho_S(0)\otimes \rho_B^{\,{\rm fid}}$. 
 This is the primary conclusion that emerges.\\
 
\noindent
{\em Remark on convexity\,:}\\
\noindent
Let us view this from a slightly different position. Since
 there is no conceivable 
manner in which a linear map $U_{SB}(t)$ acting on elements of $\Omega^{SB}$ 
 could be prevented from acting on 
convex sums (indeed, on the linear span) of such elements, we may assume---without loss of 
of generality---$\Omega^{SB}$ to be convex and ask,  
consistent with  the SL theorem: 
  What are the possible choices for  
 the collection $\Omega^{SB}$ to be {\em convex and at the same time consist entirely of states 
of vanishing quantum discord}. 
 One possibility comprises elements of the form 
$\rho_{SB}(0) = \rho_{S}(0)\otimes \rho_B^{\,{\rm fid}}$ for a fixed bath state 
 $\rho_B^{\,{\rm fid}}$  and arbitrary system state $\rho_{S}(0)$. 
This case of $\overline{\Omega^{SB}} = \Lambda_S \otimes \rho_B^{\rm fid}$ is recognized to be simply the folklore  case. 
The second one consists of elements of the form 
$\rho_{SB}(0) = \sum_{j}p_j |j \rangle \langle j| \otimes \rho_{Bj}(0)$,  
 for a {\em fixed (complete) set of orthonormal pure states} 
$\{|j \rangle \langle j|\}$. This  
 case  restricted to {\em  mutually commuting density operators} of the system  seems to be the one studied by Rodriguez-Rosario et al.\,\cite{rosario08},
  but the very notion of CP itself is unlikely to make much sense in this non-quantum case of  
 {\em classical state space} (of dimension $d_S-1$ rather than 
$d_S^{\,2}-1$),  the honorific `a recent breakthrough' notwithstanding. \\

\noindent
The stated goal of SL  was to give 
a {\em complete characterization}
 of possible initial correlations that lead to CP maps. 
It is possibly in view of the (erroneous) belief that there was 
a large class of permissible initial correlations out there 
within the SL framework, and that that class now stands  
fully characterized by the SL theorem, that
a large number of recent papers tend to list complete characterization of   
CP maps among the principal achievements of quantum discord\,\cite{modiprl,brukner10,acin10,breuer10,maniscalco10,fanchini11,discreview11,majenz13}.     
Our result implies, with no irreverence whatsoever to quantum discord, 
 that characterization of CP maps may not yet be 
 rightfully paraded as one of the principal achievements of quantum discord.

The SL theorem has influenced an enormous number 
of authors, and it is inevitable that those results of these 
authors which make essential use of the sufficiency part of the SL theorem need recalibration in the light of our result.

There are other, potentially much deeper, implications of our finding. 
 Our analysis---strictly  within the SL  
framework---has shown that this framework brings one exactly back to the 
  folklore scheme itself, as if it were a {\em fixed point}.  
 This is not at all a negative result for two reasons. First,  
 it shows that quantum discord is no `cheaper' than entanglement as far as 
 complete positivity of QDP is concerned. Second, and more importantly, the
fact that the folklore product-scheme survives  attack 
 under this powerful, well-defined, and fairly general SL framework demonstrates  
 its, perhaps unsuspected, {\em robustness}. In view of the fact that 
 this scheme has been at the heart of most applications 
 of quantum theory to real situations, virtually in every area of physical 
science, and even beyond, its robustness the SL framework has
helped to establish is likely to prove to be of far-reaching significance. 







\chapter{Correlations for two-qubit X-states}
\section{Introduction}
The study of correlations in bipartite systems has been invigorated over the last couple of decades or so. Various measures and approaches to segregate the classical and quantum contents of correlations have been explored. Entanglement has continued to be the most popular of these correlations owing to its inherent potential advantages in performing quantum computation and communication tasks\,\cite{horo-rmp}.  More recently, however, there has been a rapidly growing interest in the study of correlations from a more direct measurement perspective\,\cite{modi-rmp,celeri-rmp}, and several measures to quantify the same have been considered. Among these measures, quantum discord and classical correlation have been attracting much attention\,\cite{dutta13,terno11,caves-1bit,winter-ops-disc11,dutta-ops-disc11,lanyon08}, and have lead to several interesting results\,\cite{dakic12,piani11,robust13,gu12}. 
There has also been recent studies of different aspects of correlations like their evolution in various systems\,\cite{maziero09,mazzola10,auccprl11,galve13}, including non-markovian environments\,\cite{werlang09,fanchinipra10,wang10,franco13} and its role in spin systems\,\cite{dill08,werlang10}. Methods of witnessing quantum discord in the theoretical\,\cite{dattanull10,terno10,chitambar11,huang11,chrus12,modiprl10,dakicprl10,gessner13,chrus10,gessner11,adesso-indicator11} and experimental\,\cite{auccprl11b,xu10,girolami12,passante11,girolami13} domain have also been explored. 

In this Chapter, we undertake a comprehensive analysis of the problem of computation of  correlations in the two-qubit system, especially  the so-called $X$-states of two-qubit system\,\cite{saba13}; this class of states has come to be accorded a distinguished status in this regard. The problem of $X$-states has already been considered in\,\cite{luo,ali,wang11,chen11,du12pra,quesada12,huang13} and that of more general two-qubit states in\,\cite{james11,adesso11,zambrinipra,zambriniepl,nassajour13}. The approach which we present here exploits the very {\em geometric nature} of the problem, and helps to clarify and correct some  issues regarding computation of correlations in $X$-states in the literature, and many new insights emerge.  It may be emphasised that the geometric methods used here have been the basic tools of (classical) polarization optics for a very long time, and involve constructs like Stokes vectors, Poincar\'{e} sphere, and Mueller matrices\,\cite{simon-mueller1,simon-mueller2,simon-mueller3,simon-mueller4,simon-mueller5}.  

In Section\,\ref{rausection} we compare our analysis and results with those of the well known work of Ali, Rau, and Alber\,\cite{ali}. We show that their famous theorem that the optimal POVM for $X$-states is always a von Neumann projection either along x or along the z direction holds numerically for the entire manifold of $X$-states except for a very tiny region. Perhaps surprisingly, however, {\em their symmetry-based proof of that theorem seems to make an a priori assumption equivalent to the theorem itself}.

\section{Mueller-Stokes formalism for two-qubit states}
We begin with a brief indication as to why the Mueller-Stokes formalism of classical optics is possibly the most appropriate one for handling quantum states post measurement. In classical polarization optics the state of a light beam is represented by a $2 \times 2$ complex positive matrix $\Phi$ called the {\em polarization matrix}\,\cite{neillbook}. The intensity of the beam is identified with ${\rm Tr}\,\Phi$, and so the  matrix $({\rm Tr}\,\Phi)^{-1} \Phi$ (normalized to unit trace) represents the actual {\em state} of polarization. The polarization matrix $\Phi$ is thus analogous to the density matrix of a qubit, the only distinction being that the trace of the latter needs to assume unit value. Even this one little difference is gone when one deals with {\em conditional quantum states} post measurement\,: the probability of obtaining a conditional state becomes analogous to intensity $={\rm Tr}\, \Phi$ of the classical context. 

The Mueller-Stokes formalism itself arises from the following simple fact\,: any  $2\times 2$ matrix $\Phi$ can be invertibly associated with a  four-vector $S$, called the Stokes vector, through
\begin{align}
\Phi = \frac{1}{2} \sum_{k=0}^3 S_k \sigma_k, ~~ S_k = {\rm Tr}(\sigma_k \Phi).
\end{align} 
This representation is an immediate consequence of the fact that the Pauli triplet $\sigma_1,\,\sigma_2,\,\sigma_3$ and $\sigma_0 = 1\!\!1$, the unit matrix, form a complete orthonormal set of (hermitian) matrices. 

Clearly, hermiticity of the polarization matrix $\Phi$ is {\em equivalent} to reality of the associated four-vector $S$ and ${\rm Tr}\,\Phi = S_0$. Positivity of $\Phi$ reads $S_0 >0$, $S_0^2 -S_1^2-S_2^2-S_3^2 \geq 0$ corresponding, respectively, to the pair ${\rm Tr}\, \Phi>0$, ${\rm det}\, \Phi \geq 0$. Thus positive $2\times 2$ matrices (or their Stokes vectors) are in one-to-one correspondence with points of the {\em positive branch of the solid light cone}. Unit trace (intensity) restriction corresponds to the section of this cone at unity along the `time' axis, $S_0=1$. The resulting three-dimensional unit ball ${\cal B}_3 \in {\cal R}^3$ is the more familiar Bloch (Poincar\'{e}) ball, whose surface or boundary ${\cal P} = {\cal S}^2$ representing pure states (of unit intensity) is often called the  Bloch (Poincar\'{e}) sphere. The interior points correspond to mixed (partially polarized) states. 
 
Optical systems which map Stokes vectors {\em linearly} into Stokes vectors have been of particular interest in polarization optics. Such a linear system is represented by a $4 \times 4$ real matrix $M$, the Mueller matrix\,\cite{simon-mueller1,simon-mueller2,simon-mueller3,simon-mueller4,simon-mueller5}\,:
\begin{align}
M\,: S^{\rm in} \to S^{\rm out} = M S^{\rm in}.
\label{mspol}
\end{align}
It is evident that a (physical) Mueller matrix should necessarily map the positive solid light cone into itself. {\em It needs to respect an additional subtle restriction}, even in classical optics. \\

{\bf Remark}\,: The Mueller-Stokes formulation of classical polarization optics traditionally assumes plane waves. It would appear, within such a framework, one need not possibly place on a Mueller matrix any more demand than the requirement that it map Stokes vectors to Stokes vectors. However, the very possibility that the input (classical) light could have its polarization and spatial degree of freedoms intertwined in an inseparable manner, leads to the additional requirement that the Mueller matrix acting `locally' on the polarization indices alone map such an entangled (classical) beam into a physical beam. Interestingly, it is only recently that such a requirement has been pointed out \,\cite{simon-mueller4,simon-mueller5}, leading to a full characterization of Mueller matrices in classical polarization optics. $\blacksquare$

To see the connection between Mueller matrices and two-qubit states unfold naturally, use a single index rather than a pair of indices to label the computational basis two-qubit states $\{|jk \rangle\}$ in the familiar manner\,: $(00,01,10,11) = (0,1,2,3)$. Now note that a two-qubit density operator $\hat{\rho}_{AB}$ can be expressed in {\em two distinct ways}\,:
\begin{align}
\hat{\rho}_{AB} &= \sum_{j,k=0}^3 \rho_{jk} |j \rangle \langle k| \nonumber\\
&= \frac{1}{4} \sum_{a,b=0}^3 M_{ab} \,\sigma_a \otimes \sigma_b^*,
\label{mmatrix}
\end{align}
the second expression simply arising from the fact that the sixteen hermitian  matrices $\{\sigma_a \otimes \sigma_b^*  \}$ form a complete orthonormal set of $4 \times 4$ matrices. Hermiticity of operator $\hat{\rho}_{AB}$ is equivalent to reality of the  matrix $M = ((M_{ab}))$, but the same hermiticity is equivalent to  $\rho = ((\rho_{jk}))$ being a hermitian matrix. \\

\noindent
{\bf Remark}\,: It is clear from the defining equation\,\eqref{mmatrix} that the numerical entries of the two matrices $\rho,\,M$ thus associated with a given two-qubit state $\hat{\rho}_{AB}$ be related in an invertible linear manner. This linear relationship has been in use in polarization optics for a long time\,\cite{simon-mueller1,simon-mueller3,simon-mueller4}  and, for convenience, it is reproduced in explicit form in the Appendix. $\blacksquare$

Given a bipartite state $\hat{\rho}_{AB}$, the reduced density operators $\hat{\rho}_A,\,\hat{\rho}_{B}$ of the subsystems are readily computed from the associated $M$\,:
\begin{align}
\hat{\rho}_A &= {\rm Tr}[\hat{\rho}_{AB}] = \frac{1}{2} \sum_{a=0}^3 M_{a0}\, \sigma_a, \nonumber\\
\hat{\rho}_B &= {\rm Tr}[\hat{\rho}_{AB}] = \frac{1}{2} \sum_{b=0}^3 M_{0b}\, \sigma_b^*.
\label{partial}
\end{align} 
That is, the leading column and leading row of $M$ are precisely the Stokes vectors of reduced states $\hat{\rho}_A,\,\hat{\rho}_B$ respectively. 

It is clear that a generic  POVM element is of the form 
$\Pi_j = \frac{1}{2} \sum_{k=0}^3 S_k \sigma_k^*$. We shall call $S$  the Stokes vector of the POVM element $\Pi_j$. Occasionally one finds it convenient to write it in the form $S = (S_0, \mathbf{S})^T$ with the `spatial' $3$-vector part highlighted. 
The Stokes vector corresponding to a rank-one element has components that satisfy the relation $S_1^2 + S_2^2+ S_3^2 =S_0^2$.
Obviously, rank-one elements  are  light-like and rank-two elements are strictly time-like. 
One recalls that similar considerations apply to the density operator of a qubit as well. 

The (unnormalised) state operator post measurement (measurement element $\Pi_j$) evaluates to
\begin{align}
\rho_{\pi_j}^A&= {\rm Tr}_B[\hat{\rho}_{AB}\, \Pi_j  ] \nonumber\\
&= \frac{1}{8} {\rm Tr}_B\left[\, \left(\sum_{a,b=0}^3 M_{ab}\, \sigma_a \otimes \sigma_b^*\right)\,\left( \sum_{k=0}^3 S_k \sigma^*_k \right)\,\right]\nonumber \\
&=\frac{1}{8}  \sum_{a,b=0}^3 \sum_{k=0}^3 M_{ab} \, S_k \,\sigma_a {\rm Tr}( \sigma_b^* \sigma^*_k )\nonumber\\
&= \frac{1}{4} \sum_{a=0}^3 S^{\,'}_a \sigma_a,
\label{msa}
\end{align}  
where we used  ${\rm Tr}( \sigma_b^* \sigma^*_k ) = 2 \delta_{bk}$ in the last step. \\

\noindent
{\bf Remark}\,: It may be noted, for clarity, that we use Stokes vectors to represent both measurement elements and states. For instance,  Stokes vector $S$ in Eq.\,\eqref{msa} stands for a measurement element $\Pi_j$ on the B-side, whereas $S^{\,'}$ stands for (unnormalised) state of subsystem $A$. $\blacksquare$ 

The Stokes vector of the resultant state in Eq.\,\eqref{msa} is thus given by 
 $S^{\,'}_a= \sum_{k=0}^3 M_{ak} S_k$, which may be written in the suggestive form
\begin{align}
S^{\rm out} = M S^{\rm in}.
\label{ms}
\end{align}
Comparison with \eqref{mspol} prompts one to call $M$ the {\em Mueller matrix associated with two-qubit state} $\hat{\rho}_{AB}$. 
We repeat that the conditional state $\rho_{\pi_j}^A$ need not have unit trace and so needs to be normalised when computing entropy post measurement. To this end, we write $\rho_{\pi_j}^A = p_{j} \hat{\rho}_{\pi_j}$, where 
\begin{align}
 p_{j}   = \frac{S^{\rm out}_0}{2}, ~~ \hat{\rho}_{\pi_j} = \frac{1}{2} (1\!\!1 + (S_0^{\rm out})^{-1} \,\mathbf{S}^{\rm out}.\boldsymbol{\sigma}).
\end{align}
It is sometimes convenient to write the Mueller matrix $M$ associated with a given state $\hat{\rho}_{AB}$  in the block form 
\begin{align}
M = \left( 
\begin{array}{cc}
1 & \boldsymbol{\xi}^T \nonumber\\
\boldsymbol{\lambda} & \Omega
\label{genm}
\end{array}
\right), ~~ \boldsymbol{\lambda},\,\boldsymbol{\xi} \in {\cal R}^3.
\end{align}
Then the input-output relation \eqref{ms} reads 
\begin{align}
S_{0}^{\rm out} = S_0^{\rm in} + \boldsymbol{\xi} \cdot \mathbf{S^{\rm in}}, ~~~ \mathbf{S^{\rm out}} &= S_0^{\rm in}\,\boldsymbol{\lambda} + \Omega\, \mathbf{S^{\rm in}},
\end{align}
showing in particular that the probability of the conditional state $S^{\rm out}$ on the A-side depends on the POVM element precisely through $1 + \boldsymbol{\xi}\cdot {\bf S}^{\rm in}$.\\

\noindent
{\bf Remark}\,: The linear relationship between two-qubit density operators $\rho$ (states) and Mueller matrices (single qubit maps) we have developed in this Section can be usefully viewed as an instance of the Choi-Jamiokowski isomorphism. $\blacksquare$\\

\noindent
{\bf Remark}\,: We have chosen measurement to be on the B qubit. Had we instead chosen to compute correlations by performing measurements on subsystem $A$ then, by similar considerations as detailed above\,\eqref{msa}, we would have found $M^T$ playing the role of the Mueller matrix $M$. $\blacksquare$ 


\section{$X$-states and their Mueller matrices}
$X$-states are states whose density matrix $\rho$ has non-vanishing entries only along  the diagonal and the anti-diagonal. That is, the numerical  matrix $\rho$ has the `shape' of $X$. A general $X$-state can thus be written, to begin with, as
\begin{align}
\rho_X = \left( \begin{array}{cccc}
\rho_{00} & 0 &0 & \rho_{03} e^{i \phi_2} \\
0& \rho_{11} & \rho_{12} e^{i \phi_1} &0 \\
0&\rho_{12} e^{-i\phi_1} & \rho_{22} &0\\ 
\rho_{03} e^{-i \phi_2} & 0 & 0& \rho_{33}
\end{array}
\right),
\end{align}
where the $\rho_{ij}$'s are all real nonnegative.  One can get rid of the phases (of the off-diagonal elements) by a suitable local unitary transformation
$U_A \otimes U_B$. This is not only possible, but {\em also  desirable} because the quantities of interest, namely mutual information, quantum discord and classical correlation, are all invariant under local unitary transformations. Since it is unlikely to be profitable to carry around a baggage of irrelevant parameters, we shall indeed remove $\phi_1,\,\phi_2$ by taking $\rho_X$ to its canonical form $\rho_X^{\rm can}$.
We have  
\begin{align}
\rho_X \to \rho_X^{\rm can} =U_A \otimes U_B \, \rho_X \, U_A^{\dagger} \otimes U_B^{\dagger},
\end{align}
 where
\begin{align}
\rho_X^{\rm can}  &= \left( \begin{array}{cccc}
\rho_{00} & 0 &0 & \rho_{03}  \\ 
0& \rho_{11} & \rho_{12}  &0 \\
0&\rho_{12} & \rho_{22} &0\\ 
\rho_{03}  & 0 & 0& \rho_{33}
\end{array}
\right);\nonumber\\
U_A &= {\rm diag}(e^{-i(2\phi_1+ \phi_2)/4},e^{i \phi_2/4}),\nonumber\\ U_B &= {\rm diag}(e^{i( 2\phi_1-\phi_2)/4},e^{i \phi_2/4}).
\label{xcan}
\end{align}

\noindent
{\bf Remark}\,: We wish to clarify that $X$-states thus constitute, in the canonical form, a (real) 5-parameter family ($\rho_{00}+\rho_{11}+\rho_{22}+\rho_{33} = m_{00}=1$);  it can be lifted,  using  local unitaries $U_A,\,U_B \in SU(2)$ which have three parameters each, to a $11$-parameter subset in the $15$-parameter state space (or generalized Bloch sphere) of two-qubit states\,: they are all local unitary equivalent though they may no more have a `shape' $X$.  

With this canonical form, it is clear that the Mueller matrix for the generic $X$-state $\rho_X^{\rm can}$  has the form
\begin{align}
M = \left( \begin{array}{cccc}
1 & 0&0&m_{03}\\
0&m_{11}&0&0\\
0&0&m_{22} &0\\
m_{30}&0&0&m_{33}
\end{array}\right), 
\label{mueller}
\end{align} 
where 
\begin{align}
m_{11} &= 2(\rho_{03} + \rho_{12}), ~~m_{22} =  2(\rho_{03} - \rho_{12}), \nonumber \\
m_{03} &= \rho_{00} + \rho_{22} - (\rho_{11} + \rho_{33}), \nonumber \\
m_{33} &= \rho_{00} + \rho_{33} - (\rho_{11} + \rho_{22}), \nonumber\\
m_{30} &= \rho_{00} + \rho_{11} - (\rho_{22} + \rho_{33}),
\end{align}
as can be read off from the defining equation \eqref{mmatrix} or Eq.\,\eqref{app} in the Appendix. We note that the Mueller matrix of an $X$-state has a `sub-X' form\,: the only nonvanishing off-diagonal entries are $m_{03}$ and $m_{30}$ ($m_{12} = 0 = m_{21}$). In our computation later we will sometimes need the inverse relations
\begin{align}
\rho_{00} &= \frac{1}{4} (m_{00}+m_{03}+m_{30}+m_{33}),\nonumber\\\rho_{11}&=\frac{1}{4} (m_{00}-m_{03}+m_{30}-m_{33}),\nonumber\\ 
\rho_{22} &= \frac{1}{4} (m_{00}+m_{03}-m_{30}-m_{33}),\nonumber\\\rho_{33}&=\frac{1}{4} (m_{00}-m_{03}-m_{30}+m_{33}),\nonumber\\
\rho_{03} &= \frac{1}{4} (m_{11}+m_{22}),~~\rho_{12}=\frac{1}{4} (m_{11}-m_{22}).
\end{align}

The positivity properties of $\rho^{\rm can}_X$, namely $\rho_{00}\, \rho_{33} \geq \rho_{03}^2$, $\rho_{11}\, \rho_{22} \geq \rho_{12}^2$  transcribes to the following conditions on the entries of its Mueller matrix\,:
\begin{align}
(1+m_{33})^2 &- (m_{30}+m_{03})^2 \geq (m_{11}+m_{22})^2 \label{cp1}\\
(1-m_{33})^2 &- (m_{30}-m_{03})^2 \geq (m_{11}-m_{22})^2.
\label{cp2}
\end{align}

\noindent
{\bf Remark}\,: As noted earlier the requirements \eqref{cp1}, \eqref{cp2} on classical optical Mueller matrix \eqref{mueller} was noted for the first time in Refs.\,\cite{simon-mueller4,simon-mueller5}. These correspond to complete positivity of $M$ considered as a positive map (map which images the solid light cone into itself), and turns out to be equivalent to positivity of the corresponding two-qubit density operator. $\blacksquare$

By virtue of the direct-sum block structure of $X$-state density matrix, one can readily write down its (real) eigenvectors. We choose the following order 
\begin{align}
|\psi_0 \rangle &=\, c_{\alpha} |00\rangle + s_{\alpha} |11\rangle,~~|\psi_1 \rangle = \,c_{\beta} |01\rangle + s_{\beta} |10\rangle, \nonumber\\
|\psi_2 \rangle &= -s_{\beta} |01\rangle + c_{\beta} |10\rangle, \, |\psi_3 \rangle = -s_{\alpha}  |00\rangle + c_{\alpha}  |11\rangle,
\label{spec1}
\end{align}
where $c_{\alpha},\,s_{\alpha}$ denote respectively $\cos{\alpha}$ and $\sin{\alpha}$. And (dropping the superscript `can') we have the spectral resolution  
\begin{align}
\hat{\rho}_X = \sum_{j=0}^3 \lambda_j |\psi_j \rangle \langle \psi_j|,
\label{spec2}
\end{align} 
\begin{align}
c_{\alpha} &= \sqrt{\frac{1+\nu_1}{2}}, ~ c_{\beta} = \sqrt{\frac{1+\nu_2}{2}},\nonumber\\
\nu_1 &= \frac{\rho_{00}-\rho_{33}}{\sqrt{4\rho_{03}^2 +( \rho_{00} - \rho_{33})^2}} = \frac{m_{30}+m_{03}}{\sqrt{(m_{11}+m_{22})^2 +(m_{30}+m_{03})^2}},\nonumber\\
\nu_2 &= \frac{\rho_{11}-\rho_{22}}{\sqrt{4\rho_{12}^2 +( \rho_{11} - \rho_{22})^2}} = \frac{m_{30}-m_{03}}{\sqrt{(m_{11}-m_{22})^2 +(m_{30}-m_{03})^2}};\nonumber\\
\lambda_{0 \,{\rm or}\, 3} &= \frac{\rho_{00}+\rho_{33}}{2} \pm \frac{\sqrt{(\rho_{00}-\rho_{33})^2 + 4\,\rho_{03}^2}}{2}\nonumber\\
&= \frac{1+m_{33}}{4} \pm \frac{\sqrt{(m_{11}+m_{22})^2 +(m_{30}+m_{03})^2}}{4},\nonumber\\
\lambda_{1\, {\rm or}\, 2} &= \frac{\rho_{11}+\rho_{22}}{2} \pm \frac{\sqrt{(\rho_{11}-\rho_{22})^2 + 4\,\rho_{12}^2}}{2}\nonumber\\
&= \frac{1-m_{33}}{4} \pm \frac{\sqrt{(m_{11}-m_{22})^2 +(m_{30}-m_{03})^2}}{4}.
\label{eigen}
\end{align}
 
While computation of $S^A_{\rm min}$ will have to wait for a detailed consideration of the manifold of conditional states of $\hat{\rho}_{AB}$, the other entropic quantities can be evaluated right away. Given a qubit state specified by Stokes vector $(1,\mathbf{S})^T$, it is clear that its von Neumann entropy equals
\begin{align}
S_2(r)=-\left[\frac{1+r}{2}\right]\, {\ell og}_2 {\left[\frac{1+r}{2}\right]} -\left[\frac{1-r}{2}\right]\, {\ell og}_2 {\left[\frac{1-r}{2}\right]},
\end{align}
where $r$ is the norm of the three vector $\mathbf{S}$, or the distance of $\mathbf{S}$ from the origin of the Bloch ball.
Thus from Eq.\,\eqref{partial} we have
\begin{align}
S(\hat{\rho}_A) &= S_2(|m_{30}|), ~S(\hat{\rho}_B)=S_2(|m_{03}|), \nonumber\\
S(\hat{\rho}_{AB}) &\equiv S(\{\lambda_j \})= \sum_{j=0}^3 -\lambda_j {\ell og}_2\,(\lambda_j), 
\label{ents}
\end{align}
where $\lambda_j$, $j=0,1,2,3$ are the eigenvalues of the bipartite state $\hat{\rho}_{AB}$ given in Eq.\,\eqref{eigen}. The mutual information thus assumes the value
\begin{align}
I(\hat{\rho}_{AB}) = S_2(|m_{30}|) + S_2(|m_{03}|) - S(\{\lambda_j \}).
\label{mi}
\end{align}

\section{Correlation ellipsoid\,: Manifold of conditional states}
We have seen that the state of subsystem $A$ resulting from measurement of any POVM element on the B-side of $\hat{\rho}_{AB}$ is the Stokes vector resulting from
the action of the associated Mueller matrix on the Stokes vector of the POVM element. 
In the case of rank-one measurement elements, the `input' Stokes vectors correspond to points on the (surface ${\cal S}^2 = {\cal P}$  of the) Bloch ball. Denoting the POVM elements as
$S^{\rm in}=(1,x,y,z)^T$, $x^2+y^2+z^2=1$, we ask for the collection of corresponding normalized or conditional states. By Eq.\,\eqref{ms} we have 
\begin{align}
S^{\rm out} &= M S^{\rm in} = \left( \begin{array}{c} 1+m_{03} z\\ m_{11} \,x \\ m_{22} \, y\\ m_{30} + m_{33} \,z \end{array}\right) 
\to \left( \begin{array}{c} 1 \\ \frac{m_{11} \,x}{1+m_{03} z} \\ \frac{m_{22} \, y}{1+m_{03} z}\\ \frac{m_{30} + m_{33} \,z}{1+m_{03} z} \end{array}\right),
\label{outstates}
\end{align}
It is clear that, for $S_0^{\rm in}=1$,  $S_0^{\rm out} \neq 1$ whenever $m_{03} \neq 0$ and the input is {\em not} in the x-y plane of the Poincar\'{e} sphere.
It can be shown that the sphere $x^2+y^2+z^2=1$ at the `input' is mapped to the ellipsoid 
\begin{align}
\frac{x^2}{a_x^2} + \frac{y^2}{a_y^2} +\frac{(z-z_c)^2}{a_z^2} = 1
\end{align}
of normalized states at the output, the parameters of the ellipsoid being fully determined by the entries of $M$\,: 
\begin{align}
a_x &= \frac{|m_{11}|}{\sqrt{1-m_{03}^2}}, ~~~ a_y= \frac{|m_{22}|}{\sqrt{1-m_{03}^2}}, \nonumber \\ 
a_z &= \frac{|m_{33} - m_{03}m_{30}|}{1-m_{03}^2}, ~~~z_c = \frac{m_{30} - m_{03} m_{33}}{1-m_{03}^2}.
\label{ellipsoid}
\end{align}

{\bf Remark}\,: This ellipsoid of all possible (normalized) conditional states associated with a two-qubit state is something known as a steering ellipsoid\,\cite{demoor01,du-geom,du12pra,rudolph13}. It degenerates into a single point if and only if the state is a product state. It captures in a geometric manner correlations in the two-qubit state under consideration, and correlation is the object of focus of the present work. For those reasons, we prefer to call it the {\em correlation ellipsoid} associated with the given two-qubit state. While measurement elements $\Pi_j$ are mapped to points of the ellipsoid, measurement elements $a \Pi_j$ for all $a>0$ are mapped to one and the same point of the correlation ellipsoid. Thus, in the general case, each point of the ellipsoid corresponds to a `ray' of measurement elements. In the degenerate case, several rays map to the same point. $\blacksquare$

The x-z section of the correlation ellipsoid is pictorially depicted in Fig.\,\ref{ellipsoida}. It is clear that the geometry of the ellipsoid is determined by the four parameters $a_x, a_y,a_z,z_c$ and $z_c$ could be assumed nonnegative without loss of generality. The fifth parameter $m_{30}$ specifying the z-coordinate of the image ${\rm I}$ of the maximally mixed state on the B side is not part of this geometry. 

Having thus considered the passage from a two-qubit $X$-state to its correlation ellipsoid, we may raise the converse issue of going from the correlation ellipsoid to the associated $X$-state. To do this, however, we need the parameter $z_I=m_{30}$ as an input in addition to the ellipsoid itself. Further, change of the signature of $m_{22}$ does not affect the ellipsoid in any manner, but changes the states and correspondingly the signature of ${\rm det} M$. Thus, the signature of ${\rm det} M$ needs to be recorded as an additional binary parameter. It can be easily seen that the nonnegative $a_x,a_y,a_z,z_c$ along with $z_I$ and ${\rm sgn}\, ({\rm det} M)$ fully reconstruct the $X$-state in its canonical form \eqref{mueller}. Using local unitary freedom we can render $m_{11},\,m_{33}-m_{03} m_{30}$ and $z_c$ nonnegative so that ${\rm sgn}(m_{22}) = {\rm sgn}\, ({\rm det} M)$; $z_I = m_{30}$ can assume either signature. It turns out to be convenient to denote by $\Omega^+$ the collection of all Mueller matrices with ${\rm det} M \geq 0$ and by $\Omega^-$ those with ${\rm det} M \leq 0$. The intersection corresponds to Mueller matrices for which ${\rm det} M=0$, a measure zero set. Further, in our analysis to follow we assume, without loss of generality
\begin{align}
a_x \geq a_y.
\label{axgeqay}
\end{align}

\noindent
{\bf Remark}\,: Every two-qubit state has associated with it a unique correlation ellipsoid of (normalized) conditional states. An ellipsoid centered at the origin needs six parameters for its description\,: three for the sizes of the principal axes and three for the orientation of the ellipsoid as a rigid body. For a generic state, the centre $C$ can be shifted from the origin to vectorial location $\vec{r}_c$, thus accounting for three parameters, and ${\rm I}$ can be located at $\vec{r}_I$ anywhere inside the ellipsoid, thus accounting for another three. The three-parameter local {\em unitary} transformations on the B-side, having no effect whatsoever on the geometry of the ellipsoid, (but determines which points of the input Poincar\'{e} sphere go to which points on the surface of the ellipsoid), accounts for the final three parameters, adding to a total of $15$. For $X$-states the shift of $C$ from the origin {\em needs to be} along one of the principal directions and ${\rm I}$ is {\em constrained to be located on this very principal axis}. In other words, $\vec{r}_c$ and $\vec{r}_I$ become one-dimensional rather than three-dimensional variables rendering $X$-states a 11-parameter subfamily of the 15-parameter state space. {\em Thus $X$-states are distinguished by the fact that $C$, ${\rm I}$, and the origin are collinear with one of the principal axes of the ellipsoid}. This geometric rendering pays no special respect to the shape $X$, but is manifestly invariant under local unitaries as against the characterization in terms of `shape' $X$ of the matrix $\rho_{AB}$ in the computation basis. {\em The latter characterization is not even invariant under local unitaries!} $\blacksquare$

\begin{figure}
\begin{center}
\scalebox{1}{\includegraphics{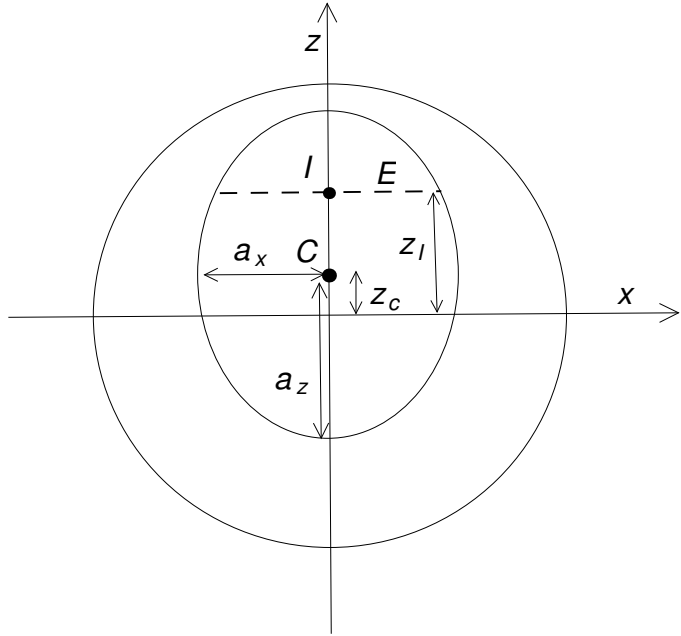}}
\end{center}
\caption{Showing the x-z cross-section of the correlation ellipsoid associated with a general $X$-state. The point ${\rm I}$ is the image of input state as identity, $C$ the center of the ellipsoid, and $E$ the image of the equatorial plane of the input Bloch sphere. 
 \label{ellipsoida}} 
\end{figure}

\section{Optimal measurement \label{optsec}}
In this Section we take up the central part of the present work which is to develop a provably optimal scheme for computation of the quantum discord for any $X$-state of a two-qubit system. Our treatment  is {\em both comprehensive and self-contained and, moreover, it is geometric in flavour}. We begin by exploiting symmetry to show, without loss of generality, that the problem itself is one of {\em optimization in just a single variable}. The analysis is entirely based on the output or correlation ellipsoid associated with a two-qubit state $\hr{AB}$, and we continue to assume that measurements are carried out on the B-side. 

The single-variable function under reference will be seen, on optimization, to divide the manifold of possible correlation ellipsoids into two subfamilies. For one subfamily the optimal measurement or POVM will be shown to be a von Neumann measurement along either x or z, {\em independent of the location (inside the ellipsoid) of ${\rm I}$, the image of the maximally mixed input}. For the other subfamily, the optimal POVM will turn out to be either a von Neumann measurement along x or a three-element POVM, {\em depending on the actual location of ${\rm I}$ in the ellipsoid}. There exists no $X$-state for which the optimal measurement requires a four-element POVM, neither does there exist an $X$-state for which the optimal POVM is von Neumann neither along x nor z.

For the special case of the centre $C$ of the ellipsoid coinciding with the origin $z=0$ of the Poincar\'{e} sphere ($z_c=0$), it will be shown that the optimal measurement is always a von Neumann measurement along x or z, {\em irrespective of the location of $z_I$ in the ellipsoid}. While this result may look analogous to the simple case of Bell mixtures earlier treated by Luo\,\cite{luo}, it should be borne in mind that these centred $X$-states form a much larger family than the family of Bell mixtures, for in the Luo scenario {\em ${\rm I}$ coincides with $C$ and hence  with the origin}, but we place no such restriction of coincidence. Stated differently, in our case $a_x,a_y,a_z$ and $z_I$ are independent variables. 

As we now turn to the analysis itself it is useful to record this\,: the popular result that the optimal POVM requires no more than four elements plays {\em a priori} no particular role of help in our analysis; it is for this reason that we shall have no occasion in our analysis to appeal to this important theorem\,\cite{ariano05,zaraket04}. \\

\noindent
{\bf Proposition}\,: The optimal POVM needs to comprise  rank-one elements.\\
\noindent
{\em Proof}\,: This fact is nearly obvious, and equally obvious is its proof. Suppose $\omega_j$ is a rank-two element of an optimal POVM and $\hat{\rho}_j^A$ the associated conditional state of subsystem $A$. Write $\omega_j$ as a positive sum of rank-one elements $\omega_{j1},\,\omega_{j2}$ and let $\hat{\rho}_{j1}^A,\,\hat{\rho}_{j2}^A$ be the conditional states corresponding respectively to $\omega_{j1},\,\omega_{j2}$.  It is then clear that $\hat{\rho}_j = \lambda \hat{\rho}_{j1}^A + (1-\lambda) \hat{\rho}_{j2}^A$, for some $0 < \lambda <1$. Concavity of the entropy function immediately implies $S(\hat{\rho}_j^A) > \lambda S(\hat{\rho}_{j1}^A) + (1-\lambda) S(\hat{\rho}_{j2}^A)$, in turn implying through \eqref{samin} that the POVM under consideration could not have been optimal. It is clear from the nature of the proof that this fact applies to all Hilbert space dimensions, and not just $d=2$. $\blacksquare$\\

\noindent 
{\bf Remark}\,: Since a rank-one POVM element $|v \rangle \langle v|$ is just a point $\mathbf{S}$ on (the surface of) the Bloch (Poincar\'{e}) sphere ${\cal P}$, a four element rank-one POVM is a quadruple of points $\mathbf{S^{(j)}}$ on ${\cal P}$, with associated probabilities $p_j$. The POVM condition $\sum_j p_j |v_j \rangle \langle v_j| = 1\!\!1$ demands that we have to solve the pair
\begin{align}
p_1 + p_3 &+p_3 +p_4 =2,\nonumber\\
&\sum_j p_j\,\mathbf{S^{(j)}} =0.
\label{povmcondition}
\end{align}
Once four points $\mathbf{S^{(j)}}$ on ${\cal P}$ are chosen, the `probabilities' $\{p_j \}$ are {\em not independent}. To see this, consider the tetrahedron for which  $\mathbf{S^{(j)}}$ are the vertices. If this tetrahedron does not contain the origin, then $\sum_j p_j \mathbf{S^{(j)}} =0$ has no solution with nonnegative $\{p_j \}$. If it contains the origin, then there exits a solution and the solution is `essentially' unique by Caratheodory theorem. 

The condition $\sum_j p_j =2$ comes into play in the following manner. Suppose we have a solution to $\sum_j p_j\,\mathbf{S^{(j)}} =0$. It is clear that $p_j \to p_j^{\,'} = a p_j$, $j=1,2,3,4$, with no change in $\mathbf{S^{(j)}}$'s, will also be a solution for any ($j$-independent) $a>0$. It is this freedom in choosing the scale parameter $a$ that gets frozen by the condition $\sum_j p_j =2$, rendering the association between tetrahedra and solutions of the pair \eqref{povmcondition} unique.

{\em We thus arrive at a geometric understanding of the manifold of all (rank-one) four-element POVM's, even though we would need such POVM's only when we go beyond $X$-states. This is precisely the manifold of all tetrahedra with vertices on ${\cal P}$, and containing the centre in the interior of ${\cal P}$.} We are not considering four-element POVM's whose $\mathbf{S^{(j)}}$ are coplanar with the origin of ${\cal P}$, because they are of no use as optimal measurements. It is clear that three element rank-one POVM's  are similarly characterized, again by the Caratheodory theorem, by triplets of points on ${\cal P}$ coplanar with the origin of ${\cal P}$, with the requirement that the triangle generated by the triplet contains the origin {\em in the interior}. Further, it is trivially seen in this manner that  2-element rank-one POVM's are von Neumann measurements determined by pairs of antipodal $\mathbf{S^{(j)}}$'s on ${\cal P}$, i.e., by `diameters' of ${\cal P}$. $\blacksquare$

The correlation ellipsoid of an $X$-state (as a subset of the Poincar\'{e} sphere) has a ${\cal Z}_2 \times {\cal Z}_2$ symmetry generated by reflections respectively about the x-z and y-z planes. We shall now use the product of these two reflections---a $\pi$-rotation or inversion about the z-axis---to simplify, without loss of generality, our problem of optimization. \\

\noindent
{\bf Proposition}\,: All elements of the optimal POVM have to necessarily correspond to Stokes vectors of the form $S_0(1,\, \sin{\theta},\,0,\,\cos{\theta})^T$. \\
\noindent
{\em Proof}\,: Suppose ${\cal N} = \{\omega_1,\,\omega_2,\,\cdots,\,\omega_k \}$ is an optimal POVM of rank-one elements (we are placing no restriction on the cardinality $k$ of ${\cal N}$, but rather allow it to unfold naturally from the analysis to follow). And let $\{\hat{\rho}_1^A,\,\hat{\rho}_2^A,\,\cdots,\,\hat{\rho}_k^A \}$ be the corresponding conditional states, these being points on the boundary of the correlation ellipsoid. Let $\tilde{\omega}_j$ and $\tilde{\hat{\rho}}_j^A$ represent, respectively, the images of $\omega_j$, $\hat{\rho}_j^A$ under $\pi$-rotation about the z-axis (of the input Poincar\'{e} sphere and of the correlation ellipsoid). It follows from symmetry that $\widetilde{{\cal N}} = \{\tilde{\omega}_1,\,\tilde{\omega}_2,\,\cdots,\,\tilde{\omega}_k \}$ too is an optimal POVM. And so is also ${\cal N}\, \overline{\bigcup} \,\widetilde{{\cal N}}$, where we have used the decorated symbol $\overline{\bigcup}$ rather than the set union symbol $\bigcup$ to distinguish from simple union of sets\,: if $S_0(1\!\!1 \pm \sigma_3)$ happens to be an element $\omega_j$ of ${\cal N}$, then $\tilde{\omega}_j = w_j$ for this element, and in that case this $\omega_j$ should be `included' in ${\cal N}\, \overline{\bigcup}\, \widetilde{{\cal N}}$ {\em not once but twice} (equivalently its `weight' needs to be doubled). The same consideration holds if ${\cal N}$ includes $\omega_j$ and $\sigma_3 \, \omega_j\,\sigma_3$.

Our supposed to be optimal POVM thus comprises pairs of elements $\omega_j,\,\tilde{\omega}_j$ related by inversion about the z-axis\,: $\tilde{\omega}_j = \sigma_3\,\omega_j\,\sigma_3$. Let us consider the associated conditional states $\hat{\rho}_j^A,\,\tilde{\hat{\rho}}_j^A$ on the (surface of the) correlation ellipsoid. They have identical z-coordinate $z_j$. The section of the ellipsoid (parallel to the x-y plane) at $z=z_j$ is an ellipse, with major axis along $x$ (recall that we have assumed, without loss of generality $a_x \geq a_y$), and $\hat{\rho}_j^A$ and $\tilde{\hat{\rho}}_j^A$ are on opposite ends of a line segment through the centre $z_j$ of the ellipse. Let us {\em assume} that this line segment is not the major axis of the ellipse $z=z_j$. That is, we assume $\hat{\rho}_j^A$, $\tilde{\hat{\rho}}_j^A$ are not in the x-z plane. 

Now slide (only) this pair along the ellipse smoothly, keeping them at equal and opposite distance from the z-axis until both reach opposite ends of the major axis of the ellipse, the x-z plane. It is clear that during this process of sliding $\hat{\rho}_j^A$, $\tilde{\hat{\rho}}_j^A$ recede away from the centre of the ellipse and hence away from the centre of the Poincar\'{e} sphere itself. As a result $S(\hat{\rho}_j^A)$ decreases, thus improving the value of $S^A_{\rm min}$ in \eqref{samin}. This would have proved that the POVM ${\cal N}$ is not optimal, unless our  assumption that $\hat{\rho}_j^A$, $\tilde{\hat{\rho}}_j^A$ are not in the x-z plane is false. This completes proof of the proposition. $\blacksquare$

This preparation immediately leads to the following important result which forms the basis for our further analysis.

\begin{theorem}\,:
The problem of computing quantum discord for $X$-states is a problem of convex optimization on a plane or optimization over a single variable.   
\end{theorem}
\noindent
{\em Proof}\,: We have just proved that elements of the optimal POVM come, in view of the ${\cal Z}_z \times {\cal Z}_2$ symmetry of $X$-states,  in pairs $S_0(1,\,\pm \sin{\theta},\,0,\,\cos{\theta})^T$ of Stokes vectors $\omega_j,\,\tilde{\omega}_j$ with $0 \leq \theta \leq \pi$. The corresponding conditional states come in pairs $\hat{\rho}_j^A$, $\tilde{\hat{\rho}}_j^A = 1/2(1\!\!1 \pm x_j \sigma_1 + z_j \sigma_3)$. The two states of such a pair of conditional states are at the same distance 
\begin{align}
r(z_j) = \sqrt{z_j^2 + a_x ^2 -(z_j -z_c)^2 a_x^2/a_z^2}
\label{rz}
\end{align}
from the origin of the Poincar\'{e} sphere, and hence they have the same von Neumann entropy 
\begin{align}
f(z_j) &= S_2(r(z_j)),\nonumber\\
S_2(r)&= -\frac{1+r}{2}\,{\ell og}_2\,\frac{1+r}{2}-\frac{1-r}{2}\,{\ell og}_2\,\frac{1-r}{2}.
\label{fz}
\end{align}

Further, continuing to assume without loss of generality $a_x \geq a_y$, our convex optimization is not over the three-dimensional ellipsoid, but effectively a {\em planar} problem over the x-z elliptic section of the correlation ellipsoid, and hence the optimal POVM cannot have more that three elements. Thus, the (Stokes vectors of the) optimal POVM elements on the B-side necessarily have the form,
\begin{align}
\Pi^{(3)}_{\theta} &= \{ 2 p_0(\theta) (1,0,0,1)^T, \,2 p_1(\theta) (1, \pm \sin{\theta}, 0, -\cos{\theta} )^T\}, \nonumber\\
 p_0(\theta) &= \frac{\cos{\theta}}{1+ \cos{\theta}}, ~~ p_1(\theta) = \frac{1}{2(1+\cos{\theta})}, ~ 0 \leq \theta \leq \pi/2.
\label{scheme}
\end{align}
The optimization itself is thus over the single variable $\theta$. $\blacksquare$\\

\noindent
{\bf Remark}\,: It is clear that $\theta=0$ and $\theta = \pi/2$ correspond respectively to von Neumann measurement along z and x, and no other von Neumann measurement gets included in $\Pi_{\theta}$. Every $\Pi_{\theta}$ in the open interval $0 < \theta < \pi/2$ corresponds to a {\em genuine} three-element POVM. 

The symmetry considerations above do allow also three-element POVM's of the form 
\begin{align}
\widetilde{\Pi}^{(3)}_{\theta} &= \{ 2 p_0(\theta) (1,0,0,-1)^T,\, 2 p_1(\theta) (1, \pm \sin{\theta}, 0, \cos{\theta} )^T\}, ~~
 0 \leq \theta \leq \pi/2,
\end{align} 
but such POVM's lead to local maximum rather than a minimum for $S^A$, and hence are of no value to us.   $\blacksquare$

\section{Computation of $S^A_{\rm min}$ \label{saminsec}}
A schematic diagram of the 3-element  POVM $\Pi^{(3)}_{\theta}$ of Eq.\,\eqref{scheme} is shown in Fig.\,\ref{opta}. The Bloch vectors of the corresponding conditional states $\hr{1}^A,\, \hr{2}^A,\, \hr{3}^A$ at the output are found to be of the form 
\begin{align}
(0,\,0,\,z_c+a_z)^T,\,&(x(z),\,0,\,z)^T,\,(-x(z),\,0,\,z)^T, \nonumber\\
~~ &x(z) = \frac{a_x}{a_z}(a_z^2 - (z-z_c)^2)^{1/2}.
\end{align}
For these states denoted $1,2,3$ in Fig.\,\ref{opta} the weights should be chosen  to realize as convex sum the state ${\rm I}$ (the image of the maximally mixed input) whose Bloch vector is  $(0,0,z_I)^T$. The von Neumann measurements along the z or x-axis correspond respectively to $z= z_c-a_z$ or $z_I$. 
\begin{figure}
\begin{center}
\scalebox{1.5}{\includegraphics{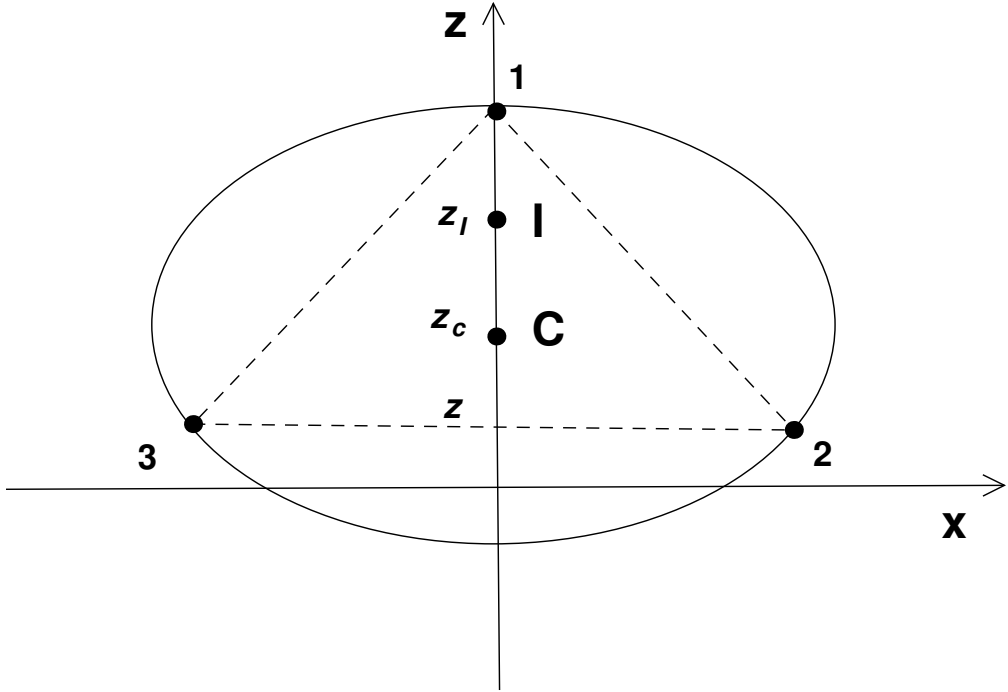}}
\end{center}
\caption{Showing the conditional states corresponding to the 3-element measurement scheme of\,\eqref{scheme}.} 
\label{opta}
\end{figure}
Using Eqs.\,\eqref{rz},\,\eqref{fz}, the expression for $S^A(z)$ is thus given by 
\begin{align}
S^A(z)= p_1(z) \, f(z_c+a_z) + p_2(z)\,f(z),\nonumber\\
p_1(z) = \frac{z_I-z}{z_c+a_z-z}, ~~ p_2(z) = \frac{z_c+a_z-z_I}{z_c+a_z-z}. 
\end{align}
The minimization of $S^A(z)$ with respect to the single variable $z$ should give $S^A_{\rm min}$. It may be noted in passing that, for a given ${\rm I}$ or $z_I$, the three-element POVM parametrized by $z$ is of no value in the present context for $z > z_I$.

For clarity of presentation, we begin by considering a specific example  $(a_z,\,z_c,\,z_{I}) = (0.58,\,0.4,0.6)$. The relevant interval for the variable $z$ in this case is $[z_c-a_z,\, c+a_z]=[-0.18,0.98]$, and we shall examine the situation as we vary $a_x$ for fixed $(a_z,z_c,z_{I})$. The behaviour of $S^A(z)$ for this example is depicted in Fig.\,\ref{sap}, wherein each curve in the $(z,\,S^A(z))$ plane corresponds to a chosen value of $a_x$, and the value of $a_x$ increases as we go down Fig.\,\ref{sap}. For values of $a_x \leq a_x^V(a_z,z_c)$, for some $a_x^V(a_z,z_c)$ to be detailed later, $S^A(z)$ is seen to be a monotone increasing function of $z$, and so its minimum $S^A_{\rm min}$ obtains at the `lower' end point $z = z_c-a_z=-0.18$, hence the optimal POVM corresponds to the vertical projection or von Neumann measurement along the z-axis. The curve marked $2$ corresponds to $a_x = a^V_x(a_z,z_c)$ [which equals $0.641441$ for our example].

Similarly for values of $a_x \geq a_x^H(a_z,z_c)$, $S^A(z)$ proves to be a monotone decreasing function of $z$, its minimum therefore obtains at the `upper' end point which is $z_I$ and not $z_c+a_z$ [recall that the three-element POVM makes no sense for $z>z_I$]; hence the optimal POVM corresponds to horizontal projection or von Neumann measurement along x-axis. It will be shown later that both $a^V_x(a_z,z_c)$, $a^H_x(a_z,z_c)$ do indeed depend only on $a_z,z_c$ and not on $z_I$. Both are therefore properties of the ellipsoid\,: all states with one and the same ellipsoid will share the same $a^V_x(a_z,z_c)$, $a^H_x(a_z,z_c)$.
\begin{figure}
\begin{center}
\scalebox{1.6}{\includegraphics{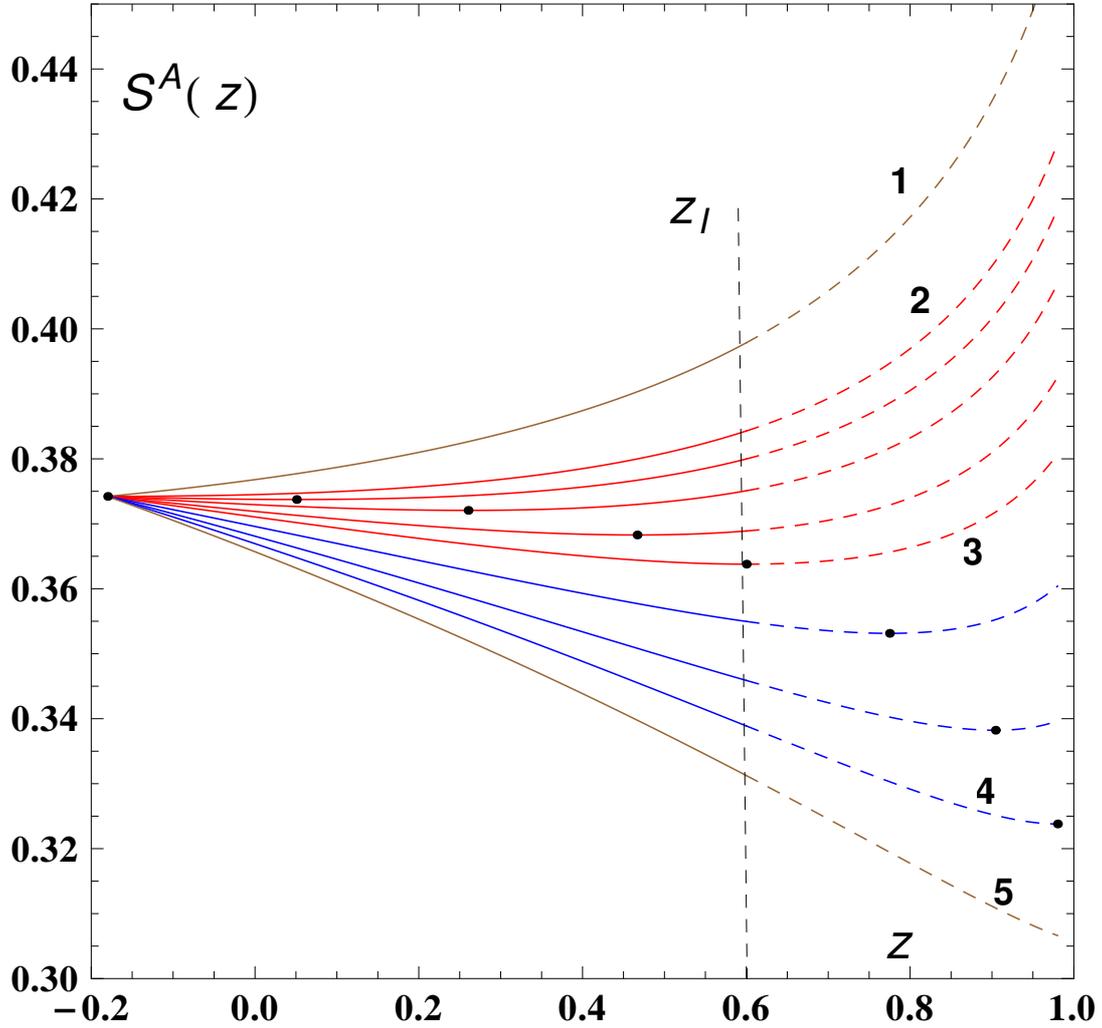}}
\end{center}
\caption{Showing $S^A(z)$ for various values of $a_x$ labelled in increasing order. The line marked $z_I$ denotes $z=z_I$. A three-element POVM scheme results for values of $a_x \in (a_x^V(a_z,z_c), a^H_x(a_z,z_c))$ [the curves (2) and (4)]. For values of $a_x \leq a^V(a_z,z_c)$ [example of curve (1)], the von Neumann projection along the z-axis is the optimal and for values of $a_x \geq a^H(a_z,z_c)$ [example of curve (5)], the von Neumann projection along the x-axis is the optimal. The optimal $z=z_0$ is obtained by minimizing $S^A(z)$ (marked with a dot). $S^A(z)$ for $z>z_I$ is not meaningful and this region of the curves is shown with dashed lines. In this example $(a_z,z_c) = (0.58,0.4,0.6)$. For $z_I = 0.6$, a three-element POVM results for (red) curves between (2) and (3).} 
\label{sap}
\end{figure} 

Thus, it is the region $a^V_x(a_z,z_c) < a_x < a^H_x(a_z,z_c)$ of values of $a_x$ that needs a more careful analysis, for it is only in this region that the optimal measurement could possibly correspond to a three-element POVM. Clearly, this region in the space of correlation ellipsoids is distinguished by the fact that $S^A(z)$ has a minimum at some value $z = z_0$ in the open interval $(z_c-a_z,\,z_c+a_z)$. For $a_x = a^V_x(a_z,z_c)$ this minimum occurs at $z_0 = z_c- a_z$, moves with increasing values of $a_x$ gradually towards $z_c+a_z$, and reaches $z_c+a_z$ itself as $a_x$ reaches $a^H_x(a_z,z_c)$. 

Not only this qualitative behaviour, but also the exact value of $z_0(a_z,z_c,a_x)$ is independent of $z_I$, as long as $z_c \neq 0$. Let us evaluate $z_0(a_z,z_c,a_x)$ by looking for the zero-crossing of the derivative function $dS^A(z)/dz$ depicted in Fig.\,\ref{firstderi}. We have 
\begin{align}
\frac{d\, S^A(z)}{dz} &= (z_c+a_z-z_I)\,G(a_z,a_x,z_c;z),\nonumber\\
G(a_z,a_x,z_c;z) &= \frac{1}{(z_c+a_z-z)^2}\left([(z_c+a_z-z)(a_x^2(z-z_c)/a_z^2-z) X(z)]\right.\nonumber\\
 &~~~~~~~~~~~~~~~~~~~~~~~~~~~~~~~~~~~~ \left. - [f(z_c+a_z)-f(z)]\right),\nonumber\\
X(z) &= \frac{1}{2r(z)} {\ell og}_2\left[\frac{1+r(z)}{1-r(z)}\right],
\label{Gexp}
\end{align} 
\begin{figure}
\begin{center}
\scalebox{1.5}{\includegraphics{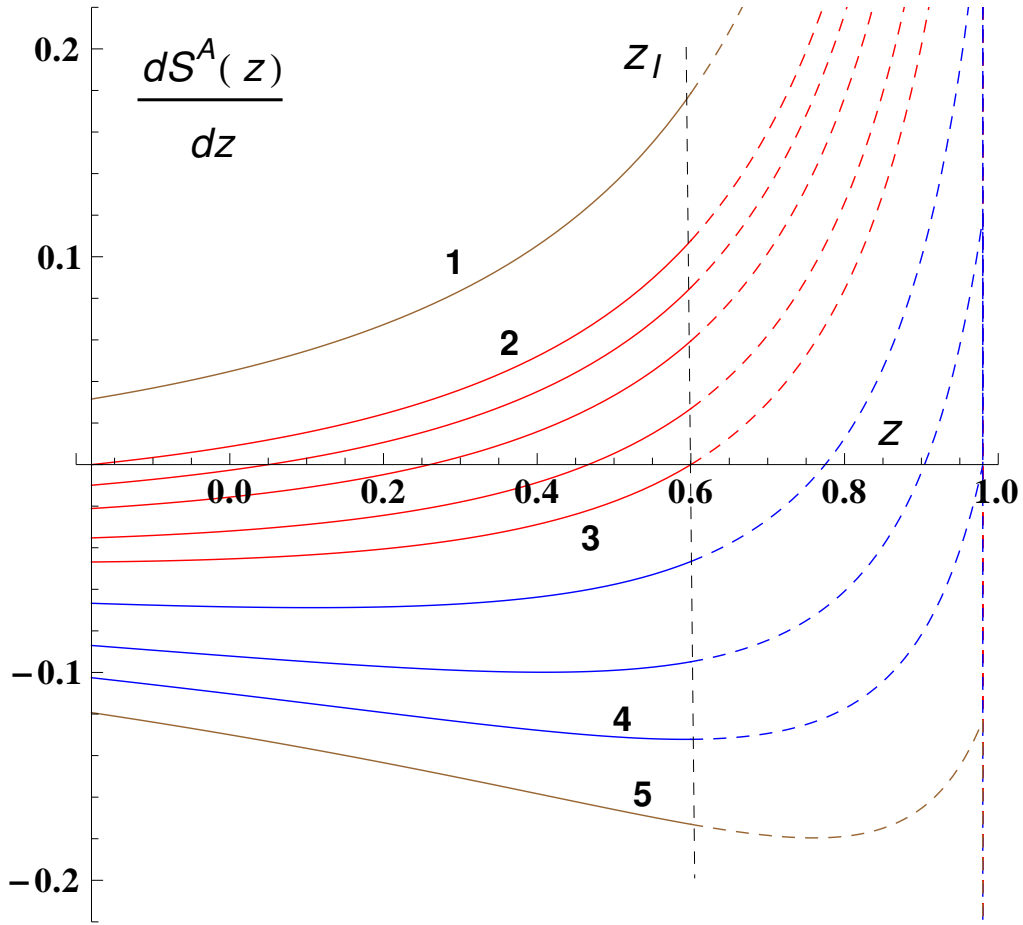}}
\end{center}
\caption{Showing $dS^A(z)/dz$ for various values of $a_x$ labelled in increasing order. A root exits  for values of $a_x \in (a_x^V(a_z,z_c), a^H_x(a_z,z_c))$ [between curves (2) and (4)]. For values of $a_x \leq a^V(a_z,z_c)$ and  $a_x \geq a^H(a_z,z_c)$, there is no $z_0$ [examples are curves (1) and (5)].} 
\label{firstderi}
\end{figure}
and we need to look for $z_0$ that solves $G(a_z,a_x,z_c;z_0)=0$. The reader may note that $z_I$ {\em does not enter} the function $G(a_z,a_x,z_c;z_0)$ defined in Eq.\,\eqref{Gexp}, showing that $z_0$ is indeed independent of $z_I$ as claimed earlier\,: {\em $z_0$ is a property of the correlation ellipsoid; all states with the same correlation ellipsoid have the same $z_0$}.

Let us focus on the two curves $a^V_x(a_z,z_c)$, $a^H_x(a_z,z_c)$ introduced earlier and defined through 
\begin{align}
a^V_x(a_z,z_c)\,:~  G(a_z,a_x^V,z_c;z_c-a_z) =0,\nonumber\\
a^H_x(a_z,z_c)\,:~ G(a_z,a_x^H,z_c;z_c+a_z) =0.
\label{bounds}
\end{align}
The  curve $a^V_x(a_z,z_c)$ characterizes, for a given $(a_z,z_c)$, the value of $a_x$ for which the first derivative of $S^A(z)$ vanishes at $z= z_c-a_z$ (i.e., $z_0 = z_c - a_z$), so that the vertical von Neumann projection is the optimal POVM for all $a_x \leq a^V_x(a_z,z_c)$. Similarly, the  curve $a^H_x(a_z,z_c)$ captures the value of $a_x$ for which the first derivative of $S^A(z)$ vanishes at $z= z_c+a_z$.  
Solving for the two curves in terms of $a_z$ and $z_c$ we obtain, after some algebra, 
\begin{align}
a^V_x(a_z,z_c) &= \sqrt{\frac{f(|z_c-a_z|)-f(z_c+a_z)}{2 X(|z_c-a_z|)} - a_z(z_c-a_z)},\nonumber\\
a^H_x(a_z,z_c) &= \frac{(z_c+a_z)}{2[Y(z_c+a_z)-X(z_c+a_z)]}\nonumber\\
& \hspace{2cm}\times  \, \left[(z_c-a_z) X(z_c+a_z) + 2a_zY(z_c+a_z) - \sqrt{w}\,\right],\nonumber\\
Y(z) &= \frac{1}{[{\ell n}\,2\,](1-r(z)^2)},\nonumber\\
~~w &= X(z_c+a_z)[(z_c-a_z)^2 X(z_c+a_z) + 4 a_z z_c Y(z_c+a_z)].
\label{ahav}
\end{align}
\begin{figure}
\begin{center}
\scalebox{1.6}{\includegraphics{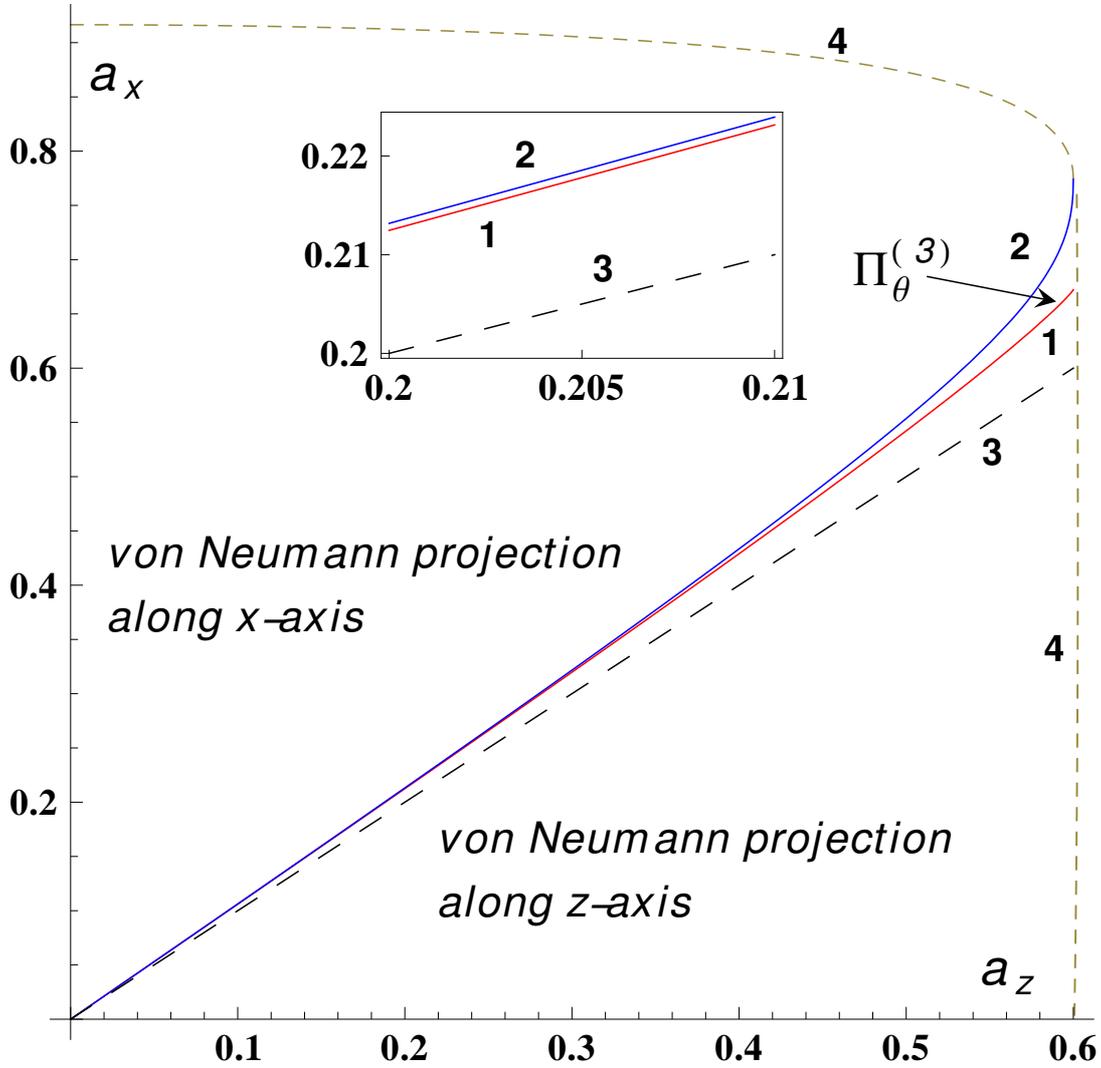}}
\end{center}
\caption{Showing the various measurement schemes across a slice (the $a_x-a_z$ plane) of  parameter space (of correlation ellipsoids) with $z_c=0.4$.  We see that only for a tiny wedge-shaped region marked $\Pi^{(3)}_{\theta}$, the region between $a^V(a_z,z_c)$ (1) and $a^H(a_z,z_c)$ (2), one can expect a 3-element POVM. Region above (2) corresponds to a von Neumann measurement along the x-axis and the region below curve (1) corresponds to a von Neumann measurement along the z-axis. Curves marked (4) depict the boundary of allowed values for $a_z,a_x$. The curve (3) is the line $a_z = a_x$. The inset shows curves (1), (2) and (3) for $a_x \in [0.2,0.21]$.} 
\label{wedge}
\end{figure}
These curves are marked $(1)$ and $(2)$ respectively  in Fig.\,\ref{wedge}. Two aspects are of particular importance\,:
\begin{enumerate}[(i)]
\item $a_x^H(a_z,z_c) \geq a^V_x(a_z,z_c)$, the inequality saturating if and only if $z_c=0$. In particular these two curves never meet, the appearance in Fig.\,\ref{wedge} notwithstanding. It is to emphasize this fact that an inset has been added to this figure. The straight line $a_x = a_z$, marked (3) in Fig.\,\ref{wedge}, shows that $a_x^V(a_z,z_c) \geq  a_z$, the inequality saturating if and only if $z_c=0$.
\item It is only in the range $a_x^V(a_z,z_c) < a_x < a_x^H(a_z,z_c)$ that we get a solution $z_0$
\end{enumerate}
\begin{align}
G(a_z,a_x,z_c;z_0) =0, ~~ z_c-a_z < z_0 < z_c+ a_z
\end{align}
corresponding to a {\em potential} three-element optimal POVM for {\em some} $X$-state corresponding to the ellipsoid under consideration; and, clearly, the optimal measurement for the state will {\em actually} correspond to a three-element POVM only if 
\begin{align}
z_I > z_0. 
\end{align}
If $z_I \leq z_0$, the optimal measurement corresponds to a von Neumann projection along the x-axis, and never the z-axis.

We also note that the range of values of $a_x$ for a fixed $(a_z,z_c)$ where the potential three-element POVM can exist increases with increasing $a_z$. 
Let us define a parameter $\delta$ as $\delta(a_z,z_c) = a^H_x(a_z,z_c) - a^V_x(a_z,z_c)$ which captures the extent of region bounded by curves (1) and (2) in Fig.\,\ref{wedge}. This object is shown in Fig.\,\ref{shifts}. 
\begin{figure}
\begin{center}
\scalebox{1.5}{\includegraphics{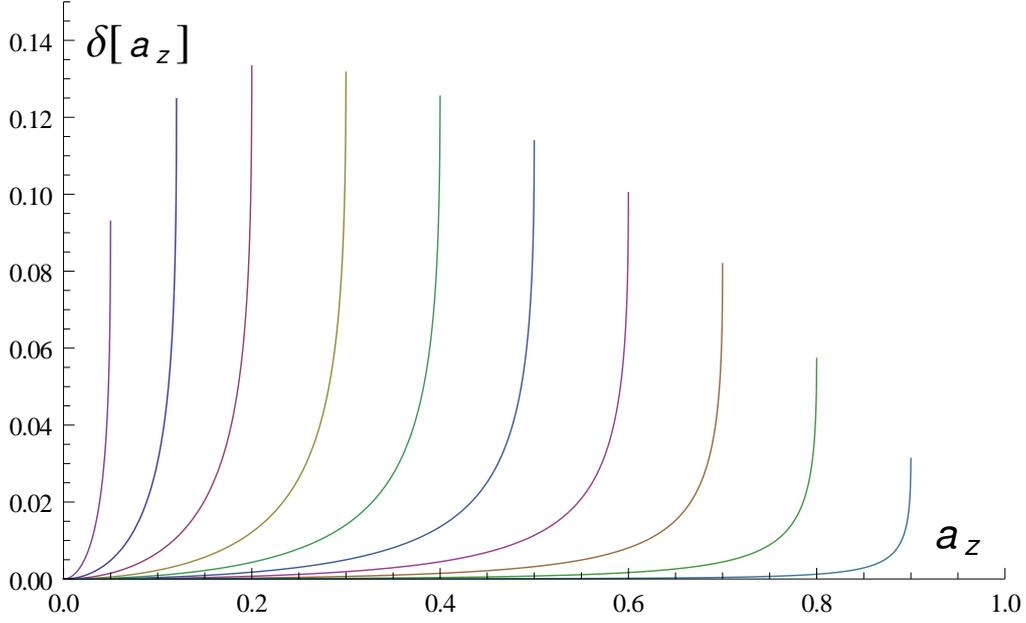}}
\end{center}
\caption{Showing $\delta(a_z)$ for decreasing values of $z_c$ from left to right. The first curve corresponds to $z_c=0.95$ and the last curve to $z_c=0.1$. We see that size of the `wedge'-shaped region (Fig.\,\ref{wedge}) first increases and then decreases with increasing $z_c$.} 
\label{shifts}
\end{figure}
We see that the range of values of $a_x$ for which a three-element POVM exists first increases with increasing $z_c$ and then decreases.\\

{\bf \em An example}\,:
We now evaluate quantum discord for a one-parameter family of states we construct and denote by $\hr{}(a)$. The Mueller matrix associated with $\hr{}(a)$ is chosen as\,:
\begin{align}
M(a) = \left[\begin{array}{cccc}
1 & 0 & 0 & y\\
0& a (1-y^2)^{1/2} & 0 & 0 \\
0& 0 & 0.59 (1-y^2)^{1/2} & 0\\
0.5 & 0 & 0 & 0.58 + 0.4 y
\end{array}
\right],
\end{align}
where $a \in [0.59,0.7]$ and $y=0.1/0.58$. The ellipsoid parameters for our class of states is given by $(a_x,a_y,a_z,z_c,z_I) = (a,0.59,0.58,0.4,0.5)$. The class of states differ only in the parameter $a_x$ which changes as the parameter $a$ is varied in the chosen interval. Using the optimal measurement scheme outlined above and in the earlier Section, we compute $S^A_{\rm min}$ and quantum discord. The values are displayed in Fig.\,\ref{example1} in which $S^A_{\rm min}$ is denoted by curve (1) and quantum discord by curve (4). An over-estimation of quantum discord by restricting to von Neumann measurements along x or z-axis is shown in curve (3) for comparison with the optimal three-element POVM. The point E denotes the change in the measurement from z-axis projection to a three-element POVM and point F denotes a change from the three-element POVM scheme to the x-axis von Neumann measurement.

\begin{figure}
\begin{center}
\scalebox{1.2}{\includegraphics{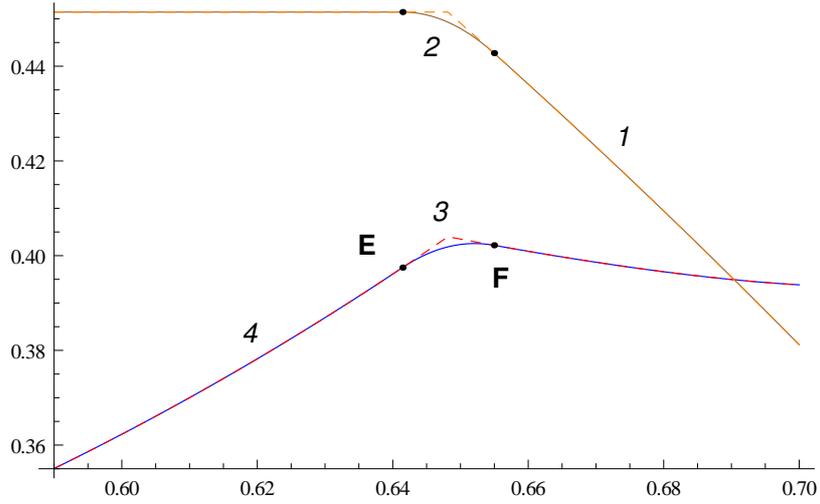}}
\end{center}
\caption{
Showing $S^A_{\rm min}$ [curve (1)] and quantum discord [curve (4)] for a one-parameter family of states $\hr{}(a)$. The ellipsoid corresponding to $\hr{}(a)$  has parameters $(a_x,a_y,a_z,z_c,z_I) = (a,0.59,0.58,0.4,0.5)$ where $a  \in [0.59,0.7]$. Point E denotes the change of the optimal measurement from a von Neumann measurement along the z-axis to a three-element POVM, while point F denotes the change of the optimal measurement from a three-element POVM to a von Neumann measurement along the x-axis, with increasing values of the parameter $a$. The curve (3) (or (2))  denotes the over(under)-estimation of quantum discord (or $S^A_{\rm min}$) by restricting the measurement scheme to a von Neumann measurement along the z or x-axis. This aspect is clearly brought out in the following Fig.\,\ref{example1b}.} 
\label{example1}
\end{figure}

\begin{figure}
\scalebox{1.2}{
\includegraphics{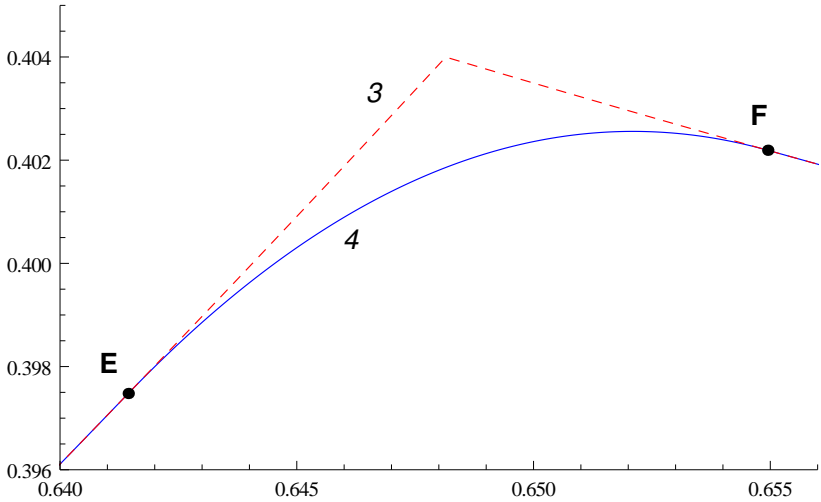}}
\caption{
Showing quantum discord [curve (4)] with increasing $a_x$ when there is a  transition of the measurement scheme from the von Neumann measurement along the z-axis to the three-element scheme (E) and finally to the von Neumann measurement along the x-axis (F). The over-estimation of quantum discord is depicted in curve (3) where the measurement scheme is restricted to one of von Neumann measurements along the z or x-axis. We see a gradual change in quantum discord in contrast to the sharp change in the restricted case.} 
\label{example1b}
\end{figure}

This transition is clearly shown in Fig.\,\ref{example1b}. We see that had one restricted to only the von Neumann measurement along the z-axis or the x-axis, one would obtain a `kink' in the value for quantum discord. Whereas, the optimal three-element scheme returns a gradual change in the value of quantum discord as we change the parameter $a$. The curves (3) and (4) will only merge for the value $z_c=0$ and this aspect will be detailed in a later Section. This behaviour of quantum discord is generic to any non-zero value of the center $z_c$ of the correlation ellipsoid. \\

\noindent
{\bf \em  Purification and  EoF}\,:
The Koashi-Winter theorem or relation\,\cite{koashi04} shows that the classical correlation  of a given bipartite state $\hat{\rho}_{AB}$ is related to the entanglement of formation of the `complimentary' state $\hat{\rho}_{CA}$. 
That is, 
\begin{align}
C(\hat{\rho}_{AB}) = S(\hat{\rho}_A) - E_F(\hat{\rho}_{CA}).
\end{align}
Comparing with the definition of $S^A_{\rm min}$ in Eq.\,\eqref{samin}, we see that
\begin{align}
 S^A_{\rm min}(\hat{\rho}_{AB}) = E_F(\hat{\rho}_{CA}).
\end{align}
In other words, the Koashi-Winter relation connects the (minimum average) conditional entropy post measurement of a bipartite state $\hat{\rho}_{AB}$ to the 
entanglement of formation of its complimentary state $\hat{\rho}_{CA}$ defined through purification of $\hat{\rho}_{AB}$ to pure state $|\phi_{C:AB}\rangle$.

The purification can be written as $|\phi_{C:AB} \rangle = \sum_{j=0}^3 \sqrt{\lambda_j}  |e_j \rangle \otimes |\psi_j\rangle$, $\{|e_j\rangle\}$ being orthonormal vectors in the Hilbert space of subsystem C. Now, the complimentary state $\hat{\rho}_{CA}$ results when subsystem $B$ is simply discarded\,:
\begin{align}
\hat{\rho}_{CA} &= {\rm Tr}_B [|\phi_{C:AB}\rangle \langle \phi_{C:AB}|] \nonumber\\
& = \sum_{j,k=0}^3 \sqrt{\lambda_j \lambda_k} |e_j \rangle \langle e_k| \otimes {\rm Tr}[|\psi_j\rangle \langle \psi_k|].
\end{align}
It is easy to see that for the case of the two qubit $X$-states, the complimentary state belongs to a $2 \times 4$ system. Now that $S^A_{\rm min}$ is determined for all $X$-states by our procedure, using Eqs.\,\eqref{spec1},\,\eqref{spec2},\,\eqref{eigen}  one can immediately write down the expressions for the entanglement of formation for the complimentary states corresponding to the entire 11-parameter family of $X$-states states using the optimal measurement scheme outlined in Sec.\,\ref{optsec}. We note in passing that examples of this connection for the particular cases of states such as rank-two two-qubit states and Bell-mixtures have been earlier studied in\,\cite{du-rank2,yan11}.

\section{Invariance group beyond local unitaries}
Recall that a measurement element (on the B side) need not be normalized. Thus in constructing the correlation ellipsoid associated with a two-qubit state $\hr{AB}$, we gave as input to the Mueller matrix associated with $\hr{AB}$ an arbitrary four-vector in the positive solid light cone (corresponding to an arbitrary $2 \times 2$ positive matrix), and then normalized the output Stokes vector to obtain the image point on the correlation ellipsoid. It follows, on the one hand, that all measurement elements which differ from one another by positive multiplicative factors lead to the same image point on the correlation ellipsoid. On the other hand it follows that  $a\,\hr{AB}$ has the same correlation ellipsoid as $\hr{AB}$, for all $a>0$. As one consequence, it is not necessary to normalize a  Mueller matrix to $m_{00}=1$ as far as construction of the correlation ellipsoid is concerned. 

The fact that construction of the correlation ellipsoid deploys the entire positive solid light cone of positive operators readily implies that {\em the ellipsoid inherits all the symmetries of this solid light cone}. These symmetries are easily enumerated. Denoting by $\psi_1,\,\psi_2$ the components of a vector $|\psi\rangle$ in Bob's Hilbert space ${\cal H}_B$, a nonsingular linear transformation 
\begin{align}
J\,: \begin{pmatrix} \psi_1 \\ \psi_2 \end{pmatrix} \to \begin{pmatrix} \psi_1^{\,'} \\ \psi_2^{\,'} \end{pmatrix}
= J \begin{pmatrix} \psi_1 \\ \psi_2 \end{pmatrix}
\end{align}
on ${\cal H}_B$ corresponds on Stokes vectors to the transformation $|{\rm det} J|\,L$ where $L$ is an element of the Lorentz group $SO(3,1)$, and the factor $|{\rm det} J|$ corresponds to `radial' scaling of the light cone. Following the convention of classical polarization optics, we may call $J$ the {\em Jones matrix} of the (non-singular) local filtering\,\cite{simon-mueller1,simon-mueller3,simon-mueller4}. When $({\rm det} J)^{-1/2} J =L$ is polar decomposed, the positive factor corresponds to pure boosts of $SO(3,1)$ while the (local) unitary factor corresponds to the `spatial' rotation subgroup $SO(3)$ of $SO(3,1)$\,\cite{simon-mueller1,simon-mueller3,simon-mueller4}. It follows that restriction of attention to the section $S_0=1$ confines the invariance group from $SO(3,1)$ to $SO(3)$. 

The positive light cone is mapped onto itself also under inversion of all `spatial' coordinates\,: $(S_0, {\bf S}) \to (S_0, -{\bf S})$. This symmetry corresponds to the Mueller matrix $T={\rm diag}(1,-1,-1,-1)$, which is equivalent to $T_0={\rm diag}(1,1,-1,1)$, and hence corresponds to the transpose map on $2\times 2$ matrices. In contradistinction to $SO(3,1)$, $T_0$ acts directly on the operators and cannot be realized or lifted as filtering on Hilbert space vectors; indeed, it cannot be realized as any physical process. Even so, it remains a symmetry of the positive light cone and hence of the correlation ellipsoid itself. 

The full invariance group ${\cal G}$ of a correlation ellipsoid thus comprises two copies of the Lorentz group and the one-parameter semigroup of radial scaling by factor $a>0$\,:
\begin{align}
{\cal G} = \{ SO(3,1), \, TSO(3,1) \approx SO(3,1)T,\,a\}.
\label{largesym}
\end{align} 
All Mueller matrices $MM_0$ with $M_0 \in {\cal G}$ and fixed $M$ correspond to one and the same correlation ellipsoid. In what follows we examine briefly the manner in which these invariances could be exploited for our purpose, and we begin with $SO(3,1)$.

The Jones matrix $J= \exp[\mu \sigma_3/2] = {\rm diag}(e^{\mu/2},e^{-\mu/2})$ corresponds to the Lorentz boost 
\begin{align}
M_0(\mu) = c_{\mu} \left[ 
\begin{array}{cccc}
1 & 0 & 0 & t_{\mu}\\
0& (c_{\mu})^{-1} & 0 & 0\\
0&0&(c_{\mu})^{-1}&0\\
t_{\mu} & 0&0&1
\end{array}
\right]
\label{mboost}
\end{align} 
along the third spatial direction on Stokes vectors. Here $c_{\mu},\,t_{\mu}$ stand respectively for $\cosh{\mu}$ and ${\rm tanh}\,{\mu}$. To see the effect of this boost on the correlation ellipsoid, consider a Mueller matrix of the form\,\eqref{mueller} with $m_{03}=0$ so that $m_{11} =a_x$, $m_{33} =a_z$, $m_{22} = \pm a_y$ and $z_c=m_{30}$. Absorbing the scale factor $c_{\mu}$ in Eq.\,\eqref{mboost} into the solid light cone, we have 
\begin{align}
M M_0(\mu) =  
\left[ 
\begin{array}{cccc}
1 & 0 & 0 & t_{\mu}\\
0&  m_{11}/c_{\mu} & 0 & 0\\
0&0&m_{22}/c_{\mu}&0\\
m_{30} + m_{33} \,t_{\mu} & 0&0&m_{33} + m_{30} \,t_{\mu}
\end{array}
\right].
\end{align} 
With the help of \eqref{axgeqay} we immediately verify that $a_x,\,a_y,\,a_z,$ and $z_c$ associated with $MM_0(\mu)$ are exactly those associated with $M$ with no change whatsoever, consistent with the fact that we expect $M$ and $MM_0(\mu)$ to have the same correlation ellipsoid. Only $z_I$, the image of identity, changes from $m_{30}$ to $m_{30} + m_{33}\,t_{\mu}$\,: as $t_{\mu}$ varies over the permitted open interval $(-1,1)$, the point ${\rm I}$ varies linearly over the open interval $(z_c-a_z, \,z_c+a_z)$. Thus, {\em it is the Lorentz boost on the B side which connects states having  one and the same correlation ellipsoid, but different values of $z_I$}. 

As an illustration of this connection, we go back to Fig.\,\ref{wedge} and consider a correlation ellipsoid in the interior of the wedge region between curves (1) and (2) of Fig.\,\ref{wedge}. We recall that a point in this region is distinguished by the fact that for states corresponding to this point the optimal POVM could {\em potentially} be a three-element POVM, but whether a three element POVM or a horizontal projection actually turns out to be the optimal one for a state requires the value of $z_I$ as additional information on the state, beyond the correlation ellipsoid. The behaviour of classical correlation, quantum discord, and mutual information as the Lorentz boost on the B side sweeps  $z_I$ across the full interval $(z_c-a_z,z_c+a_z)$ is presented in Fig.\,\ref{boost}. We repeat that the entire Fig.\,\ref{boost} corresponds to one fixed point in Fig.\,\ref{wedge}.\\
\begin{figure}
\begin{center}
\scalebox{1.4}{\includegraphics{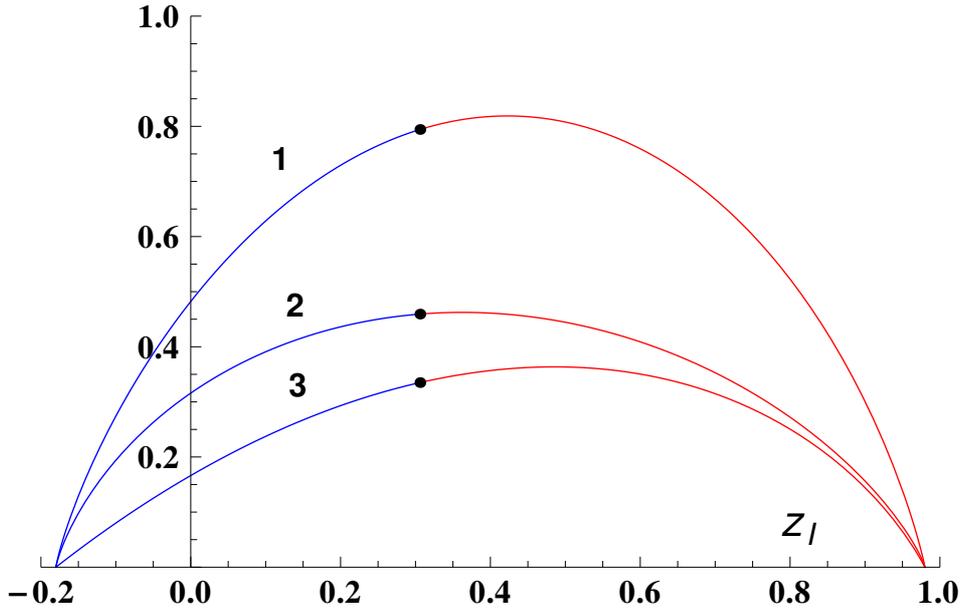}}
\end{center}
\caption{Showing mutual information (1), quantum discord (2) and classical correlation (3) as a function of $z_I$ for the ellipsoid parameters $(a_z,z_c,a_x)=(0.58,0.4,0.65)$ and $z_0=0.305919$. For $z_I \leq  z_0$, the optimal measurement is a von Neumann measurement along the x-axis, and for $z_I>z_0$ the optimal measurement is a three-element POVM.} 
\label{boost}
\end{figure}

\noindent
{\bf Remark}\,: Any entangled two-qubit pure state can be written as 
\begin{align}
|\psi\rangle_{AB} = (1\!\!1 \otimes J)\,|\psi_{\rm max}\rangle,
\end{align}
where the Jones matrix $J$ is non-singular and $|\psi_{\rm max}\rangle$ is a Bell state. Since the associated $SO(3,1)$ does not affect the correlation ellipsoid, the ellipsoid corresponding to $|\psi\rangle_{AB}$ is the same as that of the Bell state, and thereby it is the full Bloch sphere. Hence, $S^A_{\rm min}$ trivially evaluates to zero. So we see that for all two-qubit pure states $I(\hr{AB})=2C(\hr{AB})=2D(\hr{AB})=2E(\hr{AB})$.\,$\blacksquare$ \\

\noindent
{\bf Remark}\,: It is useful to make two minor observations before we leave the present discussion of the role of $SO(3,1)$. First, it is obvious that a bipartite operator $\hr{AB}$ is positive if and only if its image under any (nonsingular) local filtering $J$ is positive. This, combined with the fact that the location of $z_I$ inside the correlation ellipsoid can be freely moved around using local filtering, implies that the location of $z_I$ has no role to play in the characterization of positivity of $\hr{AB}$ given in \eqref{cp1}, \eqref{cp2}. Consequently, in forcing these positivity requirements on the correlation ellipsoid we are free to move, without loss of generality, to the simplest case corresponding to $z_I = z_c$ or $m_{03}=0$. 

Secondly, since determinant of an $SO(3,1)$ matrix is positive, we see that local filtering does not affect the signature of ${\rm det} M$,
and hence it leaves unaffected the signature of the correlation ellipsoid itself\,: $\Omega^+$ and $\Omega^-$ remain separately invariant under $SO(3,1)$. $\blacksquare$

The case of the spatial inversion $T$, to which we now turn our attention, will prove to be quite different on both counts. 
It is clear that the effect of $T_0\,: M \to M T_0$ on an $X$-state Mueller matrix is to transform $m_{22}$ to $-m_{22}$, leaving all other entries of $M$ invariant. Since the only way $m_{22}$ enters the correlation ellipsoid parameters in \eqref{ellipsoid} is through $a_y=|m_{22}|$, it follows that the correlation ellipsoid itself is left invariant, but its signature gets reversed\,: ${\rm det}\, MT_0=-{\rm det} M$. This reversal of signature of the ellipsoid has important consequences. 

As explained earlier during our discussion of the role of $SO(3,1)$ we may assume, without loss of generality, $z_I = z_c$ or, equivalently, $m_{03}=0$. The positivity conditions \eqref{cp1}, \eqref{cp2} then read as the following requirements on the ellipsoid parameters\,:
\begin{align}
(1+a_z)^2 - z_c^2 \geq (a_x + a_y)^2, \label{x1}\\
(1-a_z)^2 - z_c^2 \geq (a_x - a_y)^2 \label{x2},
\end{align}
in the case $M \in \Omega^+$, and 
\begin{align}
(1+a_z)^2 - z_c^2 \geq (a_x - a_y)^2, \label{x3}\\
(1-a_z)^2 - z_c^2 \geq (a_x + a_y)^2 \label{x4},
\end{align}
in the case $M \in \Omega^-$. But \eqref{x3} is manifestly weaker than \eqref{x4} and hence is of little consequence. The demand that $M T_0$ too correspond to a physical state requires 
\begin{align}
(1+a_z)^2 - z_c^2 \geq (a_x - a_y)^2, \label{x5}\\
(1-a_z)^2 - z_c^2 \geq (a_x + a_y)^2 \label{x6},
\end{align}
in the case $M \in \Omega^+$, and  
\begin{align}
(1+a_z)^2 - z_c^2 \geq (a_x + a_y)^2, \label{x7}\\
(1-a_z)^2 - z_c^2 \geq (a_x - a_y)^2 \label{x8},
\end{align}
in the case of $M \in \Omega^-$.

Now, in the case of $M \in \Omega^+$, \eqref{x5} is weaker than \eqref{x1} and hence is of no consequence, but \eqref{x6} is stronger than and subsumes {\em both} \eqref{x1} and \eqref{x2}. In the case $M \in \Omega^-$ on the other hand both \eqref{x7} and \eqref{x8} are weaker than \eqref{x4}. These considerations establish the following\,:
\begin{enumerate}[1.]
\item  If $M \in \Omega^-$, its positivity requirement is governed by the single condition \eqref{x4} and, further, $M T_0$ certainly corresponds to a physical state in $\Omega^+$.
\item If $M \in \Omega^+$, then $MT_0 \in \Omega^-$ is physical if and only if the additional condition \eqref{x6} which is the same as \eqref{x4} is met.
\end{enumerate}
Since $T_0$ is the same as  partial transpose on the B side, we conclude that a correlation ellipsoid corresponds to a separable state if and only if \eqref{x4} is met, and it may be emphasised that this statement is independent of the signature of the ellipsoid. Stated differently, those correlation ellipsoids in $\Omega^+$ whose signature reversed version are not present in $\Omega^-$ correspond to entangled states. In other words, the set of entangled $X$-states constitute precisely the $\Omega^-$ complement of $\Omega^+$.

Finally, the necessary and sufficient condition $(1-a_z)^2 - z_c^2 \geq (a_x + a_y)^2$ for separability can be used to ask for correlation ellipsoid of maximum volume that corresponds to a separable state, for a given $z_c$. In the case $z_c=0$, it is easily seen that the maximum volume obtains for $a_x = a_y =a_z = 1/3$, and evaluates to a fraction $1/27$ of the volume of the Bloch ball. For $z_c \neq 0$, this fractional volume $V(z_c)$ can be shown to be
\begin{align}
V(z_c) = \frac{1}{54} (2-\sqrt{1+3z_c^2})^2(1+\sqrt{1+3z_c^2}),
\end{align}
and corresponds to 
\begin{align}
a_x = a_y = \frac{[(2-\sqrt{1+3z_c^2})(1+\sqrt{1+3z_c^2})]^{1/2}}{3\sqrt{2}},~~ a_z =\frac{2-\sqrt{1+3z_c^2}}{3}.
\end{align}
 It is a monotone decreasing function of $z_c$. Thus $\Omega^-$ has no ellipsoid of fractional volume $> 1/27$.\\

 \noindent
{\bf Remark}\,: It is clear that any $X$-state whose ellipsoid degenerates into an elliptic disc necessarily corresponds to a separable state. This sufficient separability condition may be contrasted with the case of discord wherein the ellipsoid has to doubly degenerate into a line segment for nullity of quantum discord to obtain  \\

\section{Comparison with the work of Ali, Rau, and Alber \label{rausection}}
In this Section we briefly contrast our approach and results with those of the famous work of Ali, Rau, and Alber (ARA)\,\cite{ali}, whose principal claim comprises two parts\,:
\begin{enumerate}[C1\,:]
\item Among all von Neumann measurements, either the horizontal or the vertical projection always yields the optimal classical correlation and quantum discord.
\item  The values thus computed remain optimal even when general POVM's are considered. 
\end{enumerate}
\noindent
As for the second claim, the main text of ARA simply declares ``The Appendix shows how we may generalize to POVM  to get final compact expressions that are simple extensions of the more limited von Neumann measurements, thereby {\em yielding the same value} for the maximum classical correlation and discord.'' The Appendix itself  seems not to do enough towards validating this claim. It begins with ``Instead of von Neumann projectors, consider more general POVM. For instance, choose three orthogonal unit vectors mutually at $120^{\circ}$,
\begin{align}
\hat{s}_{0,1,2} = [\hat{z},(-\hat{z} \pm \sqrt{3} \hat{x})/2],
\label{raueq}
\end{align} 
and corresponding projectors $\cdots$.'' [It is not immediately clear how `orthogonal' is to be reconciled with `mutually as $120^{\circ}$' ]. Subsequent reference to their Eq.\,(11) possibly indicates that ARA have in mind two more sets of such {\em three orthogonal unit vectors mutually at $120^{\circ}$} related to\,\eqref{raueq} through $SU(2)$ rotations. In the absence of concrete computation aimed at validating the claim, one is left to wonder if the second claim (C2) of ARA is more of an  assertion than deduction. 

We now know, however,  that the actual situation in respect of the second claim is much more subtle\,: the optimal three-element POVM is hardly of the {\em three orthogonal unit vectors mutually at $120^{\circ}$} type and, further, when a three-element POVM is required as the optimal one, there seems to be no basis to anticipate that it would yield `the same value for the maximum classical correlation and discord'. 

Admittedly, the present work is not the first to discover that ARA is not the last word on quantum discord of $X$-states. Several authors have pointed to examples of $X$-states which fail ARA\,\cite{zambrinierrata,zambrinipra,zambriniepl,wang11,chen11,adesso11,huang13}. But these authors have largely been concerned with the second claim (C2) of ARA. In contradistinction, our considerations below focuses on the first one (C1). 
In order that it be clearly understood as to what the ARA claim (C1) is not,  we begin with the following three statements\,:
\begin{enumerate}[S1\,:]
\item If von Neumann projection proves to be the optimal POVM, then the projection is either along the x or z direction.
\item von Neumann projection along the x or z direction always proves to be the optimal POVM.
\item von Neumann projection along either the x or z direction proves to be the best among all von Neumann projections. 
\end{enumerate}
\noindent
Our analysis has confirmed  that the first statement (S1) is absolutely correct. We also know that the second statement (S2) is correct except for a very tiny fraction of states corresponding to the wedge-like region between curves (1) and (2) in Fig.\,\ref{wedge}. 

The first claim (C1) of ARA corresponds, however, to neither of these two but to the third statement (S3). 
We begin with a counter-example to prove that this claim (S3) is non-maintainable. The example corresponds to the ellipsoid parameters $(a_x,a_y,a_z,z_c) = (0.780936,0.616528,0.77183,0.122479)$. These parameters, together with $z_I=0.3$, fully specify the state in the canonical form, and the corresponding Mueller matrix 
\begin{align}
M = \left( \begin{array}{cccc}
1&0&0&0.23\\
0&0.76&0&0\\
0&0&0.6&0\\
0.3&0&0&0.8
\end{array}\right).
\label{ex1}
\end{align}
The parameter values verify the positivity requirements. Further, it is seen that $M \in \Omega^+$ and corresponds to a nonseparable state. The x-z section of the correlation ellipsoid corresponding to this example is depicted in Fig.\,\ref{figx3}.
\begin{figure}
\begin{center}
\scalebox{1}{\includegraphics{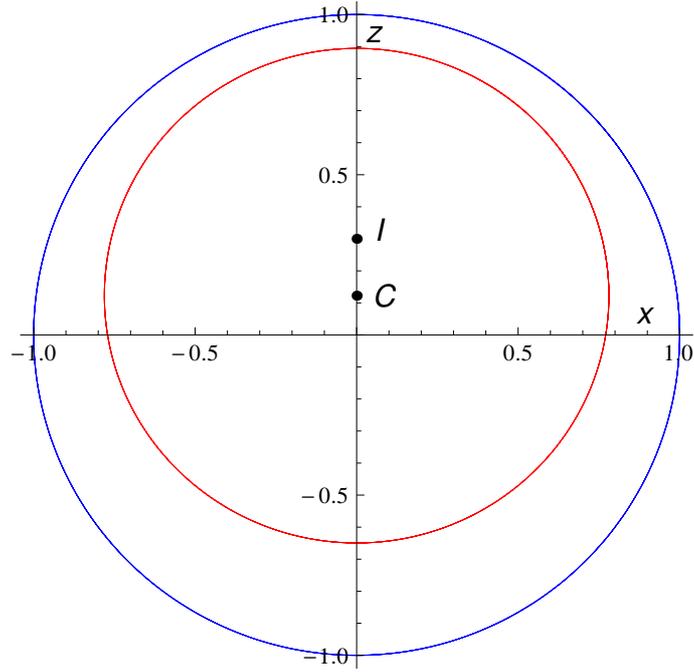}}
\end{center}
\caption{Showing the x-z cross-section of the ellipsoid associated with the Mueller matrix in\,\eqref{ex1}.
 \label{figx3}} 
\end{figure} 

Let us denote by $S^A_{vN}(\theta)$ the average conditional entropy post von Neumann measurement $\Pi_{\theta}$ parametrized by angle $\theta$\,:
\begin{align}
\Pi_{\theta} = \left\{ (1,\sin{\theta}, 0 , \cos{\theta})^T, \, (1,-\sin{\theta}, 0 ,-\cos{\theta})^T \right\}, ~~ 0 \leq \theta \leq \pi/2.
\end{align} 
It is clear that the output states are at distances $r(\theta),\,r^{\,'}(\theta)$ with respective conditional probabilities $p(\theta),\,p^{\,'}(\theta)$\,: 
\begin{align}
r(\theta) &= \frac{\sqrt{(m_{11} \,\sin{\theta})^2+(m_{30} + m_{33} \,\cos{\theta})^2}}{1+m_{03}\,\cos{\theta}},\nonumber\\
r^{\,'}(\theta) &= \frac{\sqrt{(m_{11} \,\sin{\theta})^2+(m_{30} - m_{33} \,\cos{\theta})^2}}{1-m_{03}\,\cos{\theta}},\nonumber\\
p(\theta) &= \frac{1+m_{03}\,\cos{\theta}}{2}, ~~ p^{\,'}(\theta) = \frac{1-m_{03}\,\cos{\theta}}{2}.
\end{align}
Thus $S^A_{vN}(\theta)$ evaluates to 
\begin{align}
S^A_{vN}(\theta) = \frac{1}{2} &\left[ S_2(r(\theta))+S_2(r^{\,'}(\theta)) \right. \nonumber\\
&~~~+ \left. m_{03}\,\cos{\theta}\,(S_2(r(\theta))-S_2(r^{\,'}(\theta)) )\right].
\label{von}
\end{align}
The behaviour of $S^A_{vN}(\theta)$ as a function of $\theta$ is shown in Fig.\,\ref{figx4}, and it is manifest that the optimal von Neumann obtains {\em neither at $\theta=0$ nor at $\pi/2$, but at $\theta=0.7792$ radians}. More strikingly, it is not only that neither $\theta=0$ or $\pi/2$  is the best, but both are indeed the worst in the sense that von Neumann projection along any other direction returns a better value!
\begin{figure}
\begin{center}
\scalebox{1.5}{\includegraphics{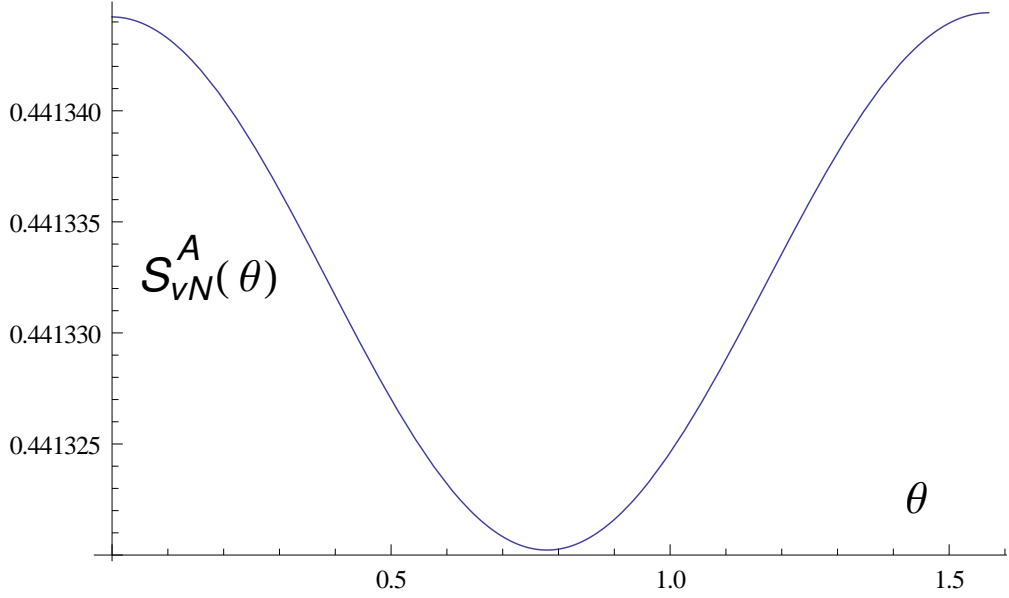}}
\end{center}
\caption{Showing the conditional entropy $S^A_{vN}(\theta)$ resulting from von Neumann measurements for the example in\,\eqref{ex1}.
 \label{figx4}} 
\end{figure} 

We know from our analysis in Section\,\ref{saminsec} that if the von Neumann measurement indeed happens to be the optimal POVM, it can not obtain for any angle other than $\theta=0$ or $\pi/2$. Thus, the fact that the best von Neumann for the present example corresponds to neither angle is already a sure signature that a three-element POVM is the optimal one for the state under consideration. Prompted by this signature, we embed the state under consideration in a one-parameter family with fixed $(a_z,a_y,z_c)=(0.616528,0.77183,0.122479)$ and $z_I$, and $a_x$ varying over the range $[0.7803,0.7816]$. The results are shown in Fig.\,\ref{figx5}.  
\begin{figure}
\begin{center}
\scalebox{1.6}{\includegraphics{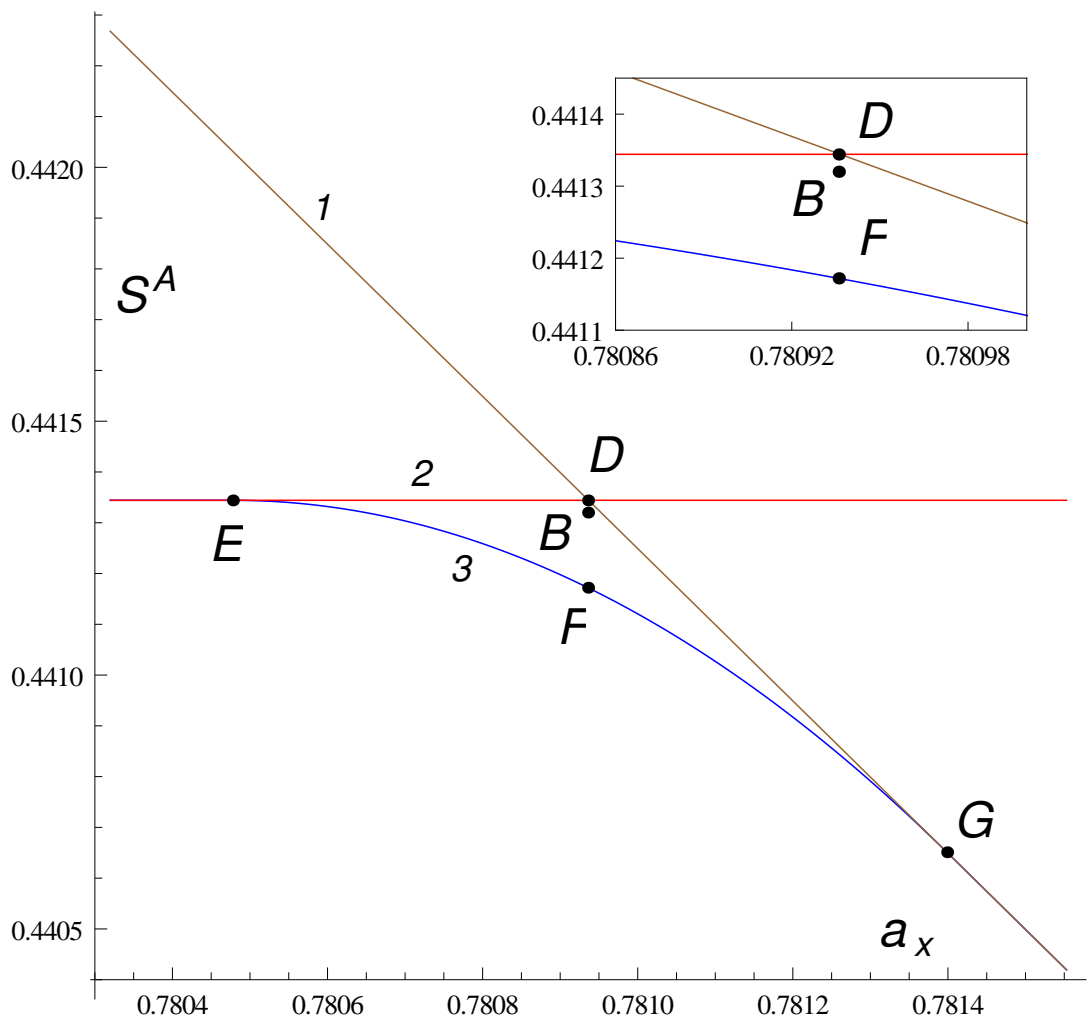}}
\end{center}
\caption{Showing the entropy variation with respect to variation of $a_x$ in $[0.78032,0.781553]$, with $(a_y,a_z,z_c)$ fixed at $(0.616528,0.77183,0.122479)$. Curve (1) depicts $S^A_{vN}$ for von Neumann measurement along $\theta=\pi/2$, the constant line [curve (2)] to a von Neumann measurement along $\theta=0$, and curve (3) to $S^A_{\rm min}$ resulting from the three element POVM scheme. The example in (\eqref{ex1}) corresponds to $a_x=0.780936$. The inset compared the various schemes of the example in \eqref{ex1}. D refers to the a measurement restricted only to the von Neumann projection along z or x-axis, B to the best von Neumann projection, and F to the optimal three-element measurement. 
 \label{figx5}} 
\end{figure} 
Curve (1) and curve (2) correspond respectively to the horizontal and vertical von Neumann projections, whereas curve (3) corresponds to the optimal three-element POVM. We emphasise that curve (3) is not asymptotic to curves (1) or (2), but joins them at $G$ and $E$ respectively. Our example of Eq.\,\eqref{ex1} embedded in this one-parameter family is highlighted by points $A,\, B,\,F$. This example is so manufactured that $S^A_{vN}$ computed by the horizontal projection equals the value computed by the vertical projection, and denoted by point $D$. The point $B$ corresponds to $S^A_{vN}$ evaluated using the best von Neumann, and $F$ to the one computed by the (three-element) optimal POVM. It may be noted that $D$ and $B$ are numerically quite close as highlighted by the inset. A numerical comparison of these values is conveniently presented in Table\,\ref{tablex1}.  
\begin{table}
\centering
\begin{tabular}{|c|c|c|c|}
\hline
Scheme&Elements& Optimal value & $S^A$\\
\hline
von Neumann & $\sigma_x$  or  $\sigma_z$ & equal & $0.441344$ \\
\hline
von Neumann & $\Pi^{vN}_{\theta}$, ~~$\theta \in [0,\pi/2]$ & $\theta_{\rm opt} = 0.779283$ & $0.44132$ \\
\hline
~~3-element POVM ~~&~~$\Pi^{(3)}_{\theta}$,~~ $\theta \in [0,\pi/2]$~~ & ~~$\theta_{\rm opt} =1.02158$ ~~&~~ $0.441172$~~\\
\hline
\end{tabular}
\caption{Showing a comparison of the von Neumann and the three-element POVM schemes for the example in Eq.\,\eqref{ex1}.\label{tablex1}}
\end{table}

It is seen from Fig.\,\ref{figx5} that vertical von Neumann is the optimal POVM upto the point E (i.e. for $a_x \leq 0.780478$), from E all the way to G the three-element POVM $\Pi^{(3)}_{\theta}$  is the optimal one, and beyond G ($a_x \geq 0.781399$) the horizontal von Neumann is the optimal POVM. The continuous evolution of the parameter $\theta$ in $\Pi^{(3)}_{\theta}$ of Eq.\,\eqref{scheme} as one moves from E to G is shown in Fig.\,\ref{figx6}. Shown also is the continuous manner in which the probability $p_0(\theta)$ in Eq.\,\eqref{scheme} continuously varies from $0.5$ to zero as $a_x$ varies over the range from E to G.

\begin{figure}
\begin{center}
\scalebox{1.6}{\includegraphics{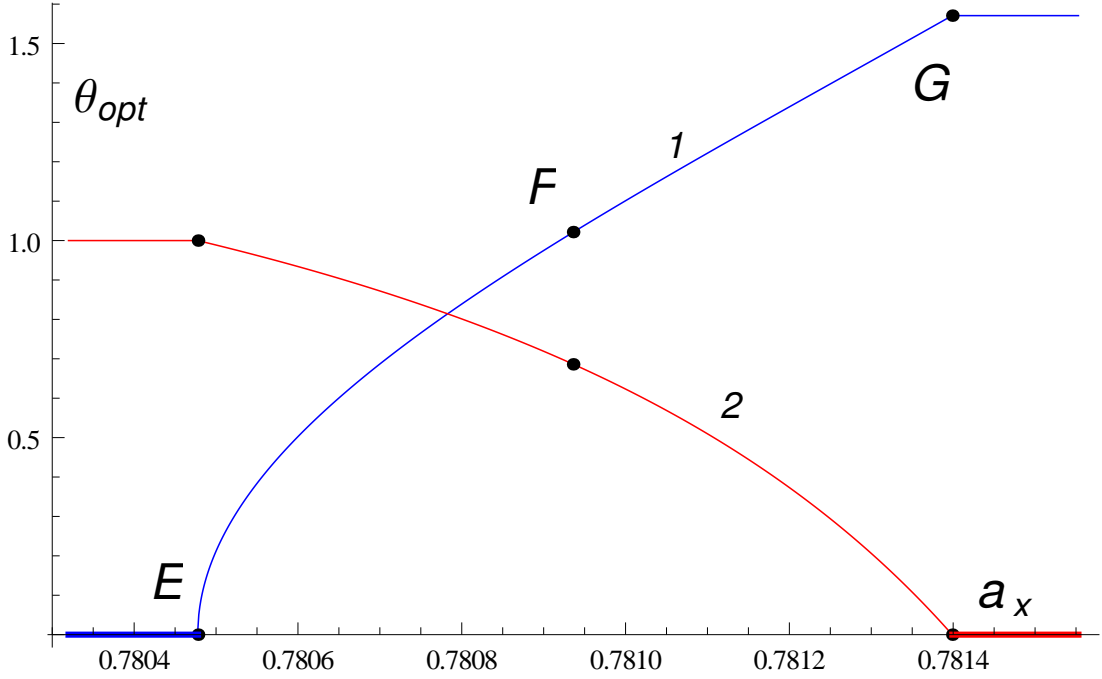}}
\end{center}
\caption{Showing the optimal $\theta=\theta_{\rm opt}$ of $\Pi^{(3)}_{\theta}$ [curve(1)] resulting in $S^A_{\rm min}$ depicted as curve (3) in Fig.\,\ref{figx5}. Curve (2) shows the probability (scaled by a factor of 2) $2p_{0}(\theta_{\rm opt})$ of the conditional state corresponding to input POVM element $(1,0,0,1)^T$. 
 \label{figx6}} 
\end{figure}

In order to reconcile ARA's first claim (S3 above) with our counter-example we briefly reexamine their very analysis leading to the claim. As we shall see the decisive stage of their argument is symmetry-based or group theoretical in tenor. It is therefore unusual that they carry around an extra baggage of irrelevant phase parameters, not only in the main text but also in the reformulation presented in their Appendix\,: the traditional first step in symmetry-based approach is to transform the problem to its simplest form (often called the canonical form) without loss of generality. Their analysis beings with parametrization of von Neumann measurements as [their Eq.\,(11)]
\begin{align}
B_i = V \Pi_i V^{\dagger}, ~~ i=0,1
\end{align} 
where $\Pi_i = \proj{i}$ is the projector on the computation state $|i\rangle \in \{|0\rangle, |1\rangle \}$ and $V \in SU(2)$. With the representation $V = t 1\!\!1 + i \vec{y}.\vec{\sigma}$, $t^2+ y_1^2+y_2^2+y_3^2=1$ they note that three of these four parameters $t,y_1,y_2,y_3$ are {\em independent}. Inspired by their Ref.\,[15] (our Ref.\,\cite{luo}), ARA recast  $t,y_1,y_2,y_3$ into four new parameters $m,n,k,\ell$ and once again emphasize that $k,m,n$ are three {\em independent} parameters describing the manifold of von Neumann measurements. \\

\noindent
{\bf Remark}\,: It is obvious that every von Neumann measurement on a qubit is fully specified by a pure state, and hence the manifold of von Neumann measurements can be no larger than ${\cal S}^2$, the Bloch sphere. Indeed, this manifold is even `smaller'\,: it coincides with the real projective space ${\cal RP}^2 = {\cal S}^2/{\cal Z}_2$ of diameters in ${\cal S}^2$, since a pure state and its orthogonal partner  define {\em one and the same} von Neumann measurement. In any case, it is not  immediately clear in what sense could this two-manifold  be described by three `independent' parameters. $\blacksquare$\\

\noindent
{\bf Remark}\,: We should hasten to add, for completeness, that ARA introduce subsequently, in an unusually well cited {\em erratum}\,\cite{alierratum}, another identity 
\begin{align}
m^2 + n^2 = klm
\label{araeq}
\end{align}
 which they claim to be independent of  $t^2+ y_1^2+y_2^2+y_3^2=1$, and hence expect it to reduce the number of independent variables parametrizing the manifold of von Neumann measurements from three to two. To understand the structure of this new identity, define two complex numbers $\alpha = t-iy_3$, $\beta=y_1+iy_2$. Then $k = |\alpha|^2$, $\ell = |\beta|^2$, $m = ({\rm Re} \alpha \beta)^2$, and $n= ({\rm Re} \alpha \beta)\, ({\rm Im} \alpha \beta)$ so that the ARA identity Eq.\,\eqref{araeq} reads 
\begin{align}
({\rm Re} \alpha \beta)^4 + ({\rm Re} \alpha \beta)^2 \,({\rm Im} \alpha \beta) = |\alpha\beta|^2 ({\rm Re} \alpha \beta)^2,
\label{araeqb}
\end{align}
showing that it is indeed independent of $k+ \ell=1$ as claimed by ARA. Indeed, it is simply the Pythagorean theorem $|z|^2 = ({\rm Re} z)^2 +({\rm Im} z)^2 $ valid for any complex number $z$  puffed up to the appearance of an eighth degree real homogeneous form. It is unlikely that such an {\em universal} identity, valid for any four numbers, would ever aid in reducing the number of independent parameters. Not only ARA, but also the large number of works which cite this erratum, seem to have missed this aspect of the ARA identity\,\eqref{araeq}.\,$\blacksquare$

Returning now to the clinching part of the ARA analysis, after setting up the expression for the conditional entropy as a function of their independent variables $k,m,n$ they correctly note that it could be minimized ``by setting equal to zero its partial derivatives with respect to $k,m$ and $n$.'' Rather than carrying out this step, however, they prefer a short cut in the form of a symmetry argument. They `{\em observe}' that the problem has a symmetry (this is the symmetry of inversion about the z-axis which we used in Section\,\ref{optsec} to simplify the optimization problem), and then use the {\em unusual symmetry argument} that if a problem has a symmetry its solution {\em ought to be} invariant under that symmetry. Obviously, one knows in advance that the only von Neumann projections that are invariant under the symmetry under consideration are the vertical or z-projection and the horizontal projection, the latter meaning x or y-projection according as $a_x > a_y$ or $a_y > a_x$. This version of symmetry argument is unusual, since the familiar folklore version reads\,: {\em if a problem has a symmetry, its solution ought to be covariant (and not necessarily invariant) under the symmetry}. In any case, unless  the ARA version of symmetry argument 
be justified as arising from some {\em special aspect} of the problem under consideration,  its deployment would amount to assuming a priori that either z or x-projection is the best von Neumann; but then this assumption is precisely the claim S3 ARA set out to prove as the very central result of their work. \\

\noindent
{\bf Remark}\,: The ARA version of symmetry argument would remain justified if it were the case that the problem is expected, from other considerations, to have a {\em unique} solution. This happens, for instance, in the case of  {\em convex optimization}. But von Neumann measurements {\em do not form a convex set} and hence the ARA problem of optimization over von Neumann measurement is not one of convex optimization. Thus demanding a unique solution in their case would again amount to an a priori assumption equivalent to the theorem they set out to prove. 

\section{$X$-states with vanishing discord}
Many authors have considered methods to enumerate the zero discord $X$-states\,\cite{dakicprl10,huang11,chrus12}. 
Our analysis below is directly based on the very definition of vanishing discord and hence is elementary; more importantly, it leads to an {\em exhaustive} classification of these states, correcting an earlier claim. Any generic two-qubit state of vanishing quantum discord can be written as\,\cite{modi-rmp} 
\begin{align}
\hat{\rho}_{AB} = U_A |0\rangle \langle 0| U_A^{\dagger} \otimes p_1 \hat{\rho}_{B1} + U_A |1\rangle \langle 1| U_A^{\dagger} \otimes p_2 \hat{\rho}_{B2}, 
\label{zerod}
\end{align}
with $p_1,\,p_2 \geq0$, $p_1 + p_2 =1$, the measurements being assumed performed on subsystem A. We may write
\begin{align}
p_1 \hat{\rho}_{B1} &= \begin{bmatrix}
a_1 & b_1 \\
b_1^* & c_1
\end{bmatrix}, ~~
p_2 \hat{\rho}_{B2} = \begin{bmatrix}
a_2 & b_2 \\
b_2^* & c_2
\end{bmatrix}, \nonumber\\~~ 
U_A &= \begin{bmatrix}
\alpha & \beta\\
-\beta^* & \alpha^*
\end{bmatrix} \in\,SU(2).
\end{align}
Clearly, the reduced state of subsystem B is $p_1 \hat{\rho}_{B1} + p_2 \hat{\rho}_{B2}$, and that of $A$ equals $ U_A\, (p_1 |0\rangle \langle 0| + p_2 |1 \rangle \langle 1|)\,U_A^{\dagger}$. We now combine this nullity condition with the demand that the state under consideration be an $X$-state in the canonical form\,\eqref{xcan}.  From the off-diagonal blocks of $\hat{\rho}_{AB}$ we immediately see that $a_1 = a_2$ and $c_1 =c_2$. 
${\rm Tr}\,\hat{\rho}_{AB} = a_1 + c_1 + a_2 + c_2 = 1$ implies $a_1 + c_1 = 1/2= a_2+c_2$. 
Vanishing of the $01$ and $23$ elements of $\hat{\rho}_{AB}$ forces the following constraints\,:
\begin{align}
&|\alpha|^2 b_1 + |\beta|^2 b_2 = 0,\nonumber \\
&|\alpha|^2 b_2 + |\beta|^2 b_1 = 0.
\end{align}     
These imply in turn that either $|\alpha| = |\beta| = 1/\sqrt{2}$ or $b_1 = b_2=0$. The first case of $|\alpha| = |\beta| = 1/\sqrt{2}$ forces $b_2 = -b_1$, and we end up with a two-parameter family of zero discord states 
\begin{align}
\hat{\rho}^A(a,b) &= \frac{1}{4}\left( 
\begin{array}{cccc}
1+a& 0 &0 & b\\
0& 1 -a & b & 0\\
0&b&1+a &0\\
b&0&0&1-a
\end{array}
\right)\nonumber\\
&= \frac{1}{4} \left[\sigma_0 \otimes \sigma_0 + a \sigma_0 \otimes \sigma_3 + b \sigma_1 \otimes \sigma_1\right]. 
\label{onewaya}
\end{align}
Positivity of $\hat{\rho}^A(a,b)$ places the constraint $a^2 + b^2 \leq 1$, a disc in the $(\sigma_0 \otimes \sigma_3, \sigma_1 \otimes \sigma_1)$ plane.  The case $b=0$ corresponds to the product state 
\begin{align}
\hat{\rho}^{AB}(a) = \frac{1}{4} 1\!\!1 \otimes \begin{bmatrix}  1+a & 0 \\ 0& 1-a \end{bmatrix}.
\end{align}

If instead the measurement was performed on the $B$ subsystem, then it can be easily seen that  similar arguments can be used to arrive at the zero discord states 
\begin{align}
\hat{\rho}^B(a,b) &= \frac{1}{4}\left( 
\begin{array}{cccc}
1+a & 0 &0 & b\\
0& 1+a & b & 0\\
0&b&1-a&0\\
b&0&0&1-a
\end{array}
\right),\nonumber\\
& = \frac{1}{4}\left[ \sigma_0 \otimes \sigma_0 + a \sigma_3 \otimes \sigma_0 + b \sigma_1 \otimes \sigma_1\right]. 
\label{onewayb}
\end{align}
Positivity again constraints $a,b$ to the disc $a^2 + b^2 \leq 1$ in the $(\sigma_3 \otimes \sigma_0, \sigma_1 \otimes \sigma_1)$ plane. 

The intersection between these two comprises the one-parameter family of $X$-states
\begin{align}
\hat{\rho}_{AB} = \frac{1}{4}\left[ \sigma_0 \otimes \sigma_0 + b \sigma_1 \otimes \sigma_1\right], ~~ -1 \leq b \leq 1. 
\label{onewayint}
\end{align}
But these are {\em not the only} two-way zero discord states, and this fact is significant in the light of\,\cite{chrus12}.  To see this, note that in deriving the canonical form \eqref{onewaya} we assumed $\beta \neq 0$. So we now consider the case $\beta=0$, so that \eqref{zerod} reads
\begin{align}
\hat{\rho}_{AB} = p_1 |0\rangle \langle 0| \otimes \hat{\rho}_{B1} + p_2 |1\rangle \langle 1| \otimes \hat{\rho}_{B2}.
\end{align}
The demand that this be an $X$-state forces $\hat{\rho}_{AB}$ to be diagonal in the computational basis\,:
\begin{align}
\hat{\rho}_{AB}(\{p_{k \ell} \}) = \sum_{k,\ell=0}^1 p_{k \ell}\, |k\rangle \langle k| \otimes |\ell \rangle \langle \ell|.
\label{twoway}
\end{align}
It is manifest that all $X$-states of this three-parameter family, determined by probabilities $\{p_{k\ell} \}$, $\sum p_{k\ell} =1$ and worth a {\em tetrahedron} in extent have vanishing quantum discord and, indeed, vanishing two-way quantum discord.  

The intersection of \eqref{onewaya} and \eqref{onewayb} given in  \eqref{onewayint} is not really outside the tetrahedron \eqref{twoway} in the canonical form because it can be diagonalized by a local unitary $U_A \otimes U_B$, $U_A= U_B = \exp [-i \pi \sigma_2/4]$\,:
\begin{align}
\sigma_0 \otimes \sigma_0 + b \sigma_1 \otimes \sigma_1 \to  \sigma_0 \otimes \sigma_0 + b \sigma_3 \otimes \sigma_3. 
\end{align}
Stated differently, the family of strictly one-way zero discord $X$-states in the canonical form is not a disc, but a disc with the diameter removed. \\

\noindent
{\bf Remark}\,: Strictly speaking, this is just an half disc with diameter removed, as seen from the fact that in \eqref{onewaya}, \eqref{onewayb}, and \eqref{onewayint} the two states $(a,b)$, $(a,-b)$ are local unitarily equivalent under $U_A \otimes U_B = \sigma_3 \otimes \sigma_3$. $\blacksquare$

We now consider the correlation ellipsoids  associated with these zero discord states.  For the one-way zero discord states in \eqref{onewayb} the non-zero Mueller matrix entries are  
\begin{align}
m_{30} = a, ~~ m_{11} = b. 
\end{align}
This ellipsoid is actually a symmetric line segment parallel to the  x-axis, of extent $2b$, translated by extent $a$, perpendicular to the line segment (i.e., along z)\,: $\{(x,y,z) = (x,0,a)|-b\leq x \leq b \}$;  it is symmetric under reflection about the z-axis. For measurements on the A side we have from \eqref{onewaya} 
\begin{align}
m_{03} = a, ~~ m_{11} = b,
\end{align}
and we get the same line segment structure (recall that now we have to consider $M^T$ in place of $M$). 

For the two-way zero discord states\,\eqref{twoway} we have  
\begin{align}
m_{03} & = p_{00}-p_{01}+p_{10} - p_{11},\nonumber\\
m_{30} &= p_{00}+p_{01}-p_{10}- p_{11},\nonumber\\
m_{33} &= p_{00}-p_{01}-p_{10} + p_{11},
\end{align}
corresponding to a point in the tetrahedron. We note that the associated correlation ellipsoid is a line segment of a diameter {\em shifted along the diameter itself}. That is, the line segment is radial. While the extent of the line segment and the shift are two parameters, the third parameter is the image ${\rm I}$ of the maximally mixed input, which does not contribute to the `shape' of the ellipsoid, but does contribute to the shape. This three parameter family should be contrasted with the claim of\,\cite{chrus12} that an `$X$-state is purely classical if and only if $\hat{\rho}_{AB}$ has components along $\sigma_0 \otimes \sigma_0, \sigma_1 \otimes \sigma_1 $' implying a one-parameter family.

\section{States not requiring an optimization}
We now exhibit a large class of states for which one can write down {\em analytic expression} for quantum discord by inspection, without the necessity to perform explicit optimization over all measurements.  We begin by first giving a geometric motivation for this class of states. 
Consider $X$-states for which the associated correlation ellipsoid is {\em centered at the origin}: 
\begin{align}
z_c = &~~\frac{m_{30} - m_{03}m_{33}}{1-m_{03}^2} =0, \nonumber\\
{\rm i.e.},~~  &~m_{30} = m_{03}m_{33}.
\end{align}
This implies on the one hand that only two of the three parameters $m_{03},\,m_{30},\,m_{33} $ are independent. On the other hand, it implies that the product of $m_{03}, m_{30}, m_{33}$ is necessarily positive and thus, by local unitary, all the three can be assumed to be positive without loss of generality. Let us take $m_{03} = \sin{\theta} > 0$, then we have $m_{30} = m_{33}\, \sin{\theta}$. 
So, we now have in the canonical form a four-parameter family of Mueller matrices
\begin{align}
M(\gamma_1,\,\gamma_2,\,\gamma_3;\,\theta) = \begin{bmatrix}
1 & 0 & 0 & \sin{\theta} \\
0& \gamma_1 \cos{\theta} & 0 & 0 \\
0& 0& \gamma_2 \,\cos{\theta} & 0 \\
\gamma_3 \, \sin{\theta} & 0 & 0 & \gamma_3
\end{bmatrix},
\label{center}
\end{align}
and correspondingly a three-parameter family of correlation ellipsoids centered at the origin, with principal axes $(a_x,a_y,a_z) = (\gamma_1, |\gamma_2|, \gamma_3)$, and $z_I = \gamma_3 \, \sin{\theta}$. We continue to assume $|\gamma_2| \leq \gamma_1$.\\

\noindent
{\bf Remark}\,: Note that we are not considering the case of Bell-diagonal states, which too correspond to ellipsoids centered at the origin. In the Bell-diagonal case, the point {\rm I} is located at the origin and, as an immediate consequence, $S^A_{\rm min}$ is entirely determined by the major axis of the ellipsoid. In the present case, $z_I = \gamma_1 \,\cos{\theta} \neq 0$, and $S^A_{\rm min}$ does depend on $z_I$. Indeed, the case of Bell-diagonal states corresponds to $\sin{\theta}=0$, and hence what we have here is a one-parameter generalization. $\blacksquare$

The  four parameter family of density matrices corresponding to Eq.\,\eqref{center} takes the form 
{\footnotesize
\begin{align}
\rho(\gamma_1,\,\gamma_2,\,\gamma_3;\,\theta)  =  \frac{1}{4}
\begin{bmatrix}
(1+\gamma_3)(1+\sin{\theta}) & 0 & 0 & (\gamma_1 + \gamma_2)\,\cos{\theta}\\
0 & (1-\gamma_3)(1-\sin{\theta}) & (\gamma_1 - \gamma_2)\,\cos{\theta}& 0 \\
0&(\gamma_1 - \gamma_2)\,\cos{\theta} & (1-\gamma_3)(1+\sin{\theta})&0\\
(\gamma_1 + \gamma_2)\,\cos{\theta} & 0 & 0 & (1+\gamma_3)(1-\sin{\theta})
\end{bmatrix}.
\label{circ}
\end{align}
}
The first CP  condition \eqref{cp1} reads
\begin{align}
 ~~(1+\gamma_3)^2 &- \sin^2{\theta}  (1+\gamma_3)^2 \geq (\gamma_1 + \gamma_2)^2 \,\cos^2{\theta},\nonumber\\
{\rm i.e.},  &~~  \gamma_1 + \gamma_2 - \gamma_3 \leq 1,
\end{align} 
while the second CP condition \eqref{cp2} reads
\begin{align}
~~(1-\gamma_3)^2 &- \sin^2{\theta}  (1-\gamma_3)^2 \geq (\gamma_1 - \gamma_2)^2 \, \cos^2{\theta}\nonumber\\
{\rm i.e.}, &~~ \gamma_1 + \gamma_3 -  \gamma_2 \leq 1.  
\end{align}
Recalling that $|\gamma|_2 \leq \gamma_1$, these two conditions can be combined into a {\em single CP condition} 
\begin{align}
\gamma_1 + |\gamma_3 -  \gamma_2|\leq 1.
\end{align}
Having given a full characterization of the centered $X$-states, we note a special property of these states in respect of quantum discord. \\

{\bf Remark}\,:
As seen from Eq.\,\eqref{ahav}, the optimal POVM for a centered $X$-state is a von Neumann measurement since the two curves $a^H(a_z,0)$ and $a^V(a_z,0)$ equal $a_z$. Therefore, the measurement is along $x$ or $z$  according as $a_x > a_z$ or $a_z > a_x$. 

{\bf Circular states}\,: This special case corresponds to $a_x = a_z$, i.e., setting $\gamma_3 = \gamma_1$ in Eq.\,\eqref{center}.
For this class of states, every von Neumann measurement in the x-z plane (indeed, every POVM with all the measurement elements lying in the x-z plane) is equally optimal. In other words, $I$  plays no role in determining the optimal POVM for a centered $X$-state.

The four eigenvalues of $\hat{\rho}(\gamma_1,\,\gamma_2;\,\theta)$ of \eqref{circ} are
\begin{align}
\{\lambda_j \} &= \frac{1}{4} \left\{ 1 + \epsilon \gamma_1 \pm y \right\},\nonumber\\
y &=  \sqrt{(1 + \epsilon \gamma_1)^2 \cos^2{\theta} + (\gamma_1 + \epsilon \gamma_2 )^2 \sin^2{\theta}} ,
\label{eigencirc}
\end{align} 
$\epsilon$ being a signature.

We can explicitly write down the various quantities of interest in respect of the circular states $\hat{\rho} (\gamma_1,\gamma_2;\theta)$.  First, we note that the conditional entropy post measurement is simply the entropy of the output states that are on the circle, and hence 
\begin{align}
S^A_{\rm min}= S_2(m_{33})=S_2(\gamma_1).
\end{align}
By Eqs.\,\eqref{ents} and \eqref{mi}, we have 
\begin{align}
I(\hat{\rho}(\gamma_1,\,\gamma_2;\,\theta))&= S_2(\gamma_1\,\sin{\theta}) + S_2(\sin{\theta}) - S(\{\lambda_j\}),\nonumber\\
C(\hat{\rho}(\gamma_1,\,\gamma_2;\,\theta)) &= S_2(\gamma_1\,\sin{\theta}) - S_2(\gamma_1),\nonumber\\
D(\hat{\rho}(\gamma_1,\,\gamma_2;\,\theta))&= S_2(\sin{\theta}) + S_2(\gamma_1) - S(\{\lambda_j\}),
\label{corrcirc}
\end{align}
where $\lambda_j \equiv \lambda_j(\gamma_1,\,\gamma_2;\,\theta)$ are given in Eq.\,\eqref{eigencirc}. 
Finally, we note that with the local unitary freedom, this 3-parameter class of states can be lifted to a 9-parameter using local unitaries.\\

\noindent
{\bf Spherical states}: The correlation ellipsoid corresponding to these states is a sphere with $z_c=0$. They can be obtained as a subset of circular states by setting $\gamma_1 = |\gamma_2|$. The expressions for the correlation are the same as those of circular states as given in Eq.\,\eqref{corrcirc}. We note that, the spherical states form a 2-parameter family of states inside the set of $X$-states. We can lift this family to a seven parameter family of states, the five parameters coming from the local unitary transformations. One parameter was however lost from the degeneracy $m_{11} = |m_{22}|$ for spherical states. \\

{\bf Bell mixtures}\,:
The next example of a convex combination of the Bell-states was considered in\,\cite{luo}. We can write the state as 
$\hat{\rho} = \sum_{j=1}^4 p_j |\phi_j \rangle \langle \phi_j|$, i.e.,
\begin{align}
\hat{\rho} = \frac{1}{2} \left( \begin{array}{cccc}
p_1 + p_2 & 0 & 0 & p_1 - p_2 \\
0& p_3+p_4 & p_3 -p_4 &0 \\
0& p_3 - p_4 & p_3 +p_4 & 0\\
p_1 - p_2 &0&0& p_1 + p_2
\end{array}
\right).
\end{align} 
The corresponding Mueller matrix is diagonal with 
\begin{align}
m_{11} &= p_1 + p_3 - (p_2 + p_4), \nonumber\\
m_{22} & =p_1 + p_4 -(p_2 + p_3), \nonumber\\
m_{33} &= p_1 + p_2 - (p_3 + p_4).
\end{align}
The correlation ellipsoid has $z_c=0$, and  more importantly, $z_I=0$. The optimal measurement is then a von Neumann projection along the direction of the longest axis length of the ellipsoid. \\

{\bf Linear states}\,:
Another example of states for which the quantum discord can be immediately written down are states for which x-z cross-section of the correlation ellipsoid is a line segment along the x-axis. We denote these states as linear states and they are obtained by setting $a_z=0$. We have
\begin{align}
m_{33} = m_{03} m_{30}.
\end{align} 
As before we see that only two of them can have a negative value which can be dropped by a local unitary transformation. This gives us a four parameter family of states for which the optimal measurement is the horizontal entropy. We make the following choice\,:
\begin{align}
M(\gamma_1,\gamma_2,\gamma_3,\theta) = \left[ 
\begin{array}{cccc}
1 & 0 & 0 & \sin{\theta}\\
0& \gamma_1 \cos{\theta} & 0 & 0\\
0 & 0& \gamma_2 \cos{\theta} &0\\
\gamma_3 & 0 & 0 & \gamma_3 \sin{\theta}
\end{array}
\right],
\end{align}
where we assume $|\gamma_2| < \gamma_1$. Then the CP conditions demand that $\gamma_1 + |\gamma_2| \leq \sqrt{1-\gamma_3^2}$. So we have
$|\gamma_2| \leq {\rm min}\, (\gamma_1, \sqrt{1-\gamma_3^2} - \gamma_1)$.

\section{Conclusions}  
We  develop an optimal scheme for computation of the quantum discord for any $X$-state of a two-qubit system. Our treatment itself is  both comprehensive and self-contained and, moreover, it is geometric in flavour. We  exploit symmetry to show, without loss of generality, that the problem itself is one of {\em optimization over just a single variable}. The analysis is entirely based on the output or correlation ellipsoid. 

The optimal measurement is shown to be a three-element POVM. Further, it emerges that the region where the optimal measurement comprises three elements is a tiny wedge shaped one in a slice of the parameter space. On either side of this wedge shaped region one has a von Neumann measurement along the z or x axis as the optimal measurement. 

Not all parameters of a two-qubit $X$-state influence the correlation ellipsoid. The parameters that influence and those which do not influence play very different roles.  The correlation ellipsoid has an invariance group which is much larger that the group of local unitary symmetries and comprises of three components. These symmetries are the Lorentz group, another copy of the Lorentz group obtained by the action of spatial inversion, and finally, a scale factor.  An appreciation of this larger invariance turns out to be essential to the simplification of the present analysis. We bring out in a transparent manner how the various parameters of the ellipsoid affect the optimal measurement scheme and also provide many examples to demonstrate the same. We also bring out the role played by the partial transpose test at the level of the correlation ellipsoid in respect of entanglement. 

Having set up and studied the properties of the optimal measurement, we clearly underline the fact that the region where the assertion of Ali {\em et al.} is numerically misplaced is really tiny. But the $X$-states in this tiny region have {\em the same symmetry} as those outside, perhaps implying that if the symmetry argument of Ali {\em et al.} is misplaced it is likely to be so everywhere, and not just in this region. We bring out all the above aspects with a useful example. 

Finally, we provide numerous examples of states for which the quantum discord can be computed without an explicit optimization problem. These include states with vanishing discord and states whose correlation ellipsoid is centered at the origin.

\subsection*{Appendix\,: Matrix elements of $\rho$ and $M$}
Matrix elements of $\rho_{AB}$ in terms of the Mueller matrix elements is given by\,: 
{\scriptsize
\begin{align*}
\rho_{AB} = \frac{1}{4}
\left[
\begin{array}{c|c|c|c}
m_{00} + m_{03} + m_{30} + m_{33} &m_{01} + i m_{02} + m_{31} + i m_{32} & m_{10} - i m_{20} + m_{13} - i m_{23} & m_{11}+i m_{12} - i m_{21}+m_{22} \\
\hline
m_{01} - i m_{02} + m_{31} - i m_{32}&m_{00} - m_{03} + m_{30} - m_{33}&m_{11}-i m_{12} - i m_{21}-m_{22}&m_{10} - i m_{20} - m_{13} + i m_{23}\\
\hline
m_{10} + i m_{20} + m_{13} + i m_{23}&m_{11}+i m_{12} + i m_{21}-m_{22}&m_{00} + m_{03} - m_{30} - m_{33}&m_{01} + i m_{02} - m_{31} - i m_{32}\\
\hline
m_{11}-i m_{12} + i m_{21}+m_{22}&m_{10} + i m_{20} - m_{13} - i m_{23} &m_{01} - i m_{02} - m_{31} + i m_{32}&m_{00} - m_{03} - m_{30} + m_{33}
\end{array} 
\right],
\end{align*}}
and that of $M$ in terms of ${\rho}_{AB}$ by\,:
{\scriptsize 
\begin{align}
M = \left[
\begin{array}{c|c|c|c}
1& \rho_{01} + \rho_{10} + \rho_{23} + \rho_{32} & -i[(\rho_{01} - \rho_{10}) + (\rho_{23} + \rho_{32})] & (\rho_{00}  - \rho_{11}) + (\rho_{22} - \rho_{33})\\
\hline
\rho_{02} + \rho_{20} + \rho_{13} + \rho_{31} & \rho_{03} + \rho_{30} + \rho_{12} + \rho_{21} & -i [(\rho_{03} - \rho_{12}) + (\rho_{21} - \rho_{30})] & \rho_{02} + \rho_{20} - (\rho_{13} + \rho_{31}) \\ 
\hline
i [(\rho_{02} -\rho_{20})+ (\rho_{13} - \rho_{31})] & i [(\rho_{03} + \rho_{12}) - (\rho_{21} + \rho_{30})] & \rho_{30} + \rho_{03} - (\rho_{12} + \rho_{21}) & i[(\rho_{02} - \rho_{20}) - (\rho_{13} - \rho_{31})] \\
\hline
\rho_{00} + \rho_{11} -( \rho_{22} + \rho_{33}) & \rho_{01} + \rho_{10} -( \rho_{23} + \rho_{32}) & -i[(\rho_{01} -\rho_{10}) - (\rho_{23} - \rho_{32})] & \rho_{00} + \rho_{33} -( \rho_{11} + \rho_{22})
\end{array}
\right]. 
\label{app}
\end{align}
}

\chapter{Robustness of non-Gaussian entanglement}
 
\section{Introduction}
Early developments in quantum information technology of continuous 
variable (CV) systems largely concentrated on Gaussian states and 
Gaussian operations\,\cite{adessoreview,wangreview,andersenreview,samreview,raulreview,patibook}. 
The Gaussian setting has 
proved to be a valuable resource in continuous variable quantum information processes with current optical technology\,\cite{lvovskyreview,opttech1,opttech2}.
These include teleportation\,\cite{tele1,tele2,tele3}, cloning\cite{cloning1,cloning2,cloning3,cloning4,cloning5}, dense
coding\,\cite{densecoding1,densecoding2,densecoding3}, quantum
cryptography\,\cite{crypt1,crypt2,crypt3,crypt4,crypt5,crypt6,crypt7,crypt8,crypt9}, and quantum computation\,\cite{qcomp1,qcomp2,qcomp3}. 

The symplectic group of linear 
canonical transformations\,\cite{simon94,simon95} is available as a handy and 
powerful tool in this Gaussian scenario, leading to an elegant
classification of permissible Gaussian processes or 
channels\,\cite{wolfeisert}.

The fact that states in the non-Gaussian sector could offer advantage 
for several quantum information tasks has resulted more recently in 
considerable interest in non-Gaussian 
states and operations, both experimental\,\cite{ng-exp1,ng-exp2,ng-exp3} and 
theoretical\,\cite{ng-theor1,ng-theor2,ng-theor3,ng-theor4,ng-meas1,ng-meas2,ng-meas3,ng-meas4,ng-meas5}. The potential use of non-Gaussian
states for quantum information processing tasks have been explored\,\cite{ng-prep1,ng-prep2,ng-prep3,ng-prep4,ng-prep5,ng-prep6,ng-prep7}.
 The use of non-Gaussian resources for
teleportation\,\cite{ng-tele1,ng-tele2,ng-tele3,ng-tele4}, 
entanglement distillation\,\cite{ng-ent1,ng-ent2,ng-ent3}, and its use in quantum
networks\,\cite{ng-qnet} have been studied. So there 
has been interest to explore the essential differences between
Gaussian states and non-Gaussian states as resources for performing
these quantum information tasks.


Since  noise is unavoidable in any actual realization of these 
information processes\,\cite{kimble,noise1,noise2,noise3,noise4,noise5,noise6}, robustness of entanglement and other 
nonclassical effects against noise becomes an important consideration. 
Allegra et. al.\,\cite{allegra} have thus studied the evolution of 
what they call {\em photon number entangled states} (PNES) (i.e., 
two-mode states of the form $|\psi\rangle = \sum\,c_n\,|n,\,n\rangle$) 
in a  {\em noisy} attenuator environment. They conjectured based on  
numerical evidence that, for a given energy, Gaussian entanglement is 
more robust than the non-Gaussian ones. Earlier Agarwal et. 
al.\,\cite{agarwal091} had shown that entanglement of the NOON state is 
more robust than Gaussian entanglement in the {\em quantum limited} 
amplifier environment. More recently, Nha et. al.\,\cite{carmichael10} have 
shown that nonclassical features, including entanglement, of several 
non-Gaussian states survive a {\em quantum limited} amplifier 
environment much longer than Gaussian entanglement. Since the 
 conjecture of Ref.\,\cite{allegra} refers to the noisy environment  
 and  the analysis in Ref.\,\cite{agarwal091,carmichael10} to the noiseless or 
quantum-limited case, the conclusions of the latter do not necessarily
amount to 
refutation of the conjecture of Ref.\,\cite{allegra}. Indeed, Adesso has 
argued very recently\,\cite{adesso} that the well known 
extremality\,\cite{extremality1,extremality2} of Gaussian states implies 
proof and rigorous validation of the conjecture of 
Ref.\,\cite{allegra}.     

In this Chapter, we employ the  
Kraus representation of bosonic Gaussian channels\,\cite{kraus10} to study 
analytically the behaviour of non-Gaussian states in {\em noisy} 
attenuator and amplifier environments. Both NOON states and a simple 
form of PNES are considered. Our results show conclusively that the 
conjecture of Ref.\,\cite{allegra} is too strong to be maintainable.

\section{Noisy attenuator environment}
Under evolution through a noisy
attenuator channel ${\cal 
  C}_{1}(\kappa,a), \, \kappa \leq 1$, an input state ${\hr{}}^{\rm in}$
with characteristic 
function (CF) $\chi_{W}^{\rm{in}}(\xi)$ goes to state ${\hr{}}^{\rm
  out}$ with CF 
\begin{eqnarray}
\chi^{\rm{out}}_{W}(\xi)= \chi^{\rm{in}}_{W}(\kappa \xi)
 \,e^{-\frac{1}{2} (1- \kappa^2 
  + a )|\xi|^2},
\label{r1}
\end{eqnarray}
where $\kappa$ is the attenuation 
parameter\,\cite{caruso06,holevo07}. 
 In this notation, 
quantum
limited channels\,\cite{carmichael10} correspond to $a=0$, and so the 
parameter 
$a$ stands
for the {\em additional Gaussian noise}. 
Thus, ${\hr{}}^{\rm in}$ is taken under the
two-sided symmetric action of  ${\cal C}_{1}(\kappa,
a)$ to ${\hr{}}^{\rm out}= {\cal C}_{1}(\kappa, a) \otimes {\cal
  C}_{1}(\kappa, a) \, ({\hr{}}^{\rm in})$
with CF 
\begin{align}
\chi^{\rm{out}}_{W}(\xi_1, \xi_2) = 
\chi^{\rm{in}}_{W}(\kappa \xi_1, \kappa \xi_2)
 \,e^{-\frac{1}{2} (1- \kappa^2 +a)(|\xi_1|^2 + |\xi_2|^2)}.
\label{r2}
\end{align}
To test for separability of ${\hr{}}^{\rm out}$ we may
 implement the partial transpose test on ${\hr{}}^{\rm
  out}$ in the Fock basis or on
$\chi^{\rm{out}}_{W}(\xi_1, \xi_2)$. {\em The choice could depend on 
the state}.

Before we begin with the analysis of the action of the noisy channels on  two-mode states, a few definitions we require are in order\,:
\begin{definition}[Critical or threshold noise]\,:
$\alpha_0(\hr{})$ is the threshold noise with the property that $\Phi(\kappa,\alpha)[\hr{}] = [C_j(\kappa,\alpha) \otimes C_j(\kappa,\alpha)][\hr{}]$ remains entangled for $\alpha < \alpha_0$ and becomes separable for $\alpha >\alpha_0$ for a given $\kappa$; $j=1,2$ according as the attenuator or amplifier channel
\end{definition}

\vspace{0.3cm}
\begin{definition}[Robustness of entanglement]\,:
Entanglement of $\hr{1}$ is more robust than that of $\hr{2}$ if $\alpha_0(\hr{1}) > \alpha_0(\hr{2})$, for a given $\kappa$.
\end{definition}

\vspace{0.3cm}
\begin{definition}[Critical noise for a set of states]\,:
The critical noise for a set of states ${\cal A} = \{\hr{1},\hr{2},\cdots \}$ is defined as 
$\alpha_0({\cal A})= {\rm max}\,(\alpha_0(\hr{1}),\, \alpha_0(\hr{2}),\,\cdots )$. In this case, the value $\alpha_0({\cal A})$ renders all the states of the set ${\cal A}$ separable for $\alpha \geq \alpha_0({\cal A})$,  for a given $\kappa$.
\end{definition}

\subsection{Action on Gaussian states}
Consider first the Gaussian case, and in particular 
the two-mode squeezed state
$|\psi(\mu)\rangle = {\rm sech}\mu \sum_{n=0}^{\infty} \tanh^n \mu
|n,n \rangle$ with variance matrix $V_{\rm sq}(\mu)$. 
Under the
two-sided action of noisy attenuator channels ${\cal C}_{1}(\kappa,
a)$, 
 the output two-mode Gaussian state ${\hr{}}^{\rm{out}}(\mu) 
= {\cal C}_{1}(\kappa, a) \otimes {\cal
  C}_{1}(\kappa, a) \, (\,|\psi(\mu)\rangle \langle\psi(\mu)|\,)$
has variance matrix  
\begin{eqnarray}  
V^{\rm out}(\mu) &=& \kappa^2 V_{\rm sq}(\mu) + (1-\kappa^2 +
a){1\!\!1}_{4}, \,\,\, \nonumber \\
V_{\rm sq}(\mu)&=& \left(\begin{array}{cc}
c_{2\mu}{1\!\!1}_{2} & s_{2\mu}\sigma_3 \\
s_{2 \mu}\sigma_3 & c_{2 \mu} {1\!\!1}_{2}
\end{array} \right), 
\label{r3}
\end{eqnarray}
where 
 $c_{2\mu}= \cosh 2\mu$, $s_{2 \mu}= \sinh 2\mu$. 
Note that our variance matrix differs from that
of some authors by a factor 2; {\em in particular, the
variance matrix of  vacuum is the unit matrix in our notation}. 
 Partial transpose test\,\cite{simon00} shows that
${\hr{}}^{\rm out}(\mu)$ is separable iff $a \geq
\kappa^2 (1 - e^{-2 \mu})$. The `additional noise' $a$
required to render ${\hr{}}^{\rm out}(\mu)$ separable is an 
increasing function of the squeeze (entanglement) parameter $\mu$ 
and saturates at $\kappa^2$. 
In particular, $|\psi(\mu_1)
\rangle$, $\mu_1 \approx 0.5185$ corresponding to one ebit of
entanglement 
is rendered separable when $a \geq \kappa^2 (1- e^{-2 \mu_1})$.
 For $a \geq \kappa^2$, ${\hr{}}^{\rm out}(\mu)$ is separable,
independent of the initial squeeze parameter $\mu$. {\em Thus  
$a = 
\kappa^2$
is the additional noise that renders separable all Gaussian states.}

\subsection{Action on non-Gaussian states}
Behaviour of non-Gaussian entanglement may be handled directly in the 
Fock basis using 
 the recently developed Kraus
representation of Gaussian channels\,\cite{kraus10}.
  In this basis 
quantum-limited attenuator ${\cal C}_1(\kappa;0), \, \kappa \leq 1$ 
and 
quantum-limited amplifier ${\cal C}_2(\kappa;0), \, \kappa \geq 1$ 
are described, respectively, by Kraus operators displayed in Table\,\ref{tablej2}\,:
\begin{eqnarray}
B_{\ell}(\kappa) &=& \sum_{m=0}^{\infty}
\sqrt{{}^{m+\ell} C_{\ell}}\, 
(\sqrt{1-\kappa^2}\,)^{\,\ell}\, {\kappa}^{m}  
| m \rangle \langle m+\ell|,\nonumber\\ 
A_{\ell}(\kappa) &=& \frac{1}{\kappa} \sum_{m=0}^{\infty} 
\sqrt{{}^{m+\ell} 
  C_{\ell}} 
(\sqrt{1-\kappa^{-2}}\,)^{\,\ell}\frac{1}{\kappa^m}
| m +\ell \rangle \langle m |,
\label{r4}
\end{eqnarray}
$\ell =0,1,2,\cdots$. In either case, the noisy channel ${\cal
  C}_j(\kappa;a), \, j=1,2$ can be 
realized in the form ${\cal C}_2(\kappa_2;0) \circ {\cal
  C}_1(\kappa_1;0)$, so that the Kraus operators
for the noisy case is simply  the product set
$\{A_{\ell^{\,'}}(\kappa_2) B_{\ell}(\kappa_1)  \} $. 
Indeed, the
composition rule ${\cal C}_2(\kappa_2;0) \circ {\cal
  C}_1(\kappa_1;0) = {\cal C}_1(\kappa_2\kappa_1;2(\kappa_2^2-1))$ 
or ${\cal C}_2(\kappa_2\kappa_1;2\kappa_2^2(1-\kappa_1^2))$ 
according as $\kappa_2\kappa_1 \leq 1$ or $\kappa_2\kappa_1 \geq 1$
implies that the noisy attenuator ${\cal C}_1(\kappa;a), 
\kappa \leq 1$ is realised by the choice $\kappa_2 = \sqrt{1 + a/2}
\geq 1, \, \kappa_1 = \kappa/\kappa_2 \leq \kappa \leq 1$, and the noisy
amplifier 
${\cal  C}_2(\kappa;a), \, \kappa \geq 1 $ by 
$\kappa_2 = \sqrt{\kappa^2 +
  a/2} \geq \kappa \geq 1, \, \kappa_1 = \kappa/\kappa_2 
\leq 1$\,\cite{kraus10}.   
 {\em Note that one goes from realization of ${\cal 
C}_1(\kappa;a),\,\kappa\le 1$ to that of
  ${\cal C}_2(\kappa;a),\,\kappa\ge 1$ simply by replacing $(1+a/2)$ 
by 
$(\kappa^2 +
  a/2)$; this fact will be exploited later.}

Under the action of ${\cal C}_j(\kappa;a) = {\cal C}_2(\kappa_2;0) \circ
{\cal C}_1(\kappa_1;0),\,j=1,2$, by Eq.\,\eqref{j71},  the elementary operators 
$|m\rangle\langle n|$ 
go to 
\begin{align}
&{\cal C}_{2}(\kappa_2; 0) \circ {\cal C}_{1}(\kappa_1; 0) \left( |m
  \rangle \langle n| \right) \nonumber \\ &= 
\kappa_2^{-2} \sum_{j=0}^{\infty} \sum_{\ell=0}^{\text{min}(m,n)}
\left[ {}^{m-\ell+j}C_j\,
  {}^{n-\ell+j}C_j\,{}^mC_{\ell}\,{}^nC_{\ell}\right]^{1/2} 
\, (\kappa_2^{-1} \kappa_1)^{(m+n-2\ell)}\,  
\nonumber \\ 
&~~~\,\times  (1-\kappa_2^{-2})^j
\,(1-\kappa_1^2)^{\ell}  \, |m-\ell +j \rangle \langle n-\ell +j|.
\label{r5}
\end{align}
Substitution of $\kappa_2 = \sqrt{1+a/2}, \kappa_1 = \kappa/\kappa_2$
gives realization of ${\cal C}_1(\kappa;a), \kappa \leq 1$ while
$\kappa_2 = \sqrt{\kappa^2+a/2}, \kappa_1 = \kappa/\kappa_2$ gives
 that of ${\cal C}_2(\kappa;a), \, \kappa \geq 1$. 

\subsubsection{NOON states}
As our first non-Gaussian example we study the NOON state. Various  aspects of the experimental generation of NOON states\,\cite{noon1,noon2,noon3,noon4,noon5,noon6,noon7} and its usefulness in measurements\,\cite{noon8,noon9,noon10} has been well studied. 

A NOON state $| \psi
\rangle = \left(|n0\rangle + |0n \rangle  \right)/\sqrt{2} $ has density matrix  
density matrix 
\begin{align}
{\hr{}} &= \frac{1}{2} \left( |n\rangle \langle n| \otimes |0
  \rangle \langle 0| + |n\rangle \langle 0| \otimes |0
  \rangle \langle n| \right. \nonumber \\
& ~~~+|0\rangle \langle n| \otimes |n
  \rangle \langle 0| + |0\rangle \langle 0| \otimes |n
  \rangle \langle n| \left.  \right).
\label{r6}
\end{align}
 The output state 
${\hr{}}^{\rm out} = {\cal C}_{1}(\kappa ; a) \otimes {\cal
  C}_{1}(\kappa ; a)( \hr{})$ can be
detailed in the Fock basis through use of Eq.\,(\ref{r5}).

To test for inseparability, we
project ${\hr{}}^{\rm out}$ 
 onto the $2 \times 2$ subspace spanned by the four bipartite vectors
$\{|00\rangle,\, |0n \rangle, \, |n,0 \rangle, \, |n,n \rangle \}$, and
test for entanglement in this subspace;  this
simple test proves sufficient for our purpose! The matrix
elements of interest are\,:  
$\hat{\rho}^{\rm out}_{00,00}$, $\hat{\rho}^{\rm out}_{nn,nn}$, and $\hat{\rho}^{\rm 
out}_{0n,n0} = \hat{\rho}^{\rm
  out\,*}_{n0,0n}$.  Negativity of
$\delta_{1}(\kappa, a) \equiv {\hat{\rho}}^{\rm out}_{00,00} {\hat{\rho}}^{\rm
  out}_{nn, nn} - |{\hat{\rho}}^{\rm out}_{0n, n0}|^2 $ will prove for 
$\hat{\rho}^{\rm out}$  not only NPT entanglement, but
also one-copy distillability \cite{horodecki97,bennett96}. 

To evaluate ${\hat{\rho}}^{\rm out}_{00,00}$, ${\hat{\rho}}^{\rm out}_{0n,n0}$, and
${\hat{\rho}}^{\rm out}_{nn,nn}$, it suffices to
evolve the four single-mode operators $|0\rangle\langle0|$, $|0 \rangle
\langle n|$, $|n \rangle \langle 0|$, and $|n \rangle \langle n|$
through the noisy attenuator ${\cal C}_1(\kappa;a)$ using
Eq.\,\eqref{r5}, and then project the output to one of these
operators. For our purpose we need only the 
following single mode matrix elements\,:
\begin{align}
x_1 &\equiv \langle n| {\cal C}_1(\kappa;a) (|n\rangle \langle n|)| n
\rangle \nonumber\\ 
&=
(1+a/2)^{-1} \sum_{\ell=0}^{n} \left[ {}^n C_{\ell} \right]^2 [\kappa^2 
(1+a/2)^{-2}]^{\ell} \nonumber\\
&\times~ [(1-\kappa^2(1+a/2)^{-1}) (1-(1+a/2)^{-1})]^{n -\ell},\nonumber\\
x_2 &\equiv \langle 0| {\cal C}_1(\kappa;a) (|n\rangle \langle n|)| 0
\rangle \nonumber \\
&= (1+a/2)^{-1} [1-\kappa^2(1+a/2)^{-1}]^n,
\label{r7a}
\end{align}
\begin{align}
 x_3 &\equiv \langle 0| {\cal C}_1(\kappa;a) (|0\rangle \langle 0|)| 0 \rangle \nonumber\\
&= (1+a/2)^{-1},\nonumber\\
x_4 &\equiv \langle n| {\cal C}_1(\kappa;a) (|0\rangle \langle 0|)| n
\rangle \nonumber \\
&=(1+a/2)^{-1}[1-(1+a/2)^{-1}]^n,\nonumber\\
x_5 &\equiv \langle n| {\cal C}_1(\kappa;a) (|n\rangle \langle 0|)| 0 \rangle \nonumber\\
&=\kappa^n (1+a/2)^{-(n+1)},\nonumber \\
 &\equiv \langle 0| {\cal C}_1(\kappa;a) (|0\rangle \langle n|)| n\rangle^{*}  
\label{r7b}
\end{align}
One finds ${\hat{\rho}}^{\rm out}_{00,00}= x_2 x_3$, ${\hat{\rho}}^{\rm
  out}_{nn,nn} = x_1 x_4$, and ${\hat{\rho}}^{\rm out}_{0n,n0} = x^2_5/2$,
and therefore 
\begin{align} 
\delta_1(\kappa, a)= x_1x_2x_3 x_4 -(|x_5|^2/2)^2, 
\label{r8}
\end{align} 
Let $a_1(\kappa)$ be the solution to $\delta_1(\kappa,a)=0$. This
means that entanglement of our NOON state survives all values of noise
$a < a_1(\kappa)$. The curve labelled $N_5$ in Fig.\,\ref{figr1}
 shows, in the $(a,\,\kappa)$ space,  $a_1(\kappa)$ for
the NOON state with $n=5$\,:  entanglement of $(| 50 \rangle + |05
\rangle)/\sqrt{2} $ survives all noisy 
attenuators below $N_5$. The straight line denoted $g_{\infty}$
corresponds to $a = \kappa^2$\,: channels above this line break
entanglement of all Gaussian states, even the ones with arbitrarily
large entanglement. The line $g_1$ denotes $a
= \kappa^2(1- e^{-2\mu_1})$, where $\mu_1 = 0.5185$ 
 corresponds to 1 ebit of Gaussian entanglement\,:
Gaussian entanglement $\leq 1$ ebit does not survive any of the
channels above this line. The region $R$ (shaded-region) of channels above $g_{\infty}$
but below $N_5$ are distinguished in this sense \,: {\em no Gaussian
entanglement survives the channels in this region, but the} NOON {\em state 
$(| 50
\rangle + | 0 5 \rangle)/\sqrt{2} $ does}.  
\begin{figure}
\begin{center}
\scalebox{0.55}{
\includegraphics{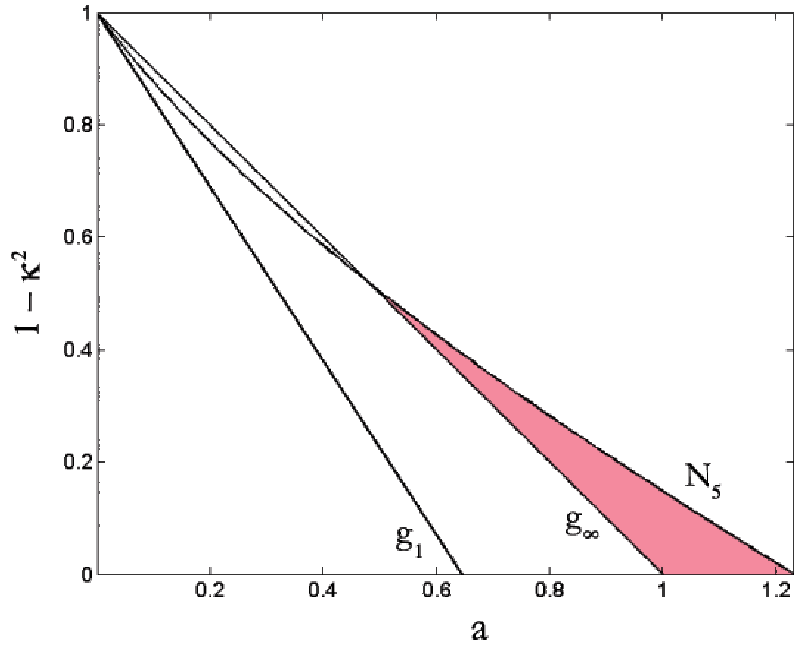}}
\end{center}
\caption{
Comparison of the robustness of the entanglement of a NOON
state with that of two-mode Gaussian states under the 
two-sided action of symmetric noisy attenuator. 
\label{figr1}} 
\end{figure}

\subsubsection{PNES states}
As a second non-Gaussian example we study the PNES 
\begin{align}|\psi \rangle  = \left(|00 \rangle + | nn \rangle
\right)/\sqrt{2}
\label{r8b}
\end{align} with density matrix 
\begin{align}
{\hat{\rho}} = &\frac{1}{2} \left( |0\rangle \langle 0| \otimes |0
  \rangle \langle 0| + |0\rangle \langle n| \otimes |0
  \rangle \langle n| \right. \nonumber \\
& ~~+|n\rangle \langle 0| \otimes |n
  \rangle \langle 0| + |n\rangle \langle n| \otimes |n
  \rangle \langle n| \left.  \right).
\label{r9}
\end{align}
The output
state ${\hat{\rho}}^{\rm out} = {\cal C}_{1}(\kappa;
a)\otimes {\cal C}_{1}(\kappa; a) \, \left( {\hat{\rho}} \right)$ can be 
detailed in the Fock basis through use of Eq.\,(\ref{r5}).

Now to test for entanglement of ${\hat{\rho}}^{\rm out}$, we project
again ${\hat{\rho}}^{\rm out}$ onto the $2 \times 2$ subspace spanned by the 
vectors 
$\{ |00\rangle,\, |0n \rangle, \, |n,0 \rangle, \, |n,n \rangle\}$,
and see if it is (NPT) entangled in this subspace. Clearly, it suffices
to evaluate the matrix elements ${\hat{\rho}}^{\rm 
  out}_{0n,0n}$, ${\hat{\rho}}^{\rm out}_{n0,n0}$, and 
${\hat{\rho}}^{\rm out}_{00,nn}$, for  if
$\delta_{2}(\kappa, a) \equiv {\hat{\rho}}^{\rm
  out}_{0n,0n} {\hat{\rho}}^{\rm out}_{n0,n0}-
|{\hat{\rho}}^{\rm out}_{00,nn} |^2$
is negative  then ${\hat{\rho}}^{\rm out}$ is NPT
entangled, and one-copy distillable. 

\begin{figure}
\begin{center}
\scalebox{0.55}{
\includegraphics{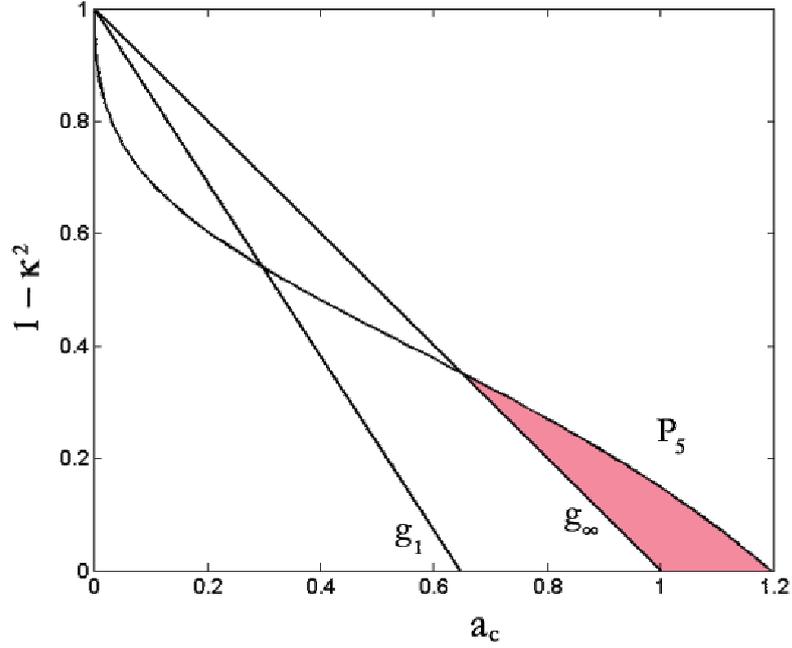}}
\end{center}
\caption{Comparison of the robustness of the entanglement of a PNES
state with that of two-mode Gaussian states under the action of
two-sided symmetric noisy attenuator. 
\label{figr2}} 
\end{figure}

Once again, the matrix elements listed in  \eqref{r7a} and \eqref{r7b} prove 
sufficient to determine $\delta_2(\kappa,a)$:  ${\hat{\rho}}^{\rm
  out}_{0n,n0} = {\hat{\rho}}^{\rm out}_{n0,n0} = (x_1x_2 + x_3x_4)/2$,
and ${\hat{\rho}}^{\rm out}_{00,nn} = |x_5|^2/2$, and so 
\begin{align}
\delta_2(\kappa,a) = ((x_1 x_2 + x_3x_4)/2)^2 - (|x_5|^2/2)^2. 
\label{r10}
\end{align}
Let $a_2(\kappa)$ denote the solution to $\delta_2(\kappa,a) =0$. That
is, entanglement of our PNES survives all $a \leq
a_2(\kappa)$. 
 This $a_2(\kappa)$ is shown as the curve labelled $P_5$ in 
Fig.\,\ref{figr2} for 
the
PNES $(|00 \rangle + | 55 \rangle )/\sqrt{2}$. The lines $g_1$
and $g_{\infty}$ have the same meaning as in Fig.\,\ref{figr1}. The 
region $R$ (shaded-region) above
$g_{\infty}$ but below $P_5$ corresponds to channels
$(\kappa,a)$ under whose action all two-mode Gaussian states are
rendered separable, while entanglement of the non-Gaussian PNES $(|00
\rangle + | 55 \rangle )/\sqrt{2}$ definitely survives.

\section{Noisy amplifier environment} 
We turn our attention now to the amplifier environment. Under the
symmetric two-sided 
action of a noisy amplifier channel ${\cal C}_{2}(\kappa; 
a), \, \kappa \geq 1$, the  two-mode CF 
$\chi^{\rm{in}}_{W}(\xi_1,\xi_2)$ is  
taken to
\begin{eqnarray}
\chi_W^{\rm{out}}(\xi_1,\xi_2) = \chi^{\rm{in}}_W(\kappa \xi_1, \kappa
\xi_2)\,e^{-\frac{1}{2}(\kappa^2-1+a)(|\xi_1|^2 + |\xi_2|^2)}. 
\label{r11}
\end{eqnarray}

\subsection{Action on Gaussian states}
In particular, the two-mode
squeezed vacuum state $|\psi(\mu) \rangle$ with variance matrix
$V_{\rm sq}(\mu)$ is taken to a Gaussian state with variance matrix
\begin{eqnarray}
V^{\text{out}}(\mu) =\kappa^2 V_{\text{sq}}(\mu) + 
(\kappa^2-1 + a) {1\!\!1}_{4}.
\label{r12}
\end{eqnarray} 
The partial transpose test\,\cite{simon00} readily shows that 
the output
state is separable when $a \geq 2-\kappa^2 (1+e^{-2\mu})$: 
the additional noise $a$ required to render the output Gaussian state 
separable  increases with the squeeze or entanglement parameter $\mu$ 
and saturates at $a =  2- \kappa^2$: for $a \geq 2- \kappa^2$ 
the output state is separable
for every Gaussian input. The noise required to render 
the two-mode squeezed state $|\psi(\mu_1)
\rangle $ with 1 ebit of
entanglement 
($\mu_1\approx 0.5185$) separable is $a = 2-\kappa^2(1+e^{-2\mu_1}) $.

\begin{figure}
\begin{center}
\scalebox{0.55}{
\includegraphics{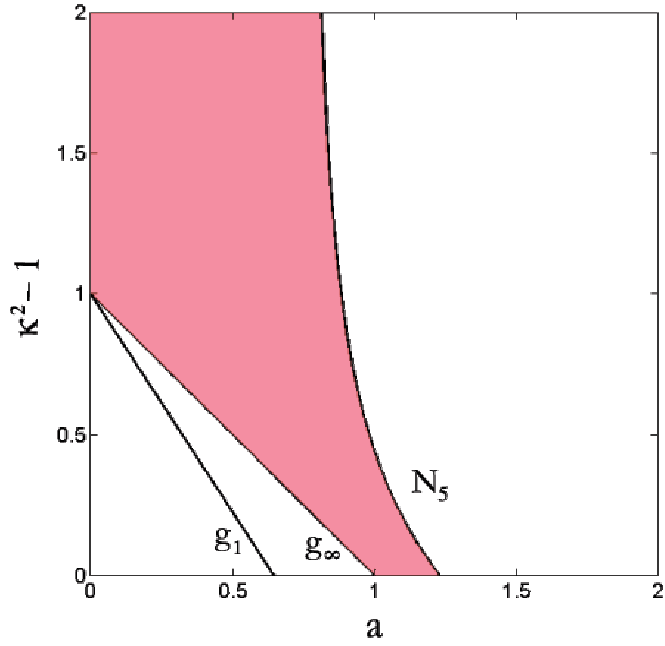}}
\end{center}
\caption{Comparison of the robustness of the entanglement of a NOON
state with that of all two-mode Gaussian states under the action of
two-sided symmetric noisy amplifier.
\label{figr3}} 
\end{figure}

\subsection{Action on non-Gaussian states}
As in the beamsplitter case, we now consider the action of the noisy amplifier channel on our choice of non-Gaussian states.

\subsubsection{NOON states}
Now we examine the behaviour of the NOON state $(|n0 \rangle + | 0n 
\rangle)/\sqrt{2}$
under the symmetric action of noisy amplifiers ${\cal
  C}_2(\kappa;a),\, \kappa \geq 1$.  Proceeding exactly
as in the attenuator case, we know that ${\hat{\rho}}^{\rm{out}}$ is definitely
entangled if $\delta_3(\kappa,a) \equiv {\hat{\rho}}^{\rm 
out}_{00,00} {\hat{\rho}}^{\rm
  out}_{nn, nn} - |{\hat{\rho}}^{\rm out}_{0n, n0}|^2  $ is negative. As
remarked earlier the expressions for ${\cal
   C}_1(\kappa;a), \, \kappa \leq 1$ in Eqs.\,\eqref{r7a} and \eqref{r7b} are 
valid 
for ${\cal
   C}_2(\kappa;a), \, \kappa \geq 1$ {\em {provided}} $1 + 
a/2$ is
replaced by $\kappa^2 + a/2$. For clarity we denote by $x^{\,'}_j$ the
expressions resulting from $x_j$ when ${\cal C}_1(\kappa;a), \, \kappa \leq 1$
replaced by ${\cal 
  C}_2(\kappa;a), \, \geq 1$ and $1 + a/2$ by $\kappa^2 + a/2$. For
instance, $x_5^{\,'} \equiv \langle n| {\cal C}_2(\kappa;a) (\,|n\rangle \langle 0|\,)| 0 \rangle =
\kappa^n (\kappa^2+a/2)^{-(n+1)} $ and $\delta_3(\kappa;a) =
x_1^{\,'} x_2^{\,'} x_3^{\,'} x_4^{\,'} - (|x_5^{\,'}|^2/2)^2$. 

Let $a_3(\kappa)$ be the solution to $\delta_3(\kappa,a)=0$. This is
represented  in Fig.\,\ref{figr3} by the curve marked $
N_5$, for the case of NOON state $(|05 \rangle + |50 \rangle
)/\sqrt{2}$. This curve is to be compared  with the line $a = 2 - \kappa^2$, denoted
$g_{\infty}$, above which no Gaussian entanglement survives, and with
the line $a = 2 - \kappa^2(1 + e^{-2\mu_1}), \, \mu_1= 0.5185$,
denoted $g_1$, above which no Gaussian entanglement $\leq 1$ ebit
survives. In particular, {\em the region $R$ (shaded-region) between $g_{\infty}$ and 
$N_5$
corresponds to noisy amplifier channels against 
which entanglement of the} NOON {\em state $( |05 \rangle   + |50\rangle
)/\sqrt{2}$ is robust, whereas no Gaussian entanglement survives}. 

\subsubsection{PNES states}
Finally, we consider the behaviour of the PNES $( |00 \rangle + |nn 
\rangle )/\sqrt{2}$ in this noisy amplifier environment. The
output, denoted ${\hat{\rho}}^{\rm{out}}$, is certainly entangled if
$\delta_4(\kappa,a) \equiv {\hat{\rho}}^{\rm out}_{0n,0n} {\hat{\rho}}^{\rm 
  out}_{n0, n0} - |\hat{\rho}^{\rm out}_{00, nn}|^2  $ is
negative. Proceeding as in the case of the attenuator, and
remembering the connection between $x_j$'s and the corresponding 
$x_j^{\,'}$'s, we have $\delta_4(\kappa,a) = ((x_1^{\,'}
x_2^{\,'} + x_3^{\,'}x_4^{\,'})/2)^2 - (|x_5^{\,'}|^2/2)^2 $. 

\begin{figure}
\begin{center}
\scalebox{0.55}{
\includegraphics{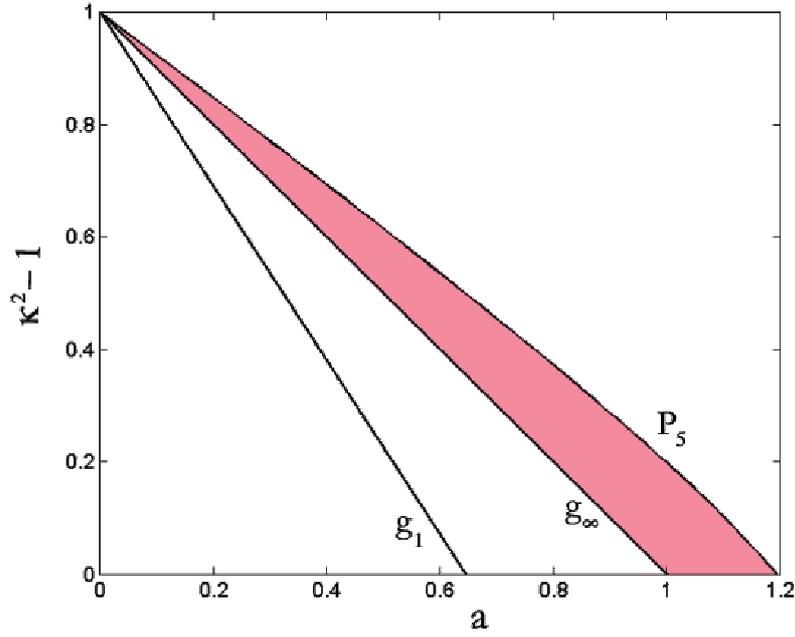}}
\end{center}
\caption{
Comparison of the robustness of the entanglement of a PNES
state with that of all two-mode Gaussian states under the action of 
 two-sided symmetric noisy amplifier.  
\label{figr4}} 
\end{figure}

The curve denoted $P_5$ in Fig.\,\ref{figr4} represents $a_4(\kappa)$
forming solution to \\$\delta_4(\kappa,a)=0$,  for the case of the PNES 
$(|00 \rangle
+ | 55\rangle )/\sqrt{2}$. The lines
$g_{\infty}$ and $g_1$ have the same meaning as in Fig.\,\ref{figr3}. 
The  
region $R$ (shaded-region) between $g_{\infty} $ and $P_5$ signifies the robustness of
our PNES\,: {\em for every $\kappa \geq 1$,
  the} PNES {\em is 
seen to 
endure more noise than Gaussian states with arbitrarily large
entanglement.}

\section{Conclusion}
We conclude with a pair of remarks.
First, our conclusion following\\ Eq.\,\eqref{r3} and Eq.\,\eqref{r12} that entanglement of two-mode 
squeezed (pure) state $|\psi(\mu)\rangle$ does not survive, for any value of $\mu$, 
channels $(\kappa,\,a)$ which satisfy the inequality $|1-\kappa^2| +a \geq 1$ applies 
to {\em all} Gaussian states. Indeed, for an arbitrary (pure or mixed) two-mode 
Gaussian state with variance matrix $V_G$ it is clear from
Eqs.\,\eqref{r3},\,\eqref{r12} that the  output Gaussian 
state has variance matrix 
$V^{\rm out} =\kappa^2\,V_G + (|1-\kappa^2| + a)1\!\!1_4$. Thus  
$|1-\kappa^2| + a \geq 1$ immediately implies, in view of nonnegativity of $V_G$, that  
$V^{\rm out} \geq 1\!\!1_4$, demonstrating separability of the output 
state for arbitrary Gaussian input\,\cite{simon00}. 

Secondly, Gaussian entanglement resides entirely `in' the variance matrix, and hence 
disappears when environmental noise raises the variance matrix above the vacuum or 
quantum noise limit. That our chosen states survive these environments 
shows that their entanglement resides in the higher moments, in turn
demonstrating that their entanglement is genuine non-Gaussian. Indeed,
the variance matrix of our PNES and NOON states for $N=5$ is six times
that of the vacuum state.

 Thus our result is likely to add further
impetus to the avalanching interest in the relatively new
`non-Gaussian-state-engineering'  in the context of realization of
distributed quantum communication networks.



\chapter{Nonclassicality breaking channels}

\section{Introduction}
Two notions that have been particularly well explored in the context of
quantum information of continuous variable states are 
{\em nonclassicality}\,\cite{ecg63,glauber63} and
{\em entanglement}\,\cite{schrodinger35,schrodinger36}. 
The `older' notion of entanglement has become one of renewed interest 
in recent decades for its central role and applications in (potential as well as demonstrated)
quantum information processes\,\cite{keyl02,horo-rmp}, while the concept of nonclassicality, which emerges
directly from the {\em diagonal representation}\,\cite{ecg63,glauber63}
had already been well explored in the quantum optical context\,\cite{kimble92,davidovich96,kolobov99},
even before the emergence of the present quantum information era.
A fundamental distinction between these two notions may be noted\,: 
{\em While nonclassicality can be defined even for states of a
single mode of radiation, the very notion of entanglement requires two or more parties}.
Nevertheless, it turns out that the two notions are not entirely independent of one another; they are rather
intimately related\,\cite{simon00,asboth05,ivan11,ivan12,kim02}. 
In fact, nonclassicality is a prerequisite
for entanglement\,\cite{ivan11,ivan12,kim02}. Since a nonclassical bipartite state
whose nonclassicality  can be removed by local unitaries could not be entangled,
one can assert, at least in an intuitive sense, that {\em entanglement is nonlocal
nonclassicality}.

An important aspect in the study of nonclassicality and entanglement
is in regard of their evolution under the action of a channel. 
A noisy channel acting on a state can degrade its nonclassical
features\,\cite{kim92,agarwal93,leuchs05,loudon97,biswas07,spagnolo09,dodonov11,meng12,filip13}. 
Similarly, entanglement can be degraded by channels acting  
locally on the constituent parties or modes\,\cite{scheel01,serafini05,poon07,song08,carmichael10,agarwal091,robust,barbosa10,agarwal10b,horodecki03, holevo08}.    
We have seen earlier\,\eqref{i115}, that {\em entanglement breaking} channels are those that render 
every  bipartite state separable by action on one of the subsystems\,\,\cite{horodecki03,holevo08,shirokov05}.

In this Chapter, we address the following issue\,: {\em which channels
possess the property of ridding every input state of its nonclassicality?}
Inspired by the notion of entanglement breaking channels, we may call
such channels {\em nonclassicality breaking channels}. The close
connection between nonclassicality and entanglement alluded to
earlier raises a related second issue\,: {\em what is the connection,
if any, between entanglement breaking channels and nonclassicality breaking
channels?} To appreciate the nontriviality of the second issue, it suffices
to simply note that the very definition of entanglement breaking refers
to bipartite states whereas the notion of nonclassicality breaking
makes no such reference. We show that both these issues
can be completely answered in the case of bosonic Gaussian channels\,:
nonclassicality breaking channels are enumerated, and it is shown that the set of all
nonclassicality breaking channels is essentially the same as the set of all entanglement breaking channels. 

We hasten to clarify the caveat `essentially'. Suppose a channel $\Gamma$ is nonclassicality breaking as well as entanglement breaking, and let us follow the action of this channel with a local unitary ${\cal U}$. The composite ${\cal U}\,\Gamma$ is clearly entanglement breaking. But local unitaries can create nonclassicality, and so ${\cal U}\,\Gamma$ need not be nonclassicality breaking. We say $\Gamma$ is {\em essentially nonclassicality breaking} 
if there exists a fixed unitary ${\cal U}$ dependent on $\Gamma$ but independent of the input state 
on which $\Gamma$ acts, so that ${\cal U}\,\Gamma$ is nonclassicality breaking. We may stress that this definition is not vacuous, for  given a collection of states {\em it is generically the case that there is no single unitary which would render the entire set nonclassical}. [This is not necessarily a property of the collection\,: given a nonclassical mixed state $\rho$, it is possibly not guaranteed that there exists an unitary ${\cal U}$ such that $\hat{\rho}^{\,'} = {\cal U}\, \hat{\rho} \,{\cal U}^{\dagger}$ is classical.] It is thus reasonable to declare the set of entanglement breaking  channels to be the same as the set of nonclassicality breaking channels if at all the two sets indeed turn out to be the same, modulo this `obvious' caveat or provision. 

We recall that  Gaussian channels are 
 physical processes that map Gaussian states to
 Gaussian states and their systematic 
 analysis was presented in\,\cite{wolfeisert,holevo07,caruso06,holevo01,kraus10,caruso08,wolf07,gio04,caruso06b,wolf08}.

\section{Nonclassicality breaking channels}

Any density operator $\hat{\rho}$ representing some state of a single mode of 
radiation field can always be expanded as
\begin{eqnarray}
\hat{\rho} = \int\frac{d^2\alpha}{\pi}\, {\phi}_{\rho} (\alpha) | \alpha \rangle \langle \alpha |,
\label{n4}
\end{eqnarray}
where ${\phi}_{\rho} (\alpha)= W_{1}(\alpha;\rho)$ is the diagonal `weight' function, 
$| \alpha \rangle$ being the
coherent state. This {\em diagonal representation} is made possible because of the 
over-completeness property
of the coherent state `basis'\,\cite{ecg63,glauber63}. The diagonal representation\,(\ref{n4})
enables the evaluation, {\em in a classical-looking manner},
of ensemble averages of normal-ordered operators, and this is
important from the experimental point of view\,\cite{mandelbook}.


An important notion that arises from the diagonal representation 
is the \\{\em classicality-nonclassicality divide}. If $\phi_{\rho}(\alpha)$
associated with density operator $\hat{\rho}$ is pointwise nonnegative
over ${\cal C}$, then the state is a convex sum, or ensemble,
of coherent states. Since coherent states are the most elementary 
of all quantum mechanical states exhibiting classical behaviour, any state
that can be written as a convex sum of these elementary classical states
is deemed classical. We have,
\begin{eqnarray}
{\phi}_{\rho} (\alpha) \geq 0 \,\,\,\,{\rm for}\,\,\,{\rm all} \,\,\,\alpha
\in {\cal C}\,\Leftrightarrow\, \hat{\rho}\,\,{\rm is}\,\,{\rm classical}.
\label{n5}
\end{eqnarray}      
Any state which cannot be so written is declared to be  nonclassical. Fock states $| n\rangle\langle n|$,
whose diagonal weight function
$\phi_{|n \rangle\langle n|}(\alpha)$ is the 
${\rm n^{th}}$ derivative of the delta function, are examples of nonclassical states. [All the above considerations generalize from one mode to $n$-modes in a painless manner, with $\alpha, \,\xi \in\, {\cal R}^{2n} \sim {\cal C}^n$.]

This classicality-nonclassicality divide leads to the following natural definition, inspired by the notion of entanglement breaking (See Fig.\,\ref{fign0b})\,:
\\

\noindent
{\bf Definition}\,: A channel $\Gamma$ is said to be {\em nonclassicality breaking}
if and only if the output state $\hat{\rho}_{\rm out}= \Gamma (\hat{\rho}_{\rm in})$ is classical
{\em for every} input state $\hat{\rho}_{\rm in}$, i.e., if and only if the diagonal
function of every output state is a genuine probability distribution (See Fig.\,\ref{fign0a}).
\\

\begin{figure}
\begin{center}
\scalebox{0.55}{\includegraphics{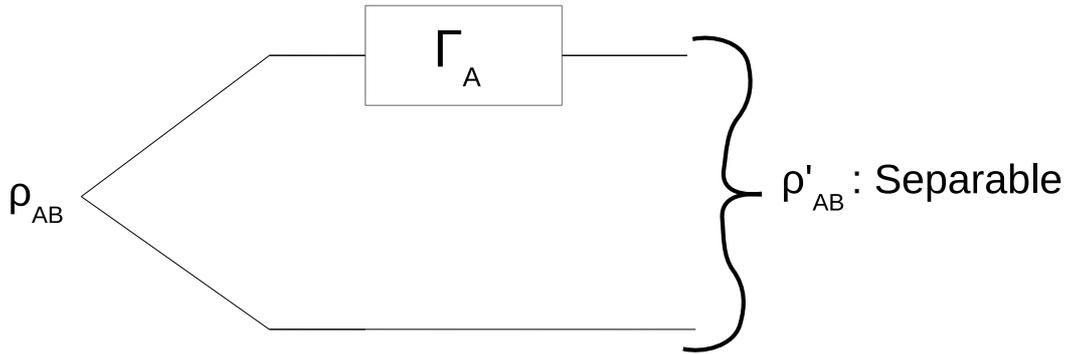}}
\end{center}
\caption{A schematic diagram depicting the notion of entanglement breaking channels. \label{fign0b}}
\end{figure} 

\vspace{0.2cm}

\begin{figure}
\begin{center}
\scalebox{0.55}{\includegraphics{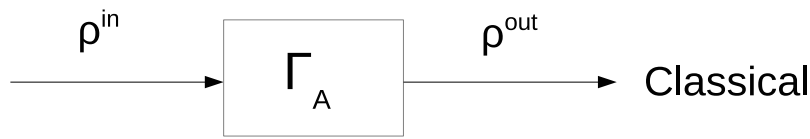}}
\end{center}
\caption{Showing the notion of nonclassicality breaking channels.\label{fign0a}}
\end{figure}

\section{Nonclassicality-based canonical forms for Gaussian channels}
The canonical forms for Gaussian channels have been described
by Holevo\,\cite{holevo07,caruso06} and Werner and Holevo\,\cite{holevo01}.
Let ${\cal S}$ denote an element of the symplectic group $Sp(2n,\,R)$ of 
linear canonical transformations and ${\cal U}({\cal S})$
the corresponding unitary (metaplectic) operator\,\cite{simon94}. One often 
encounters situations wherein the aspects one is looking for
are invariant under local unitary operations, entanglement
being an example. In such cases a Gaussian channel $\Gamma$ is `equivalent'
to ${\cal U}({\cal S}^{'})\,\Gamma\,{\cal U}({\cal S})$, for
arbitrary symplectic group elements ${\cal S}$, ${\cal S}^{'}$.
The orbits or double cosets of equivalent channels in this sense are the ones
classified and enumerated by Holevo and 
collaborators\,\cite{holevo07,caruso06,holevo01} and recalled in Table\,\ref{tablej1}.

While the classification of Holevo and collaborators is entanglement-based, as just noted,
the notion of nonclassicality breaking has {\em a more restricted invariance}.
A nonclassicality breaking Gaussian channel $\Gamma$
preceded by any Gaussian unitary ${\cal U}({\cal S})$ is nonclassicality breaking
if and only if $\Gamma$ itself is nonclassicality breaking.
In contradistinction, the nonclassicality breaking aspect of $\Gamma$
and ${\cal U}({\cal S})\,\Gamma$ [$\Gamma$ followed the Gaussian unitary
${\cal U}({\cal S})$] are not equivalent in general; they are equivalent
if and only if ${\cal S}$ is in the 
intersection $Sp(2n,\,R) \cap SO(2n,\,R) \sim U(n)$ of `symplectic phase
space rotations' or passive elements\,\cite{simon94,simon95}. In the single-mode case
this intersection is just the rotation group $SO(2) \subset Sp(2,\,R)$ described in Eq.\,\eqref{j41}. We thus
need to classify single-mode Gaussian channels $\Gamma$ into
orbits or double cosets ${\cal U}({\cal R})\,\Gamma\,{\cal U}({\cal S})$,
${\cal S} \in Sp(2,\,R)$, ${\cal R} \in SO(2) \subset Sp(2, R)$.
Equivalently, we need to classify $(X, Y)$ into orbits 
$({\cal S}\,X\,{\cal R},\,{\cal R}^{T}\,Y\,{\cal R})$. It turns out that
there are three distinct canonical forms, and the type into which a given pair $(X,Y)$ belongs is  fully determined by ${\rm det}\,X$.\\

\noindent
{\bf First canonical form}\,: ${\rm {\bf det}}\, {\bf X > 0}$.\\
A real $2 \times 2$ matrix $X$ with ${\rm det}\,X =\kappa^2 > 0$
is necessarily of the form $\kappa\,{\cal S}_{X}$ for some
${\cal S}_{X} \in Sp(2,\,R)$. Indeed we have ${\cal S}_{X}=({\rm det}\,X)^{-1/2}\,X$
Choose ${\cal R} \in SO(2)$ so as to
diagonalise $Y > 0$\,: ${\cal R}^{T}\,Y\,{\cal R}={\rm diag}(a, b)$.
With such an ${\cal R}$, the choice ${\cal S}= {\cal R}^{T}{\cal S}_{X}^{-1} \in Sp(2,\,R)$
takes $(X,\,Y)$ to the canonical form $(\kappa{1\!\!1}, \,{\rm diag}(a, b))$,
where $\kappa =\sqrt{{\rm det}\,X} > 0$, and $a,\,b$ are the eigenvalues of $Y$.\\

\noindent
{\bf Second canonical form\,:} {${\rm{\bf det}} \, {\bf X <0}$}.\\
Again choose ${\cal R}$ so that ${\cal R}^{T} Y {\cal R}={\rm diag}(a, b)$.
Since ${\rm det}\, X < 0$, $X$ is necessarily of the form $\kappa\,{\cal S}_{X}\,\sigma_3$,
for some ${\cal S}_{X} \in Sp(2,\, R)$\,: ${\cal S}_{X}=({\rm det}\, X\sigma_3)^{-1/2}X\sigma_3$.
Since ${\cal R}\,\sigma_3\,{\cal R}=\sigma_3$ for
every ${\cal R} \in SO(2)$, it is clear that the choice ${\cal S}={\cal R}\,{\cal S}_{X}^{-1} \in Sp(2,\,R)$
takes $(X,\,Y)$ to the canonical form $(\kappa\,\sigma_3,\,{\rm diag}(a,b))$ in this
case, with $\kappa=\sqrt{{\rm det}\,X\sigma_3}$, and the parameters $a,\,b$ being the
eigenvalues of $Y$.\\

\noindent
{\bf Third canonical form\,:} ${\rm{\bf det}}\,{\bf X=0}$.\\ Let $\kappa$ be the singular
value of $X$; choose ${\cal R}', \,\,{\cal R} \in SO(2)$ such that 
${\cal R}' \,X \,{\cal R}={\rm diag}(\kappa, 0)$. It is clear that the choice
${\cal S}_{X}={\rm diag}(\kappa^{-1}, \kappa)\,{\cal R}'^{\,T} \in Sp(2,\,R)$ along
with ${\cal R} \in SO(2)$ takes $(X,\, Y)$ to the canonical form 
$({\rm diag}(1,0),\,Y_{0}={\cal R}^{T}\,Y\,{\cal R})$. $Y_{0}$ does not, of course,
assume any special form. But if $X=0$, then ${\cal R} \in SO(2)$  can be chosen so as to
diagonalise $Y$\,: in that case $Y_{0}=(a,b),\,\,a,\,b$ being the eigenvalues of $Y$.

\section{Nonclassicality breaking Gaussian channels}
Having obtained the nonclassicality-based canonical forms of $(X,\,Y)$, 
we now derive the necessary and sufficient 
conditions for a single-mode Gaussian channel to be nonclassicality breaking.
We do it for the three canonical forms in that order. \\

\noindent
{\bf First canonical form\,:}\,${\bf \bm{(}X,\,Y\bm{)}\bm{=}
\bm{\left(}\bm{\kappa}{1\!\!1},\,{\rm {\bf diag}}\bm{(}a,b\bm{)}\bm{\right)}}$.\\
There are three possibilities\,: $\kappa=1$, $\kappa< 1$, and $\kappa>1$. We begin with
$\kappa =1$; it happens that the analysis  extends quite easily to the other two cases
and, indeed, to the other two canonical forms as well. 
The action on the
normal-ordered characteristic function in this case is
\begin{align}
&\chi_{N}^{\rm in}(\xi_1, \,\xi_2; \rho) \rightarrow {\chi}_{N}^{\rm out}(\xi_1, \xi_2;\rho)\nonumber\\
&=
\exp\left[-\frac{a\,\xi_1^2}{2}-\frac{b\,\xi_2^2}{2}  \right] 
\chi_{N}^{\rm in}(\xi_1,\,\xi_2; \rho).
\label{n13}
\end{align} 
[For clarity, we shall write the subscript of $\chi$ explicitly as $N$, $W$, or $A$ in place of 1, 0,  or -1]. It should be appreciated that {\em for this class of Gaussian channels} ($\kappa=1$) the above 
input-output relationship
holds even with the subscript $N$ replaced by $W$ or $A$ uniformly. Let us assume
$a,\,b > 1$ so that $a =1 + \epsilon_1$, $b= 1+ \epsilon_2$ with $\epsilon_1 , \, \epsilon_2 >0$.
The above input-output relationship can then be written in the form 
\begin{eqnarray*}
{\chi}_{N}^{\rm out}(\xi_1, \xi_2; \rho)=
\exp\left[-\frac{\epsilon_1\,\xi_1^2}{2}-\frac{\epsilon_2\,\xi_2^2}{2}  \right] 
\chi_{W}^{\rm in}(\xi_1,\,\xi_2; \rho).
\label{n14}
\end{eqnarray*}
Note that the subscript of $\chi$ on the right hand side is now $W$ and not $N$.

Define $\lambda > 0$ through $\lambda^2 = \sqrt{\epsilon_2 /\epsilon_1}$, and rewrite
the input-output relationship in the suggestive form
\begin{align}
{\chi}_{N}^{\rm out}(\lambda\xi_1, \lambda^{-1}\xi_2; \rho)&= \exp\left[-\frac{1}{2}(\sqrt{\epsilon_1\epsilon_2}\,\xi_1^2-
\sqrt{\epsilon_1\epsilon_2}\,\xi_2^2)  \right]
 \nonumber\\
&~~~~~\times  \chi_{W}^{\rm in}(\lambda\xi_1,\,\lambda^{-1}\xi_2; \rho).
\label{n15}
\end{align}
But $\chi_{W}^{\rm in}(\lambda\xi_1,\,\lambda^{-1}\xi_2; \rho)$ is simply the
Weyl-ordered or Wigner characteristic function of a (single-mode-)
squeezed version of $\hat{\rho}$, for every $\hat{\rho}$. If ${\cal U}_{\lambda}$
represents the unitary (metaplectic) operator that effects this squeezing
transformation specified by squeeze parameter $\lambda$, we have
\begin{eqnarray}
\chi_{W}^{\rm in}(\lambda\xi_1,\,\lambda^{-1}\xi_2; \rho)=
\chi_{W}^{\rm in}(\xi_1,\,\xi_2; {\cal U}_{\lambda}\,\rho\,{\cal U}_{\lambda}^{\dagger}),
\label{n16}
\end{eqnarray}
so that the right hand side of the last input-output relationship, {\em in the
special case} $\epsilon_1 \epsilon_2 =1$, reads
\begin{eqnarray}
\chi_{W}^{\rm out}(\lambda\xi_1,\,\lambda^{-1}\xi_2; \rho)=
\chi_{A}^{\rm in}(\xi_1,\,\xi_2; {\cal U}_{\lambda}\,\rho\,{\cal U}_{\lambda}^{\dagger}).
\label{n17}
\end{eqnarray}
This special case would transcribe, on Fourier transformation, to
\begin{eqnarray}
&&\phi^{\rm out}(\lambda \alpha_1,\,\lambda^{-1} \alpha_2; \rho)=
Q^{\rm in}(\alpha_1,\,\alpha_2; {\cal U}_{\lambda}\,\rho\,{\cal U}_{\lambda}^{\dagger})\nonumber \\
&&~~~~=\langle \alpha | {\cal U}_{\lambda}\,\hat{\rho}\,{\cal U}^{\dagger}_{\lambda}|\alpha\rangle
\geq 0,\,\, \forall\,\,\alpha,\,\,\forall\,\,\hat{\rho}.
\label{n18}
\end{eqnarray}
That is, the output diagonal weight function evaluated at $(\lambda\alpha_1, \,\lambda^{-1}\alpha_2 )$ equals the input
$Q$-function evaluated at $(\alpha_1, \,\alpha_2)$, and hence is nonnegative for all $\alpha \in {\cal C}$. 
Thus the output state is classical
for every input, and hence the channel is nonclassicality breaking. It is clear that if
$\epsilon_1 \epsilon_2 > 1$,
the further Gaussian convolution corresponding to the additional multiplicative factor
$\exp \left[-(\sqrt{\epsilon_1 \epsilon_2}-1) (\xi_{1}^{2}+\xi_{2}^{2})/2\right]$
in the output characteristic function will only render the output state even more strongly
classical. We have thus established this {\em sufficient condition}
\begin{eqnarray}
(a-1)(b-1)\geq 1,
\label{n19}
\end{eqnarray} 
or, equivalently,
\begin{align}
\frac{1}{a}+ \frac{1}{b} \leq 1.
\label{n20}
\end{align}

Having derived a sufficient condition for nonclassicality breaking, we 
derive a necessary condition by looking at the
signature of the output diagonal weight function {\em for a particular
input state} evaluated at {\em a particular phase space} point at the
output. Let the input be the Fock state $|1\rangle\langle1|$, the first excited
state of the oscillator. Fourier transforming the input-output relation (\ref{n13}),
one readily computes the output diagonal weight function to be 
\begin{eqnarray}
&&{\phi}^{\rm out}(\alpha_1,\,\alpha_2; |1 \rangle\langle 1|) =   \frac{2}{\sqrt{ab}}\,
\exp\left[-\frac{2 \alpha_{1}^{2}}{a}
-\frac{2 \alpha_{2}^{2}}{b}\right]\,\nonumber\\
&& \hspace{2cm} \times \left(1+ \frac{4(\alpha_1 + \alpha_2)^2}{a^{2}} - \frac{1}{a} 
-\frac{1}{b}\right). 
\label{n21}
\end{eqnarray}  
An obvious necessary condition for nonclassicality breaking is that this
function should be nonnegative everywhere in  phase space. Nonnegativity at the single phase space point $\alpha =0$
gives the necessary condition ${1}/{a} + {1}/{b} \leq 1$ which is, perhaps
surprisingly, the same as the sufficiency condition established earlier! That is, {\em the  sufficient condition (\ref{n19}) is also a necessary condition for nonclassicality breaking}.
Saturation of this inequality corresponds to the boundary wherein the channel
is `just' nonclassicality breaking. {\em The formal resemblance in this case
with the law of distances in respect of imaging by a thin convex lens} is unlikely to miss  the reader's attention.


The above proof for the particular case of classical noise channel $(\kappa=1)$ gets easily extended to noisy
beamsplitter (attenuator) channel $(\kappa < 1)$ and noisy
amplifier channel $(\kappa > 1)$. The action of the channel 
$(\kappa1\!\!1,\,{\rm diag}(a, b))$ on the
normal-ordered characteristic function follows from that on the Wigner
characteristic function given in\,(\ref{j51})\,: 
\begin{eqnarray}
{\chi}_{N}^{\rm out}(\xi; \rho)&=& \exp \left[ -\frac{\tilde{a}\, \xi_1^2}{2}
-\frac{\tilde{b}\,\xi_2^2}{2} \right]\chi_{N}^{\rm in}(\kappa\,\xi;\rho),  \nonumber \\
\tilde{a}&=& a+ \kappa^2 -1,\,\,\,\tilde{b}= b+\kappa^2 -1.  
\label{n22}
\end{eqnarray}
This may be rewritten in the suggestive form 
\begin{align}
{\chi}_{N}^{\rm out}(\kappa^{-1}\xi; \rho) = 
\exp \left[ -\frac{\tilde{a}\, \xi_1^2}{2\kappa^{2}}
  -\frac{\tilde{b}\, \xi_2^2}{2\kappa^{2}} \right]
\chi_{N}^{\rm in}(\xi; \rho).
\label{n23}
\end{align}  
With this we see that the right hand side of \eqref{n23} to be the same as right hand side of
\eqref{n13} with
${\tilde{a}}/{\kappa^2}$, ${\tilde{b}}/{\kappa^2}$
replacing $a, \,b$. 
The case $\kappa \not= 1$ thus gets essentially reduced
to the case $\kappa=1$, the case of classical noise channel,
analysed in detail above. This leads to the following {\em necessary
and sufficient condition for nonclassicality breaking} 
\begin{align}
&\frac{1}{a+\kappa^2-1} + \frac{1}{b+\kappa^2-1} \leq
\frac{1}{\kappa^2} \nonumber\\
&~ \Leftrightarrow ~ (a-1)(b-1) \geq  \kappa^4,
\label{n24}
\end{align}  
for all $\kappa>0$, thus completing our analysis of the first canonical form.\\

\noindent
{\bf Second canonical form\,:}\,\,${\bf \bm{(}X,Y\bm{)}\bm{=}\bm{(}\bm{\kappa\,\sigma_3},
\,{\rm {\bf diag}}\bm{(}a,b\bm{)}\bm{)}}$.
The noisy phase conjugation channel with canonical form $(\kappa\,\sigma_3, \,{\rm diag}(a,b))$
acts on the \\normal-ordered characteristic
function in the following manner, as may be seen from its action on the Weyl-ordered
characteristic function (\ref{j51})\,:
\begin{eqnarray}
{\chi}_{N}^{\rm out}(\xi;\rho)= 
\exp \left[ -\frac{\tilde{a}\, \xi_1^2}{2}
  -\frac{\tilde{b}\, \xi_2^2}{2} \right]
\chi_{N}^{\rm in}(\kappa\,\sigma_3\,\xi; \rho),
\label{n25}
\end{eqnarray} 
with $\tilde{a}= a+ \kappa^2 -1,\,\,\,\tilde{b}= b+\kappa^2 -1$ again, and $\kappa \,\sigma_3\,\xi$
denoting the pair $(\kappa\,\xi_1, -\kappa\,\xi_2)$.
As in the case of the noisy amplifier/attenuator channel, we rewrite it in the
form
\begin{align}
{\chi}_{N}^{\rm out}(\kappa^{-1}\,\sigma_3\, \xi; \rho) = 
\exp \left[ -\frac{\tilde{a}\,\xi_1^2}{2\kappa^{2}}
  -\frac{\tilde{b}\,\xi_2^2}{2\kappa^{2}} \right]
\chi_{N}^{\rm in}(\xi; \rho), 
\label{n26}
\end{align}  
the right hand side of \eqref{n26} has the same form as (\ref{n13}), leading to the
{\em necessary and sufficient nonclassicality breaking condition}
\begin{align}
\frac{1}{\tilde{a}}+ \frac{1}{\tilde{b}} \leq \frac{1}{\kappa^2}
~~\Leftrightarrow~~
(a-1)(b-1) \geq  \kappa^4.
\label{n27}
\end{align}  

\noindent
{\bf Remark}\,: We note in passing that in exploiting the `similarity' of
Eqs.\,(\ref{n23}) and (\ref{n26}) with Eq.\,(\ref{n13}),
we made use of the following two elementary facts\,: (1) An
invertible linear change of variables 
$[f(x) \rightarrow f(A\,x),\,\,{\rm det}\,A \not= 0]$
on a multivariable function $f(x)$ reflects as a
corresponding linear change of variables in its Fourier
transform\,; (2) A function $f(x)$ is pointwise
nonnegative if and only if $f(A\,x)$ is pointwise nonnegative
for every invertible $A$. In the case of (\ref{n23}), the linear change $A$
corresponds to uniform scaling, and in the case of (\ref{n26})
it corresponds to uniform scaling followed or preceded by 
mirror reflection. 
\vskip 0.1cm

\noindent
{\bf Third canonical form\,:} {\bf Singular $\bm{X}$}.
Unlike the previous two cases, it proves to be convenient to begin with the Weyl or symmetric-ordered 
characteristic function in this case of singular $X$\,:
\begin{eqnarray} {\chi}_{W}^{\rm out}(\xi; \rho) = \exp \left[-\frac{1}{2}
  \xi^T \,Y_{0}\, \xi \right] \chi_{W}^{\rm in}(\xi_1 , 0; \rho).
\label{n28}
\end{eqnarray}
Since we are dealing with symmetric ordering, $\chi_{W}^{\rm in}(\xi_1 , 0; \rho)$
is the Fourier transform of the marginal distribution of the first quadrature
(`position' quadrature) variable. Let us assume that the input $\hat{\rho}$
is a (single-mode-) squeezed Gaussian pure state, squeezed in the position
(or first) quadrature. For arbitrarily large squeezing, the state approaches a position
eigenstate and the position quadrature marginal approaches the Dirac delta
function. That is $\chi_{W}^{\rm in}(\xi_1 , 0; \rho)$ approaches a constant.
Thus, the Gaussian\\ $\exp \left[- (\xi^T \,Y_{0}\, \xi)/2 \right]$
is essentially the Weyl-characteristic function of the output state, and hence corresponds
to a classical state if and only if
 \begin{eqnarray}
Y_{0} \geq 1\!\!1,\,\,\,{\rm or}\,\,\,a,\,\,b\geq 1,
\label{n29}
\end{eqnarray}   
$a,\,b$ being the eigenvalues of $Y$.

We have derived this as a {\em necessary condition for nonclassicality breaking},
taking as input a highly squeezed state. It is clear that 
for any other input state the phase space distribution of the output state
will be a convolution of this Gaussian classical state with the position
quadrature marginal of the input state, rendering the output state more strongly
classical, and thus proving that the condition (\ref{n29}) is {\em also a sufficient
condition for nonclassicality breaking}.  

In the special case in which $X=0$ identically, we have the following
input-output relation in place of (\ref{n28})\,:
\begin{eqnarray}
{\chi}_{W}^{\rm out}(\xi; \rho) = \exp \left[-\frac{1}{2}
  \xi^T \,Y\, \xi \right] \chi_{W}^{\rm in}(\xi=0; \rho).
\label{n30}
\end{eqnarray}
Since $\chi_{W}^{\rm in}(\xi=0; \rho)=1$ independent of $\hat{\rho}$, the output
is an input-independent {\em fixed state}, and $\exp \left[-\frac{1}{2}
  \xi^T \,Y\, \xi \right]$ is its Weyl-characteristic function. But we know that this
fixed output is a classical state if and only if $Y \geq 1\!\!1$. In other
words, {\em the condition for nonclassicality breaking is the same for all
singular $X$, including vanishing $X$}. 

We conclude our analysis in this Section with the following, perhaps redundant,
remark\,: Since our canonical forms are nonclassicality-based,
rather than entanglement-based, if the nonclassicality breaking property
applies for one member of an orbit or double coset, it applies to the
entire orbit.

\section{Nonclassicality breaking {\em vs} entanglement breaking}

We are now fully equipped to explore the relationship between nonclassicality breaking 
Gaussian channels and entanglement breaking channels.
In the case of the first canonical form the nonclassicality breaking condition reads $(a-1)(b-1)\geq \kappa^4$, the entanglement breaking condition reads $ab \geq (1+\kappa^2)^2$, while the complete positivity condition reads $ab \geq (1-\kappa^2)^2$. These conditions are progressively weaker, indicating that the family of channels which meet these conditions are progressively larger. For the second canonical form the first two conditions have the same formal expression as the first canonical form, while the complete positivity condition has a more stringent form $ab \geq (1+\kappa^2)^2$. For the third and  final canonical form, the nonclassicality breaking condition requires both $a$ and $b$ to be bounded from below by unity, whereas both the entanglement breaking and  complete positivity conditions read $ab \geq 1$. Table\,\ref{table1} conveniently places these conditions side-by-side. In the case of  first canonical form, (first row of Table\,\ref{table1}), the complete positivity condition itself is vacuous for $\kappa=1$, the classical noise channels. 

\begin{table*}
\centering
\begin{tabular}{|c|c|c|c|}
\hline
Canonical form & NB & EB & CP \\
\hline
$(\kappa\,{1\!\!1},\, {\rm diag}(a,b))$  &   $(a-1)(b-1) \geq \kappa^4$ 
& $ab\geq(1+\kappa^2)^2$  & $ab\geq (1-\kappa^2)^2$\\
\hline
$(\kappa\,\sigma_3,\,{\rm diag}(a,b))$ & $(a-1)(b-1) \geq \kappa^4$ 
& $ab\geq(1+\kappa^2)^2$  & $ab\geq (1+\kappa^2)^2$\\
\hline
$({\rm diag}(1,0),\,Y)$, &$a,\, b \geq 1$, $a$, $b$ being & $ab\geq1$ & $a b \geq 1$ \\
& eigenvalues of $Y$ & &  \\
$({\rm diag}(0,0),\, {\rm diag}(a,b))$ &$a,\,b \geq 1$ & $ab\geq1$ & $ab \geq 1$ \\
\hline
\end{tabular}
\caption{A comparison of the nonclassicality breaking (NB) condition, the entanglement breaking (EB)
condition, and the complete positivity (CP) condition for the three canonical classes of channels.\label{table1}}
\end{table*}

\begin{figure*}
\centering
\scalebox{0.5}{\includegraphics{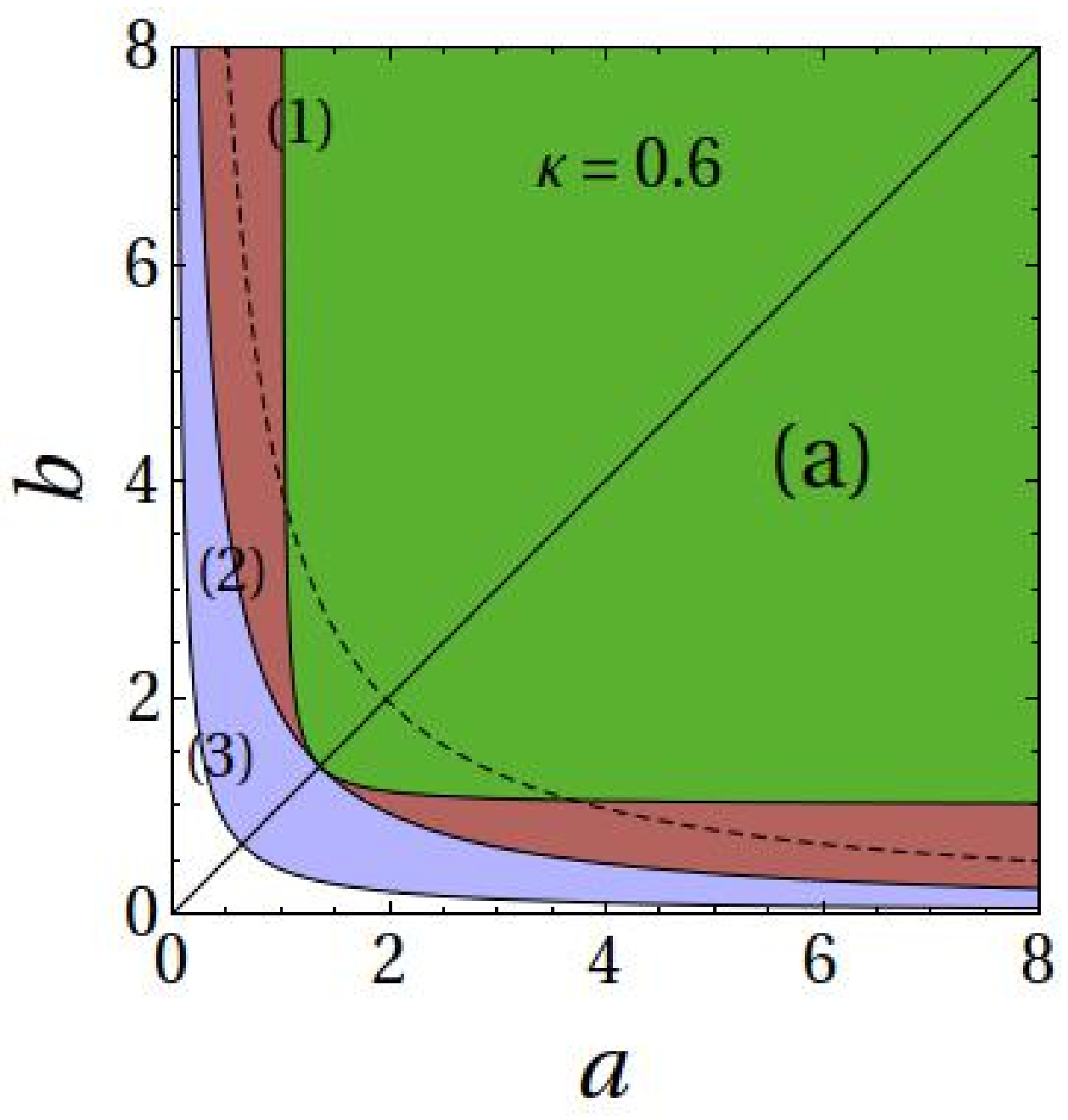}}~~ 
\scalebox{0.5}{\includegraphics{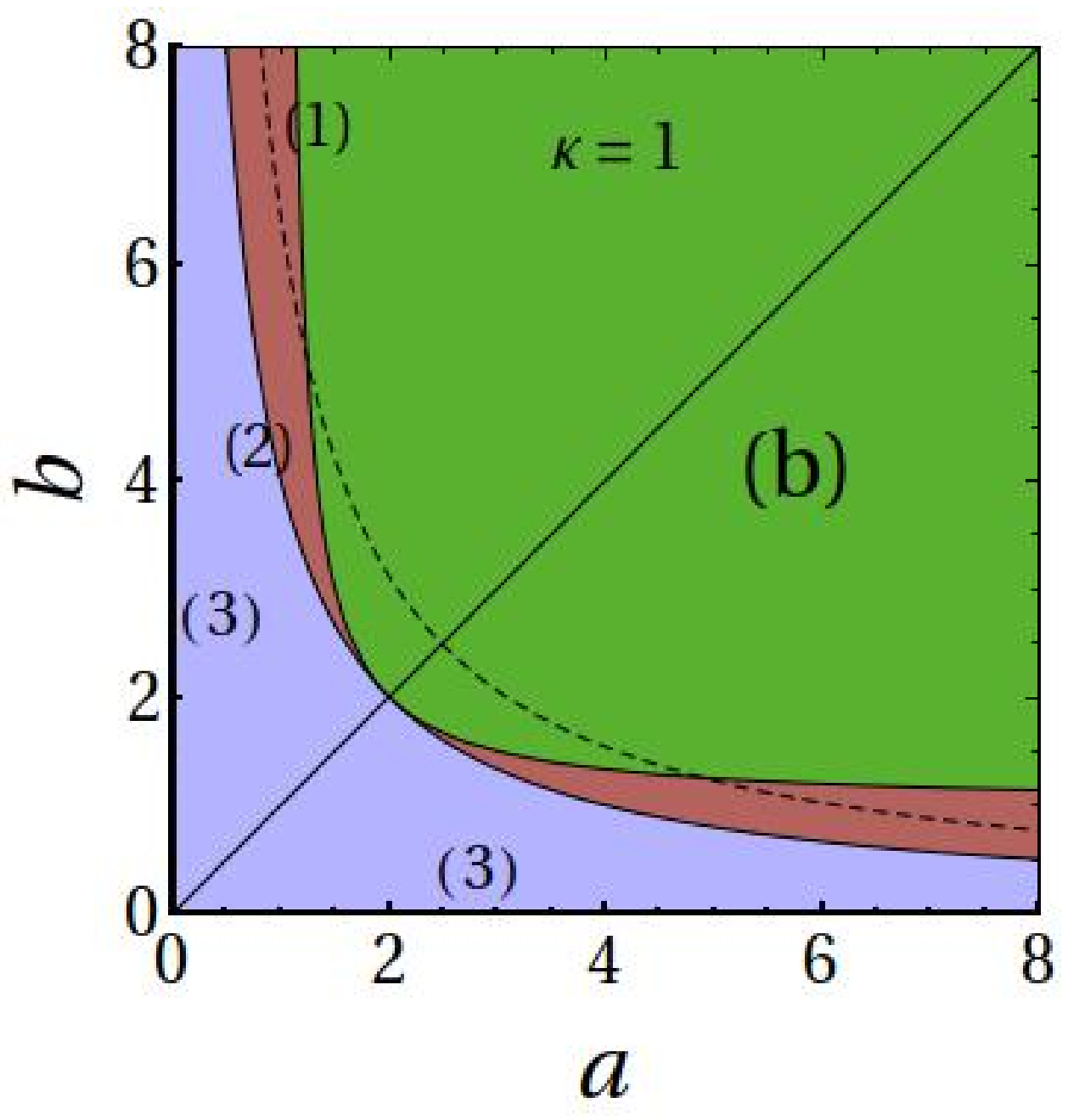}}\\
\scalebox{0.5}{\includegraphics{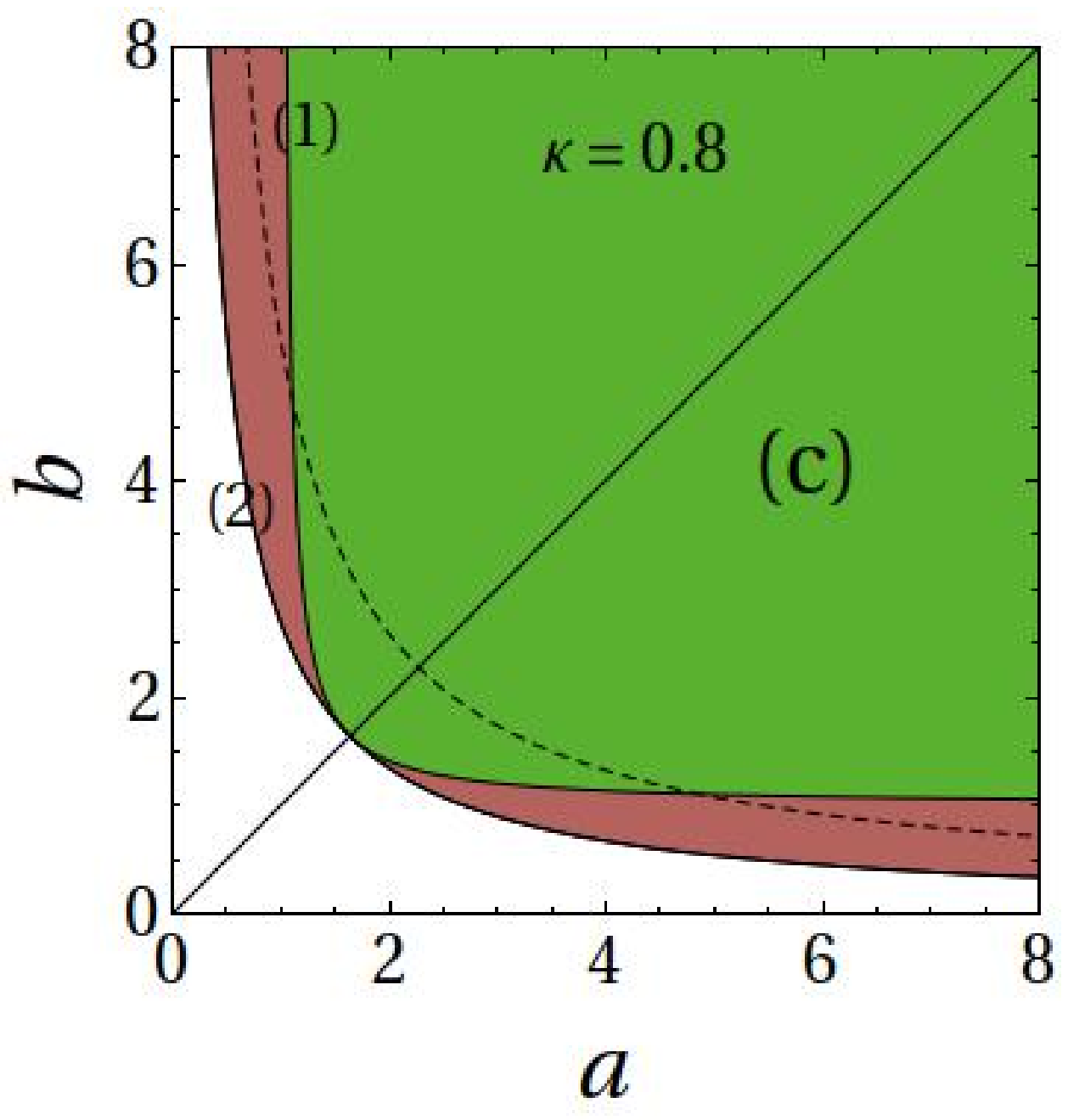}} ~~
\scalebox{0.5}{\includegraphics{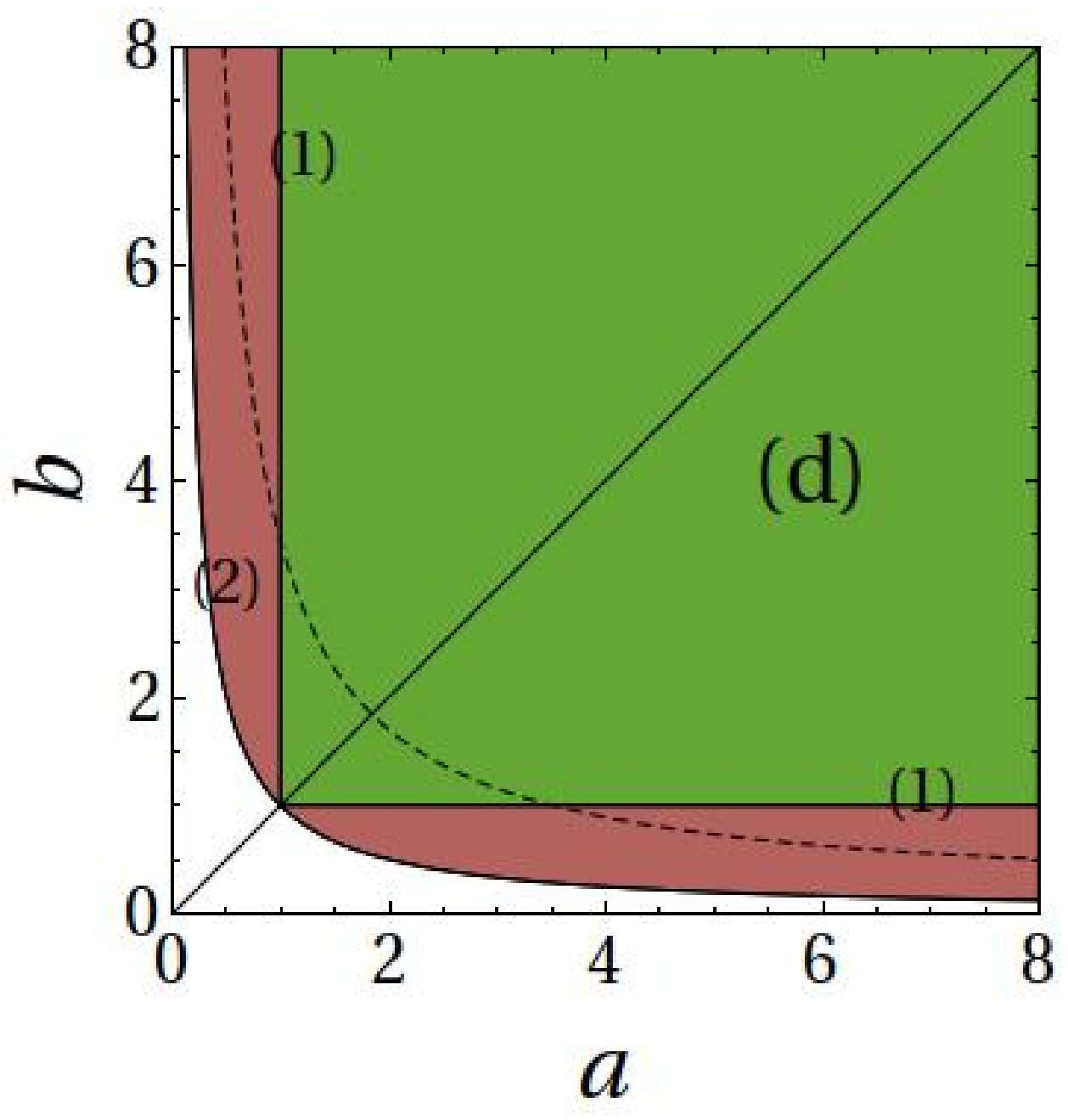}} 
\caption{
\baselineskip=12pt
Showing a pictorial comparison of the nonclassicality breaking condition, the entanglement breaking condition, and the complete positivity condition in the channel parameter space $(a,b)$, for fixed ${\rm det}\, X$. Curves (1), (2), and (3) correspond to saturation of these conditions in that order. Curve (3) thus corresponds to quantum-limited channels. Frame (a) refers to the first canonical form $(\kappa1\!\!1,\,{\rm diag}(a,b))$,
frame (c) to the second canonical form $(\kappa\,\sigma_3,\,{\rm diag}(a,b))$,
and frame (d) to the third canonical form, singular $X$. Frame (b) 
refers to the limiting case $\kappa=1$, classical noise channel.   
In all the four frames, the region to the right of (above) curve (1) corresponds
to nonclassicality breaking channels; the region to the right of (above) curve (2) corresponds
to entanglement breaking channels; curve (3) depicts the CP condition, so the
region to the right of (above) it alone corresponds to physical channels. The region to the left (below) curve (3) is unphysical as channels. In frames
(c) and (d), curves (2) and (3) coincide. In frame (b), 
curve (3) of (a) reduces to the $a$ and $b$ axis shown in bold. In frames (a) and 
(c), curves (1) and (2) meet at the point $(1+\kappa^2, 1+\kappa^2)$, in frame (b) they meet  at $(2,2)$, and in frame (d) at $(1,1)$. 
The region between (2) and (3) corresponds to the set of channels which are not entanglement breaking. That in frame (c) and (d) the two curves coincide proves that this set is vacuous for the second and third canonical forms. That in every frame the nonclassicality breaking region is properly contained in the entanglement breaking region proves that a nonclassicality breaking channel is certainly an entanglement breaking channel. The dotted curve in each frame indicates the orbit
of a generic entanglement breaking Gaussian channel under the action of a local unitary squeezing after the channel action. That the orbit of every entanglement breaking channel passes through the
nonclassicality breaking region, proves that the nonclassicality in all the output
states of an entanglement breaking channel can be removed by a fixed unitary squeezing, thus showing that every entanglement breaking channel is `essentially' a nonclassicality breaking channel. 
\label{fign1}}
\end{figure*}

This comparison is rendered pictorial in Fig.\,\ref{fign1}, in the channel parameter plane $(a,b)$, for fixed values of ${\rm det} X$. Saturation of the nonclassicality breaking condition, the entanglement breaking condition, and the complete positivity condition are  marked $(1)$, $(2)$, and $(3)$ respectively in all the four frames. Frame (a) depicts the first canonical form for $\kappa=0.6$ (attenuator channel). The case of the amplifier channel takes a qualitatively similar form in this pictorial representation. As $\kappa \to 1$,  from below ($\kappa <1$) or above ($\kappa >1$), curve $(3)$ approaches the straight lines $a=0, \, b=0$ shown as solid lines in Frame (b) which depicts this limiting $\kappa=1$ case (the classical noise channel). Frame (c) corresponds to the second canonical form (phase conjugation channel) for $\kappa=0.8$ and Frame (d) to the third canonical form. It may be noticed that in Frames (c) and (d) the curves (2) and (3) merge, indicating and consistent with that fact that channels of the second and third canonical forms are aways entanglement breaking.

It is clear that the nonclassicality breaking condition is stronger than the entanglement breaking condition. 
Thus, a nonclassicality breaking
channel is necessarily entanglement breaking\,: But there are channel 
parameter ranges wherein the channel is entanglement breaking,
though not nonclassicality breaking.
The dotted curves in Fig.\,\ref{fign1} represent orbits of a generic entanglement breaking channel $\Gamma$, fixed by the product $ab$ ($\kappa$ having been already fixed), when $\Gamma$ is followed up by a variable local unitary squeezing ${\cal U}(r)$. To see that the orbit of every entanglement breaking channel passes through the nonclassicality breaking region,
it suffices to note from Table\,\ref{table1} that the nonclassicality breaking boundary has $a=1$, $b=1$ as asymptotes whereas the entanglement breaking boundary has $a=0$, $b=0$ as the asymptotes.  
That is, for every entanglement breaking channel there exists a particular
value of squeeze-parameter $r_{0}$, depending only on the channel
parameters and not on the input state, so that the entanglement breaking
channel $\Gamma$ followed by unitary squeezing of extent $r_0$ always results
in a nonclassicality breaking channel ${\cal U}(r_0)\,\Gamma$. It is in this precise sense 
that nonclassicality breaking channels and entanglement breaking channels
are essentially one and the same. 

Stated somewhat differently, if at all the output of an entanglement breaking channel is nonclassical, the nonclassicality is of a `weak' kind in the following sense. Squeezing is not the only form of nonclassicality. Our result
not only says that the output of an entanglement breaking 
channel could at the most have a squeezing-type nonclassicality,
it further says that the nonclassicality of {\em all} output states can be
removed by a {\em fixed} unitary squeezing transformation. This is depicted schematically in Fig.\,\ref{fign2}.

\begin{figure}
\scalebox{0.55}{
\includegraphics{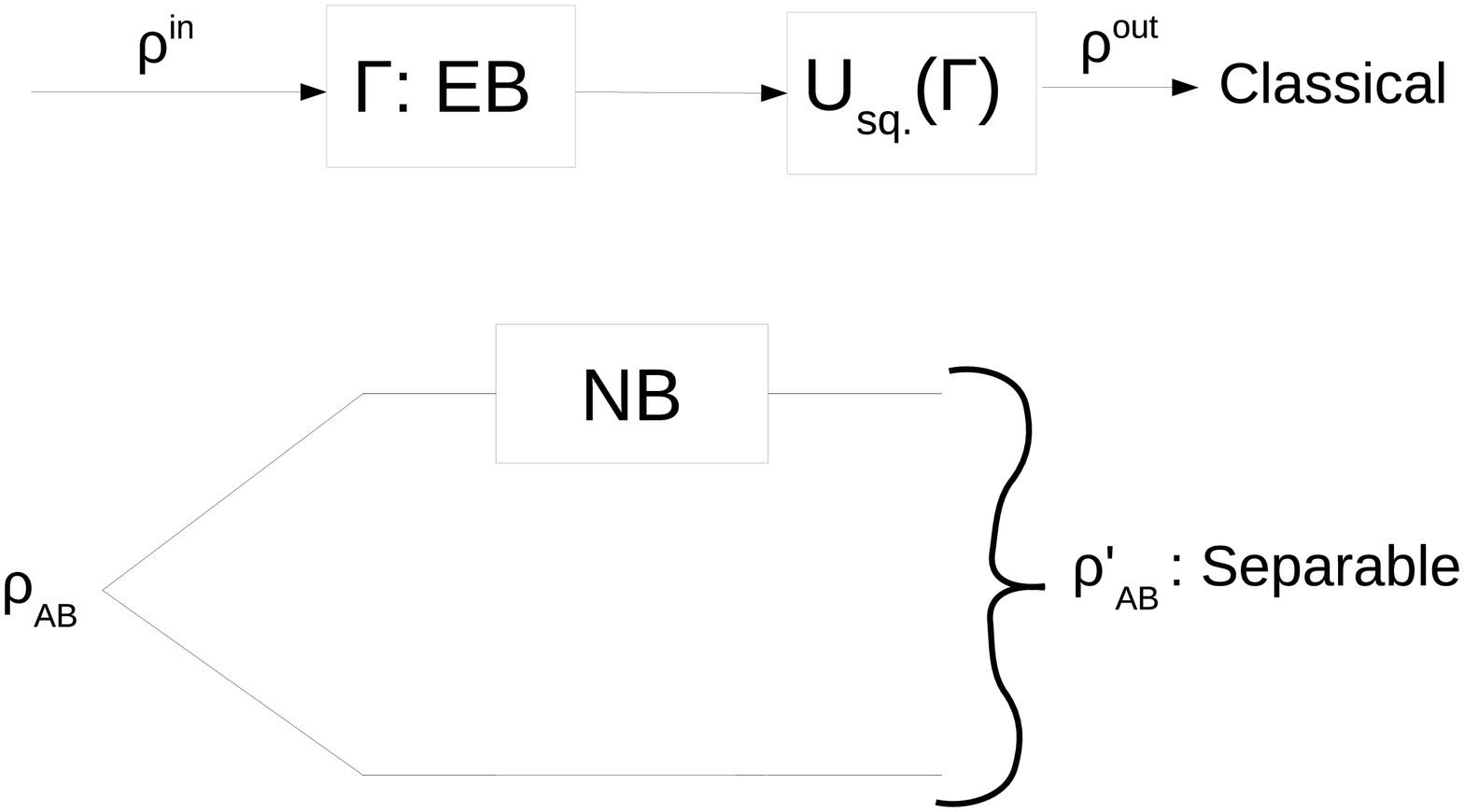}}
\caption{Showing the relationship between nonclassicality breaking and entanglement breaking channels established in the present Chapter. The output state corresponding to any input to an entanglement breaking channel is rendered classical by a single squeeze transformation that depends only on the channel parameters and independent of the input states. In other words, an entanglement breaking channel followed by a given squeeze transformation renders the original channel nonclassicality breaking. In contrast, every nonclassicality breaking channel is also entanglement breaking.
\label{fign2}}
\end{figure} 


\section{Conclusions}
We have explored the notion of nonclassicality breaking and its relation to entanglement breaking. We have shown that the two notions are effectively  equivalent in the context of bosonic Gaussian channels, even though at the level of definition the two notions are quite different, the latter requiring reference to a bipartite system. Our analysis shows that some nonclassicality could survive an entanglement breaking channel, but this residual nonclassicality would be of a particular weaker kind. 

The close relationship between entanglement and nonclassicality has been studied by several authors in the past\,\cite{simon00,asboth05,ivan11,ivan12,kim02,carmichael10,agarwal091,robust,barbosa10,agarwal10b}. It would seem that our result brings this relationship another step closer.

Finally, we have presented details of the analysis only in the case of single-mode bosonic Gaussian channels. We believe the analysis is likely to generalize to the case of $n$-mode channels in a reasonably straight forward manner.


\chapter{Conclusions}
We now provide an overall summary to the primary results of this thesis. 

In Chapter 2, we studied the role played by initial correlations of bipartite quantum systems on the subsystem dynamics. Working within the Shabani-Lidar framework, the principal result that emerges as a result of our analysis is that, for the system dynamics to be a completely positive map, or in other words, a physical evolution, the only allowed initial system bath states are (tensor) product states. This brings us back to the well known Stinespring dilation for realization of completely positive maps. Our analysis solely rested on two very reasonable assumptions of the set of initial system bath states. This demonstrated robustness of the folklore scheme could be of much importance in the study of open quantum systems. 

In Chapter 3, we studied the computation of correlations for two-qubit $X$-states, namely, classical correlation, quantum discord, and mutual information. We exploit the geometric flavour of the problem and obtain the optimal measurement scheme for computing  correlations. The optimal measurement turned out to be an optimization problem over a single real variable and this gave rise to a three-element POVM. We studied the region in the parameter space where the optimal measurement requires three elements and the region where the optimal measurement is a von Neumann measurement along x or z-axis. We further bring out clearly the role played by the larger invariance group (beyond local unitaries) in respect of the correlation ellipsoid and exploit this notion for simplifying the computation of correlations. We then immediately draw many new insights regarding the problem of computation of correlations and provide numerous concrete examples to detail the same.  

In Chapter 4, we studied the robustness of non-Gaussian entanglement. The setup involved the evolution of Gaussian and non-Gaussian states under symmetric local noisy channel action. The noisy channels we are concerned with are the noisy attenuator and the noisy amplifier channels. This problem has consequences for protocols in quantum networks involving continuous variable systems. In this physical setting it was recently conjectured that Gaussian states are more robust than non-Gaussian states with regard to robustness of entanglement against these noisy environments. This conjecture is along the lines of other well established extremality properties enjoyed by Gaussian states. We demonstrate simple examples of non-Gaussian states with 1 ebit of entanglement which are more robust than Gaussian states with arbitrary large entanglement. Thereby proving that the conjecture is too strong to be true. The result will add to the growing list of plausible uses of non-Gaussian quantum information alongside the Gaussian-only toolbox.

In Chapter 5, we explore the connection between nonclassicality and entanglement in continuous variable systems and in particular the Gaussian setting. The nonclassicality of a state is inferred from its Sudarshan diagonal function. Motivated by the definition of entanglement breaking channels, we define nonclassicality breaking channels as those channels which guarantee that the output is classical for any input state. We classify Gaussian channels that are nonclassicalitybreaking under the restricted double cosetting appropriate for the situation on hand. We show that all nonclassicalitybreaking channels are entanglementbreaking channels. This is a surprising result in light of the fact that a nonclassicalitybreaking channel requires only one mode whereas the very definition of an entanglementbreaking channel requires two modes. We further show that the nonclassicality of the output states of an entanglementbreaking channel are of a weak type. In the sense that a suitable squeeze transformation, independent of the input state, can take all these output states to classical states. The study reveals another close connection between nonclassicality and entanglement. 

A natural future direction to explore from this study is that of the role played by these channels as a resource in quantum communication, namely, the capacity problem. The capacity of a channel is the rate at which information can be reliably sent across many uses of the channel. In the quantum setting, there are many variants of capacities depending on the available resources and the tasks to be accomplished.  These quantities are well understood only for a handful of channels and is thus an interesting scope for further studies in both the finite and the continous variable systems.

}

\bibliography{lidar-ref,robust-ref,ncbc-ref,intro-ref,xstates-ref}
\end{document}